%% file: these.tex
\documentclass[a4paper,12pt,twoside]{book}
\usepackage{amsmath} 
\usepackage{amssymb} 
\usepackage{amsthm}  

\usepackage{hyperref}
\usepackage{fancyhdr}
\usepackage{enumitem}
\usepackage[utf8]{inputenc}
\usepackage{setspace}
\usepackage{textcomp}
\usepackage[retainorgcmds]{IEEEtrantools}
\usepackage{url}
\usepackage{minitoc}
\usepackage{feynmp}
\usepackage[pdftex]{graphicx}
\usepackage{tikz}
\usetikzlibrary{shapes,arrows}

\usepackage[a4paper,twoside, hmarginratio={1:1}]{geometry}
\usepackage[a4paper]{meta-donnees}
\include{abbreviations}

\begin{document}
\begin{fmffile}{fgraphs}
\dominitoc

\pagestyle{empty}
\include{couverture}

\newpage
\null\newpage
\frontmatter
\pagestyle{plain}
\null
\include{ack}

\newpage\null\newpage

\tableofcontents
\newpage\null\newpage

\mainmatter
\setcounter{page}{1}
\include{chapter0}

\include{chapter1}

 \include{chapter2}

\include{chapter4}

\include{chapter3}

  \include{chapter5}

  \include{chapter6}

  \include{chapter7}

 \include{chapter8}

 \include{chapter9}

 \addcontentsline{toc}{chapter}{Conclusion}
 \include{chapter10}

\newpage
\begin{spacing}{0.6}
\addcontentsline{toc}{chapter}{Bibliography}

\input{these.bbl}
\end{spacing}

\newpage
\renewcommand{\thesection}{\Alph{section}}
\setcounter{section}{0}

 \include{appendix}

\end{fmffile}
\end{document}

%% file: abbreviations.tex
\newcommand{\1}{\frac{1}{2}}
\newcommand{\oo}[1]{\overline{#1}}
\renewcommand{\L}{\mathcal{L}}
\def\O{\mathcal{O}}
\def\Dslash{D\!\!\!\! / \,\,\/}
\def\kslash{k\!\!\! / \,\/}
\def\dslash{\partial\!\!\! / \,\/}
\def\Geff{\Gamma}
\def\Geffq{\Gamma_{|2}{}}

\def\CC{\mathbb{C}}

\def\hc{\ +\text{h.c.}}
\def\D{\mathcal{D}}
\def\A{\mathcal{A}}
\def\P{\mathcal{P}}
\def\G{\mathcal{G}}
\def\W{W}
\def\M{\mathcal{M}}
\renewcommand{\t}[1]{\tilde{#1}}
\newcommand{\grandO}[1]{O\mathopen{}\left(#1\right)}
\def\mio{m_{i\,0}}
\def\mjo{m_{j\,0}}
\def\xvec{\overrightarrow{x}}
\def\PL{\frac{1+\gamma_5}{2}}
\def\PR{\frac{1-\gamma_5}{2}}
\def\Bloop{B_2\left(\frac{m_{\tilde{q}}^2}{m_i^2},\frac{m_i^2}{m_j^2}\right)}

\def\tb{t_\beta}
\def\ma{M_{A^0}}
\def\mz{M_Z}
\def\mw{M_W}
\def\mh{m_h}
\def\mH{m_H}
\def\Oh{\Omega h^2}
\def\mchi{m_{\tilde{\chi}^0_1}}

\def\Mf{M_{\tilde{f}}}
\def\vv{\mathbf{v}}
\def\ggg{\mathbf{g}}
\def\wp{\Gamma_\phi}

\def\AA{\mathbf{A}}
\def\chii{\tilde{\chi}_1^0}

\def\hp{\hat{h}}
\def\cha{\tilde{\chi}_1^+}
\def\chjm{\tilde{\chi}_j^-}
\def\chip{\tilde{\chi}_i^+}
\def\Hp{H^+}
\def\Hm{H^-}

\def\sw{s_w}
\def\cw{c_w}
\def\sb{s_\beta}
\def\cb{c_\beta}
\def\sdb{s_{2\beta}}
\def\cdb{c_{2\beta}}
\def\sa{s_\alpha}
\def\ca{c_\alpha}

\def\dZ{\delta Z}

\def\dA{\delta A}
\def\dB{\delta B}
\def\dP{\delta P}
\def\dM{\delta M}
\def\dm{\delta m}
\def\dU{\delta U}
\def\dV{\delta V}
\def\dX{\delta X}
\def\dY{\delta Y}

\def\gpg{g_{\phi gg}}
\def\gpw{g_{\phi WW}}
\def\gpz{g_{\phi ZZ}}
\def\gpv{g_{\phi VV}}
\def\gpga{g_{\phi \gamma\gamma}}
\def\gpt{g_{\phi \bar tt}}
\def\gpb{g_{\phi \bar bb}}
\def\gpl{g_{\phi \bar\tau\tau}}
\def\gpp{g_{\phi\phi\phi}}
\def\ghb{g_{h \bar bb}}

\def\ghg{g_{h gg}}
\def\ghga{g_{h \gamma\gamma}}
\def\chisff{\tilde{\chi}_1^0\tilde{f}f}

\def\s0{\sigma_0}
\def\seff{\sigma_{\textrm{eff}}}
\def\sol{\sigma_{\textrm{one-loop}}}

\def\DU{\Delta_{\rm eff}}
\def\DNU{\Delta_{NU}}

\def\DA{\Delta_{\alpha}}
\def\Bsmu{B_s\to\bar\mu\mu}
\def\Bsg{B\to X_s\gamma^*}
\def\Rgam{R_{\gamma\gamma}}
\def\Rz{R_{ZZ}}
\def\RtauE{R_{\bar\tau\tau}^\text{excl 95\%}}
\def\Rgamh{R_{\gamma\gamma\ h}}

\def\lanHEP{\texttt{lanHEP}}
\def\FA{\texttt{FeynArts}}
\def\FC{\texttt{FormCalc}}
\def\HD{\texttt{HDecay}}
\def\LT{\texttt{LoopTools}}
\def\SP{\texttt{SuSpect}}
\def\CH{\texttt{CalcHEP}}
\def\CO{\texttt{CompHEP}}
\def\MG{\texttt{MadGraph}}
\def\MA{\textit{Mathematica}}
\def\FO{\textit{Fortran}}
\def\momegas{\texttt{micrOmegas}}
\def\SL{\texttt{SloopS}}
\def\darksusy{\texttt{DarkSusy}}
\def\PY{\texttt{PYTHIA}}

\def\MSSM{\text{MSSM}}
\def\SM{\text{SM}}
\def\RD{relic density}

\def\vev{vacuum expectation values}
\def\mhmax{m_{h\ \text{max}}}

\def\OS{On-Shell}
\def\CUV{C_\text{UV}}

\def\fb{\ fb$^{-1}$}
\def\add{\text{add}}
\def\eff{\text{eff}}
\def\tot{\text{tot}}
\def\excl{\text{excl 95\%}}

\def\looop{\text{loop}}

\def\Moller{M\o ller}
\def\fb{\ fb$^{-1}$}
\def\pvalue{\textit{p}-value}

\newcommand{\neuto}{{\tilde{\chi}}_1^0}
\newcommand{\beqn}{\begin{eqnarray}}
\newcommand{\eeqn}{\end{eqnarray}}
\newcommand{\ra}{\rightarrow}
\newcommand{\mneuto}{m_{{\tilde{\chi}}_1^0}}
\newcommand{\neutt}{{\tilde{\chi}}_2^0}
\def\tgb{\tan \beta}

\definecolor{Zgris}{rgb}{0.87,0.85,0.85}
\newsavebox{\BBbox}
\newenvironment{DDbox}[1]{
\begin{lrbox}{\BBbox}\begin{minipage}{\linewidth}}
{\end{minipage}\end{lrbox}\noindent\colorbox{Zgris}{\usebox{\BBbox}} \\
[.5cm]}

\tikzstyle{decision} = [diamond, draw, fill=gray!20, aspect=4,
    text width=5cm, text badly centered, node distance=4cm, inner sep=0pt]
\tikzstyle{block} = [rectangle, draw, fill=white!20,
    text width=5cm, text centered, rounded corners, minimum height=4em]
\tikzstyle{line} = [draw, very thick, color=black!100, -latex']
\tikzstyle{cloud} = [draw, ellipse,line width=2pt,fill=white!20, node distance=2.5cm,
    minimum height=2em]
\tikzstyle{input}=[draw,rounded corners=3pt,text width=4cm,fill=blue!25]
\tikzstyle{inputb}=[draw,rounded corners=3pt,text width=4cm,fill=gray!25,dashed]
\tikzstyle{outputa}=[draw,text width=3cm,fill=yellow!50]
\tikzstyle{outputb}=[draw,rounded corners=6pt,text width=3cm,text=red,thick,dotted]
\tikzstyle{outputc}=[draw,diamond,aspect=4,text width=3cm,fill=yellow!50,text=red,thick]
\tikzstyle{suite}=[->,>=stealth',thick,rounded corners=4pt]
\tikzstyle{transversal}=[->,>=stealth',thick,dashed]
\tikzstyle{autre}=[<->,>=latex,very thick,dotted]

%% file: couverture.tex





\Specialite{Physique Th\'eorique}
\Arrete{7 ao\^ut 2006}
\Auteur{Guillaume DRIEU LA ROCHELLE}
\Directeur{Fawzi BOUDJEMA}
\Laboratoire{\small du Laboratoire d'Annecy-le-Vieux de Physique Th\'eorique}
\EcoleDoctorale{Ecole doctorale de Physique de Grenoble}         
\Titre{\Large Effective Approaches in and Beyond the MSSM : applications to Higgs Physics and Dark Matter observables}
\Depot{12 juillet 2012}




\Jury{
\UGTPresident{Dr, Genevi\`eve Belanger}{Directeur de Recherche, LAPTh Annecy}

\UGTRapporteur{Dr, Abdelhak Djouadi}{Directeur de Recherche, LPT Orsay}      
\UGTRapporteur{Dr, Ben Matthew Gripaios}{Lecturer, University of Cambridge}      
 
\UGTExaminateur{Dr, Rohini Godbole}{Professor, CHEP Bangalore}     
\UGTExaminateur{Dr, Christophe Grojean}{Directeur de Recherche, CEA Saclay}     

\UGTDirecteur{Dr, Fawzi Boudjema}{Directeur de Recherche, LAPTh Annecy}       

}

\MakeUGthesePDG    

%% file: ack.tex
\chapter*{Acknowledgements}

First I would like to thank my referees Abdelhak Djouadi and Ben Gripaios for their careful reading of my manuscript and their acute comments and suggestions. I also thank Rohini Godbole for the many discussions we had and her presence at my defence.\\

Je remercie également les personnes qui ont fait de mon séjour au LAPTh une véritable porte d'entrée sur la recherche en physique des particules. En premier lieu mon directeur de thèse Fawzi, qui a non seulement pris le temps et la patience (et il en a fallu) de m'apprendre une partie de son savoir, mais m'a surtout permis de confronter mes différents travaux avec la réalité de la recherche et m'a ainsi enseigné les codes d'une communauté si ... particulière. Mes remerciements vont également à la ``dream administrative team'', à savoir Dominique, Nathalie, Véronique et Virginie dont la compétence, l'enthousiasme et un sourire omniprésent m'ont laissé un souvenir que je ne suis pas près d'oublier. Je remercie l'ensemble des chercheurs, en particulier Geneviève qui, en plus de son rôle de tutrice hors pair, n'as pas manqué une occasion de m'aider, de répondre à mes questions et de me diriger vers moult conférences et autres écoles d'été. Je remercie aussi Eric P., que j'ai abusivement sollicité pour la moindre 
question concernant la théorie quantique des champs, sans jamais arriver à le prendre en défaut. Je n'oublie pas non plus le service informatique LAPP/LAPTh pour sa réactivité, avec un clin d'oeil à Matthieu.\\
J'ai eu la chance de pouvoir passer 6 mois de ma thèse au CERN et je dois ce passage, ainsi que les travaux que j'y ai entamé à Christophe Grojean, que je remercie donc pour m'avoir accueilli et m'avoir lancé sur de nouveaux projets.\\

Je ne peux manquer de remercier tout ceux qui ont fait de mes trois années à Annecy un tumulte foisonnant d'activités et de grands moments. Je pense d'abord à mon premier colocataire Timur, que je remercie entre autres pour m'avoir fait réaliser dès la première année de thèse qu'on ne saurait vivre que pour son travail, fut-il des plus intéressants. Je remercie aussi les couples du cercle très fermé des joueurs gourmands et bon vivants d'Annecy, à savoir Caroline et Olivier, Armand et Iro, Loïc et Bénédicte et les petits derniers Michael et Marie pour de folles soirées, où l'adrénaline du jeu le dispute au plaisir des papilles. Puis aux nombreux (autres) jeunes du labo avec lesquels j'ai tant partagé : mon grand frère de thèse Guillaume pour de nombreuses conférences et francs moments de rigolade ensemble, Florent pour m'avoir enseigné l'escalade, Louis pour ne m'avoir jamais laissé boire seul, Sean pour avoir ruiné sans vergogne toute mes réserves de martini, Daniel pour m'avoir appris des jeux de dé 
chiliens, Dimitra, Dudu, Florian et Timothée pour avoir contribué à faire de la promo 2009 un très grand cru (mythique?) de thésards d'exception, Gilles et Nukri pour leur sérieux et leur assiduité aux (nombreuses) pauses bayfoot-esques, Laurent pour sa légendaire concision littéraire, Jonathan pour sa contribution assidue à l'animation du bureau, Maud pour les after de la place Sainte-Claire et enfin les squattages récurrents de Lisa et Guilhem sans lesquels ce manuscrit aurait été nettement moins 
fourni en fautes d'orthographe. Je remercie ma colocataire Cécile pour avoir bien voulu croire que j'étais vivable pendant près de deux ans, et avoir contribué de sa bonne humeur à l'atmosphère unique du fameux 7 avenue de Novel. Je remercie enfin le lac d'Annecy ainsi que les installations de première nécessité du LAPP, à savoir la table de ping-pong et le babyfoot, sans lesquelles cette thèse aurait probablement eu un aspect fini, soigné, perfectionné, en un mot propre et aurait par là même tant contrasté avec mon style naturel.\\

I'd like to thank the PhD Students I met while at CERN in particular Ennio, Ahmad, Jean-Claude, Sandeepan and the whole team for the amazing ambiance!\\

Je remercie ma famille pour son soutien fort et constant tout au long de ces trois années.\\

Cette thèse est dédiée à mon moi âgé de 11 ans, qui en 1999 se lamentait à l'envie sur la durée excessive des études en France et la certitude qu'on aurait découvert le Higgs bien avant qu'il ait eu le temps d'arriver jusqu'en thèse. Bref.

%% file: chapter0.tex
\chapter*{Introduction}
\addcontentsline{toc}{chapter}{Introduction}

Despite the numerous successes of the Standard Model of particle physics, it is believed that the complete picture of particle physics could be larger, as a unified theory for instance, and thus many efforts have been devoted to the development of theories of new physics. Supersymmetry is one of the most popular extensions since in addition to a solution of the naturalness issue, it provides a viable dark matter candidate. This last sector being all the more important now that recent experimental measurements have significantly increased our knowledge about dark matter properties, in particular the experimental determination of the relic density has reached the accuracy of a few percent. When applied to the Minimal Supersymmetric Standard Model (the MSSM, which is the simplest supersymmetric extension of the Standard Model), this constraint will thus shed light on the one-loop structure of the model. The MSSM is however much more liberal with unconstrained parameters than the Standard Model is, and the full 
one-loop computation of the relic density tends to be too long to be carried out throughout this large parameter space. In this thesis I have thus explored the opportunity of accounting for those loop corrections through a set of effective couplings. This effective approach has the advantage of keeping the simplicity of a tree-level computation while encoding at the same time genuine loop features such as the non-decoupling of heavy particles. Complementary to those constraints are the observables related to the LHC, which started taking data shortly after the beginning of my PhD in fall 2009. The Higgs sector of the MSSM is tightly constrained and this results in a certain fine-tuning of the model, which led to the creation of many models beyond the MSSM (such as the Next-to-Minimal Supersymmetric Standard Model). Arguing for a more general approach, I have decided in this thesis to use again the effective approach but with a different aim : while the effective couplings in the case of dark matter are 
determined to account for the MSSM loop corrections, the effective operators we add to the Higgs sector of the MSSM are the remnants of the integration of a heavy extra spectrum. Though based on distinct aims, these two implementations show the different advantages of an effective field theory. In the first case the effective operators are parametrising the effect of an unknown UV (UltraViolet) completion, whereas in the second we assume this UV completion to be the MSSM.

The introduction of the new operators in the Higgs sector was motivated by the naturalness issue of the Standard Model Higgs. Indeed if we believe that there may be new particles at the Planck scale, then those particles will shift the running Higgs mass, which should stay of the order of the electroweak scale, up to the Planck scale. Supersymmetry provides a mechanism to evade this effect by ensuring that bosonic and fermionic corrections to the Higgs running mass cancel together. It requires however that the top superpartners (the stops) are light to be efficient. Those light stops are known to generate only moderate loop enhancements to the lightest Higgs mass. Given that the tree-level lightest Higgs mass comes from the quartic coupling which purely stems from gauge interaction, it is bounded by the electroweak scale $\mz$. As such, since its loop corrections will be moderate, a natural MSSM is doomed to have the lightest Higgs about the LEP bound (114 GeV) at best. This can be cured by assuming 
some extra physics at a heavy scale and introducing operators with MSSM superfields up to a given order in the expansion over the powers of the heavy scale, which is precisely the effective field theory approach. The interesting point of the effective approach is that it allows one to keep a generic framework towards supersymmetry since the kind of extra physics that may be realised is not specified and as such this framework, called in the literature the BMSSM (for Beyond the MSSM), accounts for many different non-minimal supersymmetric realisations subject only to the requirement that the extra physics be sufficiently heavy. The practical implementation of the BMSSM with the usual tools for phenomenology was a first obstacle since the model deals with both a non-standard K\"ahler potential and a non-standard superpotential. The issue goes even beyond the simple derivation of the Feynman rules associated to the new operators, since for some of the loop-induced processes of the Higgs such as the decay to 
photons, the loop form factor is extended by the effective operators and leads to more diagrams than in the MSSM computation. By extending different tools such as \lanHEP\ and \HD\ for our purposes, we were eventually able to generate the full Higgs phenomenology in an efficient way, which allowed us to probe the reach of the BMSSM physics. A first consequence was to raise the lightest Higgs mass up to 250 GeV. This is first alleviating the fine-tuning issue of the MSSM since the large loop corrections are not needed any more, but it does also bring the lightest Higgs in a different observable region than the MSSM from the point of view of experiments. Indeed a Higgs boson in the 150-250 GeV range is in the sensitive zone of the $WW$ and $ZZ$ searches, as compared to a lighter boson that is best probed by the $\gamma\gamma$ channel.\\

In the meantime, the LHC started collecting data. This put quickly severe constraints on a moderately heavy (say 150 to 400 GeV) Higgs. But those Higgs searches are mostly dedicated to the Standard Model Higgs boson and it is not straightforward to derive the implications in the MSSM and even less in the BMSSM where the Higgs phenomenology appears to be much richer. A certain effort has thus been devoted to the interpretation of Standard Model Higgs searches in non-Standard Model frameworks. Interestingly, a significant part of our findings were totally uncorrelated to supersymmetry : some issues of the reinterpretation are common to any of the BSM (Beyond Standard Model) theories. This is the case of the paradigm between exclusive and inclusive cross-sections : most experimental results are given as functions of the Standard Model inclusive cross-section but they have been obtained by comparing the exclusive cross-section to the data. And since the ratio between inclusive and exclusive cross-sections is 
model dependent, those results are hence model dependent. The case is best supported by the diphoton analysis done by CMS : they recasted explicitly their results in the model of a fermiophobic Higgs, which shifted the  limits by up to 50\%. The information needed to do such a recasting is twofold : one needs the efficiencies of each production mode of the Higgs for each sub-channel of the analysis on the one hand, and on the other the separate exclusion bounds for each sub-channel. It turns out however that those quantities are not available publicly, which is not specific to the diphoton channel but happens for nearly all searches. This has led us to test some approximations, such as the estimation of the efficiencies by a simplified \PY\ simulation for instance, in the recasting of the experimental results delivered by ATLAS and CMS collaborations. Using those approximations, we were able to constrain the BMSSM phenomenology based on the analysis of the 2\fb\ datatset and we found that the light Higgs 
mass had to be less than 150 GeV, hence ruling out the high masses (150-250 GeV) obtained prior to the LHC. However we have also found out that a $\mh<150$ GeV in the general BMSSM framework could be quite elusive, both at LEP and the LHC.\\

Shortly after having put those stringent bound on the Higgs parameter space, the ATLAS and CMS collaborations both reported in mid-December 2011 some excesses with the total 2011 dataset (5\fb), which were pointing to a signal at 125 GeV. Although the value of the mass is fully acceptable in the Standard Model and possible in the MSSM, it was also noted that the excesses seemed to have non-standard strengths. Given the small amount of data collected the discussions on the would-be signal strength are however even more speculative than on the existence of a Higgs boson itself at such a mass. Nonetheless, it is crucial to be ready in case a signal would emerge that is incompatible with the Standard Model properties. Turning to the BMSSM framework, we found that the Higgs phenomenology of the BMSSM could generically produce a non standard kind of signal. We have then, for the sake of the exercise, tried to reproduce the different signal strengths derived by the experiments. With the 5\fb\ dataset, the excesses 
in 
the diphoton channel are roughly twice the Standard Model Higgs expectation for both collaborations. Such an enhancement is frequent in the BMSSM, where the branching fraction to $b$ quarks is not as constrained as in the MSSM and can thus be lowered, enhancing thus other branching ratios such as the diphoton final state. Even more interesting is the claim from the CMS collaboration that the signal in the diphoton plus dijet final state could be significantly higher than the diphoton rate. This kind of signature can be readily obtained in the BMSSM : indeed since it is based on a natural MSSM spectrum, that is to say with light top superpartners, those states can alter the gluon fusion and the decay to photons altogether. We have shown in particular that a hierarchy between weak gauge boson final states, diphoton final state and diphoton plus dijet final state could be reproduced. Such a feature would be a crucial point in making the distinction between minimal and non-minimal supersymmetry.\\

We have then enlarged the focus of our work to include other experimental constraints on new physics that are the flavour physics and dark matter. Although the contribution to B physics observables is quite well known in the MSSM case, we have shown that the inclusion of higher-order operators led to a modification of the penguin amplitude in the branching ratio of $\Bsmu$, an observable known to put severe constraints on supersymmetry. Using the equations of motion, we were able to obtain the full BMSSM prediction and our analysis showed an interesting interplay with the Higgs physics. On the one hand it disfavoured the region with a too light CP-odd Higgs and on the other hand, when including the $\Bsg$ constraint which is affected by the stop contribution, it led to a reduction of the allowed parameter space as compared to our previous study.\\

Even if the Higgs search is certainly a powerful constraint on supersymmetry, we cannot leave aside the other motivation for supersymmetry which is the explanation for dark matter. For this we have some input from the experimental side and it turns out that one of the most powerful constraints comes from the measurement of the relic density of dark matter : from the experimental side a precision of the order of 3\% has been reached, and improvements are still expected. If one assumes a standard cosmological scenario, this translates to a very accurate determination of the  annihilation cross-section of two lightest neutralinos (which is the main MSSM candidate for dark matter) to Standard Model particles. This calls for a precision computation of the predicted cross-section and will thus require a one-loop computation. This is \textit{a priori} not an issue, since thanks to the thorough studies on the subject we know quite well how to deal with the complicated renormalisation of the MSSM and how to 
implement in practice the computations of the large number of Feynman diagrams that show up at the loop level. However it turns out that, perhaps because those developments are quite recent, or perhaps because the computation remains intricate and time-consuming even when automated, the idea of the one-loop computation of the relic density has not percolated through the whole community working on supersymmetry. It is quite amazing that all scans on the MSSM parameter space where the relic density is put as one of the most powerful constraint still stick to a tree-level computation, which underestimate largely the theoretical uncertainty. Our idea was to go once more to an effective approach. Indeed it is known that quantum corrections can be realised through effective operators entering the effective action. In particular this has allowed us to construct new effective vertices that would account for the dominant contributions to the radiative corrections and that are easier to compute than the full one-loop 
amplitude. We have in particular introduced effective vertices for the $\chii\tilde{f}f$ and the $\chii\chii Z$ vertices, that both enter the class of processes $\chii\chii\to \bar ff$. A complete study of the robustness of those operators as compared to the full one-loop computation has allowed us to determine the area of the MSSM parameter space where the effective approach would give satisfactory results and also to point out non decoupling effects in the relic density. Indeed it turns out that the loops with heavy sfermions may give a small but non-vanishing contribution, and since we are dealing with a precise measurement, we can draw conclusions on such a heavy spectrum. The situation is akin to the precision physics done at the Z pole at LEP, where the important accuracy can constrain a heavy spectrum. This feature is all the more welcomed in view of an interplay with the LHC since the latter will only probe squarks masses up to moderately heavy mass, say 1 to 2 TeV, whereas the relic density still 
keeps track of them at much higher masses.\\

\section*{Outline}
This thesis is structured as follows :
\begin{itemize}
\item Chapter 1 will present the necessary mathematical tools to compute predictions in a quantum field theory at the tree-level. In particular it will define the relations between initial fields and parameters of a theory and the physical ones.
\item Chapter 2 will then extend this knowledge to a computation at higher orders. We will in particular see the differences that appear whether those higher orders terms stem from new physics (as for the new operators to be introduced in the BMSSM) or the loop expansion (as will be the case for our study of the relic density).
\item Chapter 3 introduces the supersymmetric set-up in its minimal form, that is to say the MSSM.
\item Chapter 4 will close the introduction by showing why and how the predictions that we have to compute can be automated for a complete treatment and will focus on the implementation of our model within modern tools. We will see that they allow for an efficient implementation of supersymmetry within both the new physics and the loop expansion.
\item Chapter 5 describes the BMSSM framework from the theoretical point of view. It will dwell on the motivations for such an extension, the allowed independent operators, our practical implementation and some consistency checks. It will also expose some UV completions of the BMSSM that can yield such operators.
\item Chapter 6 will then carry on to the description of the experimental constraints to be considered, and the calculation of the predictions associated. We will furthermore focus on how the predictions can differ from the MSSM expectation in the Higgs sector.
\item Chapter 7 deals more specifically with the LHC data : we show first how one can extract relevant information from the experimental analyses and what issues arise. We then use those analyses to constrain the BMSSM model. In a first approach we consider the situation at the end of summer 2011 and deduce what exclusion bounds are drawn on the BMSSM Higgses. We move then on to the 5\fb\ dataset, released at the end of 2011 and entertain the possibility of a Higgs signal at 125 GeV. We also derive the consequence of such a signal for other channels, such as the $\bar\tau\tau$ channel and the other Higgses.
\item Chapter 8 brings us towards our second motivation that is the precise computation of the relic density. It presents the current experimental status and the picture of the predictions at the tree-level in the MSSM.
\item Chapter 9 will consider the inclusion of higher order corrections to the tree-level prediction for the relic density. We will first see what corrections are obtained in the BMSSM framework, and will then switch to the one-loop corrections of the MSSM. We will introduce our second effective approach, that is to say the effective couplings of the MSSM that can account for the dominant part of the full one-loop corrections, and show that the approach is particularly efficient in the bino case. The higgsino case will show a different picture and we will also comment on the applicability of the approach in this case.
\item We will finally draw the conclusion of our work for both sides of the effective approach. We will then show how our work can be extended, and which directions we intend to explore.
\item An appendix will detail some of the computations used in this thesis : those are the formulas for perturbative linear algebra, the detailed implementation of the \SL\ program, the formulas for statistics used by experimental collaborations and the computation of flavour observables in supersymmetry.
\end{itemize}

\chapter*{(Français) Introduction et Résumé}

Malgré le succès incontestable du Modèle Standard de la physique des particules, il est vraisemblable qu'il ne soit qu'une partie de la théorie complète de physique des particules -- comme c'est le cas des hypothèses de théories unifiées -- et ainsi de nombreux efforts ont été dédiés au développement de théories de Nouvelle Physique. La Supersymmétrie est l'une des extensions les plus populaires puisque qu'elle permet non seulement de résoudre le problème de Naturalité mais présente aussi un candidat viable de matière sombre. Ce dernier point a été particulièrement mis en avant avec les récentes mesures expérimentales qui ont permis d'affiner significativement notre connaissance des propriétés de cette matière sombre. En particulier la détermination de la densité relique de matière sombre dans l'univers est à présent réalisée avec une précision de l'ordre du pourcent. Dans le cadre du Modèle Standard Supersymmétrique Minimal (le MSSM), cette contrainte permet ainsi de tester la structure à une boucle de la 
théorie. Cependant c'est aussi un modèle présentant un très grand nombre de paramètres, comparé au Modèle Standard, et le calcul complet des observables à une boucle reste trop long pour être effectué sur l'ensemble de l'espace des paramètres. Dans cette thèse, je me suis donc intéressé à la possibilité de reproduire ces corrections à la boucle par un ensemble de couplages effectifs. L'approche effective présentant l'avantage de garder la simplicité d'un calcul effectué à l'arbre tout en conservant une trace des effets caractéristiques de boucle comme le non-découplage de certaines particules lourdes. Le LHC (Large Hadron Collider), dont les opérations ont démarrées juste après le début de ma thèse, soit à l'automne 2009, a fourni des données complémentaires aux observables de matière sombre. En effet le secteur du Higgs du MSSM est très peu flexible, ce qui a pour effet d'introduire ce que l'on appelle le problème du ``fine-tuning'', c'est à dire la nécessité d'avoir des valeurs très précises pour les 
paramètres. Afin d'y remédier, de nombreux modèles ont été créés au delà du MSSM, comme le NMSSM (en anglais Next-to-MSSM). Dans le but de suivre une approche plus générale, j'ai décidé au cours de cette thèse d'utiliser à nouveau l'approche effective, mais dans un but différent : alors que les couplages effectifs utilisés dans le cas de la matière sombre sont choisis pour reproduire le plus fidèlement possible les corrections à la boucle des particules du MSSM, les opérateurs effectifs que nous ajoutons au secteur du Higgs sont les effets à basse énergie (c'est à dire l'énergie de production du Higgs) d'une nouvelle physique à haute énergie. Bien que dédiées à deux buts différents, ces deux implémentations d'une même technique montrent bien ses différents atouts. Dans un des cas (celui du Higgs) les opérateurs effectifs permettent de paramétrer l'effet d'une physique ultraviolette inconnue, alors que dans l'autre cas cette physique ultraviolette se réduit au simple MSSM.\\

L'ajout de nouveaux opérateurs dans le secteur du Higgs a pour but premier de résoudre le problème de Naturalité du Higgs du Modèle Standard. En effet si nous considérons que de nouvelles particules vont apparaitre à l'échelle de Planck, alors ces nouvelles particules vont déplacer la masse du Higgs, qui est normalement à l'échelle électrofaible, jusqu'à l'échelle de Planck. La supersymmétrie permet de résoudre ce problème en introduisant une annulation des contributions bosoniques par les contributions fermioniques à la masse du Higgs. Pour que cette annulation soit effective il faut néanmoins que les superpartenaires du quark top (les stops) soient assez légers. Si tel est le cas, ces particules ne peuvent générer qu'une faible contribution à la masse du Higgs. Dans la mesure où cette masse est donnée à l'arbre par des couplages de jauge, ce qui implique notamment qu'elle est inférieure à l'échelle électrofaible $M_Z$, il s'ensuit que la masse du Higgs léger dans un modèle supersymmétrique naturel ne peut 
être éloignée de la borne inférieure du LEP (114 GeV) dans le meilleur des cas. Ce problème peut être résolu par l'introduction d'une nouvelle physique à une échelle élevée dont l'effet se manifeste par l'apparition d'opérateurs sur les superchamps du MSSM, qui sont supprimés par des puissances de l'échelle de la nouvelle physique, ce qui est la définition de l'approche effective. La puissance de l'approche effective réside dans le fait qu'elle reste très générique vis à vis de la supersymmétrie puisque aucune hypothèse n'est fait sur la nature de la nouvelle physique, à part que celle ci doit être suffisamment lourde. Cette approche a été baptisée BMSSM (Beyond the MSSM) dans la littérature et peut de ce fait représenter de nombreuses réalisations non minimales de la supersymmérie. En pratique l'implémentation du BMSSM avec les outils standards de la phénoménologie a été un premier obstacle puisque ce modèle inclut à la fois un potentiel de Kähler non standard et un superpotentiel non standard. La 
difficulté va au delà du calcul des règles de Feynman associées aux nouveaux opérateurs, puisque pour certains des processus à la boucle du Higgs comme la désintégration en deux photons, le facteur de forme de la boucle est modifié par les nouveaux opérateurs et produit de nouveaux diagrammes qui n'existent pas dans le pur MSSM. En étendant des outils tels \lanHEP\ et \HD\ à ce modèle nous avons pu générer la totalité de la phénoménologie du Higgs du BMSSM. La première conséquence obtenue est d'augmenter considérablement la masse du Higgs léger, qui peut maintenant atteindre 250 GeV. Cela permet aussitôt de se débarrasser du problème de naturalité puisque nous n'avons plus besoin de grandes corrections à la boucle. Cela va néanmoins amener ce Higgs dans une région différente vis à vis des expériences : en effet un Higgs dans une gamme de masse 150-250 GeV est très sensible aux recherches en WW et ZZ, contrastant avec un Higgs léger qui sera sensible au canal $\gamma\gamma$.\\

Entretemps le LHC a débuté sa prise de donnée, ce qui en très peu de temps a conduit à de fortes contraintes sur un Higgs modérément lourd (150-400 GeV). Mais ces recherches du Higgs sont pour la plupart dédiées au Higgs du Modèle Standard et il n'est pas tout à fait direct d'en tirer des conclusions pour d'autre modèles comme le MSSM et encore moins le BMSSM dont la phénoménologie du Higgs est bien plus riche. Un effort particulier a donc été consacré à la ré-interprétation des des recherches du Higgs du Modèle Standard dans des modèles non-standards. De manière intéressante une grande partie de nos découvertes ne sont pas restreintes à la supersymmétrie : certains problèmes sont communs à toutes les théories BSM (Beyond Standard Model). C'est ainsi le cas de la distinction entre sections efficaces inclusives et exclusives, puisque la quasi totalité des résultats expérimentaux sont donnés en fonction de la section efficace inclusive du Modèle Standard alors qu'ils ont été obtenus en comparant une section 
efficace exclusive avec les données. Comme le ratio entre section efficace inclusive et exclusive dépend du modèle, ces résultats sont donc dépendant du modèle. Un exemple de ce problème vient de l'analyse diphoton réalisée par CMS : les mêmes données ont été interprétées par la collaboration dans deux modèles différent : le Modèle Standard et un modèle fermiophobique, et la limite d'exclusion change de près de 50\% entre les deux cas. Pour faire une telle réinterprétation, il nous faut connaitre à la fois les efficacités de chaque mode de production du Higgs pour chaque sous-canal d'une analyse et les limites d'exclusion pour chaque sous canal séparément. Dans l'état actuel des recherches de telles quantités ne sont pas disponibles publiquement, que ce soit pour l'analyse en diphoton ou d'autres canaux. Ceci nous a amené à tester nombre d'approximations, parmi lesquelles l'estimation des efficacités via une simulation \PY\ par exemple, pour ré-interpréter les résultats d'ATLAS et de CMS. En utilisant ces 
approximations nous avons pu contraindre la phénoménologie du BMSSM en nous basant sur les données avec 2 \fb, ce qui nous a permis de trouver que le Higgs léger était dorénavant restreint à une masse plutôt légère, c'est à dire moins de 150 GeV, empêchant ainsi les hautes masses permises auparavant (150-250 GeV). Cela dit, les cas restant peuvent aussi être des cas particulièrement délicats à observer, que ce soit dans les analyses passées du LEP ou au LHC.\\

Quelque temps après avoir imposé ces limites sur la physique du Higgs les collaborations CMS et ATLAS ont déclaré vers la mi-décembre certains excès au sein des données de l'année 2011 (5 \fb), qui indiquaient un possible signal vers 125 GeV. Bien qu'une telle valeur soit tout à fait acceptable dans le Modèle Standard et possible dans le MSSM, il a aussi été noté que les différents excès semblaient avoir des couplages non-standards. Etant donné la faible quantité de données accumulée ces discussions sur les couplages d'un Higgs hypothétique restent dans le domaine de la pure spéculation, l'existence d'un Higgs à cette masse restant toujours à prouver. Cela dit, il est crucial de se préparer à la possibilité d'un signal non-standard au sein de futures données, qui serait ainsi incompatible avec le Modèle Standard. Dans le cadre du BMSSM, nous avons découvert que la phénoménologie du Higgs pouvait génériquement produire des effets de couplages non standard. Nous nous sommes alors soumis à l'exercice de tenter 
de 
reproduire les différents signaux obtenus par les expériences. Avec les 5 \fb\ de données, l'excès dans le canal diphoton est à peu près le double de la prédiction du Modèle Standard et ce pour les deux collaborations. Une telle augmentation est assez fréquente dans le cadre du BMSSM où le couplage du Higgs au quark b n'est pas autant contraint que dans le MSSM et peut ainsi être diminué. La déclaration de la collaboration CMS sur un signal dans le canal diphoton plus dijet significativement plus élevé que dans le canal diphoton est encore plus intéressante. Ce type de signature peut être facilement obtenu dans le BMSSM, puisque ce modèle est basé sur un spectre MSSM naturel, c'est à dire avec des stops légers et que ces stops peuvent altérer significativement la fusion de gluon et la désintégration du Higgs en photons. Nous avons en particulier montré qu'une hiérarchie entre les états finaux en bosons faibles, l'état final en diphoton ainsi que l'état final en diphoton et dijet peut être reproduite. Une 
telle caractéristique serait cruciale dans le but de distinguer une supersymmétrie minimale d'une réalisation non minimale.\\

Nous avons alors élargi le cadre de notre étude avec de nouvelles contraintes venant d'autres secteurs pouvant contraindre la nouvelle physique : la physique de la saveur et la matière sombre. Bien que la contribution de la physique du B est assez bien connue dans le MSSM, nous avons montré que l'ajout de nouveaux opérateurs d'ordre élevé menait à une modification de l'amplitude de diagramme pingouin dans le processus $\Bsmu$. Cette observable est connue pour mettre une contrainte forte sur la supersymmétrie. En utilisant les équations du mouvement, nous avons pu obtenir la prédiction complète du BMSSM et notre analyse a montré une interdépendance avec la physique du Higgs. D'un côté cela nous a permis de montrer que la région avec un Higgs de charge CP impaire trop léger était exclue, et de l'autre l'inclusion de l'observable $\Bsg$, qui dépend de la contribution des stops, nous a permis de réduire l'espace des paramètres par rapport à notre étude précédente.\\

Même si la recherche du Higgs est très certainement une contrainte puissante sur la supersymmétrie, il est difficile de laisser de côté notre deuxième motivation pour la supersymmétrie, qui est l'explication de la matière sombre. De ce côté nous avons déjà une certaine connaissance de par les expériences d'astrophysique et il s'avère que la contrainte principale vient de la mesure de la densité relique de matière sombre dans l'univers. Les expériences ont ainsi atteint une précision de l'ordre de 6\%, et une meilleure mesure est attendue très prochainement. Si l'on se tient à un scénario cosmologique standard, ceci nous permet de déterminer très précisément la section efficace d'annihilation de deux neutralinos les plus légers (qui sont les candidats principaux du MSSM pour la matière sombre) en particules du Modèle Standard. Ceci suggère un calcul extrêmement précis du côté théorique, et ainsi le calcul des effets à une boucle. Ceci n'est pas a priori un problème dans la mesure où, grâce aux études 
complètes 
sur le sujet, nous sommes désormais capable de mener à bien la renormalisation délicate du MSSM et d'employer en pratique des outils automatisés dédiés à cette tâche. Cependant il s'avère que, peut-être parce que ces développements sont encore récents ou parce que le calcul reste compliqué et long, même avec une automatisation l'idée du calcul à une boucle de la densité relique ne fait pas encore l'unanimité au sein de la communauté de supersymmétrie. Il est assez dérangeant de remarquer que la quasi totalité des scans de l'espace des paramètres du MSSM, où la densité relique est mise en avant comme l'une des plus fortes contraintes, restent encore au stade du calcul à l'arbre, sous-estimant ainsi grandement l'incertitude théorique. Notre idée a été de revenir à l'approche effective. En effet on sait que les corrections radiatives peuvent être écrites comme des opérateurs effectifs au sein de l'action effective. Cela nous a ainsi permis de construire des vertex effectifs pour prendre en compte la partie 
dominante des corrections de boucle, tout en restant plus rapides à évaluer que la correction totale. En particulier nous avons introduit des vertex effectifs pour $\chii\tilde{f}f$ et $\chii\chii Z$, qui sont tous les deux utilisés pour la classe de processus $\chii\chii\to\bar ff$. Une étude complète des performances de ces opérateurs comparées avec la correction à une boucle totale nous a permis de déterminer les régions du MSSM où l'approche effective nous donnait des résultats satisfaisants et aussi de mettre en exergue un effet de non-découplage de la densité relique. En effet il s'avère que les boucles avec des sfermions lourds donnent une contribution petite mais non négligeable et puisque nous avons affaire à une mesure de précision, nous pouvons tirer des conclusions sur le spectre des particules lourdes. La situation est assez similaire au cas du pôle du Z au LEP, où la précision obtenue pouvait contraindre un spectre de particules bien plus lourd. Cette conclusion est d'autant plus intéressante 
vis à vis de notre étude précédente du LHC, dans la mesure où ce dernier ne pourra tester que des masses de squarks modérées, jusqu'à 1 ou 2 TeV, alors que la densité relique gardera une trace de ces particules à plus grande échelle.\\

\section*{Résumé}
Voici le résumé des différents chapitres de la thèse :\\
\begin{itemize}
\item Le chapitre 1 présente les outils mathématiques nécessaires au calcul de prédictions dans le cadre d'une théorie quantique des champs à l'arbre. Nous y trouverons en particulier la description du modèle générique de physique des particules en fonction des représentations du groupe de Poincaré et des représentations du groupe de jauge. Nous verrons ensuite le formalisme des théories de jauge ainsi que le mécanisme de brisure du Higgs. Ces différents concepts seront appliqués dans le cas du Modèle Standard. Dans un deuxième temps nous nous intéresserons à la procédure de calcul de sections efficace dans une théorie quantique des champs générale, et détaillerons les calculs dans le cas de l'approximation à l'arbre. Nous verrons en particulier les relations entre quantités initiales (celles apparaissant dans le lagrangien) et les quantités physiques qui sont accessibles par les expériences.

\item Le chapitre 2 étend notre technique de calcul aux ordres supérieurs de l'action effective. Ces ordres supérieurs viennent d'une part des effets possible d'une nouvelle physique à un échelle plus élevée. Ces effets peuvent être caractérisés par l'apparition de nouveaux opérateurs dans le Lagrangien qui sont supprimés par l'échelle de la nouvelle physique. D'autre part ces ordres peuvent venir des particules déjà présentes dans la théorie via les corrections radiatives. Nous nous pencherons en détails sur les calculs de corrections radiatives à une boucle qui nous permettront de voir la méthode que nous utiliserons ainsi que les difficultés rencontrées. Dans les deux cas, nous aurons une modification des relations entre quantités initiales et quantités physique par rapport à celles obtenues dans le chapitre 1, modification que nous décrirons.

\item Le chapitre 3 va introduire une des extensions populaire du modèle standard : la Supersymmétrie. Nous allons tout d'abord nous pencher sur une des problèmes conceptuels du Modèle Standard qui est le problème de la Naturalité. Comme nous le verrons ce problème peut être résolu si l'on arrive à protéger par une nouvelle symétrie la masse du Higgs scalaire contre les corrections amenées par de nouvelles particules à de grandes échelles. La supersymmétrie permettant une telle protection, nous décrirons le modèle supersymmétrique le plus simple contenant le Modèle Standard, c'est le MSSM. Nous verrons comment utiliser les représentations de l'algèbre de super Poincaré pour établir le formalisme d'une théorie supersymmétrique, puis nous introduirons la forme de l'action en supersymmétrie avant de discuter le cas de la brisure douce de supersymmétrie. Nous finirons par une description des différents secteurs du MSSM.

\item Le chapitre 4 présente certains des différents outils utilisés dans la communauté de phénoménologie pour automatiser les calculs. En effet la supersymmétrie introduit de nombreuses nouvelles particules, de telle manière qu'un traitement à la main des diagrammes à une boucle n'est pas envisageable. Nous verrons donc quels sont les codes existants permettant de faciliter l'obtention de prédictions précises dans un modèle particulier. Ces codes se partagent en deux catégorie, la première concerne les codes qui permettent d'obtenir les règles de Feynman d'une théorie donnée (par exemple \lanHEP), la seconde est celle des codes dont le but est de calculer l'amplitude d'un processus donné dans une théorie donnée (par exemple \FA). Nous montrerons ainsi un exemple de l'utilisation de ces codes en vue du calcul à une boucle de processus en supersymmétrie.

\item Le chapitre 5 va introduire notre première approche effective, c'est à dire le BMSSM. Nous montrerons d'abord quels sont les nouveaux opérateurs qui peuvent être introduits dans le secteur du Higgs et quels sont ceux qui peuvent être éliminés par l'utilisation d'équations du mouvement. Puis nous passerons à notre implémentation personnelle de ce modèle via les outils présentés précédemment (c'est à dire avec \lanHEP\ et \HD) et montrerons comment nous nous assurons qu'une telle théorie, basée sur un développement perturbatif sur l'inverse de l'échelle de la nouvelle physique, reste bien dans un domaine perturbatif, ou en d'autres termes que la troncature de la série effective est justifiée. Nous conclurons le chapitre en exposant certaines théories complète de nouvelle physique qui peuvent mener à l'apparition de tels opérateurs à basse énergie et nous pourrons alors voir les relations induites sur les différents coefficients effectifs.

\item Le chapitre 6 va alors introduire la description des contraintes expérimentales sur le modèle. Cela passe d'abord par la description du calcul des prédictions associées aux expériences : dans le cadre de la recherche du Higgs il va s'agir de calculer les sections efficace de production du Higgs ainsi que de ses désintégrations. Pour ce faire nous avons modifié le code \HD\ en particulier pour les processus à la boucle comme la fusion de gluons et la désintégration en photons, dans ce cas nous décrirons explicitement quelles sont les modifications amenées par les opérateurs effectifs et quels nouveaux diagrammes peuvent apparaitre. Nous terminerons par une analyse des modifications apportées aux différent couplages des Higgs par rapport au MSSM et au Modèle Standard. Nous verrons en particulier que le domaine de découplage du MSSM est modifié par les opérateurs effectifs puisque ceux ci peuvent induire un non découplage même pour des valeurs de $\ma$ relativement grandes.

\item Le chapitre 7 se concentre quant à lui sur la ré-interprétation des données du LHC. Nous allons dans un premier temps montrer quelles sont les informations pertinentes que l'on peut extraire des résultats expérimentaux parus publiquement. Nous verrons alors les difficultés rencontrées pour adapter ces limites à des modèles non standards. D'une part il s'avère que le ratio entre sections efficaces inclusives et exclusives dépend du modèle envisagé, de telle façon que pour passer d'un modèle à un autre il est nécessaire de connaitre les efficacités de production des différents canaux utilisés par l'analyse. D'autre part les combinaisons de différent canaux sont toujours dépendantes d'un modèle particulier et dans ce sens, une fois que la combinaison est faite il est difficile de changer de modèle. Dans une première approche nous considérerons la situation telle qu'elle était à la fin de l'été 2011 et déduirons, moyennant quelques approximations, les conséquences sur la phénoménologie du Higgs dans le 
BMSSM. 
Nous passerons ensuite aux données totales de l'année 2011 c'est à dire le lot de donnée avec 5 \fb, et prendrons l'hypothèse d'un signal de Higgs à 125 GeV. Cet exercice nous permettra notamment de considérer les différentes prédictions du BMSSM dans les canaux de recherche et en particulier de prouver qu'un excès en diphoton peut être tout à fait compatible avec ce modèle.

\item Le chapitre 8 nous amènera vers notre deuxième champ d'application de la théorie effective des champs : la matière sombre. Après avoir rappelé les différentes observations pouvant être interprétées par de la matière sombre, nous nous intéresserons en particulier à la densité relique de  matière sombre. Nous montrerons d'abord comment cette quantité est reliée à la section efficace d'annihilation du candidat de matière sombre vers les particules du Modèle Standard, puis nous passerons à l'étude de cette section efficace dans le cadre du MSSM. Nous verrons alors que le candidat le plus populaire est le neutralino le plus léger, et que la valeur de la densité relique associée est très dépendante de la nature de ce neutralino. En effet des neutralinos de type bino, wino ou higgsino ne vont pas procéder par les même canaux et la densité relique résultante peut varier de manière importante.

\item Nous nous dirigerons dans le chapitre 9 vers les calculs de précision de la section efficace d'annihilation de neutralinos. Pour obtenir une précision comparable à celle des expériences (c'est à dire de l'ordre de quelque pourcents) il est nécessaire de tenir compte des corrections radiatives. A l'ordre d'une boucle les corrections radiatives du MSSM ne présentent pas de problèmes conceptuels particuliers, car de nombreuses études ont montrés comment renormaliser ce modèle de manière cohérente et comment implémenter les longs et fastidieux calculs via des codes automatiques. Cependant cela reste un calcul délicat et nous avons donc décidé d'opter pour une approche effective. Nous verrons donc les résultats de notre étude où nous introduisons des vertex effectifs $\chii\tilde{f}f$ et $\chii\chii Z$, pour pouvoir ainsi comparer les résultats effectifs avec les résultats du calcul à la boucle complet sur le processus test $\chii\chii\to\bar\mu\mu$. Comme nous le verrons, l'approche effective est 
particulièrement indiquée dans le cas d'un neutralino de type bino : dans ce cas l'accord entre les deux approches est meilleur que 2\%, alors que l'approche effective peut se targuer d'un temps de calcul plus faible de plusieurs ordres de grandeur. Nous verrons ensuite la cas d'un neutralino de type higgsino et conclurons sur les possibilités de l'approche effective dans le cadre de la densité relique du MSSM.
\end{itemize}

\adjustmtc

%% file: chapter1.tex
\chapter{Basic Physics : the building of a model}

\minitoc\vspace{1cm}

\section{Elements of gauge theory}
The description of a model of particle physics usually lies in two quantities, the first being the particle content and the second the Lagrangian. The first item is simply a list of different particles that are uniquely characterised by their Poincaré and gauge representations. We will now detail what these representations are.\\

\noindent\underline{Remark :} The notions introduced in this chapter and the following stem for the most part from the textbooks from M.Peskin and D.Schroeder (\cite{peskin}), M.Nakahara (\cite{nakahara}) and S.Weinberg (\cite{weinberg}).

\subsection{Unitary representations of the Poincaré group}
There are two quantities that will define uniquely such a representation : the first is the eigenvalue of the Casimir operator $P^2=P_\mu P^\mu$, which we denote by $m^2$ where $m$ is named the mass of the particle, and the second is the maximal weight of the representation in the Lie algebra. $sl(2,\CC)$ is characterised by two weights $(i,j)$ that take half integer values. It turns out that most of phenomenological models only use a few of those representations :
\begin{itemize}
\item $m,\ (0,0)$ : this is the trivial Poincaré representation, whose particles are called scalars, denoted $\phi$.
\item $m,\ (\1,0)\ \text{or}\ (0,\1)$ : those are the two fundamental representations, the first case yielding particles known as left-handed Weyl spinors and the second right-handed Weyl spinors. In both the massive and massless case they are two dimensional. They are denoted $\psi$.
\item $m=0,\ (\1,\1)$ : this is the massless vector representation. There are two states, labelled by helicity $\lambda=\pm 1$.
\item $m>0,\ (\1,\1)$ : this is the massive vector representation. There are three states, labelled by spin $s=-1,0,1$.
\end{itemize}

Let us focus first on the fundamental representations that are the Weyl spinors. Given a left handed Weyl spinor $\psi$ we will label its components by $\psi_\alpha$, and the right-handed representation $\chi$ will be labelled by $\chi_{\dot\alpha}$. We introduce the invariant
\begin{equation}
\psi^T\epsilon\psi\qquad\text{where}\qquad\epsilon=\begin{pmatrix}0&1\\-1&0\end{pmatrix}
\end{equation}
We will from now on use the shorthand notation
\begin{equation*}
\psi\psi=\psi^T\epsilon\psi
\end{equation*}
Both representations can be switched by taking the conjugate of the field, it is in particular conventional to denote $\bar\psi$ the conjugate of $\psi$, that is to say $\psi^*$.\\

Going now to the spin one case, one can prove that an element $X=(X^{\alpha\dot\alpha})$ in the $(\1,\1)$ representation is equivalent to a one-form $A=(A^\mu)$ by identifying
$$X^{\alpha\dot\alpha}=A^\mu\sigma_\mu^{\alpha\dot\alpha}$$
where $(\sigma_\mu)=(1,\sigma_i)$ is an invariant of the representation $(\1,0)\otimes(0,\1)\otimes(\1,\1)$. So we will from now on consider both massive and massless spin 1 representations as embedded in the one-form representation $A=(A_\mu)$. Although this representation seems to have more states than needed, we will see below that the unwanted states can be removed later in the process. This representation is fully motivated by the geometrical picture of the gauge principle that comes next.\\

\paragraph{Lagrangian :}~\\
Having worked out the invariant terms that we could write with each kind of particle, the first Lagrangian we can write is based purely on the derivative of the fields.
\begin{equation}
\L=\1|d\phi|^2+i\bar\psi\dslash\psi-\1|dA|^2
\label{eq:poincare_lag}
\end{equation}
which is computed by going to the components of each field, that is to say
$$\L=\1(\partial_\mu\phi)^\dag(\partial^\mu\phi)+i\bar\psi_{\dot\alpha}\sigma_\mu^{\dot\alpha\alpha}\partial^\mu\psi_\alpha-\frac{1}{4}(\partial_\mu A_\nu-\partial_\nu A_\mu)^\dag(\partial^\mu A^\nu-\partial^\nu A^\mu)$$
where the antisymmetry of the last term is a consequence of the identification of the spin 1 field to a one form.
We have explicitly chosen the fermion $\psi$ to be left-handed, which is always possible given that the conjugate of a right-handed fermion is left-handed and that we can equivalently write the theory for one field or its conjugate.\\

\subsection{Gauge principle}
The gauge interaction can be consistently added in this geometrical picture, described in more details in reference \cite{nakahara}. One introduces a group $G$, called the gauge group, and increases the spacetime structure by putting on each point a vector space $F$, called the fibre or the internal space, which is a representation space for $G$. This means that the total space on which the field is introduced is not $\mathbb{R}^4$ any more, but a fibre bundle $M$ which is locally equal to
$$M=\mathbb{R}^4\times F.$$
The fibre $F$ will change from one particle to another depending on its gauge quantum number. Labelling $(x,f)$ an element of $M$, the Poincaré group will act on $x$ and the gauge group on $f$. The main feature of the geometrical approach to the gauge principle is that, in a similar way that a covariant derivative has to be introduced in General Relativity to transport tangent vector fields along the spacetime, a covariant derivative is needed to describe the elements of the fibre along spacetime. To define this parallel transport one must express the tangent space to $G$ as a direct sum of a horizontal space which is invariant by the action of $G$ and a vertical space : a vector field on $M$ is said to be parallel transported if its tangent field belongs to the horizontal subspace. This separation is fully parametrised by a one-form which take values in the adjoint representation of $G$, called the connection by mathematicians and the gauge vector field by physicists :
\begin{equation}
\AA=A_{\mu\ a} T_adx^\mu
\end{equation}
where $a$ ranges from one to the rank of $G$ and $T_a$ are the representations of the generators of $G$ in the adjoint representation. Analogously to the covariant derivative of General Relativity, we will define the covariant derivative as
\begin{equation}
D=d+ig\AA
\end{equation}
where the $i$ factor is used for convenience since it will make the gauge vector field $\AA$ hermitian if the gauge group $G$ is compact (which is usually the case), or more precisely it will make the field $A_{\mu\,a}$ real and the generators $T_a$ hermitian. This covariant derivative acts on any form $\Phi$ on $M$ as
\begin{equation}
 D(\Phi)=d\Phi+ig\rho(\AA)\wedge\Phi
\end{equation}
where $\rho$ is the representation in which $\Phi$ lies and $\wedge$ is the exterior product. In particular, since $\AA$ is in the adjoint representation its covariant derivative is\footnote{Note that because of the antisymmetry of the exterior derivative we have $[\AA,\AA]=2\AA\wedge\AA$}
$$D(\AA)=d\AA+\AA\wedge\AA.$$

The basis of the bundle can also be changed by acting with an element $g$ of $G$, which is called a gauge transformation. Since $g$ can be a field over the spacetime rather than a constant this is a local gauge transformation. Under this transformation the different matter fields $\Phi$ and the gauge field $\AA$ will undergo the following change
$$\Phi\rightarrow \rho(g)\left(\Phi\right),\quad \AA\rightarrow\AA'=g^{-1}\AA g+g^{-1}dg.$$
Note that in the abelian case this is the familiar expression
\begin{equation}
A'=A+id\alpha\qquad\qquad\text{where }g=e^{i\alpha}.
\end{equation}
Since this transformation leaves physics invariant it generates a symmetry, called the gauge symmetry. In particular the covariant derivative of the gauge field $\AA$ is unchanged by a gauge transformation. Again, in a similar way as General Relativity, we can thus define a term in the action associated with the curvature : the gauge curvature, defined as
\begin{equation}
\L_\text{gauge}=-\1|\mathbf{F}|^2=-\1|D\AA|^2.
\end{equation}
In the case of an abelian gauge group, it reduces to the simple kinetic term $|dA|^2$, however for non-abelian theory there will be interactions between gauge bosons.\\

\paragraph{Gauge fixing :}~\\
Because the gauge curvature has no mass term, gauge bosons are massless spin one fields, which means that they are two dimensional representations of the Poincaré group. Our description by a four vector $A_\mu$ is thus quite redundant. Nonetheless, we can get rid of the additional states by using the gauge symmetry.

Indeed, since the action is invariant by the gauge symmetry, two different bases for fields (both matter and gauge ones) that are related by a gauge transformation describe the same physics. To get rid of the unwanted degrees of freedom of the gauge field, it is thus enough to fix the gauge, that is to say specify a configuration of the gauge field and matter fields. There is a trick, due to Fadeev and Popov, to introduce this gauge fixing directly in the Lagrangian and keep the gauge bosons as four vectors. In the abelian case, the procedure can be summarised as follows : one defines a gauge fixing condition called $G$, which usually is
$$G(A)=\frac{1}{\sqrt{\xi}}\left(\partial_\mu A^\mu-\xi x\right)$$
where $\xi$ is a real parameter and $x$ can be any combination of parameters and fields of the theory. This term $G(A)$ is then used as a Lagrange multiplier in the path integral and eventually is absorbed in the Lagrangian by the redefinition
\begin{equation}
\L\longrightarrow\L-\1|G|^2.
\end{equation}
However this procedure also brings a determinant of the gauge fixing function $\left|\frac{\delta G}{\delta A}\right|$ in the path integral, which must then be absorbed by fictitious fields called the ghost fields.
\begin{equation}
\L\xrightarrow{\text{gauge fixing}}\L-\1|G|^2+\L_\text{ghost}
\end{equation}

\paragraph{Application : The Standard Model (Part I) :}~\\
We can now apply this set-up to the Standard Model. Its gauge group is 
\begin{equation}
G=SU(3)\times SU(2)\times U(1)
\end{equation}
We introduce a gauge field and a coupling constant for each subgroup, as shown on table \ref{tab:sm_gauge}.

\begin{table}[!h]
\begin{center}
\begin{tabular}{cccc}
Group & Coupling constant & Basis & Gauge boson\\\hline
$SU(3)$ & $g_s$ & $\Lambda^a$ & $G^a$\\
$SU(2)$ & $g_1$ & $\sigma^a/2$ & $W^a$\\
$U(1)$ & $g_2$ & $1$ & $B$\\
\end{tabular}
\caption{\label{tab:sm_gauge}{\em Gauge content of the Standard Model. For each subgroup of $G$ we define a coupling constant, a basis of the associated Lie algebra ($T^a$) and the gauge fields associated, which number equals the rank of the subgroup.}}
\end{center}
\end{table}

\begin{table}[!h]
\begin{center}
\begin{tabular}{ccl}
Field &$\rho_P$&$\rho_G$\\\hline
$L$ & $(1/2,0)$ & $1\otimes 2\otimes -\1$\\
$E$ & $(1/2,0)$ & $1\otimes 1\otimes 1$\\
$Q$ & $(1/2,0)$ & $3\otimes 2\otimes \frac{1}{6}$\\
$U$ & $(1/2,0)$ & $\bar3\otimes 1\otimes -\frac{2}{3}$\\
$D$ & $(1/2,0)$ & $\bar3\otimes 1\otimes \frac{1}{3}$\\\hline
$H$ & $(0,0)$ & $1\otimes 2\otimes -\1$
\end{tabular}
\caption{\label{tab:sm_matter}{\em Matter content of the Standard Model : the quarks are introduced with the fields $Q,U$ and $D$ and the leptons with $L$ and $E$. Note that since the Standard Model has three nearly identical generations, there will be three of each field which differ only by the masses. The last field, the Higgs $H$, is the only fundamental scalar particle of the Standard Model.}}
\end{center}
\end{table}

Its matter content is shown in table \ref{tab:sm_matter} (with $\rho_P$ and $\rho_G$ denoting respectively the Poincaré representation and the gauge representation), and is mainly a set of massless fermions with the addition of a massless scalar particle, the Higgs boson. Everything is ruled by the somewhat simple Lagrangian :
$$\L=\1|DH|^2+\sum_{L,E,Q,U,D}i\bar\psi\Dslash\psi-\1\sum_{G,W,B}|DA|^2+\L_\text{gauge fix.}.$$

In particular all fields are massless and all gauge symmetries are unbroken, which brings us to the Higgs mechanism.

\subsection{The Higgs mechanism}
To trigger this mechanism the field $H$ will be ruled by an additional potential
\begin{equation}
\L_\text{Higgs}=\mu|H|^2-\lambda|H|^4.
\end{equation}
Since this potential shows a minimum at $|H|^2=v^2=\frac{\mu}{2\lambda}$ which is not at the origin, the Higgs field will develop a non vanishing vacuum expectation value
\begin{equation*}
\left<H\right>=v.
\end{equation*}
This means that the physical field is no longer $H$ but :
\begin{equation*}
H'=H-\mathbf{v}
\end{equation*}
where $\mathbf{v}$ is a vector which norm equals $v$. In the following we will take
$$\mathbf{v}=\binom{v}{0}.$$
Note that the Lagrangian still exhibits the full gauge symmetry, however it is non-linearly realised on $H'$ whereas it was linearly realised in the non physical basis (on $H$). In the physical basis the Higgs Lagrangian will be expanded as 
\begin{eqnarray*}
\L&=&|DH|^2+\mu|H|^2-\lambda|H|^4\\
\Rightarrow\quad \L&=&|D(H'+v)|^2+\mu|(H'+v)|^2-\lambda|(H'+v)|^4\\
\end{eqnarray*}
which will lead to mass terms for gauge bosons and $H'$. In particular the mass term for gauge bosons will look like
\begin{eqnarray*}
\L&\to&|g\AA\mathbf{v}|^2\\
&\to&\left|\left(\sum_{a,i}g_i\AA_{i,a}\right)\mathbf{v}\right|^2
\end{eqnarray*}
where $i$ labels the different simple subgroups of $G$ and $a$ the generators of a given subgroup. We then have to rotate $g\AA$ to another base of $G$ in which the mass matrix is diagonal.
\begin{equation}
g\AA\rightarrow g'\AA'
\end{equation}
It is useful to note at this point that the masses of gauge bosons are solely determined from the gauge couplings and the vacuum expectation value of the Higgs field.

\paragraph{Unitary gauge :}~\\
Since some gauge bosons have been turned from massless to massive, it means that they have gained a degree of freedom. The Goldstone theorem tells us that these degrees of freedom originate from the Higgs fields, indeed some states of the Higgs field, called Goldstone bosons, will decouple from the physical observables and become hence unphysical. Moreover, if the action of $G$ on the Higgs field is transitive we can always find a gauge configuration where the Goldstone fields vanish identically. This has the important consequence that we can fix entirely the gauge in the broken sector and leave the unbroken gauge free. Equivalently, one can access this gauge configuration by modifying the gauge-fixing condition, for instance in an abelian broken group
\begin{equation}
 G[A]=\partial_\mu A^\mu \xrightarrow{\text{Unitary gauge}} G[A]=\partial_\mu A^\mu+ gvG^0
\end{equation}
where $G^0$ is the Goldstone bosons associated to $A$. This gauge is called the unitary gauge.

\paragraph{Higgs mechanism for fermions}~\\
The Higgs boson is also believed to give mass to fermions in the Standard Model. First, let us recall that the way to introduce massive charged fermions is to connect a pair of two Weyl fermions of opposite charges, say $\psi_1$, $\psi_2$ in
\begin{equation}
\L_\text{mass}=m_f\psi_1\psi_2\hc
\end{equation}
and then to create a single fermion from those two Weyl fermions, called a Dirac fermion
\begin{equation}
\psi=\binom{\psi_1}{\bar\psi_2}\qquad\psi\in\ \left(\1,0\right)\oplus\left(0,\1\right)
\end{equation}
Using this new fermion, the quadratic lagrangian can be written without any mixing, hence $\psi$ actuallly is a mass eigenstate,
\begin{equation}
 \L_\psi=\bar\psi(\dslash+m_f)\psi
\end{equation}
where in this equation $\bar\psi$ stands for $\gamma_0\psi^\dag$ in order to make a Poincaré invariant.\\

The Higgs mechanism for fermions is simply the addition of the Yukawa potential to the theory, for instance 
$$\L_Y=yH\psi_1\psi_2\hc$$
which, when replacing $H\to H'+v$ will yield a mass term
$$\L_Y\to m_f\psi_1\psi_2\hc$$
with $m_f=yv$. This relates the mass term to the Yukawa coupling. In particular it will tell us that the couplings of the fermion to the physical Higgs states (the massive ones) are proportional to its mass.

\subsection{Application : The Standard Model (Part II)}
We can now upgrade our description of the Standard Model by introducing the potential
\begin{equation}
\L_{H+Y}=\mu|H|^2-\lambda|H|^4+y_LHLE+y_UH^\dag QU+y_DHQD
\end{equation}
As previously said, the Higgs will exhibit a non vanishing vacuum expectation value $v$. The first thing to do is look at the mass matrix for the gauge bosons $W$ and $B$. The expression of the gauge vector $(g\AA)$ in the $2\otimes\1$ representation is
\begin{equation*}
g\AA=\1\begin{pmatrix}g_2B+g_1W^3&g_1W^+\\g_1W^-&g_2B-g_1W^3\end{pmatrix}
\end{equation*}
so that we have
\begin{equation*}
g\AA\mathbf{v}=\frac{v}{2}\binom{g_2B+g_1W^3}{g_1W_-}.
\end{equation*}
For $|g\AA\mathbf{v}|^2$ to be diagonal, we take another gauge basis vector
\begin{equation*}
Z=\frac{1}{g_1^2+g_2^2}(g_2B+g_1W^3)
\end{equation*}
which we have normalised for the transformation to be unitary. The coupling constants $g_1,g_2$ being real, the unitary transformation is then fully specified by one angle, called the Weinberg angle $\theta_w$, defined by 
\begin{equation}
\cw=\frac{g_1}{g_1^2+g_2^2}\ ,\sw=\frac{g_2}{g_1^2+g_2^2}\qquad\left(\sw=\sin(\theta_w),\cw=\cos(\theta_w)\right).
\end{equation}
The rotation between the unbroken gauge fields $B,W_3$ and the physical ones $A,Z$ is then
\begin{equation}
\binom{B}{W_3}=R(\theta_w)\binom{A}{Z}.
\end{equation}
As foreseen, this rotation will also act on couplings and generators $(g_1\frac{\sigma_3}{2},g_2 I)$ to yield
\begin{equation*}
\binom{g_2 I}{g_1\frac{\sigma_3}{2}}=R(\theta_w)\binom{eQ}{g_zT_z}
\end{equation*}
which results in 
\begin{center}
\begin{tabular}{ll}
$Q=I+\frac{\sigma_3}{2}$ & $T_z=\frac{\sigma_3}{2}-\sw^2\,Q$\\
$e=\sw\cw$& $g_z=\frac{g_1}{\sw}$
\end{tabular}
\end{center}

The mass term then reads
\begin{equation}
|g\AA\mathbf{v}|^2=v^2g_1^2|W^+|^2+v^2(g_1^2+g_2^2)|Z|^2.
\end{equation}
The only unbroken generator of $SU(2)\times U(1)$ is then the electromagnetic charge $Q$ and its gauge boson the photon $A$. The broken gauge bosons being the $W$ and the $Z$ with masses
\begin{equation}
\mw=v^2g_1^2,\qquad\mz=v^2\frac{g_1^2+g_2^2}{2}.
\end{equation}
Since we have broken three generators, three of the Higgs fields are unphsyical and we are left with only one physical scalar field denoted $h$
\begin{equation}
H=\binom{v+h+iG^0}{G^-}
\end{equation}
where $G^{(0,-,+)}$ are the three Goldstone bosons.\\

Concerning fermions, since the $SU(2)$ is now broken we have to label the components of each doublet 
\begin{equation}
L=\binom{e_L}{\nu_L},\ Q=\binom{u_L}{d_L}
\end{equation}
Then the Yukawa potential will yield mass terms for fermions
\begin{equation}
\L_Y\to y_L\,ve_Le_R+y_Uv\,u_Lu_R+y_Dv\,d_Ld_R\hc
\end{equation}
which will allow to pair Weyl fermions together to obtain the charged leptons and quarks
\begin{equation}
e=\binom{e_L}{\bar e_R},\ u=\binom{u_L}{\bar u_R},\ d=\binom{d_L}{\bar d_R}.
\end{equation}
Note that no Dirac fermion is constructed for the neutrino. We can summarize the Standard Model in the physical basis for gauge bosons in Table \ref{tab:xx}.

\begin{table}[!h]
\begin{center}
\begin{tabular}{cccc}
Group & Coupling constant & Basis & Gauge boson\\\hline
$SU(3)$ & $g_s$ & $\Lambda^a$ & $G^a$\\
$SU(2)$ & $\left\{\begin{matrix}g_Z\\g_1\\g_1\end{matrix}\right.$ & $\begin{matrix}T_Z\\\sigma_-\\\sigma_+\end{matrix}$ & $\left.\begin{matrix}Z\\W^-\\W^+\end{matrix}\right\}$\\
$U(1)$ & $e$ & $Q$ & $A$\\
\end{tabular}
\caption{\label{tab:xx}{\em Gauge sector of the Standard Model in the physical basis, that is to say after the electroweak symmetry breaking. The three massive bosons are the $Z$ (neutral) and $W$ (charged), where the photon $A$ and the gluon $G$ stay massless. The index $a$ labels the different elements of a group with rank higher than one.}}
\end{center}
\end{table}

On the matter side, there is a novelty : the Dirac fermions do not correspond to representations of the broken $SU(2)$. In particular, their right-handed part is trivially coupled under this group, while their left part is not. This is an explicit chiral behaviour with respect to the weak interaction, which corresponds exactly to what we observe.

\section{Relating theory to observables}
\subsection{Turning ideas into predictions}
Let us switch now to a more general point of view and see how one relates a theoretical set-up with observations in a generic quantum field theory. We have seen that such a generic theory is described by a Lagrangian $\L$ :
\begin{eqnarray}
 \L&=&\L_G+\L_\text{others}\\
 \L&=&\sum_\phi|D\phi|^2+\sum_\psi i\oo{\psi}\Dslash\psi+\sum_A|DA|^2+\L_\text{others}
\end{eqnarray}
with ``others'' accounting for subtleties such as Yukawa terms and Higgs potential. However it is fair to say that quantities appearing in $\L$, namely the fields and the parameters, are usually not accessible to the experiments, and on a more general ground, to phenomenology. Indeed one only has access to observables, and we will turn now to their computation. There exists an elegant way to re-express the value of an observable $\O$ in the framework of Feynman diagrams, it is called the effective action. It is a functional $\Gamma[\Phi]$ defined by
\begin{eqnarray}
W[J]&=&-i\ln\left(\int\D\Phi\ e^{i\int d^4x\ \L+J\Phi}\right)\\
\tilde{J}&=&\left(\frac{\delta W}{\delta J}\right)^{-1}\\
\Gamma[\Phi]&=&W[\tilde{J}[\Phi]]-\int d^4x\,\Phi\tilde{J}[\Phi].\label{eq:gammaeff}
\end{eqnarray}
What the effective action actually encodes are the connected one particle irreducible diagrams, which are the building blocks to evaluate the Green correlation functions. In particular, to evaluate a process it is enough to consider Feynman diagrams at the lowest order, but with couplings obtained from the effective action. In other words it is equivalent to consider a quantum field theory with a classical Lagrangian $\L$ and a classical field theory with the effective action $\Geff$ derived from $\L$.\\

Since the physics is encoded in the effective action $\Geff$ and no more in the Lagrangian $\L$, the initial fields and couplings -- that is to say fields and couplings such as they appear in $\L$ -- will not be the states and interactions that we observe in nature. This means that if we want to use our simple Lagrangian to compute physical observables, we need to relate initial and physical parameters and fields.\\

\paragraph{Effective action at leading order :}~\\
The effective action can be developed in a perturbative expansion : this is indeed the Feynman expansion :
\begin{equation}
\Geff=\Geff^{(0)}+\epsilon\Geff^{(1)}+\epsilon^2\Geff^{(2)}+\cdots
\label{eq:gamma_eff}
\end{equation}
where $\epsilon$ stand for the couplings of the theory. Its accuracy depends on the order of the truncation. It seems that a first approach, qualitative, would be to truncate at the lowest order : this approach is called the tree-level approximation. This denomination does not mean that there are no powers of the coupling constants appearing in the coefficients of each operator but that those coefficients, which are defined by the one-point irreducible correlation functions, are restricted to the one-point irreducible diagrams with the lowest power of those constants. It can be shown that in this case the effective action reduces to the classical one, the propagators become the free propagators and the couplings are the coupling constants.
\begin{equation}
 \Geff^{(0)}=\L
\end{equation}

\subsection{Physical definition of a particle}
To define the notion of a physical field, we have to write its propagator $D$, related to the quadratic effective action $\Geffq$ in the following way :
\begin{equation}
D(k^2)=\left(\Geff_{|2}(k^2)\right)^{-1}
\end{equation}
Note that both $D$ and $\Geffq$ are matrices acting on the vector containing all fields. The first requirement is then that the propagator from a particle $X$ to a particle $Y$ vanishes on its mass-shell, called the non-mixing condition :
\begin{equation}
\forall\ X,Y\in\Phi,\qquad D_{XY}(m_X^2)=0.
\end{equation}
The second condition for a physical propagator is the requirement of a simple pole structure at $k^2=m_X^2$, this is the pole condition :
\begin{equation}
\forall\ X\in\Phi,\qquad D_{XX} \overset{m_X^2}{\sim}\frac{1}{k^2-m_X^2}.
\end{equation}
This closes our definition of a physical particle, but we still have to find a relation between initial parameters and a set of physical observables (decays, scattering amplitude, etc...).\\

The relations between initial and physical quantities will be entirely parametrised by a matrix $Z$ and a function $f$ satisfying
\begin{equation}
 \Phi_I=Z\Phi_R\qquad\&\qquad P_I=f(P_R)
\end{equation}
where $\Phi$ and $P$ denote respectively the vector containing all fields and the set of parameters, and ${}_I$ and ${}_R$ subscript denote initial and physical (renormalised) quantities. In order to compute the mixing matrix $Z$, one has to work out the propagators, which can be done as follow.

\paragraph{Scalars}~\\ 
Considering scalars first, and writing the most generic Lagrangian we have :
\begin{equation}
 \L_{|2}=\Phi_I^\dag\left(k^2+B\right)\Phi_I\qquad\rightarrow\qquad D=(k^2+B)^{-1}.
\end{equation}
The physical pole condition reads
$$D_{ii}\overset{m_i^2}{\sim}\frac{1}{k^2-m_i^2}.$$
By inverting the relation, one obtains\\
$$D^{-1}\sim k^2-M^2$$
where $M^2$ is the diagonal matrix of all squared masses and the equivalence hold on different values, $m_i^2$, throughout the matrix. Since $B$ is hermitian this condition is quickly obtained by going to the base where it is real and diagonal :
\begin{equation}
 B=P^\dag MP.
 \label{eq:diag}
\end{equation}
So we end up with the masses $M$ and mixing matrix $Z$ defined as
\begin{equation}
 \text{mass : }M,\ \text{mixing : }Z=P^{-1}\qquad\text{ where $P,M$ diagonalises $B$}.
\end{equation}

\paragraph{Fermions}~\\ 
The fermion case is a bit more involved since massive fermions can flip chirality when propagating, because of the mass term. The generic Lagrangian of a set of left-handed Weyl fermions $\psi$ is :
\begin{equation}
\L_{|2}=\psi^\dag\kslash\psi+(\psi^TB\psi\hc).
\end{equation}
As compared to the previous case the mass matrix $B$ is symmetric and not hermitian. Furthermore, because of the $\kslash$ factor, it is not straightforward to obtain the propagator. At this point, it would be tempting to use a Takagi diagonalisation\footnote{This diagonalisation is simply the usual diagonalisation but applied to a symmetric complex matrix instead of a hermitian complex matrix.} on $B$ (since it is symmetric) but this would mix fermions with possible different quantum numbers, since $B$ relates fields with opposite quantum numbers. A workaround is to decompose the $\psi$ vector into three vectors $\psi^+$, $\psi^-$ and $\psi^0$ according to the sign of the electromagnetic charge of the fermions. In the $\psi^+,\psi^-,\psi^0$ basis the matrix $B$ reads
\begin{equation*}
B=\begin{pmatrix}
&B_+&\\
B_+^T&&\\
&&B_0
\end{pmatrix}
\end{equation*}
where $B_+$ is an ordinary complex matrix (possibly not square), and $B_0$ a square symmetric matrix. The next step is to perform a singular value decomposition of $B_+$ and a Takagi diagonalisation of $B_0$. Thus we introduce three unitary matrices $P_+,P_-,P_0$ acting on $\psi^+$, $\psi^-$ and $\psi^0$ separately. 
\begin{equation}
 B_+=P_-^TXP_+,\qquad B_0=P_0^TYP_0,\qquad X,Y\text{ diagonal}
 \label{eq:SVD}
\end{equation}
The Lagrangian reads in the new basis
\begin{equation*}
\L_{|2}=\sum_{\psi^0} \binom{\psi^0}{\bar\psi^0}^\dag\begin{pmatrix}\kslash &-m\\-m&\kslash\end{pmatrix}\binom{\psi^0}{\bar\psi^0}+\sum_{\psi^+,\psi^-} \binom{\psi_1^+}{\bar\psi_2^-}^\dag\begin{pmatrix}\kslash &-m\\-m&\kslash\end{pmatrix}\binom{\psi_1^+}{\bar\psi_2^-}
\end{equation*}
where the first sum runs on neutral fermions (\textit{i.e.} Majorana fermions), and the second on charged fermions (that are paired in Dirac fermions). Note that $m$ can always be turned from negative to positive by rotating fields by a factor $i$. In this form the inversion can be performed and the propagator reads (either for Majorana or Dirac fermions) :
\begin{equation}
D=\frac{1}{k^2-m^2}\begin{pmatrix}\kslash &m\\m&\kslash\end{pmatrix}.
\end{equation}
By doing so, we have made explicit the no-mixing condition and the pole structure, and defined the mixing matrix by 
\begin{equation}
M=\binom{X}{Y},Z=\begin{pmatrix}P_+^{-1}&&\\&P_-^{-1}&\\&&P_0^{-1}\end{pmatrix},\qquad X,Y,P_-,P_+,P_0\text{ defined in eq.\ref{eq:SVD}}.
\end{equation}

\paragraph{Vector Bosons}~\\ 
In the case of vector bosons the mass matrix has once again Lorentz indices, and this will also affect the process. Indeed the Lagrangian will read
\begin{equation*}
\L_{|2}=A_\mu^\dag\left(\left(k^2g^{\mu\nu}-k^\mu k^\nu+\frac{1}{\xi} k^\mu k^\nu\right)-Bg^{\mu\nu}\right)A_\nu
\end{equation*}
where the relation between the $k^2g^{\mu\nu}$ and the $k^\mu k^\nu$ terms stems from gauge invariance. The second term, parametrised by $\xi$, comes from the gauge fixing (and, as such, spoils the relation in the first term). A first step is naturally to diagonalise the hermitian matrix $B$. This is done by the same matrices $P,M$ as in eq.\ref{eq:diag}, except that the gauge bosons being real, $P$ will be real. In the new basis $A'$ the Lagrangian reads
\begin{equation}
\L_{|2}=A^{\prime\dag}_\mu\left(\left(k^2g^{\mu\nu}-k^\mu k^\nu+\frac{1}{\xi} k^\mu k^\nu\right)-Mg^{\mu\nu}\right)A^\prime_\nu.
\label{eq:l2gauge}
\end{equation}
This is by no mean an accident that the $\xi$ terms break the gauge relation between $g_{\mu\nu}$ and $k_\mu k_\nu$ -- that is precisely where the gauge-fixing term is compulsory. Indeed it turns out that the element $(k^2g^{\mu\nu}-k^\mu k^\nu)$ is singular, so if it were not for the gauge fixing term, the quadratic Lagrangian would not be invertible. By inverting the equation \ref{eq:l2gauge}, we find that the propagator is, in the $A'$ basis,
\begin{equation}
D=\frac{-i}{k^2-M^2}\left(g^{\mu\nu}-\frac{k^\mu k^\nu}{k^2-\xi M^2}(1-\xi)\right).
\end{equation}
One may wonder how a physical propagator could have a dependence on an unphysical parameter, namely on $\xi$. This is due to the fact that the propagator of the gauge boson itself is not physical, until we add the propagators of its ghost and in the massive case its goldstone partner. Those additional terms comes from the gauge-breaking and the gauge-fixing.\\

The $Z$ matrix that relates the physical fields and the initial one is then
\begin{equation*}
M, Z=P^{-1}
\end{equation*}

\subsection{Parameters}
It is usually not difficult to express physical quantities in term of initial parameters, since most quantities are expressed as scattering amplitudes. One obtains then a relation
\begin{equation}
 P_R=g(P_I)
 \label{eq:pi2pr}
\end{equation}
where the function $g$ has the drawback of being possibly non-linear, complicating thus our task to obtain its inverse. In the simple case of the standard Model the inversion can be done exactly. In this case we have
$$P_I=\left(g_s,g_1,g_2,v,m_h,y_f\right)$$
where the last parameter stands for all Yukawa couplings. A set of physical observations in one to one correspondence to $P_I$ can be
$$P_R=\left(\alpha_s,\alpha_e,\mw,\mz,m_h,m_f\right)$$
where the first two are the strong and electromagnetic couplings. The inversion is straightforward since all quantities have very simple expressions at tree-level. In fact the only part where there is an interplay is the gauge breaking sector : there we can isolate
\begin{equation}
\left(\mw,\mz,e\right)=\left(g_1^2\frac{v^2}{2},(g_1^2+g_2^2)\frac{v^2}{2},\frac{g_1g_2}{\sqrt{g_1^2+g_2^2}}\right)
\end{equation}
which leads to 
\begin{equation}
(g_1,g_2,v)=\left(\frac{e}{s_w},ec_w,2M_W\frac{s_w}{e}\right)
\end{equation}
where we have used the shorthand notations
$$c_w=\frac{M_W}{M_Z},\quad s_w=\sqrt{1-c_w^2}.$$


%% file: chapter2.tex
\chapter{Precision Physics : predictions with accuracy}

\minitoc\vspace{1cm}

\section{Precision phenomenology : relating Lagrangian to observables with accuracy}
\subsection{The effective action at the next order}
Although some physical aspects are quite well reproduced by the tree-level approximation, some important features are still missing. Generically this missing part comes either from the Lagrangian $\L$ itself or from the truncation of the effective action. That is to say, either we are omitting new particles or new interactions that would stem from physics beyond the Standard Model, or we are neglecting important radiative corrections. Assuming that New Physics occurs at a scale significantly higher than our observables, both of these contributions will appear as perturbative expansions. The difference being that coefficients of the loop expansion are known and can be momentum dependent, whereas New Physics coefficients are \textit{a priori} unknown numbers. Including those corrections in the game, the effective action becomes
\begin{equation}
 \Geff=\Geff^{(0)}+\Geff_\text{loop}+\Geff_\text{eff}.
\end{equation}
It is important to dissociate the loop effective action $\Geff_\text{loop}$ which corresponds to the definition in eq. \ref{eq:gammaeff} where the full particle content is known and $\Geff_\text{eff}$ which accounts for effects of new particles on top of the spectrum considered and that we define precisely in the next paragraph. In other words $\Geff_\text{loop}$ must be thought as the set of diagrams involving at least one loop and no extra particles and $\Geff_\text{eff}$ as the set of diagrams involving at least one extra particle and any number of loops, possibly none.

\paragraph{Integrating out new particles}~\\
It can be shown that, when dealing with phenomena at a given scale $Q$\footnote{The notion of the scale of a process is not strictly defined : it is usually related to the momentum and masses of the particles entering the process. In this section it is sufficient to consider $Q$ as a bounding scale of the process, that is to say all masses and momenta of the ingoing and outgoing particles have norms lower than $Q$.}, particles with mass $M\gg Q$ can be removed from the theory : this is called integrating out the heavy spectrum. A full proof can be found in \cite{burgess_eft}, the idea being to separate a set of light fields $l$ from the heavy ones $h$ in the effective action:
$$\Gamma=\Gamma[l,h]$$
One uses then the property that, for observables with no heavy particles as external states, we can replace $h$ in the effective action by its stationary point $\bar h$ :
\begin{equation}
 \Gamma[l,h]\to\Gamma[l,\bar h(l)],\qquad\text{where}\quad \frac{\delta \Gamma}{\delta h}[\bar h(l)]=0
\end{equation}
Since $\bar h(l)$ depends now on the light fields only, we end up with a functional which depends only on the light fields
$$\Gamma[l,h]\longrightarrow\Gamma_\text{eff}[l].$$
When the physics for both light and heavy spectra is known one can compute exactly $\Gamma_\text{eff}$ from $\Gamma$. This is done by solving the equation of motion for $h$, which yields an equation that relates $\bar h$ to $l$. We will see some examples of this in the next sections. However, if the heavy spectrum is unknown, $\Gamma_\text{eff}$ cannot be computed. In this case, one can still write a canonical expansion as
\begin{equation}
\Gamma_\text{eff}=\Gamma[l,0]+\sum_{i,n}c_{n,i}\frac{1}{M^n}O^{(n)}_i[l]
\label{eq:eff_parametrisation}
\end{equation}
where $O_i$ are gauge and Lorentz invariant operators involving the light spectrum and $c_{n,i}$ are coefficients, remnants from the couplings of the light spectrum to the heavy one. We typically call them effective operators and effective coefficients. This parametrisation is extremely powerful : indeed it will cover all cases of possible interactions at the UV scale and the requirement of gauge invariance allows to reduce drastically the number of operators. This requirement applies more generally to any symmetries of the low energy physics, since if those symmetries are present at the low energy scale, then they must be valid up to the high scale. In that way, the more sophisticated (that is to say with the greater number of symmetries) the low energy theory will be, the more constrained will be the higher order operators. In practice, one writes all allowed operators and defines their effective coefficients as free parameters, then one compares predictions of the effective model with experiments to provide 
constraints on the effective coefficients.\\

There exist some cases where the decoupling of heavy particles does not happen and although we will not be directly concerned by such effects in our study of an effective MSSM they are quite interesting. The first one is the non-decoupling of the top quark in the Standard Model. But this is not so surprising considering that if the low-energy theory is defined to be the Standard Model without the top quark, then this theory has a non vanishing gauge anomaly, hence it is not self-consistent. So there has to be an effect of the top quark on low energy observables even if this mass is raised to a very high scale, which breaks down the decoupling. Another effect would be the consequence of a heavy fourth generation of fermions on the Standard Model Higgs self-energy. Indeed those contributions will not be suppressed by the mass of the heavy particles, as would be expected in an effective theory. But now the issue stands with the couplings, since if we believe the Yukawa couplings of the heavy fermions to be 
proportional to the mass, these couplings will grow with 
the scale, hence this UV completion escapes the scope of the effective field theory. One must then keep in mind that there may be a difference between a heavy extra-physics and an effective field theory.

\paragraph{Radiative corrections}~\\
The other term appearing in the effective action, $\Gamma_\text{loop}$, is nothing but the higher orders of the effective action $\Gamma^{(n\geq 1)}$ in the expansion over the couplings of the theory, as defined in eq.\ref{eq:gamma_eff}. They correspond to loop diagrams in the Feynman expansion and require a specific treatment, to be detailed in the next section.\\

Though both contributions to the effective action have different origins, they share the common feature of being perturbative expansions. The parameter of the expansion being $1/M$ for $\Gamma_\text{eff}$ and the coupling constant for $\Gamma_\text{loop}$. This feature will turn out to be particularly handy when doing the computation.

\subsection{Defining the perturbative expansion : the renormalisation schemes}
As anticipated in the last chapter, the appearance of the new terms will alter the computation of physical quantities, hence it will alter the relations between initial and physical parameters and fields. However for various reasons, such as the dependence on the momentum squared of the quantities appearing in $\Geff$ or interdependencies in the parameters (the fact that couplings will depend on the masses and mixing, which themselves will depend on the couplings), it is no possible to do so analytically. Doing it numerically leads directly to the appearance of very large numbers, not to say infinities, going in the calculation of $\Geff_\text{loop}$, a point that will be explained later. However there is a neat way of getting analytic expressions : since our calculation of the effective action is based on perturbation theory, we can also compute masses, mixing and couplings and even cross-sections as perturbation series, which eases a lot our task. Expressing this use of the perturbation series we now have
\begin{eqnarray}
Z&=&Z^{(0)}+Z^{(1)}+Z^{(2)}+\cdots\\
P_I&=&P_I^{(0)}+P_I^{(1)}+P_I^{(2)}+\cdots
\end{eqnarray}
This does cure the issue of interdependencies of parameters by linearising the calculation of $Z^{(i\geq 1)}$ and $P_I^{(i\geq 1)}$ and also the momentum dependence of quantities since the zeroth order value of $\Geff$ does not depend on the momentum. We will see in a moment how the computations are done in the perturbative regime.\\

We are then left with one issue : if the physical condition that we impose on the propagator can be expanded in a perturbative series (and by this token allows us to determine all orders of $Z$), it is not the case of an experimental measurement such as the $P_R$ set where we have only one value for an experimental result, not an order by order tower of values. Precisely, when truncating the expansion at order N, we have N initial quantities $P_I^{(i)}$ to determine with only one physical observable $P_R$. Hence we need to add other conditions to eq.\ref{eq:pi2pr} used at tree-level. Fortunately, since the decomposition of the parameter in a series is only a useful mathematical artefact and does not carry physical information, we can use mathematical conditions. Those are called the renormalisation conditions, in reference to the renormalisation procedure that we will see in the next section. There exists different kinds of renormalisation conditions, but since they do not convey physical information, they 
should all yield the same result. This is however true only if we write all orders. When truncating the series at order $N$, two different renormalisation schemes will give results that differ by a quantity of the order $N+1$, reflecting our truncation of the series. For simplicity I will show the case of a truncation at order one, which is called the one-loop order :
\begin{eqnarray}
Z&=&Z^{(0)}(1+\delta Z)\\
P_I&=&P_I^{(0)}+\delta P_I
\end{eqnarray}
I will only present two mainstream renormalisation schemes as an example. Note that there is no requirement to take the same scheme for all parameters.
\begin{itemize}
\item The $\overline{MS}$ (Modified Minimal Subtraction) scheme. One introduces a fictitious scale $Q$ called the renormalisation scale, and fix $\delta P_I$ and $\dZ$ to replace exactly the divergences appearing in the computation of $\Geff_\text{loop}$ by factor of $\ln Q^2$. This scheme will be presented in detail in section \ref{MS_scheme}.
\item The $OS$ (On-Shell) scheme. The physical conditions hold order by order. That means that the order one of the propagator vanish $D^{(1)}=0$ as well as the order one of $g(P_I)$ as defined in eq \ref{eq:pi2pr}, namely $g(P_I)^{(1)}=0$. This scheme is used in the work presented here.
\end{itemize}
I will now detail the calculations in the $OS$ scheme.

\subsection{New Physics corrections}
We will now see how to determine the relations between initial fields and parameters and physical fields and parameters when we include the $\Gamma_\text{eff}$ contribution to the effective action. We keep in mind that those results will appear throughout our study of the Higgs phenomenology beyond the MSSM.

\paragraph{Scalars}~\\
In the zeroth order physical basis the propagator reads :
\begin{equation}
 D=\left(k^2(1+\dA)-M^2+\dB\right)^{-1}
 \label{eq:scalar_prop_eff}
\end{equation}
where $\delta\Geffq=k^2\dA+\dB$ is the contribution from $\Geff_\eff$, so that we have omitted higher derivatives in the effective expansion. Because $\delta\Geffq$ is hermitian, so will be $\dA$ and $\dB$. The physical states are obtained in two steps, the first being the transformation $\sqrt{1+\dA}=1+\1\dA$, which yields
\begin{equation*}
 D=\left(k^2-(M^2-\dB')\right)^{-1}\qquad\qquad\text{with }\dB'=\1(M^2\dA+\dA M^2)+\dB
\end{equation*}
At this point we can perform a perturbative diagonalisation of $M^2-\dB'$ (since $\dB'$ is hermitian), which is explained in Appendix \ref{eff_diag}. This yields
\begin{eqnarray}
\dP_{ij}&=&\frac{\dB'_{ij}}{m_j^2-m_i^2}\label{eq:scalar_eff}\\
\dm_i&=&-\dB'_{ii}\nonumber
\end{eqnarray}
And the physical fields are then defined by
\begin{equation}
 \dZ=\1\dA-\dP,\quad\dm
 \label{eq:dZ_scalar_eff}
\end{equation}

Notice that because of the $\dA$ term, the $Z$ matrix is no more unitary. Moreover, as can be seen in eq.\ref{eq:scalar_eff}, the perturbative expansion does not hold for degenerate masses at tree-level ($m_i=m_j$). We will come back on this issue in chapter 5 when we will assess the accuracy of the effective expansion.\\

\paragraph{Fermions}~\\
If we go to the zeroth order basis, the quadratic action reads :
$$\Geffq=\begin{pmatrix}\kslash(1+\dA_L)&M+\dB\\M+\dB&\kslash(1+\dA_R)\end{pmatrix}$$
where $\dA_L$, $\dA_R$ and $\dB$ are nothing but the matrix elements of $\delta\Geffq$. The same trick as for scalars can be used, that is to say a transformation $\left(1+\1\dA_L,1+\1\dA_R\right)$ to get to a basis where
$$\Geffq=\begin{pmatrix}\kslash&M+\dB'\\M+\dB'&\kslash\end{pmatrix}\qquad\qquad\text{with }\dB'=\1(\dA_L^TM+M\dA_R)+\dB.$$
The ${}_{L/R}$ index reflects the fact that for Dirac fermions, the mass term matches a left-handed spinor to a right-handed one. Then $M+\dB'$ will be decomposed on the $\psi_-,\psi_+,\psi_0$ basis and by carrying out a perturbative singular valued decomposition and a Takagi diagonalisation (further decribed in Appendix \ref{eff_diag}), we obtain the diagonal basis, defined by :
\begin{eqnarray*}
\dP_{+\ ij}&=&\frac{m_i\dB'_{ji}+m_j\dB^{\prime*}_{ij}}{m_j^2-m_i^2},\quad\dP_{+\ ii}=-\frac{i}{2m_i}Im(\dB'_{ii})\\
\dP_{-\ ij}&=&\frac{m_i\dB'_{ij}+m_j\dB^{\prime*}_{ji}}{m_j^2-m_i^2},\quad\dP_{-\ ii}=-\frac{i}{2m_i}Im(\dB'_{ii})\\
\dP_0&=&\frac{Re(\dB'_{ij})}{m_j-m_i}-i\frac{Im(\dB'_{ij})}{m_j+m_i}\\
\dm&=&Re(\dB'_{ii}).
\end{eqnarray*}
The physical fermions are then obtained by
\begin{equation}
\dZ=\1\dA_{L/R}-\dP_{-/+/0},\quad\dm.
\end{equation}

\paragraph{Bosons}~\\
In the zeroth order basis, the quadratic action of a vector boson reads
$$\Geffq=\left(k^2-M^2+\delta\Geffq^T\right)\left(g_{\mu\mu}-\frac{k_\mu k\nu}{k^2}\right)+\left(\left(1-\frac{1}{\xi}\right)k^2+\delta\Geffq^L\right)\frac{k_\mu k\nu}{k^2}$$
where we have purposely separated the transverse part and the longitudinal part (respectively denoted by the indices ${}^T$ and ${}^L$) since the Ward identity will ensure that the higher orders of the longitudinal part vanish. We will then focus on the transverse part,
$$\Geffq^T=k^2(1+\dA)-M^2+\dB$$
where $\delta\Geffq^T=k^2\dA+\dB$. We recognise the very same expression as for scalars, so we can use straightaway the result of eq.\ref{eq:dZ_scalar_eff}

\subsection{Radiative corrections}
Let us now turn to the case where we include the $\Gamma_\looop$ contribution to the effective action. Once again the computation will depend on the spin of the particle. In the following we will consider only the real part of the quadratic action $\Geffq\to Re(\Geffq)$ since the imaginary part only accounts for the width of the particle and does not affect the physical conditions derived in the last chapter.

\paragraph{Scalars}~\\
The peculiarity of the loop action is that it depends on $k^2$ in a non-trivial way. However we can bypass the difficulty by linearising it around the zeroth-order masses
\begin{equation}
\delta\Geffq{}_{ij}=\delta\Geffq(m_{i\,0}^2)_{ij}+(k^2-\mio^2)\delta\Geffq'(\mio^2)_{ij}+\grandO{(k^2-\mio^2)^2}.
\end{equation}
We are then back to the case of the effective expansion $\Geff_\text{eff}$ of eq.\ref{eq:dZ_scalar_eff}, with
\begin{equation}
\dA=\delta\Geffq'(\mio^2),\quad\dB=\delta\Geffq(\mio^2)-M^2\delta\Geffq'(\mio^2)
\end{equation}
The result being then
\begin{eqnarray}
 \dZ_{ij}&=&\frac{\delta\Geffq(\mio^2)}{\mio^2-\mjo^2}\\
 \dZ_{ii}&=&\1\delta\Geffq'(\mio^2)\\
 \dm&=&-\delta\Geffq(\mio^2)
\end{eqnarray}

\paragraph{Fermions}~\\
The case of fermions is a bit more complicated since there will be factors of $\kslash$ together with $k^2$. A solution is to proceed in two steps, first write the action at $k^2=\mio^2$
\begin{equation}
\delta\Geffq{}_{ij}=\begin{pmatrix}\kslash\delta\Geffq{}_{L}(m_{i\,0}^2)_{ij}&\delta\Geffq{}_S(m_{i\,0}^2)_{ij}\\\delta\Geffq{}_S(m_{i\,0}^2)_{ij}&\kslash\delta\Geffq{}_{R}(m_{i\,0}^2)_{ij}\end{pmatrix}
\end{equation}
and use the result from the case of the effective expansion with
$$\dA=\delta\Geffq{}_{L/R}(m_{i\,0}^2),\quad\dB=\delta\Geffq{}_S(m_{i\,0}^2).$$
This will bring us in a basis where
\begin{equation}
\delta\Geffq=\begin{pmatrix}\kslash(1+(k^2-M^{\prime\,2})\dA'_L)&M+(k^2-M^{\prime\,2})\dB'\\(k^2-M^{\prime\,2})\dB'&\kslash(1+(k^2-M^{\prime\,2})\dA'_R)\end{pmatrix}
\end{equation}
with $\dA'_{L/R}=\delta\Geffq{}_{L/R}'(m_{i\,0}^2)$ and $\dB'=\delta\Geffq{}_S'(m_{i\,0}^2)$. The additional term will rescale the pole structure when the inversion of the quadratic action is performed. So it will be accounted for by an additional term to $\dZ_{ii}$, which will be the same for the left-handed and right-handed spinors of a Dirac pair. The final result is then :
\begin{eqnarray}
\dZ_{L\ ij}&=&\delta\Geffq_{L\ ij}+\frac{m_i\delta\Geffq_{S\ ji}+m_j\delta\Geffq_{S\ ij}}{m_j^2-m_i^2}\\
\dZ_{L\ ii}&=&\1\delta\Geffq_{L\ ii}+m_i(\delta\Geffq'_{L\ ii}+\delta\Geffq'_{R\ ii})+2\delta\Geffq'_{S\ ii}\\
\dm&=&\1m_i(\delta\Geffq_{L\ ii}+\delta\Geffq_{R\ ii})+\delta\Geffq_{S\ ii}
\end{eqnarray}

\paragraph{Spin $1$}~\\
The case of vector boson is pretty much similar to what we have encountered before : we will use the Ward identity to get rid of the vector indices, and then apply the result for scalar propagators obtained at this section. This closes our discussion of the physical fields at higher order in the loop and the effective expansion.

\subsection{One loop relations for parameters}
In the same manner that the relations between initial and physical fields are not the same at the tree-level or at higher orders, the relations between initial and physical parameters need to be modified. By writing the physical parameters $P_R$ as an expansion series of the initial ones $P_I$, one gets
\begin{equation*}
P_R=g(P_I)=g^{0}(P_I^0)+g^{0}{}'(\delta P_I)+\delta g(P_I^{0})
\end{equation*}
If we take the $OS$ scheme, the inversion is performed order by order. This is expressed by the two equations
\begin{equation*}
g^{0}{}_\circ f^{0}=id,\qquad g^{0}{}'\cdot\delta f+\delta g=0
\end{equation*}
where the dot reminds us that $g^0{}'$ is a matrix whose size is the number of parameters (it is simply the Jacobian of the transformation). The system is solved by
\begin{eqnarray}
f^{0}&=&g^{0\ -1}\\
\delta f&=&-g^{0}{}'\cdot\delta g.
\end{eqnarray}
So the only difficult operation, the inversion, needs only be performed once, and only at tree-level.\\

\section{Loop computations}
\begin{figure}[!h]
\begin{center}
\begin{tabular}{ccc}
\begin{fmfgraph*}(80,50)
\fmfleft{p1}
\fmfright{p2}
\fmf{dashes}{p1,v1}
\fmf{dashes}{v2,p2}
\fmf{dashes,left,tension=0.4}{v1,v2,v1}
\end{fmfgraph*}
&
\begin{fmfgraph*}(80,50)
\fmfleft{p1}
\fmfright{p2}
\fmf{fermion}{p1,v1}
\fmf{fermion,tension=0.4}{v1,v2}
\fmf{fermion}{v2,p2}
\fmffreeze
\fmf{dashes,left}{v1,v2}
\end{fmfgraph*}
&
\begin{fmfgraph*}(80,50)
\fmfleft{p1}
\fmfright{p2}
\fmf{photon}{p1,v1}
\fmf{photon}{v2,p2}
\fmf{fermion,left,tension=0.4}{v1,v2,v1}
\end{fmfgraph*}\\
Scalar & Fermion & Vector
\end{tabular}
\end{center}
\caption{\label{diag:selfs}{\em Self-Energy diagrams for the different particles : on the left is a scalar loop on a scalar propagator (as for instance for the Higgs self-correction), on the middle a scalar correction to a fermionic propagator (e.g. the Higgs to the bottom quark) and on the right a fermionic loop to a vector boson propagator (as the lepton correction to the photon propagator).}}
\end{figure}
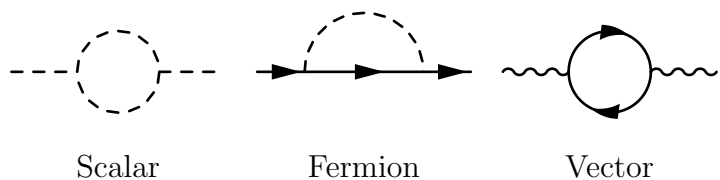

What we now need are the radiative corrections entering $\delta\Geff_\text{loop}$, that is to say the one-loop diagrams. At the quadratic level we are only interested in self-energy diagrams, shown in figure \ref{diag:selfs}. We are now entering the core of the Feynman procedure since these diagrams represent parts of the amplitude of the correlation functions, which are precisely the building blocks of the effective action.

\subsection{The loop amplitudes}
\paragraph{How to compute a loop :}~\\
The strength of the Feynman technique is that the computation of one-loop diagrams can be treated in a systematic way, that is to say independently of the number of internal propagators and the type of particles going in. Indeed the generic loop form factor can be written as 
\begin{equation}
\A=\int \frac{d^4k}{(2\pi^2)^4}\ G\times\frac{N}{D}
\end{equation}
where $G$ is the product of all couplings entering in the loop, $N$ the numerator and $D$ the product of all poles coming from propagators :
\begin{equation}
D=\prod_i(k_i^2-m_i^2)
\end{equation}
We denote here by $k_i$ and $m_i$ the momentum and mass of the particle inside the $i$th propagator. The numerator collects the additional part from fermion and vector propagator, that is to say all combinations of $\kslash_i,\ m_i,\ ,k_{\mu_i},\ g_{\mu_i\nu_i}$. It is common to use the Passarino-Veltman (see \cite{passarino_1979,consoli_1979,veltman_1980,green_1980}) basis at this point. It is the collection of elements
 \begin{equation}
 I_{m_1,k_1,...,m_n,k_n}^{\mu_1..\mu_p}=\int\frac{d^4k}{(2\pi^2)^4} \frac{k^{\mu_1}..k^{\mu_p}}{(k_1^2-m_1^2)..(k_n^2-m_n^2)}
 \end{equation}
Once the expression for elements of the basis has been worked out, one can compute any one-loop diagram just by decomposing the form factor $\A$ on the basis.
$$\A=c_0I_{m_1,k_1,...,m_n,k_n}+c_{1\mu}I_{m_1,k_1,...,m_n,k_n}^\mu+\cdots$$
Practically, we do the following : we use the Feynman rules to compute the different factors $c_0,c_{1\mu},...$ and we evaluate the Passarino-Veltman integrals $I_{m_1,k_1,...,m_n,k_n}$ with the numerical library \LT\ (\cite{feynarts}). It must be noted that this is only one way of performing a one-loop computation and many others can also be applied in the same purpose.

\paragraph{Divergences of loop amplitudes :}~\\
However a very important technical point has been left aside so far : the loop divergences. Indeed taking for instance a generic scalar amplitude
\begin{equation}
I_n(\Delta)=(-1)^{n_i}\int\frac{d^4k}{(2\pi)^4}\frac{1}{(k^2-\Delta)^n}
\end{equation}
where $\Delta$ is a function of the masses and external momenta. This integral does simply not converge when $n<3$. This puts at risk our whole procedure, because if this integral was physical -- which we naively expect since it appears in the computation of an observable -- we should be able to compute it correctly. However this does not signal a breakdown of the physical principles of particle physics, but of the decomposition of the scattering amplitude in different parts that are integrated separately. The renormalisation theory shows how a consistent decomposition can be realised.

\subsection{Renormalisation Theory}
\paragraph{Divergences are fictitious :}~\\
The breakthrough of renormalisation theory was to claim that those divergent integrals do not prevent us from accessing the physical quantities in the amplitude. To do so, the implementation of the quantum field theory has to be modified to obtain a case where loop integrals converge. This process is called the regularisation. The most used schemes are either to introduce a cut-off on the integration variables, that is
\begin{equation}
\int_0^\infty\,d^4k\longrightarrow\int_0^\Lambda\,d^4k,
\end{equation}
or to change the space time dimension. The latter, called dimensional regularisation, will be used in this thesis. The idea is that if the integration is performed in a spacetime with $4-\epsilon$ instead of 4 dimension, that is with
\begin{equation}
\int\frac{d^4k}{(2\pi)^4}\longrightarrow \int\mu^\epsilon\frac{d^{4-\epsilon}k}{(2\pi)^{4-\epsilon}}
\end{equation}
then the integrals we have seen would all converge as long as $\epsilon$ is not an integer. Note that $\epsilon$ does not have to be positive or particularly small. The parameter $\mu$ that we have introduced has the dimension of a mass, it is needed to keep the total dimension of the loop form factor. $\mu$ is called the regularisation scale since it is introduced by the regularisation procedure, and since it is another mathematical artefact, physical results should not depend on it. If we try to compute the scalar integral $I_n$ that was badly divergent in the previous paragraph within such a theory, we obtain a finite result, namely
\begin{equation}
I_n(\Delta)=\frac{\Delta^{2-\frac{\epsilon}{2}-n}}{(4\pi)^{2-\frac{\epsilon}{2}}}\frac{\Gamma(n-2+\frac{\epsilon}{2})}{\Gamma(n)},
\end{equation}
where $\Gamma$ is the gamma function
\begin{equation}
\Gamma(z)=\int_0^\infty e^{-t}t^{z-1}dt.
\end{equation}
$\Gamma$ admits poles at all negative or null integer values, and those poles are precisely the divergences of the amplitude computed above.\\

\paragraph{The full picture :}~\\
The important point of renormalisation theory is the following : if we compute an observable $\O$ for $\epsilon\neq 0$ and take the limit ($\epsilon\to 0$), then not only does this limit exists, but it is also equal to the result we wanted to compute. This theorem of renormalisation can be rephrased as 
\begin{equation}
\O=\lim_{\epsilon\to 0}\O(\epsilon).
\end{equation}
where $\O$ is the actual value of the observable and $\O(\epsilon)$ its value in the modified theory. Note that because the loop computation depend on the parameter $\mu$, the one-loop order of the parameters $\delta P_I$ (namely the counterterms) will also depend on $\mu$ so that the final result for $\O$ has no dependency on $\mu$.

\subsection{The loop action with or without divergences}
\label{MS_scheme}
In practice, the integrals of the Passarino-Veltman basis have only one kind of divergence at the one-loop order, which is often denoted $\CUV$, where UV stands for ultra-violet,
\begin{equation}
C_{UV}=\frac{1}{\epsilon}-\gamma_E+\ln 4\pi
\end{equation}
where $\gamma_E$ is the Euler-Mascheroni constant. It may be strange to include finite parts in the definition of the pole, however such finite parts do not spoil the computation since we know that in any observable the sum of the coefficients of all divergences vanishes. Those finite terms coming with any divergent integral, it is easier to put them aside in the divergent quantity $\CUV$. Since physical observable are independent of $\CUV$, this is a very stringent test of our computations : in practice we will vary $\CUV$ from 0 to $10^7$ and check that our prediction is the same in both cases. It must be noted that with each occurrence of $\CUV$ appears the term $\ln \mu^2$ which is the remnant of the dimensional regularisation. Since the observables are also independent from this scale, this can lead to another check of the result but, at least at the one-loop level, the check is identically the same as the one on $\CUV$ because those terms appear together during the loop integration.

\paragraph{$\overline{MS}$ scheme :}~\\
In the $\overline{MS}$ scheme, the parameters $\delta P_I$ are defined so that they exactly cancel the $\CUV$ part and turn the regularisation scale $\mu$ in a scale $M$ called the renormalisation scale :
\begin{equation}
 \overline{MS}\ :\ \CUV+\ln\mu^2\rightarrow\ln M^2
\end{equation}
Because of this definition, the first order of the parameters $P_I^{(0)}$ now depends on $M$, which is the main difference with the \OS\ scheme where the first order is a constant. Furthermore, since the $\ln M^2$ term now enters the zeroth order of the couplings, they may appear at higher powers in a process amplitude : for instance in a process that depends on $\alpha_S(M)^2$, there will be a squared logarithm. This feature prevents the result from being independent of $M$, since the $\dP_I$ which also contains a dependence on $\ln M$ only appear linearly. Nonetheless, this is not such a shortcoming since this dependence only reflects a higher order (two-loops and more) contribution. Hence by varying the renormalisation scale $M$ one can assess the range of the higher order corrections : this is typically used in QCD for instance.

\subsection{Infrared divergences}
Saying that it is enough to introduce a regularisation parameter, $\epsilon$, to have all loop integrals well behaved away from the critical point $\epsilon=0$ is hiding another subtlety of quantum field theory : the infrared divergences. An appropriate example is to work out a self-energy diagram where a massless boson runs in the loop. Keeping only the term associated to the divergence, we end up with a term such as
$$\A=\int\frac{d^dx}{(2\pi)^d}\frac{1}{k^4}$$
which, when decomposing the integral in the polar coordinates yields a function $F$
\begin{equation}
 F=\int_0^{+\infty}dx\,\frac{x^{d-1}}{x^4}\to\int_0^{+\infty}dx\ x^{\epsilon-1}
\end{equation}
We are led to the conclusion that the integral near the lower bound (say on $[0,1]$) converges for $\epsilon>0$ while the integral near the upper bound (so on $[1,+\infty]$) converges for $\epsilon<0$. Both diverge for $\epsilon=0$, but the trouble is that we cannot regularise them at the same time. Because the lower bound is associated to the low energy regime and the upper bound to the energy high one, we call the first an infrared-type of divergence and the second an ultraviolet-type of divergence. The correct regularisation process is then to split the integration domain into two complementary parts, and use a different regularisation factor on each, namely $\epsilon_{IR}$ and $\epsilon_{UV}$. By doing so the function computed above is correctly regularised on $\epsilon_{IR}>0,\epsilon_{UV}<0$, and can be expressed in terms of the gamma functions : its pole ends up to be
\begin{equation}
 F_{|\text{pole}}=\frac{1}{\epsilon_{UV}}+\frac{1}{\epsilon_{IR}}.
\end{equation}
Note that it is important to leave those two parameters independent since for instance the choice $\epsilon_{UV}=-\epsilon_{IR}$ would give the wrong impression that $F$ has no divergences at all.\\

\noindent The infrared divergences fall into two categories :
\begin{itemize}
 \item soft divergences : vanishing masses in the loop
 \item collinear divergences : vanishing scalar product between external and internal (with respect to the loop) momenta
\end{itemize}
Though it might not seem obvious, infrared divergences are somehow less risky for the theory. Indeed, the ultraviolet divergences cancel themselves between loop diagrams and parameters values. This is a very strong result since there is  an infinity of loop diagrams at the same loop order but only a finite number of parameters. This important theorem, which also outlines the fact that those divergences are more a mathematical caveat unavoidable in the Feynman expansion than an issue of the quantum field theory, requires that all interaction terms must be of dimension 4 or less. On the infrared side, it was proven that infrared divergences cancel themselves in between diagrams, without help from the parameters. This means in particular that we do not need to take care of the infrared when computing the physical fields and parameters at the loop level.\\

\begin{figure}[!h]
\begin{center}
\begin{fmfgraph*}(80,70)
\fmfleft{p1}
\fmfright{p2,p3}
\fmf{photon}{p1,v1}
\fmf{fermion}{p2,v2}
\fmf{fermion}{v3,p3}
\fmf{fermion,tension=0.4}{v2,v1,v3}
\fmffreeze
\fmf{photon,label=$\gamma$}{v2,v3}
\end{fmfgraph*}
\raisebox{1cm}{ + }
\begin{fmfgraph*}(80,70)
\fmfleft{p1}
\fmfright{p2,pb,p3}
\fmf{photon}{p1,v1}
\fmf{fermion}{p2,v2}
\fmf{fermion}{v3,p3}
\fmf{fermion,tension=0.4}{v2,v1,v3}
\fmffreeze
\fmf{photon,label=$\gamma$}{v2,pb}
\end{fmfgraph*}
\raisebox{1cm}{ + }
\begin{fmfgraph*}(80,70)
\fmfleft{p1}
\fmfright{p2,pb,p3}
\fmf{photon}{p1,v1}
\fmf{fermion}{p2,v2}
\fmf{fermion}{v3,p3}
\fmf{fermion,tension=0.4}{v2,v1,v3}
\fmffreeze
\fmf{photon,label=$\gamma$,label.side=left}{v3,pb}
\end{fmfgraph*}
\raisebox{1cm}{ = IR finite}
\end{center}
 \caption{\label{fig:infra} {\em Cancellation in infrared divergences : the soft divergences (due to the vanishing mass of the photon) appearing in the loop computation of the left diagram are cancelled by the soft or collinear emission of photons of the following diagrams.}}
\end{figure}
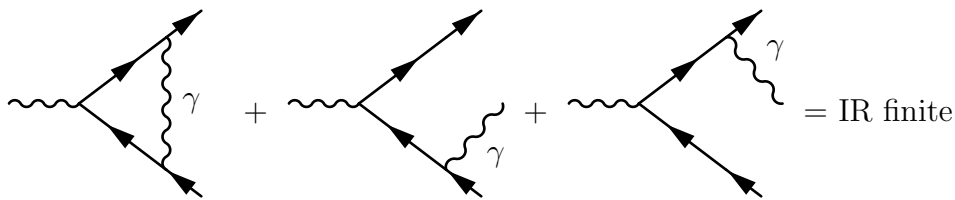

This last feature has a deep quantum mechanical explanation. Indeed for loops that are infrared divergent, one can always replace the loop by the emission of the same particle, either in a soft limit or a collinear limit, as shown in figure \ref{fig:infra}.

We know from quantum mechanics that a massless particle is undistinguishable from the vacuum in its soft limit, or that two collinear particles cannot be resolved. This is for instance the case of an electron, which cannot exist as a bare particle : it is always surrounded by photons in the soft limit. Being a lesser evil than ultraviolet divergences, infrared divergences are not to be neglected, indeed the physics in a hadron collider is plagued with them since hadrons are often charged and coloured, which results in gluon and photon emissions. Several methods have been designed to remove in an efficient way those divergences : first one has to add to the one-loop process all tree diagrams that are obtained by emitting an external photon or gluon then add all amplitude together,
\begin{equation}
\delta\A=\A_\text{loop}+\A_\text{real emission}.
\end{equation}

Depending on whether one is interested in inclusive observables (that is with any number of external photons or gluons) or in exclusive ones (where one does not allow for a photon with energy above a cut $E_\text{min}$) the additional diagrams will be evaluated fully or partially over the phase space. In any case, since the divergence lies in the low energy region, the divergences will be removed. The emission of an extra photon is called the bremstrahlung, and is in practice separated into the soft and the hard emission.
\begin{equation}
\A_\text{real emission}=\A_\text{soft}+\A_\text{hard}
\end{equation}
This separation is intrinsic to the definition of a scattering cross-section : indeed we have to separate the coherent sets of amplitudes : $\A_\text{loop}$ and $\A_\text{soft}$ are in the same coherent set, whereas $\A_\text{hard}$ is in another set since the final states are distinguishable. In fact it turns out that the soft emission does not depend on the process : it is a given divergence that only depends on the charge of the particle that is emitting the photon. The hard emission being itself process dependent, it must be integrated over the phase space, which makes it much more complicated to compute. To define the soft and the hard parts, one introduces an energy cut $k_c$. The sum of both contributions should be independent of the choice $k_c$, since we are integrating exactly the same thing on both sides, we are simply using an approximate analytic formula for the soft part and a numerical evaluation on the hard part. The quality of the approximation and numerical evaluation makes the result varying with 
$k_
c$, however it does exhibit a plateau at some point : this plateau corresponds to the numerical stability, hence the desired result. The trouble being that the location of the plateau can itself depend on the process studied.

\subsection{Effective Field Theories at the one-loop order}
\label{sec:eft_ol}
The discussion held in this chapter has purposely separated the loop action $\Geff_\looop$ from the effective expansion $\Geff_\eff$. This is however only possible if one consider only the tree-level contribution of extra particles to $\Geff_\eff$. Going to higher orders will require the possible introduction of divergences in the effective operators $c_iO_i$ in order to maintain the renormalisation procedure correct as explained in reference \cite{burgess_eft}. In other words, if one considers loop diagrams with effective vertices (that is to say vertices obtained from the $O_i$ operators), then in principle one will encounter divergences since those operators are non-renormalisable from a power counting point of view and dealing with those divergences require to know the complete theory. We will see in our study of the Higgs sector a precise example of how this question can be addressed.


%% file: chapter4.tex
\chapter{New Physics : the road to the unknown}
\minitoc\vspace{1cm}

Although the Standard Model has been able to reproduce nearly all observed phenomena to a great accuracy, many efforts have been devoted to extend it. These extensions appeal thus to the existence of New Physics. On the theoretical side, there has always been a motivation in unifying physics so it is actually not new to introduce extra particles : indeed if one tries to insert the Standard Model in a larger model, where gauge bosons emerge from one simple group and matter from one single representation, one is often led to cope with new particles and interactions. However, theory does not require those new particles to be closer to the TeV scale than to the unification scale. In other words we can build up a perfectly consistent theory which would just look like the Standard Model up to a very high scale. However, it turns out that not only do we have some evidence for new physics, but there is also realistic hope for new physics at a lower scale. This is first supported by dark matter experiments, since the 
observed relic 
density could correspond to a new particle at the electroweak scale, and also by a more theoretical point : the naturalness argument, which I will develop now.

\section{Naturalness}
At the core of the Naturalness issue lies the Hierarchy of the standard model, that is the wide separation between the electroweak scale $\sim 100$ GeV and the Planck scale at $10^{19}$ GeV. The problem arises with the Higgs particle, and in particular with its mass. The argument is the following : in the first place, quantum corrections are known to induce an energy dependence in the propagator of any particle. This is often highlighted by the introduction of the running mass $m(Q)$ : in contrast to the pole mass, which describes the pole structure of the propagator, the running mass will describe the behaviour of the propagator away from the pole, namely at the scale $Q$. In some sense it translates the fact that the effective mass of a particle will not be the same depending on the energy one is looking at. Then, because the Higgs is a scalar, its running mass will show a logarithmic dependence on the energy for each fermion present in the theory which is proportional to the squared mass of the fermion. 
To see how this happens, consider the simplified Lagrangian,
\begin{equation}
\L=|dh|^2-m_h^2|h|^2+i\bar\psi(\dslash-m_f)\psi+y\,h\bar\psi\psi
\end{equation}
where $y$ is precisely the Yukawa coupling to $\psi$. In order to compute the propagator of the Higgs field $h$ we have to regularise and renormalise the theory, which can be done in different way as we have seen in chapter 3. Using our usual tools that are dimensional regularisation and on-shell renormalisation scheme we end up with a propagator
\begin{equation}
D^{-1}(k^2)=k^2-m_h^2+\frac{3y^2}{4\pi^2}\left(m_f^2\ln\frac{k^2}{m_h^2}+\cdots\right)
\end{equation}
where we have omitted sub-leading terms. What this tells us is that the mass of the Higgs field will run with energy, as expected, and in particular its dependence on energy will show up as 
\begin{equation}
\frac{dm_h^2(Q)}{d\ln Q}=-\frac{3y^2}{4\pi^2}m_f^2.
\label{eq:fine_tuning1}
\end{equation}
Note that for a fermion field the running mass would not depend on the scalar mass, but on its own mass, namely :
\begin{equation*}
\frac{dm_f(Q)}{d\ln Q}=m_f
\end{equation*}
Such a running does not depend on the mass of any heavy particle in the loop, which means that the running mass would not be much changed with energy, and in particular stays close to the pole mass even at high energy. However turning back to the scalar field we end up with this complication : if one assumes that there is more than the Standard Model around, and that new particles will appear at higher scales, then it means that the mass of the Higgs field at the high scale is very different from the real mass. This is not in itself an issue, but what we call the naturalness problem is the fact that writing down a theory at some high scale which predicts correctly the pole mass of the Higgs boson is extremely difficult. The fine-tuning is a measure of this difficulty. Following the prescription of \cite{fine_tuning_1,fine_tuning_2}, the fine-tuning can be defined as
\begin{equation}
\Delta=\frac{d\log m_h^2}{d\log m_h^2(\Lambda_\text{NP})}.
\end{equation}
where $\Lambda_\text{NP}$ is the scale of new physics. Using eq. \ref{eq:fine_tuning1}, and assuming $m_f\sim \Lambda_{NP}$ we get
\begin{equation}
\Delta \sim \frac{\Lambda_\text{NP}^2}{m_h^2}.
\end{equation}
So that the fine-tuning is directly proportional to the hierarchy of scales. If new physics does not appear before the GUT scale, for instance, the fine-tuning will be of the order of $10^{28}$. A practical meaning of this fine-tuning is that if one expresses the theory in terms of the parameters of the unified theory (which are then representative of the high scale physics), one must have precision of 28 decimals to get the phenomenology right.\\

The naturalness, or equivalently the fine-tuning issue, has always been vigorously debated, as can be seen for instance in \cite{giudice_naturalness} and \cite{lodone_naturalness}. Indeed, one can always take the point of view that there is no new particle existing at whatever scale or that parameters of the theory at high energy have a precise value, by taking the anthropomorphic argument. Taking instead this as a hint of new particles at an observable scale, we can start extending the Standard Model, and one of its most popular extension is Supersymmetry.

\section{Supersymmetry}
Before introducing supersymmetry, let me expose briefly how one goes beyond the Standard Model. If we allow ourselves a bit of abstraction, the different parts of a quantum field theory that can be extended are not so plentiful. One can state :
\begin{itemize}
\item the matter content
\item the gauge group
\item the Poincaré group
\item Spacetime
\end{itemize}
Then one can work and improve each sector, as shown in Table \ref{tab:SM_extension}. Enhancing the matter sector means adding new states, usually fermionic but also possibly scalars. This is the case of fourth generation models (see \cite{frampton_sm4} for a review and \cite{holdom_sm4} for developments), where an entire generation of fermions is added to the Standard Model. This is also a good opportunity to introduce a right-handed neutrino which has the nice feature that it can describe the neutrino masses. Meddling with the gauge group usually calls for enlarging the structure, either to a unified group, for instance $SO(10)$ (see \cite{robinett_so10}), or with a minimalist perspective, by simply adding some extra $U(1)$ factors (see \cite{langacker_u1}). New gauge interactions will then have to be broken at a high scale, or to be hidden from Standard Model particles. The third attempt is playing with the Poincaré group. It turns out that there is a single way to enhance it : this is the super Poincaré 
group, and with it comes the whole foundation of supersymmetry. The last alternative is to change spacetime, which is usually done first by going to higher dimensions, and then by compactifying those non-observed dimensions (see \cite{cheng_extra_dim} for a review). One can also allows to warp extra dimensions, that is to allow for a non-trivial metric along the compactified dimensions, which can address the hierarchy issue (as in \cite{rs_1} for instance) or reduce the flavour hierarchy issue of the Standard Model as shown in \cite{wed}.

\begin{table}[!h]
\begin{center}
\begin{tabular}{|c|l|}
\hline
Sector & Model\\\hline
Matter & RH neutrinos, Fourth generation (\cite{frampton_sm4,holdom_sm4}), 2HDM (\cite{branco_2hdm})\\
Gauge & GUTs ($SO(10)$ \cite{robinett_so10}), Extra $U(1)$ (\cite{langacker_u1})\\
Poincaré & Supersymmetry\\
Spacetime & UED (\cite{cheng_extra_dim}), WED (\cite{wed_gherghetta}), RS (\cite{rs_1,rs_2}), Composite models (\cite{contino_composite})\\\hline
\end{tabular}
\caption{\label{tab:SM_extension} {\em Selection of some popular extensions of the Standard Model. In this table 2HDM, UED, WED, RS stands for Two Higgs Doublet Model, Unified Extra Dimensions, Warped Extra Dimensions and Randall-Sundrum models respectively.}}
\end{center}
\end{table}

\noindent\underline{Remark :} Most of the conventions and equations in this chapter stem from the review from A. Bilal (\cite{bilal}) and the compilation edited by P. West (\cite{susy}).

\subsection{Super Poincaré group}
One of the original motivation for supersymmetry was the attempt to find an extension of the Poincaré algebra. It turned out that this was a theoretical challenge, since the search was first unsuccessful. Indeed none of the usual algebras could realise this extension, to such an extent that it was finally proven by Mandula and Coleman that no Lie algebra could extend the Poincaré algebra other than in the trivial way :
\begin{equation*}
\P\to\P\oplus\G
\end{equation*}
The search then led to the discovery of superalgebras.

\subsection{Poincaré superalgebra}
The notion of superalgebra is closely related to the one of Grassman algebra. A $n$-dimensional Grassman algebra is the algebra generated by a set of $n$ anticommuting generators
\begin{equation}
\Lambda_n=Vect\{\theta_1,\cdots,\theta_n\}\qquad\forall\ i,j\ \{\theta_i,\theta_j\}=0
\end{equation}
in particular any product of $\theta$ factors containing twice the same $\theta_i$ will vanish, which ensure that the algebra has a finite dimension.\\

The minimal extension of the Poincaré algebra by a superalgebra is obtained by introducing a four dimensional Grassman algebra, where the generators are introduced as one pair, $\theta_\alpha$ transforming in the $(\1,0)$ representation of the Lorentz group, and the other, $\bar\theta_{\dot\alpha}$ i the $(0,\1)$. In order to show that the structure obtained is non-trivial, it is useful to change part of the basis as
$$P_\mu,\theta,\bar\theta\longrightarrow P,Q,\bar Q,$$
with
\begin{eqnarray}
P=(\theta\sigma^\mu\bar\theta)P_\mu\\
Q=\partial+(\sigma^\mu\bar\theta)P_\mu\\
\bar Q=\bar\partial+(\theta\sigma^\mu)P_\mu.
\end{eqnarray}
It turns out that in this case the (anti)commutation relations of the graded algebra read
\begin{eqnarray}
[P,Q]&=&[P,\bar Q]=0\\
\{Q,Q\}&=&\{\bar Q,\bar Q\}=0\\
\{Q,\bar Q\}&=&2\sigma^\mu P_\mu.
\end{eqnarray}
And the commutation relations between $Q,\bar Q$ and the Lorentz generators are also kept since $Q$ and $\bar Q$ are still in the $(\1,0)$ and $(0,\1)$ representations. This is the minimal set-up of supersymmetry, but we could extend it with other generators $Q',\bar Q'$. It is usual to denote $N$ the number of $Q,\bar Q$ pairs introduced, so that the algebra described above is the $N=1$ super Poincaré algebra.

\subsection{Super Poincaré representations}
Working out the super Poincaré representations presents some similarities with the normal case, as we will see now.

\paragraph{Massless representations}~\\
In this case the representation is labelled by an helicity number $\lambda$ but it now contains two states, one with helicity $\lambda$ and one with helicity $\lambda+\1$. Those two states are related by the $Q,\bar Q$ generators : indeed one of them raises the helicity by $\1$ while the other lowers it by the same amount. However such a representation is not really fit to describe particles realistically, since we usually have both helicity signs. So massless representations of supersymmetric theories will be a direct sum of the opposite helicities. One usually calls the lowest helicity representation the chiral representation, and the next one the vector representation. Note that the chiral representation has two states of zero helicity and two states of helicity $\1$ which turn out to give a complex scalar field and a Weyl fermion. On the other hand the vector representation has two spin 1 states and two spin $\1$ states, so we end up with a vector boson and a Weyl fermion.\\

\paragraph{Massive representations}~\\
The massive superfields are defined by a spin $s$ as in the standard case, but each of the $2s+1$ states will give rise to four different states. For instance the action of the supercharges on the $s=0$ state will give the four states of spin $(-\1,0,0,\1)$, this representation being called the massive scalar multiplet. Since it is exactly a complex scalar field and a Weyl spinor it has the same states as the massless chiral representation. The next massive multiplet is based on the two states of the $s=\1$ representation which yield  $(0,\1,\1,1)$ and $(-1,\1,\1,0)$, so we end up with a massive vector boson, a massive real scalar field and a massive Dirac fermion.

\subsection{Supersymmetry and Naturalness}
It was realised early that the contribution of a superparticle to a scalar propagator vanishes : it can be understood as an exact cancellation between the fermionic loop and the bosonic loop. No more loop corrections of this kind means that there is no behaviour of the Higgs mass proportional to any new mass scale : we can have supersymmetric particles anywhere on the spectrum and there will still be no fine-tuning. This so-called supersymmetric naturalness is perhaps one of the best motivation for supersymmetry.

\subsection{Superspace and superfields}
The question is now the following : how does one perform actual computation in superspace? To start with, we have seen that all $\theta_i^2$ term vanish, a key feature when writing Taylor-Lagrange expansions. Since these generators are in the left-handed and right-handed representations we will use the same computation rules that we used for Weyl fermions so far. In particular we can write the only non vanishing product $\theta_1\theta_2$ as $\frac{1}{2}\theta\epsilon\theta$,  so the only terms that will appear in Taylor expansions are
$$\theta,\bar\theta,\theta\theta,\bar\theta\bar\theta,\theta\theta\bar\theta\bar\theta$$
This is actually good news from the phenomenological point of view since one can express a superfield in term of fields : the generic form of a superfield $\Phi$ being
\begin{equation}
 \Phi=\phi+\theta\psi+\bar\theta\bar\psi+\theta^2f+\bar\theta^2\bar f+\theta^2 \bar\theta\bar\psi'+\bar\theta^2 \theta\psi'+\bar\theta^2\theta^2 D
\end{equation}
However such a form does not fit with the number of states of each representation previously mentioned : we thus have to apply some restrictions on the superfield whether it is a chiral one, an antichiral or a vector gauge superfield. For the vector superfield we use the Wess-Zumino gauge. Denoting $\Phi$ a chiral superfield, $\bar\Phi$ its conjugate and $e^V$ a vector superfield in this gauge we can write them as :
\begin{eqnarray*}
\Phi=&\phi+\sqrt{2}\theta\psi+i\theta\sigma^\mu\overline{\theta}\partial_\mu\phi+\theta^2 f +i\sqrt{2}\theta\sigma^\mu\oo{\theta}\;\theta\partial_\mu\psi-\frac{1}{4}\theta^2\oo{\theta}^2\square\phi\\
\overline{\Phi}=&\oo{\phi}+\sqrt{2}\oo{\theta}\oo{\psi}-i\theta\sigma^\mu\overline{\theta}\partial_\mu\oo{\phi}+\oo{\theta}^2\oo{f} -i\sqrt{2}\theta\sigma^\mu\oo{\theta}\;\oo{\theta}\partial_\mu\oo{\psi}-\frac{1}{4}\theta^2\oo{\theta}^2\square\oo{\phi}\\
e^{V}=&1+\theta\sigma^\mu\oo{\theta}2gv_\mu+i\theta^2\oo{\theta}2g\oo{\lambda}-i\oo{\theta}^2\theta2g\lambda+\theta^2\oo{\theta}^2(g\mathcal{D}+(gv)^2)
\label{eq:superfields}
\end{eqnarray*}
Note that there still seems to be more fields than needed : the $f$ field from chiral superfield and the $\D$ field from vector superfield. However they turn out not to have any dynamical dependence in the action, so they are removed from the Lagrangian by using their equations of motion and have thus no physical nature. As we will see later, using the equations of motion to modify the action is a frequent game in supersymmetry, for it can be very powerful.\\

The supersymmetric action now reads as follows :
\begin{equation}
S=\int d^4x\,d^2\theta\ \text{Tr}\mathcal{W}^a\mathcal{W}_a+\int d^4x\,d^2\theta d^2\bar\theta\ K + \int d^4x\,d^2\theta\ W + \int d^4x\,d^2\bar\theta\ \bar W
\end{equation}
where $\cal{W}$, $K$ and $W$ are functions of the superfields, $K$ being real and $W$ holomorphic, that is to say it can be derived as a complex function. $\cal{W}$ encodes the gauge interaction, it is defined by 
\begin{equation}
 \mathcal{W}=-\frac{1}{4}\bar D\bar D\left(e^{-V}De^{V}\right).
\end{equation}

$K$ is known as the K\"ahler potential and $W$ the superpotential. They can be decomposed as 
\begin{equation}
\L=\int d^2\theta\ \text{Tr}\mathcal{W}^a\mathcal{W}_a+ \int d^2\theta d^2\bar\theta\ K + \int d^2\theta\ W + \int d^2\bar\theta\ \bar W
\end{equation}
where $K$, $W$ can now be expressed as functions of fields. This transformation tells us that our supersymmetric theory, formulated on superfields living in a superspace, is equivalent to a usual quantum field theory, formulated on fields living in the usual spacetime. And this is quite an important result since most of the tools for phenomenology only deal with the usual spacetime.

\section{Supersymmetry breaking}
Since we have not been able to observe superpartners so far, it means that supersymmetry cannot be realised as a symmetry at the energy scale of the Standard Model, or in other words that it must be broken. It may seem strange to be so eager to break a symmetry right after having developed its formalism, but this is only a misconception of the role of a symmetry. A symmetry does not have to be explicitly exact at all scales to be efficient : back in the Standard Model we already had to introduce a gauge symmetry, $SU(2)\times U(1)$ to regularise the calculations and it did work even though the gauge symmetry appeared to be in a broken phase. In a similar way, the supersymmetry can still protect the Higgs mass jeopardised by the Naturalness argument, even if it is broken.\\

The breaking of supersymmetry has led to the rivalry between two schools : the exact description of the supersymmetry breaking against the effective description. We will focus in this thesis on the effective description : the idea is to break explicitly supersymmetry by including in the Lagrangian terms that include only the field with the lowest helicity for each superfield, so scalars for chiral superfields and fermions for vector superfields. We furthermore require those terms to be renormalisable, so they are split between a quadratic part (called the soft masses) and the trilinear couplings. The denomination ``soft'' of those supersymmetry breaking terms refers to the fact that they do not re-introduce the quadratic divergences (see \cite{girardello_soft}) as for the Higgs mass for instance that were cancelled in exact supersymmetry. Since we are in an effective approach we will say that each possible soft mass or trilinear coupling is a free parameter.
\begin{equation}
 \L_{SB}=\sum_{i,j}\lambda_i^\dag M_{\lambda ij}\lambda_j+\sum_{i,j}\phi_i^\dag M_{\phi ij}\phi_j+\sum_{i,j,k}A_{ijk}\phi_i\phi_j\phi_k
\end{equation}
The advantage of the method is that it keeps supersymmetry unconstrained since it does not require any further assumptions, but the inconvenience is also clear to the eye of the phenomenologist : this introduces an important number of free parameters in the model.\\

Concerning naturalness, the effect of supersymmetry breaking is the following : since superpartners have been given new mass terms, there will be a dicrepancy between the bosonic and fermionic contribution to the Higgs mass term. In particular, if we take the generic superpartner mass to be the scale of supersymmetry breaking $M_S$, the fine tuning will re-appear as the ratio of the $M_S$ scale to the electroweak scale. This means that, one has to keep $M_S$ of the order of the hundreds of GeV to keep supersymmetric fine-tuning acceptable.

\section{The Minimal Supersymmetric Standard Model}
The MSSM is the minimal supersymmetric model which includes the Standard Model, its superfield content is shown on Table \ref{tab:MSSM_matter}. Apart from an extra Higgs that has been introduced ($H_1$ and $H_2$ instead of $H$), we have simply upgraded the Standard Model field to superfields. There are two reasons for this new Higgs superfield, one is related to the generation of fermion masses through the Yukawa potential which we will see in a while and the second is related to quantum anomalies. Indeed any theory including chiral charged fermions is likely to see its gauge symmetry be broken at the quantum level, unless the fermions charges obey a specific pattern. We know in particular that the fermions of the Standard Model have charges such that the whole gauge group is preserved. The MSSM has however more fermions and in particular a single Higgs fermionic field (called the Higgsino) would spoil the hypercharge anomaly. By introducing two Higgs superfields with opposite hypercharges, the symmetry is 
then restored at the quantum level.

\begin{table}[!h]
\begin{center}
\begin{tabular}{c|l}
Superfield &Gauge Representation\\\hline
$L$ & $1_\otimes 2_\otimes -\1$\\
$E$ & $1_\otimes 1_\otimes 1$\\
$Q$ & $3_\otimes 2_\otimes \frac{1}{6}$\\
$U$ & $\bar3_\otimes 1_\otimes -\frac{2}{3}$\\
$D$ & $\bar3_\otimes 1_\otimes \frac{1}{3}$\\
$H_1$ & $1_\otimes 2_\otimes -\1$\\
$H_2$ & $1_\otimes 2_\otimes \1$\\
\end{tabular}
\caption{\label{tab:MSSM_matter} {\em Matter sector of the MSSM : as compared to the Standard Model shown in Table \ref{tab:sm_matter}, all fermionic fields have been turned into chiral superfields. Then, instead of one scalar Higss $H$ we now have two chiral superfields $H_1$ and $H_2$. Note that we still have $G=SU(3)_\otimes SU(2)_\otimes U(1)$}}
\end{center}
\end{table}

\noindent\underline{Remark :} The construction of the MSSM presented here and in particular the renormalisation of the parameters closely follows the ones in \cite{Sloops-higgspaper,baro09,boudjema_chalons1}.

\subsection{MSSM Lagrangian}
The MSSM matter potential will be decomposed into different parts : the first is the K\"ahler potential ($K$), which has a canonical form, an other is the superpotential ($W$) which encodes the Yukawa terms plus the $\mu$ term and finally there is the supersymmetry breaking potential which has a general form :

\begin{eqnarray}
K_\MSSM&=&\sum_\Phi \Phi^\dag e^V\Phi\\
\W_\text{MSSM}&=&y_LH_1LE+y_uH_2QU+y_dH_1QD+\mu H_1\cdot H_2\\
\L_{sb}&=&\sum_\text{Matter} m_{\phi}\phi^\dag\phi + \sum_\text{Gauge} m_{\tilde{A}}\tilde{A}\tilde{A} + \sum_\text{Higgs-Matter} A_{ijk} h_i\tilde{f}_j\tilde{f}_k\hc
\end{eqnarray}
In this equation $\L_{sb}$ stands for the supersymmetry-breaking lagrangian.
 
\subsection{Lagrangian in field components}
Since we are interested in the phenomenology of the MSSM, we will need the Lagrangian in terms of fields instead of a functional of the superfields. To do this, one expands all superfields as in eq. \ref{eq:superfields} and performs the integration over Grassman variables. This leads to the final result
\begin{eqnarray}
\L&=&\sum_\phi|D\phi|^2+\sum_\psi i\bar\psi\Dslash\psi+\sum_A|DA|^2\\
&&-|\partial_i W|^2-\1|\partial_n K|^2 +\left(\partial^2_{ij}W\psi_i\psi_j+i\sqrt{2}\phi_i\psi_jg\lambda\hc\right)\nonumber\\
&& +\L_{sb}.\nonumber
\end{eqnarray}
where the partial derivative $\partial_i$ means the derivative along the field $\Phi_i$ and $\partial_n$ the derivative along the gauge superfield $V_n$. Moreover it is implicit that the derivatives of $W$ and $K$ are evaluated on the scalar part of each superfields, that is to say $\phi$ for each chiral superfield $\Phi$ and $1$ for each vector superfield $e^V$, so that we have :
$$\partial_i W=\frac{\partial W}{\partial \Phi^i}(\phi,1),\quad \partial_n K=\frac{\partial K}{\partial e^{V_n}}(\phi,1),\ \cdots$$

The soft breaking terms entering $\L_{sb}$ will be soft masses for Higgses, sfermions and gauginos :
$$m_1,m_2,m_{12},\quad M_{xi}\ (x=l,r,q,u,d,\ i=1,2,3),\quad M_1,M_2,M_3$$
plus trilinear couplings for each fermion type
$$A_f,\ (f=e,\mu,\tau,u,c,t,d,s,b).$$
Let us now see what this implies for each sector of the MSSM, starting by the Higgs sector.

\subsection{Scalar Higgs sector}
It may already be clear that the Higgs sector will differ greatly from the Standard Model, indeed we have introduced another Higgs doublet but we are still breaking the same number of gauge generators, namely three, which implies that instead of the one physical particle that was left in the Standard Model, we now have 5 particles. The Higgs potential reads
\begin{eqnarray}
\L&=&\tilde{m}_1^2|h_1|^2+\tilde{m}_2^2|h_2|^2+\tilde{m}_{12}^2(h_1\cdot h_2\hc)\\
&&+\frac{g_1^2+g_2^2}{8}(|h_1|^2-|h_2|^2)^2+\frac{g_1^2}{2}|h_1^\dag h_2|^2\nonumber
\end{eqnarray}
where we have defined
\begin{equation}
\tilde{m}_1^2=m_1^2+\mu^2\qquad\tilde{m}_2^2=m_2^2+\mu^2\qquad\tilde{m}_{12}^2=m_{12}.
\end{equation}
We impose the following \vev\ :
\begin{equation}
\left<h_1\right>=\binom{0}{v_1},\ \left<h_2\right>=\binom{v_2}{0}
\end{equation}
It is not however certain than this potential generates a correct minimum. This imposes first the condition
\begin{equation}
\tilde{m}_1^2+\tilde{m}_2^2>2|\tilde{m}_{12}^2|.
\end{equation}
In order to have a global minimum away from the origin, we must ensure that the derivative of $V$ nearby the origin is negative. This condition yields
\begin{equation}
\tilde{m}_{12}^2>\tilde{m}_1^2\tilde{m}_2^2.
\end{equation}
Assuming that those two conditions are realised, the minimum is then specified by expressing that we have a critical point, that is to say that the derivative of $V$ vanishes. At this point it is useful to notice that we can rewrite the potential as 
\begin{eqnarray}
\L&=&\tilde{m}_1^2|h_1|^2+\tilde{m}_2^2|\hp_2|^2+\tilde{m}_{12}^2(h_1^\dag\hp_2+\hc)\\
&&+\frac{g_1^2+g_2^2}{8}(|h_1|^2-|\hp_2|^2)^2+\frac{g_1^2}{2}|h_1\cdot\hp_2|^2\nonumber
\end{eqnarray}
where $\hp_2=\epsilon\bar h_2$. This new field has the interesting feature of being in the same representation as $h_1$. In particular we have
\begin{equation}
 \left<h_1\right>=\binom{0}{v_1},\qquad \left<\hp_2\right>=\binom{0}{v_2}
\end{equation}
which means that if we define an angle $\beta$\footnotemark by
\begin{equation}
 \tan\beta=\frac{v_2}{v_1}
\end{equation}
\footnotetext{In all the thesis we will use the short-hand notation $\tb=\tan\beta,\ s_\beta=\sin\beta\ \text{and }c_\beta=\cos\beta$.}
then we can rotate the basis $(h_1,\hp_2)\xrightarrow{\beta}(h_a,h_b)$, where we now have
\begin{equation}
 \left<h_a\right>=\binom{v}{0},\qquad \left<h_b\right>=\binom{0}{0}
\end{equation}
and 
\begin{equation}
 v^2=v_1^2+v_2^2.
\end{equation}
That is, we have now two scalar fields, one is the Higgs of the Standard Model : $h_a$, which will hence yield three Goldstone bosons and one physical, neutral and CP even, Higgs ($h_a^0$); and another scalar field $h_b$, which has a vanishing vacuum expectation value, hence will show two neutral Higgs $h_b^0$ (CP-even) and $h_b^I$ (CP-odd) and a charged boson $h_b^+$. Since the Goldstone bosons do not mix with physical states, we know that the two latter are mass eigenstates, they are usually called $A^0$ and $H^+$. Note that since the breaking of the $SU(2)\times U(1)$ is now exactly SM-like, the relation between $v,g_1,g_2$ and $\mz,\mw$ are the same as in the Standard Model. The masses of these scalar Higgses are obtained by using the criticality conditions. Concerning $A^0$, we have
$$\ma=\frac{\partial V}{\partial |h_b|^2}$$
This does a priori depend on $\tilde{m}_1,\tilde{m}_2,\tilde{m}_{12}$ however since those are free parameters of the model, we will choose to trade one of them with $\ma$, and take $\ma$ as a parameter.\\

\noindent Looking at $H^+$, we have
\begin{eqnarray}
M_{H^+}&=&\frac{\partial V}{\partial |h_a\cdot h_b|^2}v^2+\frac{\partial V}{\partial |h_b|^2}\nonumber\\
&=&\frac{g_1^2v^2}{2}(\cb^2+\sb^2)+\ma^2\nonumber\\
M_{H^+}&=&\mw^2+\ma^2
\end{eqnarray}
And for the CP-even part, we end up with the mass matrix
\begin{equation}
M^{0\ 2}=\mz^2\begin{pmatrix}\cdb^2&-\sdb\cdb\\-\sdb\cdb&\sdb^2\end{pmatrix}+\ma^2\begin{pmatrix}0&0\\0&1\end{pmatrix}
\label{eq:even_mass_mat}
\end{equation}
which can be diagonalised to the basis $H,h$ by a further rotation, which is usually parametrised by the angle $\alpha$ defined by
\begin{equation}
 \binom{H}{h}=\begin{pmatrix}\ca&\sa\\-\sa&\ca\end{pmatrix}\binom{h_1^0}{h_2^0}
\end{equation}
where $h,H$ are respectively the light and the heavy mass eigenstates. Notice that the transformation to go from $(h_a,h_b)$ to $(H,h)$ is thus the rotation by the angle $\alpha-\beta$.\\

\paragraph{Decoupling limit}~\\
One notices that in the limit $\ma\gg\mz$ the mass matrix in eq. \ref{eq:even_mass_mat} is equivalent to
$$\begin{pmatrix}0&0\\0&\ma^2\end{pmatrix}$$
which is already diagonal. That is why in this limit, called the decoupling limit, one obtains the following approximation :
\begin{equation}
 \alpha=\frac{\pi}{2}-\beta\qquad\mh\sim\mz,\ \mH\sim\ma.
\end{equation}

\subsection{Standard Model sector}
\paragraph{Fermions :}~\\
What changes with respect to the Standard Model is the form of the Yukawa terms, since the $h_a$ scalar Higgs of the Standard Model has now been rotated by the $\beta$ angle, as described in the previous section. Fermion coming from superfields coupled to $H_1$ will exhibit a $\cb$ factor, and those coupling to $H_2$ a $\sb$ one. This is summarised by
\begin{equation}
 y_l,y_d\to\cb(y_l,y_d)\qquad y_u\to\ y_u\sb.
\end{equation}
However since the fermions masses are the same as in the Standard Model, it means that the tree level Yukawa couplings are changed as :
\begin{eqnarray*}
 y_{l,d}^\MSSM=y_{l,d}^\SM/\cb\\
 y_u^\MSSM=y_u^\SM/\sb.
 \end{eqnarray*}
It is now customary to take $\tb>1$, in order to reach a lightest Higgs mass compatible with the experiments. In this case one see that when increasing $\tb$ the Yukawa coupling to leptons and down-type quarks increases nearly linearly with $\tb$ whereas the coupling to up-type quarks is equivalent to the standard model one. This is an example of the ``$\tb$-enhanced effects'' which are common to the MSSM.

\paragraph{Gauge bosons :}~\\
The gauge bosons being broken by the same mechanism as in the Standard Model, we end up with the same physical basis. Once again the only change is with the couplings of the massive bosons to the Higgs field. Since this coupling stems from the $(\ggg\cdot \AA \vv)(\ggg\cdot \AA h)$ term, it implies that in our basis $(h_a,h_b)$ only $h_a$ couples to massive gauge bosons since $\left<h_b\right>=0$. In particular in the decoupling limit we have
\begin{equation}
 g_{h(WW,ZZ)}^\MSSM\sim g_{h(WW,ZZ)}^\SM\qquad\&\qquad g_{H(WW,ZZ)}^\MSSM\to 0
\end{equation}
In the general case the coupling to massive gauge boson will be rotated by the second rotation, parametrised by $\alpha-\beta$, which means that
\begin{equation}
 \binom{g_{H(WW,ZZ)}^\MSSM}{g_{h(WW/ZZ)}^\MSSM}=R(\alpha-\beta)\binom{g_{h(WW,ZZ)}^\SM}{0}.
\end{equation}

As in the Standard Model, we will use the gauge-fixing Lagrangian to remove mixing between gauge and Goldstone bosons. Moreover, since we have some freedom in the choice of the gauge fixing function, we will add new, non-linear terms that will help in checking the gauge invariance of our results.
\begin{eqnarray}
F^{+}&=&(\partial_{\mu}-ie\tilde{\alpha}\gamma_{\mu}-ie\frac{c_{W}}{s_{W}}\tilde{\beta}Z_{\mu})W^{\mu +}
+i\xi_{W}\frac{e}{2s_{W}}(v+\tilde{\delta}h^{0}+\tilde{\omega}H^{0}+i \tilde{\rho}A^{0}+i\tilde{\kappa}G^{0})G^{+} \, ,\nonumber\\
F^{Z}&=&\partial_{\mu}Z^{\mu}
+\xi_{Z}\frac{e}{s_{2W}}(v+\tilde{\epsilon}h^{0}+\tilde{\gamma}H^{0})G^{0} \, ,\nonumber \\
F^{A}&=&\partial_{\mu}A^{\mu} \, .
\end{eqnarray}
where $\tilde{\alpha},\cdots,\tilde{\kappa}$ are the new gauge fixing parameters, that should have no influence on the results. This kind of check is extremely useful when one is trying to separate a process amplitude into different physical parts.

\subsection{Scalar superparticles}
Neglecting the effect of generation mixing, the sfermion sector leads to a generic mass matrix of the kind
\begin{equation}
 M_{\tilde{f}}=\begin{pmatrix}\M_{LL}&\M_{LR}\\\M_{LR}&\M_{RR}\end{pmatrix}.
\end{equation}
The expression for $\M_{LL},\M_{LR},\M_{RR}$ is the following
\begin{eqnarray}
 \M_{LL}&=&M_{\tilde{f_L}}+m_f^2+\mz^2\cdb(T^3_f-Qf\sw^2)\\
 \M_{LR}&=&m_f(A_f-\mu(1/\tb,\tb))\\
 \M_{LR}&=&M_{\tilde{f_R}}+m_f^2+\mz^2\cdb Qf\sw^2
\end{eqnarray}
where the $(1/\tb,\tb)$ dependence reflects the difference between fermions coupling to $H_1$ ($\tb$-enhanced) and those coupling to $H_2$ ($\tb$-suppressed).\\

We will hence define rotation matrices to go from initial fields $\tilde{f}_L,\tilde{f}_R$ to physical fields, denoted by $\tilde{f}_1,\tilde{f}_2$, as has been explained in chapter 2 and 3.
\begin{equation}
 \binom{\tilde{f}_L}{\tilde{f}_R}=Z_{\tilde{f}}\binom{\tilde{f}_1}{\tilde{f}_2}
\end{equation}
In the simplest case of MSSM parameters, the mass matrix is real so the rotation is fully parametrised by one angle, denoted as $\theta_f$
\begin{equation}
 Z_{\tilde{f}}=R(\theta_f)
\end{equation}

One can suspect that, since the mixing angle $\theta_{\tilde{f}}$ is mostly driven by the off diagonal element which is proportional to the fermion mass, the superpartners of light fermions will show practically no mixing, and indeed, we will see that the most important $\theta$ angles correspond to the third generation : $\theta_{t},\theta_{b},\theta_{\tau}$.

\subsection{Fermionic superparticles}
The new fermions that are introduced are either the Higgsinos ($\tilde{h}_i$) that are components of the $H_i$ superfields or the gaugino $\tilde{A}_a$, components of the gauge vector superfields. As described in the procedure in chapter 2, we first have to separate those Weyl fermions in three categories $\psi^-,\psi^+,\psi^0$ :
\begin{equation}
 \psi^-\,:\,\tilde{W}^-,\tilde{h}_1^-,\qquad\psi^+\,:\,\tilde{W}^+,\tilde{h}_2^+,\qquad\psi^0\,:\,\tilde{B},\tilde{W}^3,\tilde{h}_1^0,\tilde{h}_2^0
\end{equation}
Then when looking at the mass terms, we see that they come either from
\begin{itemize}
 \item soft gaugino masses $\L\to M_a\tilde{A}^a\tilde{A}^a\hc$
 \item $\mu$ term $\L\to \mu\tilde{h}_1^-\tilde{h}_2^+\hc$
 \item gauge interaction $\L\to \sqrt{2}\sum_k h_k^\dag\, (i\ggg\cdot \tilde{\AA})\,\tilde{h}_k\ \hc$
\end{itemize}
where $g,\,\tilde{A}$ are the vectors of all coupling constants and gauge fermions fields.\\

We note that the first two mass terms are real while the last is imaginary. However, this can be cured by rotating all gauginos by a factor of $i$, so we will take the basis
\begin{equation}
\psi^-=(-i\tilde{W}^-,\tilde{h}_1^-),\qquad\psi^+=(-i\tilde{W}^+,\tilde{h}_2^+),\qquad\psi^0=(-i\tilde{B},-i\tilde{W}^3,\tilde{h}_1^0,\tilde{h}_2^0).
\end{equation}
As previously described this basis will be rotated to the mass eigenstates basis, that we write
\begin{equation}
 \tilde{\chi}^-_{i\,R},\tilde{\chi}^+_{i\,L},\tilde{\chi}^0_{j}\qquad(i=1..2,\ j=1..4)
\end{equation}
where the four charged Weyl fermions are turned into two Dirac fermions
\begin{equation}
 \tilde{\chi}^+_i=\binom{\tilde{\chi}^+_{i\,L}}{\overline{\tilde{\chi}}^-_{i\,R}}.
\end{equation}

We will call neutral fermions neutralinos and charged fermions charginos. Note that since we have not paired the neutralinos together (since they are neutral they can be massive without the need for forming a Dirac pair), they will stay as Weyl (or Majorana) fermions. Because of the smallness of gauge interactions (only electroweak sector is involved), masses and mixing of those particles will be mostly driven by the subset of parameters
\begin{equation*}
 M_1,M_2,\mu.
\end{equation*}

\subsection{Fixing the MSSM initial parameters}
We have seen how to fix the initial parameters in the case of the Standard Model. In the case of the MSSM the objective is to define most of the new parameters from masses.\\

\paragraph{Scalar superpartners $M_l,M_r,A_f$}~\\
For each lepton flavour, we generate 3 masses and one mixing angle from three parameters : we can either extract all three from the masses, or from two of the masses and the mixing angle.
\begin{equation}
M_{l\,i},M_{r\,i},A_l\longleftrightarrow \begin{matrix}m_{\tilde{l}_1},m_{\tilde{l}_2},m_{\tilde{\nu}_l}\\m_{\tilde{l}_1},m_{\tilde{l}_2},\theta_l\end{matrix}
\end{equation}
where $l=e,\mu,\tau$, $i=1,2,3$. The mixing angle is not an observable strictly speaking, but can be related to production cross sections and decay observables through interactions with weak gauge bosons for instance. The second choice may be preferable if one of the three masses is somewhat higher than the others.\\

For each quark flavour pair we generate 4 masses and two mixing angles from five parameters, so it can once again lead to different choices
\begin{equation}
M_u,M_d,M_q,A_u,A_d\leftrightarrow m_{\tilde{u}_1},m_{\tilde{u}_2},m_{\tilde{d}_1},m_{\tilde{d}_2},\theta_u,\theta_d.
\end{equation}

\paragraph{Fermionic superpartners $M_{\tilde{A}},\mu$}~\\
Leaving the gluino apart (one mass for one parameter), we generate 6 masses (4 neutralinos and 2 charginos) from three parameters. The most used schemes are two neutralinos and one chargino, or two charginos and one neutralino.
\begin{equation}
M_1,M_2,\mu\leftrightarrow m_{\tilde{\chi}^0_1},m_{\tilde{\chi}^0_2},m_{\tilde{\chi}^+_1},m_{\tilde{\chi}^+_2},\cdots
\end{equation}
We will see in particular in our dark matter study in chapter 10, how to compare the schemes where the two chargino masses and one neutralino mass are used, when switching from the lightest neutralino to another neutralino.\\

\paragraph{Higgs sector}~\\
We have seen that the Higgs sector was usually parametrised by $\ma,\tb$. The extraction of $\ma$ is pretty obvious (at least if we can produce it at colliders), but unfortunately is not straightforward at all for $\tb$. If one sticks to the $OS$ scheme one can try observables in the Higgs sector such as
\begin{itemize}
\item $\mH$
\item $A_0\to\bar\tau\tau$ decay.
\end{itemize}
However other schemes have also been used throughout the literature : in a non-$OS$ scheme, one can accommodate a definition for $\tb$ which is not directly related to an observable. For instance the Dabelstein-Chankowski-Pokorski-Rosiek scheme (DCPR) is based on the $A^0-Z$ transition that is to say $\delta\tb$ is set so that the transition $A^0-Z$ vanishes at $k^2=\ma^2$. Note that it differs from the no mixing condition which is that the transition vanishes at $k^2=\mz^2$. This definition turns out not to be gauge invariant, which will lead us to prefer schemes based on physical input as the first two.

\subsubsection*{Fixing parameters : an optional step?}
On a pessimistic perspective it seems that we have not done much : actually we have traded unknown initial parameters for unknown physical quantities since the experimental observable are still missing. This implies that for some sectors of supersymmetry one usually takes as input $P_I^{(0)}$ (that is to say the zeroth order of the initial parameter) instead of $P_R$ (which is itself the physical input used to determine $P_I$). Note that in any cases $\delta P_I$ can never be an input, since it is fixed by the renormalisation scheme. For instance one usually takes as input parameters
\begin{equation}
 P_I^{(0)}=\left(M_l,M_r,M_d,M_u,M_q,A_f\right)
\end{equation}
and then construct the masses and mixing angles.\\

\section{Supersymmetry and unexplained phenomena}
Although Supersymmetry was created on very mathematical purposes (first transgressing the ``No-Go'' theorem, then curing the unification of constants), and was also put forward by string theorists to obtain consistent theories, it would have gone completely lost in the phenomenology community if it had not been a plausible explanation to different phenomena. Among them stands the dark matter problem : if dark matter really suggests a stable new fermionic state of the order of the 100 GeV, only coupled to standard model by weak interactions, then the lightest neutralino is very likely to fit the bill. Its stable characteristic is ensured by R-parity, which was first imposed on supersymmetry to forbid operators leading to fast proton decay.\\

However another feature of supersymmetry is that it may lead to observable effects in a collider running above the TeV scale. Indeed supersymmetry often predicts a light Higgs that present some similarities with the Standard Model one. So, by getting a very high sensitivity towards the Standard Model Higgs, heavy constraints can be put on supersymmetry. On the other side, if supersymmetry has anything to do with the problems with which the Standard Model is struggling, we had better have some superpartners at the TeV scale. A totally decoupled supersymmetry (except for the light Higgs) would not be of a great phenomenological interest. A more annoying scenario would be phenomenological interesting supersymmetry with light enough superpartners but with a pathological spectrum : for instance superpartners too close in mass to be clearly seen at the LHC. Although such a scenario would be rather difficult to obtain from the theoretical point of view it lies nonetheless also in an experimental challenging region.


%% file: chapter3.tex
\chapter{Effortless Physics, or the phenomenologist's toolbox}

\minitoc\vspace{1cm}

We have recalled the general principles of calculations in a quantum field theory and the specificities of supersymmetric theories, however the hardest part is still to come. Indeed in order to get predictions for many different phenomena ranging from collider to astrophysical experiments, one has to deal with complicated and lengthy computations, and with plenty of them. Fortunately the modern phenomenologist can rely on tools to ease this task.

\section{About automation}
It is hard to deny that the most impressive results in particle physics phenomenology are approximate ones, in the sense that the associated calculations use approximate formulas instead of exact ones. But this is not an issue as such, since by their very experimental nature, observables come with an uncertainty. On the theoretical side, this comes first with the truncation of the Feynman expansion. Interestingly, the validity of the expansion itself is very uncertain since in most of the cases the series is actually divergent. However even then it seems that, by truncating the series at a specific order, one obtains a very good approximation of the exact result. This was proven in very simplified (and unfortunately, unphysical) limits, such as in zero-dimensional spaces (see \cite{greynat_0907} for instance) where the exact amplitude can be computed, and compared to the truncated series. In those cases, it was shown that there exists an order $n$ that optimizes the approximation obtained with the series, and some 
theorems were developed to evaluate the accuracy of the approximation by the series. Those 
questions are much less explored in the Standard Model framework, where not only these theorems are not existent, but the number of orders one can compute may seem ridiculous : less than three for most of the processes. But, although we may not understand exactly how the series behave, the very same expansion has proven to be extremely accurate in QED measurements with an outstanding precision of one part per $10^8$, which strongly supports this method.\\

The main difficulty arising in the computation of those Feynman expansions is their exponential complexity. Indeed it relies on the successive evaluation of all Feynman diagrams contributing to the scattering process, and this very number grows exponentially with the order of the truncation. Computations slightly involved at order $N$ will often be fully intractable at order $N+1$. This is for the dark side of the situation. On the sunny side, the Feynman procedure is general and systematic : it is in particular well suited for an automated treatment. Along this chapter we are going to see precisely how one performs this task.

\section{Automating the calculation of the effective action : a step by step approach}
The Feynman procedure is an iterative approach, relying on a succession of steps to obtain the final result. A good sketch of the situation would be the following path
\begin{enumerate}
\item Defining the model : the Lagrangian $\L$, the gauge group $G$ and the fields $\Phi$ appearing in $\L$.
\item Finding the physical particles (i.e. the mass eigenstates) and expressing initial parameters from a set of physical observables.
\item Deriving the Feynman rules. This is the list of rules to handle \begin{itemize}
	\item external particle wavefunctions
	\item coupling tensors responsible for interactions between the physical particles.
	\item propagators of the physical particles
\end{itemize}
\item Generating the Feynman diagrams according to the process considered and the Feynman rules previously obtained.
\item Computing amplitudes of each diagram, and adding them all. This requires integration tools, either for integral over internal momenta in loops, or on the kinematic variables of the final states, in case one is interested in inclusive cross-sections.
\end{enumerate} 

It must be noticed that the first three steps (1-3) are independent of the process considered and can hence be realised once and for all for a given model. Then, for each process that motivates a prediction, one only has to carry out the last two steps (4-5). This was used very early in the community of numerical techniques for particle physics : the first part can be labelled as \textbf{Derivation of Feynman rules} and the second \textbf{Computation of process amplitude}. Let us now see what kind of automated codes have been developed to tackle the two parts.

\section{Codes}
The general idea of automation is that for different parts of the computation of a prediction in a quantum field theory there exists one code which aims at providing a routine to do this part : hence the full process is basically to decompose the overall computation in different blocks, find a specific code for each block, and eventually link together all codes. In fact the possible new computation part of a new prediction is actually a very little part of the whole computation, the difficulty being precisely to extract this tiny part. This also means that there is no code that aims at giving all predictions for all models : when one wishes to compute some quantity in a given model, one has to decide which codes should be used (and in some cases improve such codes) and when turning to other observables or other models, then other codes may be needed. It is important to distinguish a code, which is a routine to perform a specific part of a calculation in a quantum field theory, from a program which is a 
bundle of different codes assembled together 
to produce a prediction. Ideally, codes ought to be well-defined\footnote{In particular the standardisation of input and output methods is a constant progress axis.} and publicly available while programs should be made on a case by case basis by the user. Following this idea, I will first present the codes that were used throughout our work and not try a comparison between codes, and then I will move to the description of a program (\SL) that was used. Let me start with a classification of some of those codes\footnote{No attempt at exhaustion is intended, the list only reflects what I have been confronted with.} :
\begin{itemize}
\item Derivation of Feynman Rules\\
\begin{tabular}{lp{10cm}}
 \lanHEP & obtains Feynman rules for a supersymmetric theory, at the one-loop order (\cite{lanhep1,lanhep2})\\
 \texttt{FeynRules} & obtains Feynman rules in supersymmetric theories (\cite{feynrules})\\
 \texttt{SARAH} & obtains Feynamn rules, spectrum and Running group evolutions in supersymmetric theories (\cite{sarah})\\
\end{tabular}
\item Computation of a process amplitude\\
\begin{tabular}{lp{10cm}}
\FA/\FC & Computes the analytical and numerical evaluation of a process at one-loop order (\cite{feynarts})\\
\HD & Computes masses and decays of Higgses in the MSSM (\cite{hdec})\\
\LT & Library of numerical evaluation of one-loop integrals (\cite{feynarts})\\
\SP & Computes loop-corrected spectrum of the MSSM (\cite{suspect})\\
\MG & Computes numerical evaluation of a process at tree-level in some models (\cite{madgraph})\\
\CH & Computes analytical and numerical tree-level cross-sections (\cite{calchep})
\end{tabular}

\end{itemize}

This must be understood as a mere selection picking elements from a wide list : depending on the type of observables (astrophysics, leptonic or hadronic collider physics, low energy physics, and so on), on possible specifications dictated by an experiment (compute an inclusive cross-section, generate events weighted by an exclusive cross-section or apply a reconstruction through the detector) and other requirements, some tools are more or less adapted. The point being that it is up to the expertise of the user to choose the right codes, hence are shown only the ones that have been the most convenient for our purposes, a choice to be soon detailed.

\subsection{Derivation of the Feynman rules}
Deriving the Feynman rules of a model can quickly become a difficult task. In supersymmetry for instance one has to deal with a Lagrangian expressed in term of potentials in the superspace, which is not strictly speaking ``Feynman rules compatible''. The other issue arises at step 2, that is relating physical particles and parameters to initial ones. When going beyond the tree-level, we have to know the Feynman rules of the specific model, since we are to compute loops. Those Feynman rules only come at the next step, for the very good reason that one needs to know the physical fields and the initial parameters to write those rules. However this is easily tackled by using the perturbative expansion, but it means that some part of the definition of masses and mixing matrices will have to be determined after the Feynman rules have been worked out.\\

Those difficulties left aside, the derivation of the Feynman rules can be automated in a fairly generic way. For each of the three codes that are available to derive the Feynman rules (\lanHEP, \texttt{FeynRules} and \texttt{SARAH}) the idea is to enter as input the Lagrangian in its simplest form, that is its most theoretical-like form, and to get in output the set of Feynman rules in a format that can be used in post-processing tools. A pleasant feature is an input very textbook-compliant : one can use directly notions such as covariant derivatives and superfields. In order to give an idea about the feasibility of the use of this machinery, I will focus on \lanHEP\ since it is the one I have learnt and developed. With a bit of training one can come up with an input file that looks like this for QED\\

\begin{DDbox}{\linewidth}
\begin{verbatim}
1  % QED at one-loop
2  % Parameters of the theory
3  parameter EE=0.3034,Me=0.0005123.
4  
5  % Fields of the theory
6  vector A:(gauge, charge 0).
7  spinor e:(mass Me, charge -1).
8
9  % Covariant derivative of U(1) gauge group
10 alias D(x)=deriv+i*charge(x)*EE*A.
11 
12 % Mixing and masses at order one.
13 infinitesimal dZAA,dZeL,dZeR,dEE,dMe.
14 transform A->A*(1+dZAA),
             e->(1+(1+gamma5)/2*dZeL+(1+gamma5)/2*dZeL)*e,
             EE->EE+dEE,Me+dMe.
15
16  lterm -1/4*(D(A)^mu^nu-D(A)^nu^mu)**2.
17  lterm anti(e)*(i*gamma*D(e)-Me)*e.
\end{verbatim}
\end{DDbox}

We introduce first the zeroth order of the initial parameters $P_I^{(0)}$ at line 3, and since we are using the \OS\ scheme, this is obtained by inverting the observable $\alpha(0),m_e$. Then on line 6-7, the fields are defined, both by their Poincaré nature (vector or spinor) and their gauge representation (here the charge). The keyword \texttt{gauge} is used for the photon to indicate that a ghost field must be added to the physical field when writing the Feynman rules of the theory. The Lagrangian is then written in the usual way on line 16-17. However this is preceded by the declaration of the parameters and mixing matrices at first order (the wave-function renormalisation) on line 13, and the subsequent rotation of fields at the next line. Their values cannot be computed at this point since it require loop integration, so \lanHEP\ keeps them as unknown parameters, and they will be computed on the fly at the next step. This is a much too simplistic choice of model from many points of view (abelian 
symmetry, no supersymmetry, no gauge 
breaking, no off-diagonal mixing between fields), however it is self consistent. A more complex one is presented in the Appendix B.

\subsection{Computation of the process amplitude}
The computation of the process is a completely different story : since there exists very different kinds of processes, there is no single code offering the complete choice of all that can be computed. However we can still find a subset of codes that aim at a more or less general purpose, namely computing the scattering cross section of a given initial state to a given final state in a given model, provided that the total number of particles involved is small (say, up to five). There are several such codes on the market :
\begin{itemize}
\item \FC/\FA
\item \CH/\CO
\item \MG
\end{itemize}
The input of those codes is precisely the Feynman rules of a given model, hence they are fairly model-independent since any model can be turned into Feynman rules. The other input is the process one wishes to compute. The core of the codes is mostly the implementation of the Feynman procedure. There are two difficult points, the first being the computation of analytic expressions for each diagram. Remember that this number grows exponentially with the order, while the precision of the result does not always converge very fast : for instance in the QCD loop expansion the ratio between amplitudes of two successive orders can be around one half or so. The second point is the integration on the final state : indeed when specifying a final state, the phenomenologist is usually not interested in a specific direction and energy of each particle but on the average, that is the cross section integrated over all the phase space. One has then to rely on an efficient numerical integration.\\

One of the codes used in this thesis is $\FA/\FC$. This code is articulated in two parts : an analytic computation of the amplitude of the process (run on an analytic software, here \MA) followed by a numerical evaluation of the expression obtained (run on a numerical software, here \FO). The first part is done with very explicit functions, to wit :\\

\begin{DDbox}{\linewidth}
\begin{verbatim}
1  (* Computation of e+,e- to mu+,mu- at one loop order *)
3  $process= {prt["e"],-prt["e"]} -> {prt["mu"],-prt["mu"]};
4
5  (* Tree-level amplitude *)
6  diagram$0 = InsertFields[ CreateTopologies[ 0 , 2->2 ], $process ];
7  amplitude$0 = CalcFeynAmp[ CreateFeynAmp[ diagram$0 ] ];
8  (* One-loop amplitude *)
9  diagram$1 = InsertFields[ CreateTopologies[ 1 , 2->2 ], $process ];
10 amplitude$1 = CalcFeynAmp[ CreateFeynAmp[ diagram$1 ] ];
11
12 WriteSquaredME [amplitude$0 , amplitude$1 , $FortranCode];
\end{verbatim}
\end{DDbox}

Here the user first defines the process and then generates the Feynman diagrams by creating all graphs with the relevant number of loops and inserting the external fields corresponding to the process (line 6 and 9). The corresponding analytic amplitudes are then calculated (line 7 and 10), and the result is written in a \FO\ routine (line 12), ready for the numerical evaluation. Note that amplitudes must be given separately at each order, since the cross section is given as a perturbative series. The computation of the one-loop amplitude will inevitably involve loop integration, and a nice feature of \FC\ is that the loop integration technique is an independent part, hence can be changed very easily by the user. For instance one can choose between unitarity cuts methods or the Passarino-Veltman decomposition that we discussed in chapter 2. Having used the last one, we will just link the \LT\ library to the \FO\ code and leave to it those loop integrals. The final part is the numerical evaluation, which is more 
or less straightforward (depending mostly on the number of particles involved, $2\to 2$ and $2\to 3$ processes being much easier to handle than higher multiplicities), at this point the only modifications one can bring along are changes of the parameters of the model and specifications on the final states, such as cuts, etc.\\

\section{Recasting observables : the Higgs example}
Computing an observable right from scratch may look natural and even easy if appropriate automated tools are available, however it is essentially limited to some processes. As an example there is no generic code for fully automated two-loop computations. And it is also no surprise that dedicated tools, such as those for the Standard Model predictions, are more accurate than generic ones. In the Higgs phenomenology for instance there is on one side the Standard Model which, being a reference model in particle physics, benefits from dedicated studies on each process and is now known with a very good precision and on the other side many models of new physics which also predict a Higgs (and usually additional particles as compared to the Standard Model), but for which the computation of productions and decays are done at lower orders. It is then all the more tempting to re-use Standard Model results since their contribution is expected in any of its extension. For instance, Higgs observables at the LHC are 
usually plagued with QCD loop contributions which have to be computed at high orders to start yielding reliable results, but in all extensions suggested by new physics QCD stays the same.\\

As a simple example, let me take a theory where the Higgs boson has a different coupling to the $b$ quarks. We can parametrise such a coupling by a parameter $\xi$ :
$$g_{h\bar bb}=\xi g_{h\bar bb}^\SM$$
If we now want to predict the cross-section $\sigma$ for the process $e^+e^-\to hZ\to\bar b bZ$, we can either compute it ourselves, but we need to include all radiative corrections from the strong sector, or we can approximate the result by
\begin{equation}
 \sigma=|\xi|^2\sigma^\SM
\end{equation}
since all diagrams where the coupling $h\bar bb$ appears are rescaled by $\xi$. The approximation only breaks up for diagrams where the coupling does not appear, or appears more than once, but in this case they would not contribute much. Hence  we achieved a precision comparable to the Standard Model without performing involved computations. Of course in practice the recasting process is rarely so straightforward. However, it can be so powerful that it is always worth considering, even if it may lead to a delicate manipulation.

\section{Application : The \SL\ program}
Having seen the description of the different codes that can be plugged in the computation of a given observable, let us now see how we can arrange them in a program. This idea is based on the work of the \SL\ collaboration that originates from the particle physics group at Annecy (N. Baro, F. Boudjema, G. Chalons, A. Semenov and myself) and aims at computing in an automated way cross-sections at the one-loop order in supersymmetric models. This is the perfect opportunity to see what I meant by the notion of program, since we are so far left with codes with identified purposes and requirements, but by no means a complete automation procedure. The first step is to define the skeleton of the program:\\

\begin{tabular}{lp{13cm}}
Name & SloopS\\\hline
Purpose & Given a supersymmetric model, computes the cross-section of a given process at the one-loop level of accuracy.\\\hline
Input &\begin{itemize}
 \item A matter sector (set of chiral superfields), a gauge sector (Gauge group and vector superfields), potentials (Kahler potential, superpotential, susy breaking potential).
 \item A set of physical parameters and physical fields.
 \item A process $X_1..X_m\to Y_1..Y_n$ ($m\leq 2$)
\end{itemize}\\\hline
Output & The integrated cross-section $\sigma$.\\\hline
Specificities &\begin{itemize}
 \item Scan compliant : the cross-section must be easily evaluated on scans over the parameter space.
 \item The process may be given with a list of restrictions, in the case where one is interested only in part of the amplitude.
 \item The program must be sufficiently modular so that each part of the calculation can be modified separately.
\end{itemize}
\end{tabular}

The program can naturally be separated in two steps : generation of the model, which has to be performed when incorporating new particles or adding new vertices, and computation of the process which is performed on a process by process basis. The first part is achieved through the writing of a set of model files and the subsequent run of \lanHEP, the second is done via the \MA\ front-end using \FA/\FC/\LT\ as a back-end.

\subsection{\SL : the global picture}
The program can now be represented diagrammatically, in the following way

\begin{figure}[!h]
\small
\begin{center}
\begin{tikzpicture}[scale=4, auto]
	\node [input] (input) at (0,3) {Susy Input\\\begin{itemize}\item Initial Superfields \item Initial Parameters \item Potentials \end{itemize}};
	\node [input,text width=3.2cm] (mass_mat) at (1.6,3) {
	\begin{itemize}
	 \item Mass Matrices
	 \item Coupling tensors
	\end{itemize}
};
	\node [input,text width=3.5cm] (tree) at (2.8,3) {Physical Fields/Parameters\\\begin{center}\textbf{Tree-level}\end{center}};
	\node [inputb] (loop) at (1,2) {Physical Fields/Parameters\\Renormalisation Scheme\\\begin{center}\textbf{1-loop}\end{center}};
	\node [input] (rules) at (2.4,2) {Feynman Rules};
	\node [inputb,text width=2cm] (process) at (0,1) {Process $X->YZ$};
	\node [outputc,text width=3.5cm] (fortran) at (1.5,1) {\FO\  routine};
	\node [outputb] (divergences) at (3,1) {\begin{center}Check UV/IR finite\end{center}};
	\node [outputa,text width=3.5cm] (sigma) at (1.5,0) {$\sigma(\text{model parameters})$};
	\draw [suite] (input) -- (mass_mat) node[midway,fill=white] {\lanHEP};
	\draw [suite] (mass_mat) -- (tree);
	\draw [-,>=stealth',thick,rounded corners=4pt] (tree) -- (2.8,2.5) node[near end,fill=white] {\lanHEP};
	\draw [suite] (2.8,2.5) -- (rules);
	\draw [suite] (2.8,2.5) -| (loop.north);
	\draw [suite] (rules) -- (fortran.north) node[midway,fill=white] {\FA/\FC};
	\draw [suite] (fortran) -- (sigma) node[midway,fill=white] {\LT/ Scan};
	\draw [transversal] (loop) -- (rules);
	\draw [transversal] (process) -- (fortran);
	\draw [autre] (fortran) -- (divergences);

\end{tikzpicture}
\end{center}
\normalsize
\caption{Overall procedure}
\end{figure}
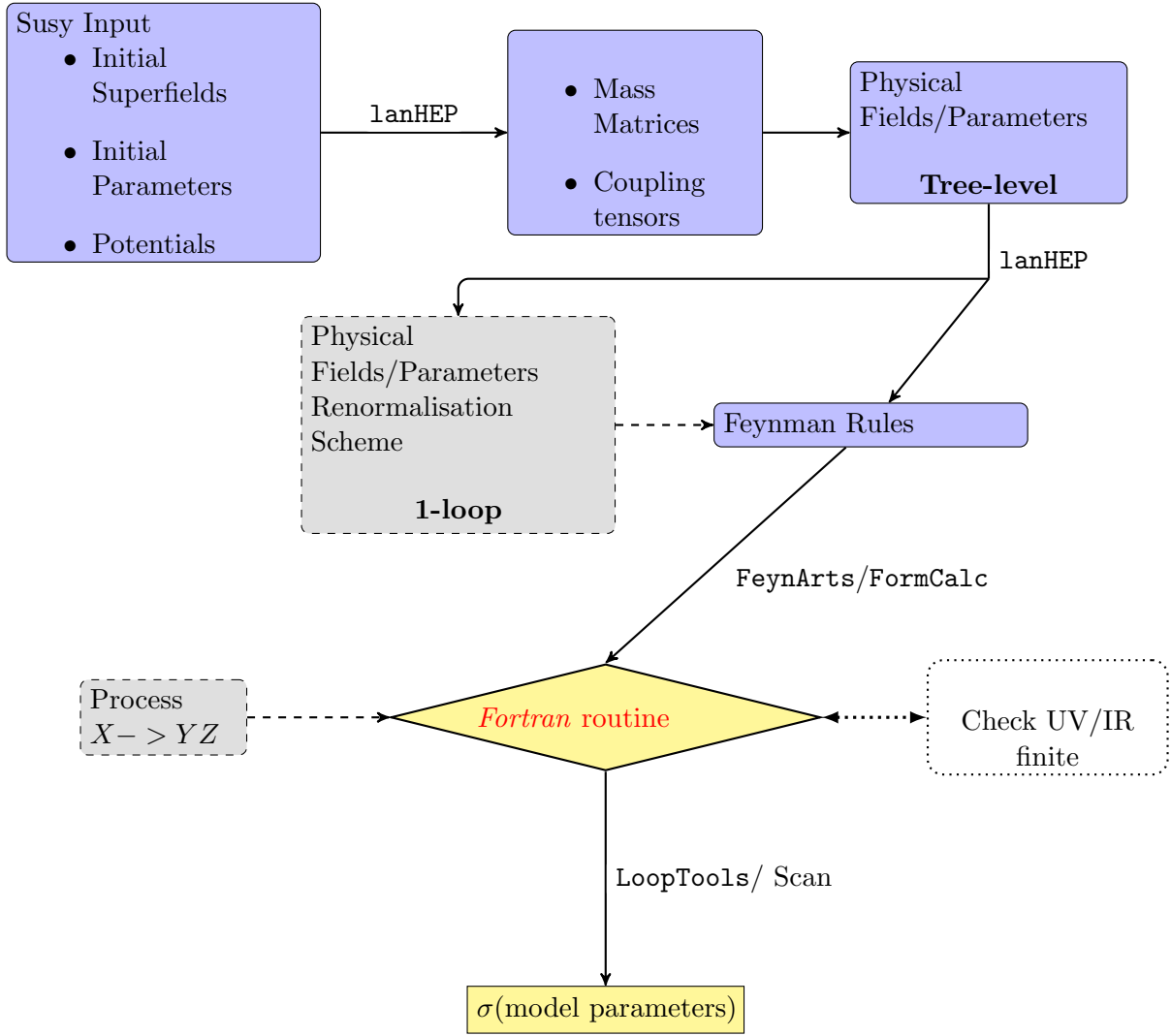


%% file: chapter5.tex
\chapter{Higgs sector : the need for non-minimal supersymmetry}

\minitoc\vspace{1cm}

\section{Hints for non minimal Higgses}
Despite its simplicity, the MSSM has lost some of its appeal over the last year (2011) with the growing tension brought by the non-observation of superparticles
at the LHC. In particular, constrained versions of the MSSM, which are probably the most simple supersymmetric theories based on the Standard Model, are more
and more disfavoured by the experimental data. Although this is often mistaken for a possibility to rule out supersymmetry itself, it actually means that if
supersymmetry is indeed realised in Nature, it may not be in a minimal form. To rephrase this, although minimal supersymmetry is going to be more and more in
trouble, probing general supersymmetry requires a more sophisticated framework than what I have discussed so far. This is particularly true in the Higgs sector,
so I will start by recalling some indications for a non-minimal Higgs sector.

\subsection{Naturalness in supersymmetry}
We have introduced supersymmetry with the motivation, among others, that it could cure the Naturalness issue of the Standard Model. Indeed in the limit of exact
supersymmetry the Higgs scalar field would not receive any quadratic corrections. The hierarchy of the MSSM was then only between the electroweak scale and the
supersymmetry breaking scale, rather than the gauge unification scale. Natural supersymmetry leads thus to the following requirement :
$$M_\text{SUSY}=\grandO{100\text{ GeV}}$$
where $M_\text{SUSY}$ stands for all supersymmetric masses that can run into the self-energy of the Higgs, that is to say the superpartner masses and $\mu$.
Precisely, since the loop contribution of superpartners is driven by their Yukawa couplings, the relevant superpartner masses are the stops and sbottoms
masses\footnote{and, at the two loops order, the gluino mass.}. As such, there is no fine-tuning in the MSSM if for instance one takes
stops and sbottoms masses and $\mu$ equal to 100 GeV. However, some trouble arises when one also considers experimental constraints : the non observation of 100
GeV charginos or squarks has already ruled out such a spectrum. It means that we have to increase our supersymmetry-breaking mass, raising thus the amount of
fine-tuning.\\

However, we can still cope with a small fine-tuning, and thus putting stops/sbottom masses together with $\mu$ around 300 GeV would bring us safe from
experimental direct searches without losing the idea of supersymmetric naturalness. The real issue does not come from the direct searches, but from an indirect
one which is precisely the light Higgs search. Indeed the experimental searches, both at LEP and now at LHC, are pushing the light Higgs to be roughly higher
than $120$ GeV. Such a high mass, in the context of the MSSM where the tree-level value is bounded by $\mz$, is only attainable by large radiative corrections.
And this requires to have heavy stops, as well as a maximal mixing between the left and right handed stops. This problem, usually referred to as the little
hierarchy problem, is exactly the tension between the concept of naturalness and the experimental constraint. On one side, we would tend to make $M_\text{SUSY}$
as close as possible to the electroweak scale, and on the other, raising the 
lightest Higgs mass drives us to a heavy spectrum. The issue is even more severe when one require the light Higgs to be compatible with the hint of signal at
$\mh=125$ GeV.\\

It must be noticed that requiring the $\mu$ parameter to be of the order of the electroweak scale is not a straightforward assumption : in the MSSM the $\mu$
term comes from the superpotential and has a priori no connection to the electroweak scale, in fact there is no reason why it would not be at the unification
scale for instance. This is known as the $\mu$-problem and can be interpreted as the need for a new mechanism to obtain this $\mu$ term. This is precisely one
of the motivation of the NMSSM where the addition of the singlet chiral superfield $S$ allows the term $\lambda SH_1H_2$ in the superpotential, which will then
be turned in the $\mu$ term by giving the $S$ field a non-vanishing vacuum expectation value. The advantage of such a mechanism is that $\lambda$ being of order
one (since it is a dimensionless coupling), the size of the $\mu$ parameter is directly linked to the vacuum expectation value of the singlet chiral superfield,
which can itself be taken at a low scale.

\subsection{Non standard signals}
Apart from the naturalness issue and leaving dark matter aside\footnote{if supersymmetry is quite decoupled, the LSP will tend to be too heavy to account
properly for dark matter in the universe. However it must be note that an LSP of the order of a couple of TeV is still possible.}, there is unfortunately
nothing wrong with a totally decoupled supersymmetry, that is with only the Standard Model spectrum accessible. This would entirely blow up the interest of
supersymmetry in phenomenology, and would make quite futile the discrimination between a minimal and a more complicated version of supersymmetry. However, this
does not forbid the possibility of being lucky and actually observing a signal in contradiction with the MSSM expectations. This perspective was particularly
alive after the release of the Higgs analyses at the LHC with the 5\fb\ dataset, which pointed to a possible non-Standard Model like, and even non-MSSM like
type of signal. By the time of the writing of this thesis, both the explanation of the 
excesses by a Higgs signal or by a statistical fluctuation are likely, and  in the signal case the alleged values for signal strengths are given with large
error bands. It is as such impossible to conclude, but given that the experimental data is now going to increase at a fast rate, it is of the utmost importance
not to content ourselves with a minimal version of supersymmetry, which would be at lost if a non-standard signal emerges.\\

Those issues have led to several expeditions in the landscape Beyond the MSSM. Most often one starts by postulating some extra physics : new matter, new gauge
structure, extra dimensions, and then incorporates the new particles and new vertices into the computation. The Next-to-Minimal Supersymmetric Standard Model
(NMSSM) (\cite{ellwanger_nmssm}) is an example of a MSSM extension that allow for a richer Higgs phenomenology. This is not much a surprise, given that the
extra physics is precisely a new Higgs superfield, although it is a singlet. Being able to enhance the Higgs mass without the contribution of heavy stops, it
does naturally alleviate the fine-tuning issue of the MSSM. From the point of view of Higgs searches at the LHC, the flexibility gain is immediate : there are
now three CP even Higgs scalar states, hence playing on the mixing between species will give a much richer structure. However one could think also of modifying
the Higgs potential and Higgs-matter-gauge relations altogether. This is 
the case with more complicated type of extra physics, such as extra $U(1)$ gauge groups among others.\\

\section{Effective Field Theories}
The drawback of each specific extension of the MSSM is that the flexibility we gain is very much dependent on our UV completion, that is to say our extra
physics, which tends to narrow the reach or our study. There is nevertheless a well-known path to escape this gloomy future, it is the Effective Field Theory
(EFT) approach. As introduced in the third chapter, the point is the following : if the extra-physics we are postulating is quite heavy compared to the scale of
the experiments, those particles will never show up in initial or final states and their contribution to the effective action can be integrated out. The key
feature is that this can be done without any assumptions on the extra physics, this use of the EFT framework is exactly a way of parametrising the unknown.

\subsection{Application to the supersymmetric Higgs sector}
In the case of supersymmetry, equation \ref{eq:eff_parametrisation} can be rewritten as
\begin{eqnarray}
 \Geff&=&\Geff_\MSSM+\frac{1}{M}\Geff^{(1)}+\frac{1}{M^2}\Geff^{(2)}+\grandO{\frac{1}{M^3}}\\
 \Geff^{(i)}&=&\sum_n c_i^nO_i^n[\Phi]
\end{eqnarray}
where $O_i$ are operators of dimension $4+i$ and $c_i$ free coefficients. An interesting point in having the MSSM as the low-energy theory as compared to the
Standard Model is that we can require the effective operators $O_i$ to be functions of superfields instead of fields.\\

Given the number of fields present in the MSSM there seems to be a large number of possibilities, even when truncating the effective expansion and requiring the
super Poincaré and gauge invariance of each operator (this is indeed shown in reference \cite{wudka_piriz_bmssm_1997}). However, since we want to investigate
what flexibility is gained on the Higgs side, it is quite clear that some operators will have no impact, whatever the value of their coefficients. A simple
approach is then to restrict all operators to involve only the Higgs superfields of the MSSM $H_1$ and $H_2$ : we will see that we end up with a reasonable set
of operators.

\subsubsection{K\"ahler and superpotential}
Since a supersymmetric Lagrangian is made of a K\"ahler potential and a superpotential, the effective operators will show up in those potentials. Requiring only
Higgs superfields, one ends up with the following operators (as shown in ref \cite{antoniadis_bmssmhiggs_0910}) in the superpotential
\begin{eqnarray}
O_A&=&\frac{1}{M}\left(H_1.H_2\right)^2
\end{eqnarray}
and in the K\"ahler potential
\begin{eqnarray}
O_B&=&\frac{1}{M}D^\alpha\left[H_2e^V_2\right]D_\alpha\left[e^V_1H_1\right]\hc\\
O_{1,2}&=&\frac{1}{M^2}\left(H_i^{\dag}e^{V_i}H_i\right)^2\\
O_3&=&\frac{1}{M^2}\left(H_1^{\dag}e^{V_1}H_1\right)\left(H_2^{\dag}e^{V_2}H_2\right)\\
O_4&=&\frac{1}{M^2}\bigl(H_1.H_2\bigr)\left(H_1^{\dag}.H_2^{\dag}\right)\\
O_{5,6}&=&\frac{1}{M^2}\left(H_1.H_2+H_1^{\dag}.H_2^{\dag}\right)\left(H_i^{\dag}e^{V_i}H_i\right)\\
O_{7,8}&=&\frac{1}{M^2}H_i^\dag\bar\Delta^2e^{V_i}\Delta^2H_i\\
O_{9,10}&=&\frac{1}{M^2}H_i^\dag e^{V_i}\Delta^\alpha W_\alpha^iH_i\\
O_{11,12}&=&\frac{1}{M^2}H_i^\dag e^{V_i}W_\alpha^i\Delta_\alpha H_i
\end{eqnarray}
where we use $\Delta_\alpha=e^{-V_i}D_\alpha e^{V_i}$ and $W_\alpha^i$ is the field strength of the vector superfield acting on $H_i$. However it turns out that
not all of these operators are independent, indeed many of them can be removed by field redefinitions and the application of the equations of motion, as we will
now see. First, the dimension 5 operator $O_B$ can be removed by field redefinitions as explained in \cite{antoniadis_bmssmhiggs_0806}. Then, by writing the
equations of motion for the on-shell Higgs superfields we get :
\begin{eqnarray}
-\frac{1}{4}\bar D^2\left(H_2^\dag e^{V_2}\right)+\mu H_1^T(i\sigma_2)&=&0\\
\frac{1}{4}\bar D^2\left(H_1^\dag e^{V_1}\right)+\mu H_2^T(i\sigma_2)&=&0
\end{eqnarray}
By plugging these results in the expression of operators $O_{7/8}$, one gets
\begin{equation}
O_7\sim 16\mu^2 H_1^\dag e^{V_1}H_1,\hspace{1cm}O_{8}\sim 16\mu^2 H_2^\dag e^{V_2}H_2
\end{equation}
which are simply shifts of the usual MSSM operators. We also obtain that operators $O_{9/10}$ will vanish after integrating by part, and that $O_{11/12}$
operators will give a null contribution because of the definition of $W_\alpha$. The final result is that the effect of operators $O_{7..12}$ is simply to shift
the wave function renormalisation of the Higgs superfields. Thus, those operators are not independent and their contribution can be dropped (a more detailed
discussion is to be found in reference \cite{antoniadis_bmssmhiggs_0910}).\\

Based on those considerations, we will hence consider the set of independent operators up to the second order of the effective expansion to be the following :
it was previouly obtained in ref \cite{carena_bmssmhiggs_0909,antoniadis_bmssmhiggs_0910}.\\
\begin{IEEEeqnarray}{rCl}
W_{\text{eff}}&=&\zeta_1\frac{1}{M}\left(H_1.H_2\right)^2\\
K_{\text{eff}}&=&a_1\frac{1}{M^2}\left(H_1^{\dag}e^{V_1}H_1\right)^2+a_2\frac{1}{M^2}\left(H_2^{\dag}e^{V_2}H_2\right)^2+a_3\frac{1}{M^2}\left(H_1^{\dag}e^{V_1}
H_1\right)\left(H_2^{\dag}e^{V_2}H_2\right)\\
&&+a_4\frac{1}{M^2}\bigl(H_1.H_2\bigr)\left(H_1^{\dag}.H_2^{\dag}\right)+\,\frac{1}{M^2}\left(H_1.H_2+H_1^{\dag}.H_2^{\dag}\right)\left(a_5H_1^{\dag}e^{V_1}
H_1+a_6H_2^{\dag}e^{V_2}H_2\right)\nonumber
\end{IEEEeqnarray}
The effective coefficients $a_i,\zeta_1$ are a priori unknown, stemming from UV physics. If we postulate further that the extra physics is weakly coupled we can
harmlessly predict $|a_i|<1$. Note that though the effects of the order-5 operator were already well-known :  they were extensively studied by \cite{brignole},
\cite{dine_bmssmhiggs_0707}, \cite{antoniadis_bmssm_0708,antoniadis_bmssmhiggs_0806,antoniadis_bmssmhiggs_0910,antoniadis_bmssmhiggs_1012},
\cite{ponton_susy_higgs_0809}, \cite{carena_bmssmhiggs_0909,carena_bmssmhiggs_1005,carena_bmssmhiggs_1111}. The precise point of the fine-tuning issue of the
MSSM was shown in (\cite{espinosa_finetuning_2004,ross_finetuning_0903,cassel_bmssmhiggs_1103}), with CP-violation (\cite{altmannshofer_bmssmcp_1107}) and the
vacuum stability (\cite{delaunay}). The effects of order-6 operators have been less appreciated throughout the literature, but have been shown independently by
\cite{carena_bmssmhiggs_0909,carena_bmssmhiggs_1111} and \cite{antoniadis_bmssmhiggs_0910} to be important.

\subsubsection{Supersymmetry breaking}
This would be the full description of the model, if we believed it to be strictly supersymmetric. However, just as the MSSM is supersymmetry broken, we can
consider our operators to be susy-broken. In the low energy side this breaking shows up in the spurion terms appearing in the coefficients
 \begin{eqnarray}
\zeta_1&\longrightarrow&\zeta_{10}+\zeta_{11}m_s\theta^2\\
a_i&\longrightarrow&
a_{i0}+a_{i1}m_s\theta^2+a_{i1}^*m_s\overline{\theta}^2+a_{i2}m_s^2\overline{\theta}^2\theta^2
 \end{eqnarray}
where we use $m_s$ as a book-keeping scale to keep all coefficients dimensionless. Since the scale $M$ of this new physics is expected to be high, we will
assume that it is approximatively supersymmetric, that is to say $m_s/M<1$. This ratio being likely to show up in the effective expansion of observables, if we
want to keep the EFT on the perturbativity side, which is typically a broadly appreciated feature, it is all the more crucial to keep this ratio low.\\

We have used the following numerical values
\begin{equation}
 M=1.5\text{ TeV},\quad m_s=300\text{ GeV}\ \Rightarrow\ \frac{m_s}{M}=0.2\qquad\left(\frac{m_s}{M}<1\right)
 \label{eq:ratio}
\end{equation}
which corresponds to a new physics scale much higher than the light Higgs production scale ($\sim 100$ GeV), but close enough to have a significant impact on
phenomenology. Raising $M$ will decrease the different deviations stemming from effective operators, but they will not stand all on an equal footing. As far as
scaling with the mass $M$ of the new physics is concerned, it is important to realise  that the leading 
corrections brought about by the new operators enter as $\mu/M$ and $m_s/M$. The fact that we take $m_s$ as in eq.~\ref{eq:ratio} tacitly assumes 
that the underlying theory is approximately supersymmetric. The fact that we take $m_s=\mu$ means that all leading $1/M$ corrections scale the same way with
$M$. 
There are also corrections  that enter as  
$(\mu/M)^2, (m_s/M)^2$ and $(m_s\mu/M^2)$. These type of corrections affect the  the quartic part of the Higgs potential. 
There are also  higher order operators (dim$>4$) in the scalar potential  that scale like $(v^2/M^2)$.  Since in our case $v \sim \mu=m_s$ ($v=246$ GeV) , all
$1/M^2$ 
effects scale in the same way with $M$. One could have for example taken larger values for $\mu=m_s$ while keeping the same ratio 
$\mu/M$ fixed, however the contribution of the small $v^2/M^2$ effect would be very small.  Therefore when $1/M$
effects are dominant it would be not too difficult  to recover the result with another 
value of $M$ from the results we will show. In general the situation is more complicated as there can be a balance 
between the $1/M$ and $1/M^2$ terms that have even an impact on the stability of the potential. 
 In any case we should warn
that lowering $M$ increases the value of the effective corrections,
similar in part to increasing the
value of the dimensionless parameters. This would put the effective
approach at risk, an $1/M$ expansion with too low $M$ being likely to
break down and many points would not pass the ``perturbativity"
criterion that we will define in section~\ref{sec:pertubativity}.\\

Concerning effective coefficients we assume the extra physics to be weakly coupled, and take thus their values in
\begin{equation}
 \zeta_{1i},a_{ij}\in[-1,1].
\end{equation}

\subsection{New Lagrangian}
We know from our phenomenological presentation of the MSSM that although the K\"ahler, the superpotential and the susy-breaking potential encode in a compact
form the physics of the model, they do not describe explicitly the dynamics at low energies, which is itself encoded in the Lagrangian expressed in terms of
field components. It turns out that the effective operators will alter significantly the derivation of this Lagrangian. This is well exemplified by the
computation of F and D terms.\\

I will use in the following equations several short-hand notations, first the index notation to denote functional derivative (as in $W_i=\partial_{\Phi_i}W$),
$i$ for indexing chiral superfields $\Phi$, $\oo{i}$ for their chiral conjugate, $n$ for indexing representations of the gauge superfield $V$.\\

\subsubsection{F terms}
The $f$ component of chiral superfields appears in the following terms in the Lagrangian.
$$\overline{f}^{\overline{i}}K_{\overline{i}j}f^j+(K_{\overline{i}jk}\oo{f}^i\psi^j\psi^k+W_if^i+ h.c.)\subset\mathcal{L}$$
Since $f$ is not a dynamical field, it will take the value that minimises such a potential. In a matrix notation, this can be seen as the quantity $X^\dag
AX+BX+(BX)^\dag$ evaluated at its minimum, which yields the canonical result
$X=-B^\dag A^{-1}B$. Coming back to the actual calculation, we get the potential\vspace{2mm}
\begin{equation}
V_F=-\left(\oo{W}^{\oo{i}}+\1 K_{\overline{i}kl}\psi^k\psi^l\right)K_{\oo{i}j}\left(W^j+\1 K_{j\oo{k}\oo{l}}\oo{\psi}^{\oo{k}}\oo{\psi}^{\oo{l}}\right)
\end{equation}
which we quickly check to yield the correct result in the standard case (in which $K_{\overline{i}kl}=0$ and $K_{\oo{i}j}=\delta_{\oo{i}j}$).\\

\subsubsection{D terms}
The part of the Lagrangian contributing to the $\mathcal{D}$ term is the following :
$$K_n\;g\mathcal{D}_n+\1\mathcal{D}^a\mathcal{D}_a \subset\mathcal{L},$$
the second term coming from the gauge Lagrangian, which is not modified by effective operators. To pursue any further we need to extract the $\mathcal{D}$ field
itself -- lying in the adjoint representation, labelled by $a$ indices -- from the $\rho_n$ representations:
$$K_n\;g\rho_n(T^a)\;\mathcal{D}_a+\1\mathcal{D}^a\mathcal{D}_a \subset\mathcal{L}$$
where $T^a$ is the basis of the gauge Lie algebra.
This  yields the
potential :
\begin{equation}
V_{\mathcal{D}}=-\1|K_n\;gT^n|^2=-\1(K_n\;gT^n_{\ a})(K_n\;gT^{n\ a})
\end{equation}
where the $T^n_{\ a}$ are the representation of the gauge matrices. Hence they also carry $SU(3)$, $SU(2)$ and $U(1)$ indices, however $K_n$ has the same
indices (since we have derived along the whole gauge superfield) hence they get contracted.

\subsubsection{Full Lagrangian}
Having seen that the F and D term already have expressions quite different from the ones of the MSSM, it is not surprising to see that the full Lagrangian
exhibits an intricate structure

\begin{eqnarray}
\mathcal{L}_K\ &= \ &
    -(\oo{\textbf{W}}^{\oo{i}}+\1 K_{\overline{i}kl}\psi^k\psi^l)K_{\oo{i}j}(\textbf{W}^j+\1
K_{j\oo{k}\oo{l}}\psi^{\oo{k}}\psi^{\oo{l}})-\1|K_n\;gT^n|^2\textbf{W}_W^{\,-1}\nonumber\\
    && +iK_{in}\;\partial_\mu\phi^i\;(gv)^{n\ \mu}
    +K_{nm}\;(gv)^{n\ \mu}\;(gv)^m_{\ \mu}+K_n\;(gv)^{n\ 2}
    +\1 K_{i\oo{j}}\;\partial_\mu\phi^i\;\partial^\mu\oo{\phi}^{\oo{j}}\nonumber\\
    && +i\frac{\sigma^\mu}{2}K_{i\oo{j}}\;\partial_\mu\psi^i\;\oo{\psi}^{\oo{j}}+\1
\textbf{W}_{ij}\;\psi^i\;\psi^j+i\frac{\sigma^\mu}{2}K_{ij\oo{k}}\;\partial_\mu\phi^i\;\psi^j\;\oo{\psi}^{\oo{k}}\nonumber\\
    && +i\sqrt{2}K_{in}\;\psi^i\;(g\lambda)^n+\sigma^\mu K_{i\oo{j}n}\;\psi^i\;\oo{\psi}^{\oo{j}}\;(gv)^n_{\ \mu}\nonumber\\
    && +\frac{1}{4}K_{ij\oo{k}\oo{l}}\;\psi^i\;\psi^j\;\oo{\psi}^{\oo{k}}\;\oo{\psi}^{\oo{l}}\nonumber\\
    && + \textbf{W}_W\,\left(-\frac{1}{4}F_{\mu\nu}F^{\mu\nu}-i\lambda\sigma^{\mu}D_{\mu}\oo{\lambda}\right)\nonumber\\
    && +\mathcal{L}_{SSB}
\label{full_lag}
\end{eqnarray}
which is pretty much unpleasant at first sight. However we shall not be afraid of such a complexity since we know how to use automated tools to handle the full
derivation. Note that a handmade treatment is still feasible, however the risk of an error is high, and it is not convenient for introducing new operators. We
decided to use \lanHEP, since it is quite handy for deriving Feynamn rules in supersymmetric theories, as discussed in chapter 5. Nevertheless it required a bit
of upgrading in order to deal properly with higher order functional derivatives, since the usual K\"ahler potentials do not exhibit that many non vanishing
derivatives. The problem stands as follows : in most of the supersymmetric theories where the particle content is fully determined, the K\"ahler potential
simply writes as
$$K=\sum_\Phi\Phi^\dag e^V\Phi.$$
This form is particularly simple for the derivation of the Lagrangian since its double derivative (along the superfields) is simply a delta function : $K_{i\bar
j}=\delta_{i\bar j}$. Similarly, the superpotential usually only needs to be evaluated on its first derivative (in the F term) or on the second derivative (for
fermion masses). Those features are however lost in the BMSSM framework where higher-order derivatives of both potentials will be needed. This was at first
sight an issue since \lanHEP\ did not include the possibility to compute such derivatives. In a collaboration with its author A.Semenov, we were able to improve
\lanHEP, which resulted eventually in a new version of \lanHEP\ which can now deal with any kind of K\"ahler potential an superpotential.\\

Before going to the discussion of the important phenomenological features, let me point out that the apparent dimension of an effective operator in the
superfield form may not be the same in the field Lagrangian. As an example let us derive the lagrangian associated to the simple superpotential
\begin{equation}
W=\mu H_1\cdot H_2+\frac{\zeta_{10}}{M}\left(H_1\cdot H_2\right)^2
\end{equation}
which has a dimension 4 operator an a dimension 5 one, suppressed by a power of $M$. The first order in $1/M$ of the Lagrangian will look like
\begin{equation}
\L_{|1/M}=\frac{\mu}{M}\zeta_{10}\,{}^{\backprime\backprime} h^4{}^{\prime\prime}\,+\frac{\zeta_{10}}{M}\,{}^{\backprime\backprime} h^2{}^{\prime\prime}\
{}^{\backprime\backprime}\tilde{h}^2{}^{\prime\prime}
\end{equation}
where the ${}^{\backprime\backprime} X^i{}^{\prime\prime}$ terms stand for generic polynomials of the field $X$ at order $i$. We already see that the scalar
term is of dimension 4 and the scale suppression factor is now a scale ratio $\mu/M$. The other term is \textit{a priori} non renormalisable, but when we shift
to the physical basis, that is to say $h^0\to h^0+v$, we will see terms appearing with dimension three and four, now with a suppression factor of $v/M$. The
point is then to show that, while there will be some non-renormalisable terms appearing in the Lagrangian $\L$, most of the terms derived from the effective
operators (introduced as function of superfields) will be renormalisable and thus will not lead to new divergences in the computation. This is in particular the
case of couplings of the Higgses to fermions which have the same Lorentz structure as in the MSSM. We will see furthermore that for the few new Lorentz
structures that may lead to actual divergences, the couplings can often be 
simplified by the 
use of the equations of 
motion. Let us now focus on the major phenomenological features of this new Higgs sector.

\section{Beyond the MSSM phenomenology : What's new?}
\subsection{Higgs scalar potential}
The Higgs scalar potential is changed because F and D terms are changed on the one hand, and because of the supersymmetry breaking part of the effective
operators on the other. We will now see the consequences of those modifications on the observable side, that is to say the masses, the mixing and the
couplings.
\subsubsection{Kinetic mixing}
The first outcome is that, because of the terms $K_{ij}\partial\bar\phi^i\cdot\partial\phi^j$ and $K_{ij}\bar\psi^i\dslash\psi^j$ in the Lagrangian, the kinetic
terms for the Higgs fields become non standard when the Higgs fields acquire non vanishing \vev. Precisely, when compared to the effective quadratic Lagrangian
worked out in eq. \ref{eq:scalar_prop_eff}, we will have a non vanishing term $\dA$. As in the MSSM, we will write
\begin{equation}
h_1\rightarrow \binom{v_1}{0}+h_1\qquad h_2\rightarrow \binom{0}{v_2}+h_2\qquad\qquad v_1,v_2>0.
\end{equation}
The symmetry breaking only occurring in a charge and CP conserving way (though relaxing such a hypothesis has been discussed in \cite{ponton_susy_higgs_0809},
we will not consider this here), there will be no mixing between charged and neutral, or CP even and odd, Higgses. The kinetic mixing will be driven by $K_2$, which is
the double derivative of the K\"ahler potential evaluated on the vacuum expectations values of the fields\footnote{And as such, $K_2$ is a dimensionless
quantity which does not depend on the physical fields} $K_2=\left(K_{i\bar j}(v_1,v_2)\right)$. $K_2$ being itself decomposed in two parts, one acting on
neutral fields $K_2^0$ and one on the charged fields $K_2^c$. The kinetic Lagrangian can then be written as
\begin{equation}
\L_k=\sum \partial_\mu\phi^{0\dag}K_2^0\partial^\mu\phi^0 + \sum \partial_\mu\phi_c^{\dag}K_2^c\partial^\mu\phi_c + \sum
\oo{\tilde{\phi}}^0K_2^0\dslash\tilde{\phi}^0+ \sum \oo{\tilde{\phi}}^cK_2^c\dslash\tilde{\phi}^c.
\end{equation}
As described in chapter 3, we obtain the physical basis by the transformations $P^0_\partial$ and $P^c_\partial$ acting respectively on neutral and charged
Higgses, defined by :
\begin{equation}
 P_\delta^0=\sqrt{K_2^0}\qquad P_\delta^c=\sqrt{K_2^c}.
\end{equation}
We now turn to the physical fields by applying the transformation :
\begin{equation}
 H^0{}'=P_\partial^0 H^0\qquad H^c{}'=P_\partial^c H^c
\end{equation}

\subsection{Higgs stabilisation}
\subsubsection{Computing the vacuum}
The consistent electroweak breaking imposes that $v_1,v_2$ describe a global minimum of the Higgs scalar potential. However in the case of an effective field
theory, this must be qualified : since the theory is valid only up to a certain energy, we can only require to have a global minimum in the domain where the
effective theory is valid. In the case under study, since we have fixed the heavy scale $M\sim 1.5$ TeV, we will impose to have a global minimum within a range
\begin{equation}
 \left<h_i\right><1\text{ TeV}.
 \label{eq:range_minimum}
\end{equation}
An issue arising now is the fact that the potential $V$ is of order 6,
\begin{eqnarray}
 \L&=&\tilde{m}_1^2|h_1|^2+\tilde{m}_2^2|h_2|^2+\tilde{m}_{12}^2(h_1\cdot h_2\hc)\nonumber\\
 &&+\frac{g_1^2+g_2^2}{8}(|h_1|^2-|h_2|^2)^2+\frac{g_1^2}{2}|h_1^\dag h_2|^2 + c_\eff^4 |h|^4 + c_\eff^6 |h|^6
\end{eqnarray}
where $|h|^4$ and $|h|^6$ stands for generic terms or order 4 and 6 and $c_\eff^4,c_\eff^6$ for generic effective coefficients with the $1/M,1/M^2$ suppression
factors implicit (not that $c^6_\eff$ must be of order $1/M^2$, but we have both orders for $c^4_\eff$). We cannot analytically determine the minimum of such a
polynomial. Instead, we will carry on the following path : first we fix the effective coefficients (in other words we take them as input), then we trade the
rest of the initial parameters to physical ones, that is to say
\begin{equation}
 g_1,g_2,v_1,v_2,\tilde{m}_1,\tilde{m}_2,\tilde{m}_{12}\longleftrightarrow \ma,\tb,e,m_W,M_Z,dV=0
\end{equation}
where $dV=0$ corresponds to imposing that $(v_1,v_2)$ is a critical point, which is a necessary condition for it to be a minimum. Note furthermore that since
$H$ and $H'$ are related by a linear transformation, this condition is the same in both bases. The system can be treated in a linear way since we know the
solution at zeroth order (it is the MSSM vacuum) and one can do the inversion perturbatively, as discussed in chapter 3.\\

Once the effective inversion has been performed, we are now dealing with a parameter point giving the correct physical values, and located on a critical point
of the Higgs scalar potential. We have now to check that it is indeed a global minimum within the range of eq. \ref{eq:range_minimum}. This is done numerically,
since we have to deal with a dimension 6 polynomial.

\subsubsection{Classifying the vacuum}

It has been noticed in the past (see \cite{delaunay}) that the vacuum localisation could have different sources. In particular there are two limits : the first
being when we set all effective coefficients to 0 and recover hence the MSSM limit. The second when we tune effective coefficients so that soft masses vanish,
in which case the breaking is susy-conserving.\\
\begin{center}
\begin{tabular}{cl}
 $c_i\rightarrow0$ & MSSM-like\\
 $m_1,m_2,m_{12}\rightarrow0$ & susy EWSB
\end{tabular}
\end{center}
That is to say that either the electroweak symmetry breaking is triggered by soft masses in the Higgs sector, or it is due to extra physics causing the
appearance of effective operators. The general case lying in between those two limits.\\

\begin{figure}[!h]
\begin{center}
\begin{tabular}{cc}
\includegraphics[scale=0.35,trim=0 0 0 0,clip=true]{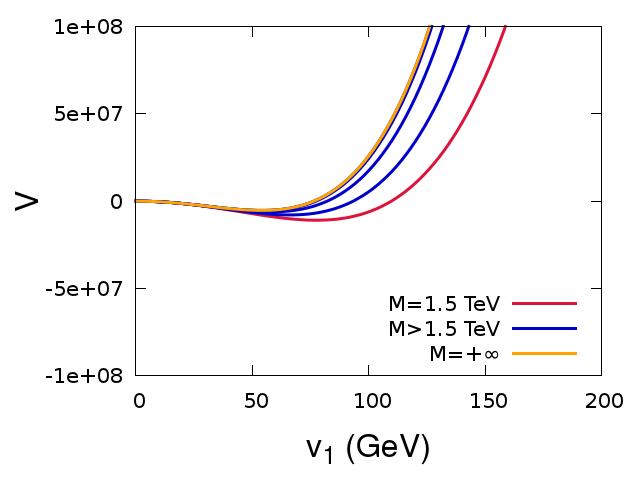}&
\includegraphics[scale=0.35,trim=0 0 0 0,clip=true]{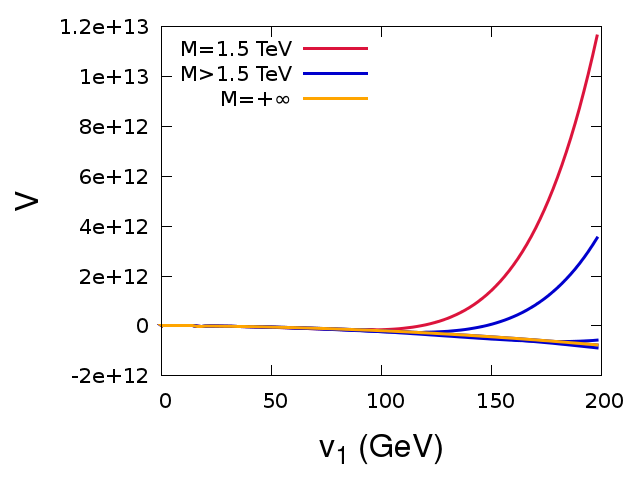}\\
MSSM-like & susy EWSB
\end{tabular}
\end{center}
\caption{\label{fig:vacuum_type} {\em We show here the form of the potential along $v_1$ (with $\tb=2$ enforced) in the initial case $M=1.5$ TeV (in red), in
the limit $1/M=0$ (in yellow) and for $M=2$ and $M=5$ TeV (in blue). On the left is an MSSM type of vacuum and on the right a susy EWSB type. Notice that in the
last case, the minimum is driven to the infinity before jumping to 0.}}
\end{figure}

The Higgs phenomenology itself is not sensitive to the kind of vacuum which is realised. However they can be told apart by looking at the behaviour of the
potential when $M$ is raised to infinity while keeping the soft masses to their initial values. The situation is the following : one starts on a given point of
the effective parameter space with a consistent electroweak symmetry breaking. Then one "freezes" the value of the MSSM parameters, in particular the soft
masses $m_1,m_2,m_{12}$, and then play on the values of the effective coefficients. The aim being to observe the behaviour of the localisation of the vacuum
along such modifications. Indeed, this localisation will \textit{a priori} evolve since we are changing effective parameters but not soft masses so the tadpoles
conditions $dV=0$ will change. What is interesting is the specific behaviour when the strength of the effective operators is lowered, that is to say, when $M$
is raised. Since the value of $M$ only enters as $1/M$ factors in the 
Lagrangian, we can even take the exact limit $M=+\infty$ by setting $1/M=0$. The behaviour of this evolution is shown in two cases in figure
\ref{fig:vacuum_type} : we show the initial point (in red), the final point $1/M=0$ (in yellow) and intermediate points in blue (corresponding to $M=2$ TeV and
$M=5$ TeV). In the case of an MSSM-like vacuum, the potential is smoothly deformed and in the limit $1/M=0$ the minimum $v_{1\ min}$ will be shifted by a
certain amount. In the susy EWSB case, the minimum of the potential will be driven to the infinity and when taking $\frac{1}{M}=0$ the potential will jump to a
configuration where either there is no minimum or it is at the origin. Note that we have chosen to look at the potential along the direction $\tb=2$, so that
the position of the minimum is fully parametrised by $v_1$, which clarifies the picture without affecting its conclusion.

\subsection{Higgs phenomenology}
\subsubsection{Mass/Mixing}
\begin{figure}[!h]
\begin{center}
\includegraphics[scale=0.4,trim=0 0 0 0,clip=true]{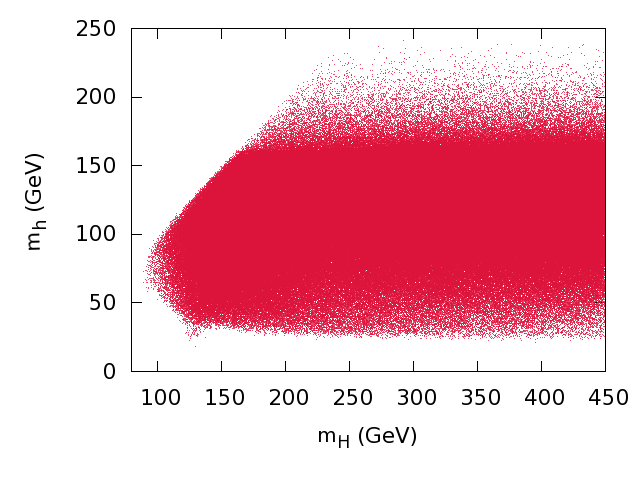}
\end{center}
\caption{\label{fig:mh_mass} {\em Reach of the BMSSM in the $\mH,\mh$ plane, before considering any experimental constraint.}}
\end{figure}

The most impressive alteration appearing in the BMSSM framework is probably the mass of the lightest Higgs $\mh$. Indeed it will have contributions from nearly
all operators, and if the major contributions add together, the effect is to raise the mass up to $\mh=250$ GeV. I show on fig \ref{fig:mh_mass} the reach of
the BMSSM in the plane $\mH,\mh$ in the $m_{h\ max}$ MSSM scenario (described in \cite{carenaheinemeyer}). Although I will specify more closely this scenario
later, it is clear that the significant rise of $\mh$ is purely due to the effective operators, in particular the dimension 5 ones.

\subsubsection{The non-decoupling}
\begin{figure}[!h]
\begin{center}
\begin{tabular}{cc}
\includegraphics[scale=0.35,trim=0 0 0 0,clip=true]{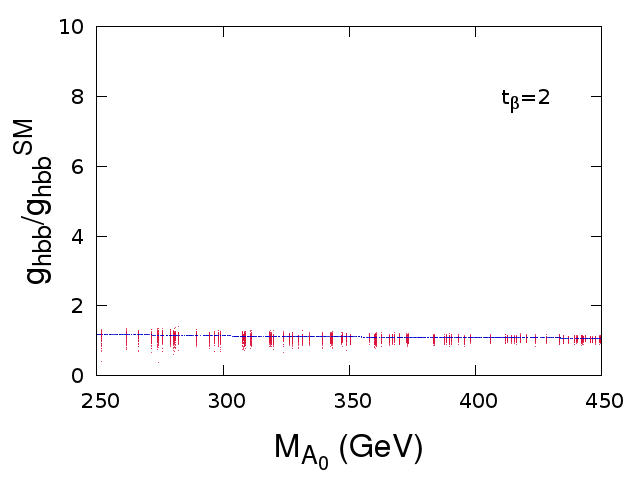}&
\includegraphics[scale=0.35,trim=0 0 0 0,clip=true]{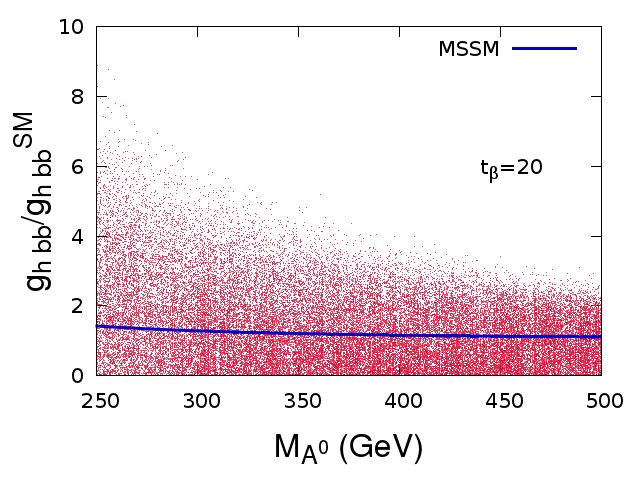}
\end{tabular}
\end{center}
\caption{\label{fig:decoupling} {\em We show the decoupling limit of the MSSM, here obtained by truncating $\ma>250$ GeV in two different $\tb$ regime : on the
left we have $\tb=2$ and on the right $\tb=20$. MSSM points are shown in blue while BMSSM are in red.}}
\end{figure}

Another feature is the fate of the decoupling limit of the MSSM. We have seen that in the limit $\ma\gg\mz$ there was a one-to-one correspondence between the
$\alpha$ angle and the $\beta$ angle. This stems from the relation
 \begin{equation}
  t_{2\alpha}=\frac{\ma^2-\mz^2}{\ma^2+\mz^2}t_{2\beta}
 \end{equation}
which in the limit $\ma\gg\mz$ writes simply as 
\begin{equation}
 \alpha=\frac{\pi}{2}+\beta+\dX+\grandO{\dX^2}
 \label{eq:decoup}
\end{equation}
where $\dX=\sdb\cdb\frac{\mz^2}{\ma^2}$, which is indeed small in the limit $\ma\gg\mz$, whatever the value of $\tb$. However the requirement for a
decoupling is stronger than $|\dX|<1$, indeed this perturbative expansion must hold when taking the sine and cosine of $\alpha$ :
\begin{eqnarray*}
 \sa&=&\cb-\sb\dX+\grandO{\dX^2}\\
 \ca&=&-\sb-\cb\dX+\grandO{\dX^2}
\end{eqnarray*}
which can be turned in 
\begin{eqnarray*}
 \sa&=&\cb(1-\tb\dX)+\grandO{\dX^2}\\
 \ca&=&-\sb\left(1+\frac{1}{\tb}\dX\right)+\grandO{\dX^2}.
\end{eqnarray*}
So, if we we want the decoupling limit to hold, we also need the condition 
\begin{equation}
\dY=\tb\dX=o\left(1\right)
\end{equation}
in the high $\tb$ limit. This will in particular be needed for the couplings of the light Higgs to down-type quarks, which are proportional to $\sa/\cb$. In
particular we have
$$g_{h\bar bb}=-\frac{\sa}{\cb}y_{b\ \SM}=-(1-\dY)y_{b\ \SM}.$$
In the MSSM the condition is typically satisfied since $\dY_\MSSM\sim-2\frac{\mz^2}{\ma^2}$ at high $\tb$. However this feature is lost in the BMSSM. In this
case we write 
$$\dY=\dY_\MSSM+\dY_\eff$$
and it turns out that $\dY_\eff$ has a $\tb$-enhanced contribution, which is the following :
\begin{equation}
\begin{split}
\delta Y_{\textrm{eff}}\sim\tb\frac{v_0^2}{\ma^2}&\left(-4a_{62}\frac{m_s^2}{M^2}+a_{60}\frac{3\mz^2-\ma^2+4\mu^2}{M^2}\right.\\
&\qquad\left.-4\frac{\mu}{M}\left(2a_{21}+a_{31}+a_{41}-2a_{50}\frac{\mu}{M}+2\zeta_{10}\right)\right)
\end{split}
\end{equation}
where $v_0$ corresponds to the Standard Model Higgs vacuum expectation value. This term will precisely blow up the decoupling in the large $\tb$ limit (apart
from the case where $A_0$ is particularly heavy). Even without speaking of the $\tb$ factor, one notices that the MSSM contribution is driven by
$\frac{\mz^2}{\ma^2}$, whereas the effective one is driven by $\frac{v_0^2}{\ma^2}$, which is substantially larger. This leads to a non-decoupling appearing already with
moderate values for $\tb$. The conclusion is hence that the decoupling region of the BMSSM is much smaller then in the MSSM : all points with moderate $\ma$,
say $\ma<500$ GeV, will be likely to show non-standard couplings for the light Higgs if $\tb$ is higher than 5. As a consequence of this fact, the coupling to
$b$ quarks can be non standard even at a relatively high $\ma$, which we show on figure \ref{fig:decoupling} where we take as an example the comparison between
$\tb=2$ and $\tb=20$.

\subsection{Perturbativity of the effective expansion}
\label{sec:pertubativity}
When we chose the effective path to account for new physics beyond the MSSM, we implicitly set a deal : all new physics effects would be correctly reproduced
if and only if the perturbative expansion in power of $1/M$ would hold. From a naive dimensional analysis point of view this was ensured by taking the high mass
$M$ one order higher than typical scales for Higgs processes. However, this point of view is not an irrefutable argument and besides, we want the truncation at
second order of the Effective Field Theory to be a good approximation of the result to all orders, so it turns out that we have to do a bit more work to be sure
that our expansion is reliable.\\

If a breaking of perturbativity occurs, it should be noticed by the fact that neglected higher order terms such as $\grandO{1/M^n}$ with
$n>2$ get non negligible. If the inclusion of these higher
order terms makes a difference, the expansion is not to be trusted
and such configurations of parameters should be discarded. These effects typically occur when the leading order
contribution is very small or then there is an accidental
suppression that makes the higher order effect important. For
 instance  when $M_{A_0}$ is low and $\tb$ moderate
 to high, then $h$ and $H$ can get nearly degenerate at
 zeroth order in the effective expansion. Hence
 any observable involving
 $\sqrt{m_{H}^2-m_{h}^2}$ will have
 an ill-defined perturbative expansion, since the
 derivatives of the square root near 0 are
 infinite. If the degeneracy is lifted by the effective operators, than the mass difference, and also the mixing will depend on the effective operators at the
leading order. In such a case the perturbative expansion is more likely to be less accurate.\\

 Our check on the accuracy of
 the effective expansion is made on the light Higgs mass
 $m_h$, after all the reason for  including $\grandO{1/M^2}$
was because there were non negligible contributions from this
order to $m_h$. The masses of the CP-even Higgses are computed
from the Higgs
 $2\times 2$ mass matrix $\mathcal{M}$, which in
 the case where the Lagrangian is truncated at
 second order, reads as
\begin{equation}
\label{mhatM2}
 \mathcal{M}_{|2}=\mathcal{M}^{(0)}+\frac{c_5}{M}\mathcal{M}^{(5,1)}+\frac{c_6}{M^2}\mathcal{M}^{(6,1)}+\frac{c_5^2}{M^2}\mathcal{M}^{(5,2)}
\end{equation}
Note that $\mathcal{M}^{(0)}$ is the MSSM loop corrected mass matrix, that is to say the mass matrix generated by the MSSM with loops (without any extra
particles entering in them, of course).
To have this concise form, we have used generic names for
effective coefficients : $c_5$ for the order 5 coefficients
$\{\zeta_{10},\zeta_{11}\}$ and $c_6$ for the $\{a_{ij}\}$. Now,
since $m_h$ is obtained by solving a quadratic equation, using
eq.~\ref{mhatM2} leads to a {\em solution}
that includes contributions up to $\grandO{1/M^4}$
\begin{equation}
 m_h=m_h^{(0)}+\frac{c_5}{M}m_h^{(5,1)}+\frac{c_6}{M^2}m_h^{(6,1)}+\frac{c_5^2}{M^2}m_h^{(5,2)}+\frac{c_5^3}{M^3}m_h^{(5,3)}+\frac{c_5c_6}{M^3}m_h^{\binom{5,1}{
6,1}}+\grandO{1/M^4}
 \label{mh_an}
\end{equation}
So our first check on the effective expansion was to ensure that
the $\grandO{1/M^3}$ terms in eq. \ref{mh_an} were small compared to
the $\grandO{1/M^2}$ terms. We have therefore imposed the condition
\begin{equation}
 \left|\frac{\frac{c_5^3}{M^3}m_h^{(5,3)}}{m_h^{(0)}+\frac{c_5}{M}m_h^{(5,1)}+\frac{c_6}{M^2}m_h^{(6,1)}+\frac{c_5^2}{M^2}m_h^{(5,2)}}\right|+\left|\frac{\frac{
c_5c_6}{M^3}m_h^{\binom{5,1}{6,1}}}{m_h^{(0)}+\frac{c_5}{M}m_h^{(5,1)}+\frac{c_6}{M^2}m_h^{(6,1)}+\frac{c_5^2}{M^2}m_h^{(5,2)}}\right|<0.1
\end{equation}
 so that points that do not pass this condition were discarded.\\

Once this algebraic test was passed, we performed another purely
numerical test based on the explicit inclusion of an operator of
$\grandO{1/M^3}$. This operator being the following
 \begin{equation}
\mathcal{O}_7=\zeta_3\left(H_1\cdot H_2\right)^3.
\end{equation}
Now the CP-even mass matrix becomes\footnote{The
reason why $c_5^3$ and $c_6c_5$ terms pop up is that the
Lagrangian is itself a non-linear function of the K\"ahler
potential and the superpotential.}
 \begin{equation}
 \mathcal{M}_{|3}=\mathcal{M}_{|2}+\frac{c_5^3}{M^3}\mathcal{M}^{(5,3)}+\frac{c_6c_5}{M^3}\mathcal{M}^{\binom{5,1}{6,1}}+\frac{c_7}{M^3}\mathcal{M}^{(7,1)},
\end{equation}
where $c_7$ stands for the new operator $\mathcal{O}_7$. To compute the shift in $m_h$, we have run again \texttt{lanHEP}
including now the new operator, and requiring the Feynman rules to
be computed at order $1/M^3$. This being done we could evaluate
numerically $\mathcal{M}_{|3}$ and compute the resulting value for
$m_h$. To do this, we had to assign a value to the $c_7$
coefficient. We choose it to be the maximum (in absolute value) of
all lower-order coefficients.
$$c_7=max\left(|\zeta_{1l}|,|a_{ij}|\right).$$ The additional
constraint was set as
\begin{equation}
\left|\frac{m_h(\mathcal{M}_{|3})-m_h(\mathcal{M}_{|2})}{m_h(\mathcal{M}_{|2})}\right|<0.1
\end{equation}
Once again, a point failing these two constraints will be
discarded. Those two checks are complementary in the sense that
the first one ensures only that we do not hit any singular point
when computing the Higgs mass, which is essential to use
perturbation theory but does not say much about  the contribution
of higher orders, while the second constraint is an explicit check
that the next order contribution is indeed small enough.

\paragraph*{Conclusion}~\\
We have now succeded in the first step of the phenomenology, which was how to break the electroweak symmetry consistently and how to relate initial parameters
to physical quantities.  We now end up with a parameter space which is made of MSSM soft breaking terms (gaugino masses, sfermions masses and trilinear
couplings), the $\mu$ parameter, $\ma$ and $\tb$ and the effective coefficients. But before going to the analysis of the predictions of such a model, let us see
what kind of extra physics would yield such operators.

\section{UV Completion}
Although the construction of the effective theory followed the path of parametrising an unknown physics, it is not uninteresting to have a look at the other
side of the effective theory, that is how does the expansion looks like when the high energy theory is known. In particular, one can see how different operators
can be generated by different species of new physics, which clarifies the need for having all coefficients uncorrelated.

\subsection{MSSM with extra Singlet}
Let us start with the simplest expansion of the MSSM, that is to say the addition of a chiral singlet superfield $S$. It modifies the
supersymmetric potentials by the following quantities,
\begin{eqnarray}
 K_\add^\tot&=&S^\dag S\\
 W_\add^\tot&=&\1 MS^2+\lambda_S SH_1\cdot H_2+\frac{\lambda'}{3}S^3
\end{eqnarray}
up to supersymmetry breaking terms, which will be introduced later. The $\tot$ index (for total) refers to the UV complete theory as opposed to the $\eff$
index. It is important to stress that this is not the usual NMSSM, for the following reasons : first there is a mass term $\1 MS^2$ in the superpotential, and
secondly the aim of the $\lambda_S SH_1\cdot H_2$ term is not to give the $\mu$ term of the MSSM since this term is already present. On the phenomenological
side it is also quite different, since we have no light state associated to the singlet superfield, after all the whole idea of the effective theory is that
particles are sufficiently massive that we can integrate them out. The reason why we have chosen such a model is that it remains the simplest UV completion of
the BMSSM and shows in a simple way how effective coefficients can be correlated. Those remarks left aside, the supersymmetric equation of motion for the
superfield $S$ can be obtained as
\begin{equation}
 -\frac{1}{4}\bar D^2\bar S+MS+\lambda_SH_1\cdot H_2+\lambda' S^2=0.
\end{equation}
If we now assume that $M$ is large compared to the typical momentum of the processes, we can use a perturbative series with $\bar D^2\bar S\gg MS$, which yield
$S$ at several orders
\begin{eqnarray}
 S^{(0)}&=&\frac{\lambda_S}{M}H_1\cdot H_2\\
 S^{(1)}&=&\frac{1}{4}\frac{\lambda_S}{M^2}\bar D^2\bar H_1\cdot\bar H_2\\
 S^{(2)}&=&\grandO{\frac{1}{M^3}}
\end{eqnarray}
By plugging this in $K_\add$ and $W_\add$, and truncating at second order we get
\begin{equation}
 K_\add^\eff=\frac{\lambda_S}{M^2}|H_1\cdot H_2|^2
\end{equation}
where we recognise the effective operator associated to $a_4$, and
\begin{equation*}
W_\add^\eff=-\1\frac{\lambda_S}{M}\left(H_1\cdot H_2\right)^2
\end{equation*}
where we recognise the effective operator associated to $\zeta_1$. The supersymmetric breaking terms are easily added since they come as multiplicative factor
with a spurion field $X=m_s\theta^2$, as shown here
\begin{eqnarray}
 K_\add^\tot&=&S^\dag S\left(1+\alpha X^\dag X\right)\\
 W_\add^\tot&=&\1 MS^2\left(1+\beta_1 X\right)+\lambda_S SH_1\cdot H_2\left(1+\beta_2 X\right)+\frac{\lambda'}{3}S^3\left(1+\beta_3 X\right)
\end{eqnarray}
which are propagated through the calculations up to the final result
\begin{eqnarray*}
 K_\add^\eff&=&\frac{\lambda_S}{M^2}|H_1\cdot H_2|^2\left(1+(\beta_1-\beta_2)(X+X^\dag)+\left((\beta_1-\beta_2)^2-\alpha\right)X^\dag X\right)\\
 W_\add^\eff&=&-\1\frac{\lambda_S}{M}\left(H_1\cdot H_2\right)^2\left(1+(\beta_1-2\beta_2)X\right).
\end{eqnarray*}

So that we have now generated our effective coefficients as
\begin{eqnarray}
\zeta_{10}&=&-\frac{1}{2}\lambda_S,\quad\zeta_{11}=-\frac{1}{2}\lambda_S(\beta_1-2\beta_2)\\
a_{40}&=&\lambda_S,\quad a_{41}=\lambda_S(\beta_1-\beta_2),\quad a_{41}=\lambda_S((\beta_1-\beta_2)^2-\alpha)
\end{eqnarray}
Note that for this particular set-up, the effective coefficients are not all uncorrelated, for instance we have $\zeta_{10}=-\1 a_{40}$.

\subsection{MSSM with triplets}
Another extension of the MSSM can be obtained by adding chiral superfields $T$ and $T'$ that belong to the triplet representation of $SU(2)$, with hypercharge
$y=1$ and $y'=-1$. This enhances the supersymmetric potentials by
\begin{eqnarray}
 K_\add^\tot&=&T^\dag e^VT+T^{\prime\,\dag}e^{V'}T'\\
 W_\add^\tot&=&MTT'+\lambda_TH_1\cdot TH_1+\lambda_{T'}H_2\cdot T'H_2.
\end{eqnarray}
To express the equations of motion we have to introduce the coordinates $T=(T_a)$ for each superfield and the $SU(2)$ generators in the fundamental
representation $(\tau_a)$ as well as the adjoint representation $(X_a)$ (so that $T=T_aX_a$) :
$$-\frac{1}{4}\left(e^V\bar D^2\bar T\right)_a+MT'_a+\lambda_TH_1\cdot\tau_aH_1=0$$
and a similar one for $T_a$, which yields at the lowest order
\begin{equation}
T_a=-\frac{\lambda_{T'}}{M}H_2\cdot\tau_aH_2,\qquad T'_a=-\frac{\lambda_{T}}{M}H_1\cdot\tau_aH_1.
\end{equation}
When going back to $K_\add$, we will use the following identity
\begin{equation}
\int d^4\theta (H\cdot\tau_aHX^a)^\dag e^{V}(H\cdot\tau_aHX^a)=
\1\int d^4\theta \left(H^\dag e^{V}H\right)^2
\end{equation}

so that we eventually end up with 
\begin{eqnarray}
K_\add^\eff&=&\1\frac{\lambda_{T'}^2}{M^2}\left(H_2^\dag e^{V_2}H_2\right)^2+\1\frac{\lambda_{T}^2}{M^2}\left(H_1^\dag e^{V_1}H_1\right)^2\\
W_\add^\eff&=&\frac{\lambda_T\lambda_{T'}}{8M^2}\left(H_1{}_\cdot H_2\right)^2
\end{eqnarray}
where we recognise the effective operators associated to $a_1$ and $a_2$, and $\zeta_1$. The supersymmetry-breaking operators can then be added in the same way
as in the previous paragraph.

\subsection{U(1)'MSSM}
Instead of introducing more matter, one can also extend the gauge group. The simplest realisation is to add a $U(1)$ factor gauged via a vector superfield
$V'$, as exemplified in reference \cite{dine_bmssmhiggs_0707}. In order to give a high mass to this field we need to add some scalar fields to break the gauge
symmetry, so we will add $\phi_+,\phi_-,\phi_0$ with charges $1,-1,0$. The additional part of the potentials reads
\begin{eqnarray}
K_\add&=&\phi_+^\dag e^{V'}\phi_++\phi_-^\dag e^{-V'}\phi_-\\
W_\add&=&\phi_0\left(\phi_+\phi_--M^2\right)
\end{eqnarray}
Note that there is also a modification of the MSSM K\"ahler potential since the Higgs superfields $H_1,H_2$ acquire a charge under $V'$ : $q_1$ and $q_2$. At
the first order in $V'$ we can write this change as
\begin{equation}
K_\MSSM\rightarrow K_\MSSM(V')=K_\MSSM(0)+V'\frac{\delta K_\MSSM}{\delta V'}(0)+\grandO{V^{\prime\,2}}
\end{equation}
Turning first to the new chiral superfields, their equations of motion yield
\begin{equation}
\begin{matrix}
 -\frac{1}{4}e^{V'}\bar D^2\bar\phi_++\phi_0\phi_-&=&0\\
 -\frac{1}{4}e^{-V'}\bar D^2\bar\phi_-+\phi_0\phi_+&=&0\\
 -\frac{1}{4}\bar D^2\bar\phi_0+\phi_+\phi_--m^2&=&0
\end{matrix}
\}
\phi=0,\ \phi_+\phi_-=M^2
\end{equation}
If we choose the solution $\phi_+=\phi_-=M$, the additional K\"ahler potential is
$$K_\add=M^2\left(e^{V'}+e^{-V'}\right)$$
The equation of motion for $V'$ will mix this term and the Higgs part of the MSSM one $K_\MSSM(V')$:
$$2M^2+\frac{\delta K_\MSSM}{\delta V'}(0)=0$$
This allows us to write eventually the effective K\"ahler potential as
\begin{eqnarray*}
K_\eff&=&M^2\left(e^{V'}+e^{-V'}\right)+V'\frac{\delta K_\MSSM}{\delta V'}(0)\\
&=&\frac{1}{4M^2}\left(\frac{\delta K_\MSSM}{\delta V'}(0)\right)^2-\frac{1}{2M^2}\left(\frac{\delta K_\MSSM}{\delta V'}(0)\right)^2\\
K_\eff&=&-\frac{1}{4M^2}\left(\frac{\delta K_\MSSM}{\delta V'}(0)\right)^2
\end{eqnarray*}
Since the derivative of $K_\MSSM$ is the following
$$\frac{\delta K_\MSSM}{\delta V'}(0)=q_1H_1^\dag e^{V_1}H_1+q_2H_2^\dag e^{V_2}H_2$$
we get to the final result
\begin{equation}
K_\eff=-\frac{1}{4M^2}\left(q_1H_1^\dag e^{V_1}H_1+q_2H_2^\dag e^{V_2}H_2\right)^2
\end{equation}
which will give contribution to the $a_1,a_2$ and $a_3$ coefficients. Once again, the supersymmetry-breaking part is easily obtained by the mutiplication with
spurions.


%% file: chapter6.tex
\chapter{Constraining a supersymmetric model}

\minitoc\vspace{1cm}

Before trying to ask our first question ``what do we predict in present and future experiments?'', it is fair to consider the more mundane issue : ``what can we
predict that is not already ruled out by past experiments?''. It is the burden of phenomenologists to go to great lengths on this point before entertaining new
trendy hypothetical searches. So, let us forget for a while what we are going to scrutinize in the future and first focus on the past. We will finally be
rewarded by the fact that, after having spent some time on those verifications, predictions in view of future measurements will be quite facilitated.

\section{Preliminary : Computing observables}
As I have stressed in chapter 5, there is no exact computation for a prediction. And since applying a constraint is merely comparing a prediction with an
observation, we are bound to tackle the issue of accuracy. We have seen that, in order to have a hint of validity, physical observables had be computed to a
certain order in the loop and effective expansion, and we have also characterised the precision on the effective side. The same question arises in the loop
expansion. Loop effects may be small in the electroweak regime, but this is a regime where measurements are extremely precise. Higgs Physics at hadron colliders
may be a rough sector, with an experimental uncertainty up  to 10 to 20\%, but then the loop effects reach often more than 50\%. This explicitly raises the
issue : at which point can we rely on the perturbative Feynman expansion? There is to my knowledge no definite answer for realistic models, and in most cases
one simply uses empirical arguments such as ``if the next order to be computed is 
small, than the perturbative expansion is safe and the result is reliable''. It would also seem that this question is model-dependent, but fortunately in the
BMSSM framework, the discussion can be much eased by noting that the most important loop contributions will be Standard Model ones, or MSSM ones.\\

\subsection{Higgs observables}
It is a well-known fact that Higgs phenomenology in supersymmetry cannot be
separated from loop computations. However, we cannot reach the
state of the art predictions obtained in the Standard Model with event
generators and multi-loop computation tools starting from scratch
with a brand new model. Besides, the accuracy would be unnecessary
since we intend to vary freely our effective coefficients in the
interval $[-1,1]$. Depending on the observable, some
radiative corrections will be computed, as we describe now. First,
let me recall that we can write any observable as a double
expansion, the loop expansion, and the effective one. We will classify
those expansions as follows :

\begin{itemize}
 \item Decorrelated expansion. We mean by this that effective
 couplings and MSSM/QCD loops do not interfere, allowing thus to
 write
 $\O=\O^{(0)}+\delta_{\textrm{loop}}\,\O+\delta_{\textrm{eff}}\,\O$,
 and we can take separately the prediction for
 $\delta_{\textrm{eff}}\,\O$ from a tree-level code and
 $\delta_{\textrm{loop}}\,\O$ from an MSSM-dedicated code. This
 will be the case of masses for instance :
 $\delta_{\textrm{eff}}\,m$ are taken from the \lanHEP\ output
 whereas $\delta_{\textrm{loop}}\,m$ are obtained from a spectrum
 calculator code, namely \texttt{Suspect} \cite{suspect}. Note that for the computation of the $\delta_\eff m$ corrections we will use the formulas derived in
chapter 3, where the zeroth order of the mass matrix will be the mass matrix with radiative corrections obtained from \texttt{Suspect}. For some observables
such as
 decay to neutralino/charginos, $\delta_{\textrm{loop}}\,\O$ will
 be neglected.
 \item Factorisable expansion. This case, which is an example of the recasting discussion held in chapter 5, arises when
 we can factorise the effective expansion from the loop expansion.
 That is to say that the scale factor between the tree-level
 amplitude of the MSSM and the tree-level amplitude in the
 effective theory is the same as the scale factor between the
 one-loop amplitudes in both theories, and so up to all orders. Hence this scale factor can safely be applied
 to the full cross-section. This is equivalent to requiring that
 both theories have the same K-factor, for the given observable.
 This is the case of most of the Higgses decays, for instance the
 partial decay width of the lightest Higgs  into $b$ fermions.
 \begin{equation}
\Gamma_{h\rightarrow\oo{b}b}=R_{g_{h\oo{b}b}\ \textrm{eff}}\times\Gamma_{h\rightarrow\oo{b}b\ \textrm{loop}}^{\text{MSSM}}\label{approx_hbb}
 \end{equation}
 where $R_{g_{h\oo{b}b}\ \textrm{eff}}$ is the ratio of the $h\oo{b}b$ coupling in the effective theory to the MSSM one and $\Gamma_{h\rightarrow\oo{b}b\
\textrm{loop}}^{\text{MSSM}}$ is the MSSM partial width.\footnote{Strictly speaking, it is the MSSM partial width obtained by replacing the MSSM masses with
effective masses.}
 \item Nested expansion. This happens when the loop contribution
 cannot be neglected, but cannot be factorised either. This
 is the case for observables such as Higgs decays to photons or gluons, which occur first at the one-loop level. In those cases, since the
 effective rescaling of the coupling $g_{h\oo{b}b}$ and
 $g_{h\oo{t}t}$ are different, the effective scale factor cannot be
 factorised. Then relations such as eq. \ref{approx_hbb} do not
 apply and the computation has to be done with the specific ratio,
 which usually means modifying some MSSM-dedicated codes such as \texttt{HDecay}\cite{hdecay}, as will
 be detailed later.
\end{itemize}

\paragraph{Higgs observables}~\\
Since they will have a special role in the discussion on the BMSSM, I will detail a bit the observables related to the Higgs. At the end of the day, the
observables we need are the masses and the product of the production cross-section by the branching ratio to the final state considered : $\sigma\times BR$.
Concerning the masses, they are computed with a reasonable accuracy on both sides of the expansion (\textit{i.e.} the loop and effective corrections), and in a reasonable time since the effective shift is an
analytic formulae, and the loop shift has to be re-evaluated only when changing MSSM parameters.
\begin{equation}
 m=m_\text{loop}+\delta m_{\text{eff}}
\end{equation}
For cross-sections and branching ratios, the experimental results
are usually given rescaled from the SM prediction, so in the case
of a factorisable expansion (with respect to the standard model
case) the loop precision is more than sufficient and the
computation is straightforward since effective ratios are given as
analytic formulas. Examples are decays to fermions or weak gauge
bosons :
\begin{IEEEeqnarray}{cc}
 \frac{\Gamma_{h\rightarrow\tau\tau}}{\Gamma_{h\rightarrow\tau\tau\SM}}=\left(\frac{g_{h\oo{\tau}\tau}}{g_{h\oo{\tau}\tau\SM}}\right)^2 \\
 \frac{\Gamma_{h\rightarrow WW}}{\Gamma_{h\rightarrow WW\SM}}=\left(\frac{g_{hWW}}{g_{hWW\SM}}\right)^2
\end{IEEEeqnarray}
This is also the case of the vector boson fusion (VBF) process,
associated vector boson production and heavy quarks associated
production.
\begin{equation}
 \frac{\sigma_{ZH}}{\sigma_{ZH\ SM}}=\left(\frac{g_{hZZ}}{g_{hZZ\ SM}}\right)^2
\end{equation}
In the case of nested expansion or MSSM factorisable one, we have
used a modified version of \texttt{HDECAY} \cite{hdecay}. This
will be used for decays where supersymmetric loop contributions are not
negligible and also for loop-induced decays. It is for instance used
to compute $\Gamma_{h \to\gamma\gamma}$. Finally, observables
where no explicit loop computation was needed have been computed
with \texttt{CalcHEP} (\cite{calchep}) : this is the case for
instance of Higgs decay to other Higgses, where the loop
correction can be reproduced by an effective potential (see
\cite{boudjema_efflambda}).\\ Following this choice, several
approximations have been made. The most important is the gluon
fusion cross-section. Since it is a nested expansion case, we started
with a modified version of \texttt{Higlu} (\cite{higlu}), an
MSSM-dedicated code. However, the integration over the parton
density functions set was an unacceptable lost in time,
considering that the effective parameter space was already
22-dimensional. Hence we used the approximation :
\begin{equation}
\sigma_{gg\rightarrow h}=\frac{\Gamma_{h\rightarrow gg}}{\Gamma_{h\rightarrow gg\ SM}}\sigma_{gg\rightarrow h\ SM}.
\label{ggh}
\end{equation}
Similarly if we wanted to consider SUSY loop corrections to the
Vector Boson Fusion (VBF), associated vector boson production and
heavy quark associated production, we could have used results
from MSSM-dedicated code, with the same issue on pdf, thus what we have done is simply to use the Standard Model result rescaled by the square of the ratio of the coupling $g_{hZZ}$. \\
About the branching ratios, using \texttt{HDECAY} did not
reproduce the best results for decays into off-shell gauge bosons,
but using a more precise tool such as \texttt{Prophecy4f}
(\cite{prophecy4f}) would have increased a lot the computation time,
since it goes up to event generation. At the end of the day,
production cross-sections were rescaled from the SM predictions
(avoiding hence the use of specific codes as \texttt{Higlu}, or
codes for VBF), and decays were rescaled from MSSM predictions
obtained with \texttt{HDecay} (avoiding the use of
\texttt{Prophecy4f}), which allowed for a large gain in computing
speed for a very moderate precision loss.\\

\begin{figure}[!h]
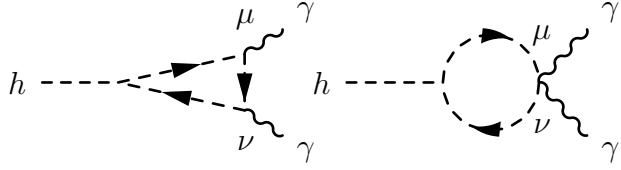

\begin{center}
\begin{tabular}{cc}
\fmfframe(1,20)(1,20){
\begin{fmfchar*}(100,40)
	\fmflabel{$h$}{pa}
	\fmflabel{$\gamma$}{pb}
	\fmflabel{$\gamma$}{pc}
	\fmfleft{pa}
	\fmf{dashes}{pa,v1}
	\fmf{scalar,tension=0.3}{v1,v2,v3,v1}
	\fmf{photon,label=$\mu$,label.side=left}{v2,pb}
	\fmf{photon,label=$\nu$}{v3,pc}
	\fmfright{pc,pb}
\end{fmfchar*}}
	&
\fmfframe(1,20)(1,20){
\begin{fmfchar*}(100,40)
	\fmflabel{$h$}{pa}
	\fmflabel{$\gamma$}{pb}
	\fmflabel{$\gamma$}{pc}
	\fmfleft{pa}
	\fmf{dashes}{pa,v1}
	\fmf{scalar,tension=0.5,left}{v1,v2,v1}
	\fmf{photon,label=$\mu$,label.side=left}{v2,pb}
	\fmf{photon,label=$\nu$,label.side=right}{v2,pc}
	\fmfright{pc,pb}
\end{fmfchar*}}
\end{tabular}
\end{center}
\caption{\label{diag:Hp_MSSM} {\em Charged Higgs contribution to the $h\to\gamma\gamma$ process in the MSSM.}}
\end{figure}

\paragraph{Higgs decay to photons :}~\\
It is not always straightforward to include the effective operators in the case of a nested expansion. This is typically the case for loop diagrams, and
specifically the decay of a Higgs scalar to two photons exhibits an interesting behaviour. To wit, we will now see an example of new Lorentz structures (in
particular derivative couplings) that are introduced by the effective operators with possible new divergences. This effect stems from the charged Higgs loops,
which are shown in figure \ref{diag:Hp_MSSM} in the MSSM case.

In the BMSSM, the triple Higgs coupling $hH^+H^-$ gets modified by the effective operators : one ends up with
\begin{eqnarray}
g_\MSSM\Hp\Hm h&\rightarrow& g_{\textrm{MSSM}}\,(1+\delta_{\textrm{A}})\,\Hp\Hm h\nonumber\\
&&\hspace*{-2cm}+\delta_B\,\left(2(\partial_{\mu}\Hp)(\partial^{\mu}\Hm) h+(\partial_{\mu}h)\,(\partial^{\mu}\Hp)
\Hm+(\partial_{\mu}h)\,(\partial^{\mu}\Hm)\Hp\right)
\label{eq:Hdecg}
\end{eqnarray}
where $\delta_A$ and $\delta_B$ are suppressed by the heavy scale $M$. We notice at once that the $\delta_A$ is an overall factor of the MSSM amplitude. Working
out the implications of $\delta_B$ can be done in two ways, either by the use of the equations of motion or by computing the new diagrams. Though the first
method tends to be faster we will still present both (more from a pedagogical viewpoint) and show how they lead to a unique result. Starting on the first method
we get
\begin{eqnarray*}
\left(2(\partial_{\mu}\Hp)(\partial^{\mu}\Hm) h\right.&&\left.+(\partial_{\mu}h)\,(\partial^{\mu}\Hp) \Hm+(\partial_{\mu}h)\,(\partial^{\mu}\Hm)\Hp\right)\\
&&=-\left((\partial^2\Hp)\Hm h+(\partial^2\Hm)\Hp h\right)\\
&&=-\left((D^2\Hp)\Hm h+(D^2\Hm)\Hp h\right)\\
&&=-2M_{\Hp}^2
\end{eqnarray*}
We have used at the next to last line the fact that, since the derivative stems from a gauge theory, it must appear in the Lagrangian as a covariant derivative.
We have used the off-shell equations of motion (since the charged Higgs is off-shell in the loop) at the last line, but we have omitted additional terms with
other Higgs fields that appear in the equation of motion but would not contribute to the process at the same order. This tells us that no new divergences
appear, and we have a simple rescaling of the MSSM coupling as 
$$g_\MSSM\to g_\MSSM(1+\delta_A-2\delta_BM_{\Hp}^2)$$

The same result can be checked by computing explicitly the different diagrams generated by the new Lorentz structures, which is done by expanding the derivative
of eq. \ref{eq:Hdecg} in the covariant derivative $D_{\mu}H^{\pm}=\left(\partial_{\mu}\pm ieA_{\mu}\right)H^{\pm}$ (leaving alone the weak part, which is of no
relevance here). We get then the following new vertices :
\begin{eqnarray}
&&\delta_B\left(2\partial_{\mu}\Hp\partial^{\mu}\Hm h+\partial_{\mu}h\,\partial^{\mu}\Hp \Hm+\partial_{\mu}h\,\partial^{\mu}\Hm\Hp\right)\nonumber\\
  &\rightarrow&\delta_B\left(2(\partial_{\mu}\Hp\partial^{\mu}\Hm h+ieA^{\mu}\partial_{\mu}\Hm\Hp h-ieA^{\mu}\partial_{\mu}\Hp\Hm h+e^2A^{\mu}A_{\mu}\Hp\Hm
h\right)\nonumber\\
  &&+\partial_{\mu}h\,\partial^{\mu}\Hp\Hm+\partial_{\mu}h\,\partial^{\mu}\Hm\Hp)\nonumber\\
  &\rightarrow&\delta_B\left(-\left[2k_+.k_-+k.k_++k.k_-\right]\Hp\Hm h+2e(k_--k_+)_{\mu}A^{\mu}\Hp\Hm h+2e^2A^{\mu}A_{\mu}\Hp\Hm h\right)\nonumber\\
  &\rightarrow&\delta_B\left((D_++D_-)\Hp\Hm h+2e(k_--k_+)_{\mu}A^{\mu}\Hp\Hm h+2e^2A^{\mu}A_{\mu}\Hp\Hm h\right)\nonumber\\
  &&+2\delta_BM_{\Hp}^2\Hp\Hm h
\end{eqnarray}
where $k_-,k_+,k$ stand for the momentum of $\Hm,\Hp,h$, respectively and $D_{\pm}=(k_{\pm}^2-M_{\Hp}^2)$. To obtain the last line we have added and subtracted
$M_{\Hp}^2$ so that the propagator appears. The last part ($2\delta_BM_{\Hp}^2\Hp\Hm h$) falls into the multiplicative factor over the MSSM amplitude
\begin{equation}
\delta_A\rightarrow\delta_A'=\delta_A-2\delta_BM_{\Hp}^2.
\label{eq:Hp_delta_A}
\end{equation}
As for the new vertices we got, they yield new diagrams shown in figure \ref{diag:Hp_BMSSM} :

\begin{figure}[!h]
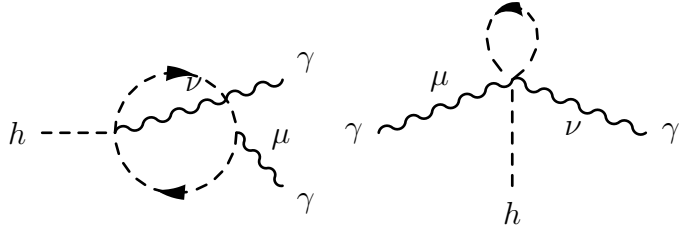

\begin{center}
\vspace{5mm}
\begin{tabular}{ccc}
\fmfframe(1,20)(1,20){
\begin{fmfchar*}(100,40)
	\fmflabel{$h$}{pa}
	\fmflabel{$\gamma$}{pb}
	\fmflabel{$\gamma$}{pc}
	\fmfleft{pa}
	\fmf{dashes}{pa,v1}
	\fmf{scalar,left,tension=0.3}{v1,v2,v1}
	\fmfright{pc,x,pb}
	\fmf{phantom}{v2,x}
	\fmffreeze
	\fmf{photon,label=$\mu$}{v2,pc}
	\fmf{photon,label=$\nu$}{v1,pb}
\end{fmfchar*}}
	&\hspace{5mm}&
\fmfframe(1,20)(1,20){
\begin{fmfchar*}(100,40)
	\fmflabel{$h$}{pa}
	\fmflabel{$\gamma$}{pb}
	\fmflabel{$\gamma$}{pc}
	\fmfbottom{pa}
	\fmf{dashes,tension=-2}{pa,v1}
	\fmf{scalar,tension=1.2}{v1,v1}
	\fmf{photon,label=$\mu$,tension=2}{pb,v1}
	\fmf{photon,label=$\nu$,tension=2}{v1,pc}
	\fmfleft{pb}
	\fmfright{pc}
\end{fmfchar*}}
\end{tabular}
\end{center}
\caption{\label{diag:Hp_BMSSM} {\em New diagrams introduced in the BMSSM in the process $h\to\gamma\gamma$.}}
\end{figure}

It seems that the computation is going to be much more involved since we now have to sum the amplitudes of all diagrams. In particular, there will be divergent
parts associated to each of them, and it is crucial that the sum of all divergent parts vanishes. We will now compute the effective contribution to the
different amplitudes related to the diagrams appearing in figure \ref{diag:Hp_MSSM} and \ref{diag:Hp_BMSSM}. For the sake of simplicity we will omit some
numerical factors that are the same in the four diagrams. We denote by $\A$ the loop integrand, that is to say the loop form factor before integration over the
internal momentum.\\

\subparagraph{Effective contribution to diagram MSSM 1}~\\
\begin{figure}[!h]
\begin{center}
\vspace{4mm}
\fmfframe(1,5)(1,5){
\begin{fmfchar*}(80,40)
	\fmflabel{$h$}{pa}
	\fmflabel{$\gamma$}{pb}
	\fmflabel{$\gamma$}{pc}
	\fmfleft{pa}
	\fmf{dashes,tension=1.3}{pa,v1}
	\fmf{scalar,tension=0.3,label=$k+p_1$,label.side=left}{v1,v2}
	\fmf{scalar,tension=0.3,label=$k$,label.side=left}{v2,v3}
	\fmf{scalar,tension=0.3,label=$k-p_2$,label.side=left}{v3,v1}
	\fmf{photon,label=$p_1^{\mu}$,label.side=left}{v2,pb}
	\fmf{photon,label=$p_2^{\nu}$}{v3,pc}
	\fmfright{pc,pb}
\end{fmfchar*}}\vspace{1cm}\\
$\mathcal{A}_1=2\delta_B\,(ie)^2(4k^{\mu}k^{\nu})\left(\frac{1}{D_0D_{p_2}}+\frac{1}{D_0D_{p_1}}\right)$
\end{center}
\end{figure}

The factor 2 comes from exchanging the external photons (which leaves the amplitude unchanged), and the terms containing $p_1^{\mu}$ or $p_2^{\nu}$ have been
removed since their products with external polarisation $\epsilon^{\mu}$ vanish. The notation $D_p$ stands for the propagator of $H^+$ evaluated at the momentum
$k+p$, where $k$ is the internal momentum running in the loop.\\

\subparagraph{Effective contribution to diagram BMSSM 1}~\\
\begin{figure}[!h]
\begin{center}
\fmfframe(1,20)(1,20){
\begin{fmfchar*}(100,40)
	\fmflabel{$h$}{pa}
	\fmflabel{$\gamma$}{pb}
	\fmflabel{$\gamma$}{pc}
	\fmfleft{pa}
	\fmf{dashes}{pa,v1}
	\fmf{scalar,left,tension=0.3,label=$k$}{v1,v2}
	\fmf{scalar,left,tension=0.3,label=$k-p_2$}{v2,v1}
	\fmfright{pc,x,pb}
	\fmf{phantom}{v2,x}
	\fmffreeze
	\fmf{photon,label=$p_1^{\mu}$}{v2,pc}
	\fmf{photon,label=$p_2^{\nu}$,label.side=right}{v1,pb}
\end{fmfchar*}}\vspace{1cm}\\
$\mathcal{A}_2=-2\delta_B\,(ie)^2(4k^{\mu}k^{\nu})\left(\frac{1}{D_0D_{p_2}}+\frac{1}{D_0D_{p_1}}\right)$
\end{center}
\end{figure}
The terms with $p_1^{\mu}p_1^{\nu}$ or $p_2^{\mu}p_2^{\nu}$ have also been removed (indeed the term $\frac{2(k^{\nu}p_2^{\mu}+k^{\mu}p_2^{\nu})}{D_0D_{p_2}}$
will yields only terms with $p_2^{\mu}p_2^{\nu}$, a similar point goes for $p_1$), leaving only the $k^{\mu}k^{\nu}$ term.

\subparagraph{Effective contribution to diagram MSSM 2}~\\
\begin{figure}[!h]
\begin{center}
\fmfframe(1,20)(1,20){
\begin{fmfchar*}(100,40)
	\fmflabel{$h$}{pa}
	\fmflabel{$\gamma$}{pb}
	\fmflabel{$\gamma$}{pc}
	\fmfleft{pa}
	\fmf{dashes}{pa,v1}
	\fmf{scalar,tension=0.5,left,label=$k+p_1$}{v1,v2}
	\fmf{scalar,tension=0.5,left,label=$k-p_2$}{v2,v1}
	\fmf{photon,label=$p_1^{\mu}$,label.side=left}{v2,pb}
	\fmf{photon,label=$p_2^{\nu}$,label.side=right}{v2,pc}
	\fmfright{pc,pb}
\end{fmfchar*}}\vspace{1cm}\\
$\mathcal{A}_3=\delta_B\,2e^2\frac{1}{D_0}$
\end{center}
\end{figure}

\subparagraph{Effective contribution to diagram BMSSM 2}~\\
\begin{figure}[!h]
\begin{center}
\fmfframe(1,20)(1,20){
\begin{fmfchar*}(100,30)
	\fmflabel{$h$}{pa}
	\fmflabel{$\gamma$}{pb}
	\fmflabel{$\gamma$}{pc}
	\fmfbottom{pa}
	\fmf{dashes,tension=-2}{pa,v1}
	\fmf{scalar,tension=1.2,label=$k$}{v1,v1}
	\fmf{photon,label=$\mu$,tension=2}{pb,v1}
	\fmf{photon,label=$\nu$,tension=2}{v1,pc}
	\fmfleft{pb}
	\fmfright{pc}
\end{fmfchar*}}\vspace{1cm}\\
$\mathcal{A}_4=-\delta_B2e^2\frac{1}{D_0}$
\end{center}
\end{figure}

One directly sees that diagrams MSSM 1 and BMSSM 1 cancel and so do diagrams MSSM 2 and BMSSM 2. Hence the $\delta_B$ contribution to the amplitude
vanishes. We can then write our final result as
\begin{equation}
\mathcal{A}_{\textrm{eff}}=\mathcal{A}_{\textrm{MSSM}}(1+\delta_A')
\end{equation}
where $\delta_A'$ was defined in eq.\ref{eq:Hp_delta_A}. This result has the interesting property of being only a rescaling of a MSSM loop amplitude, which
simplifies our task. In practice we derive the value of $\delta_A'$ from the Feynman rules obtained by \lanHEP, and we plug this into the loop form factor
computed via \HD.\\

\subsection{Divergences in an Effective Field Theory}
It would seem that the loop computation we have done for the process $h\to\gamma\gamma$ goes against the idea developed in section \ref{sec:eft_ol} that since
effective operators are non renormalisable they should not be included in a loop computation. This is however not precise enough, indeed while the operators in
term of superfields (that is to say, as they appear in the superpotential and the K\"ahler potential) are indeed non-renormalisable, it does not means that the
operators that are generated in the lagrangian of fields (defined in eq. \ref{full_lag}) are themselves non-renormalisable. Indeed it turns out that many
effective contributions are simply rescaling of the MSSM vertices, which hence do not spoil the renormalisation. The trouble arise with effective terms that
will present new Lorentz structure with additional derivatives. We have seen that in the case of the decay to two photons the gauge invariance prevented the
appearance of divergences. We will see in a while that new Lorentz 
structures also enter the chargino-chargino-Higgs vertex which is in particular used for the penguin contribution to the $\Bsmu$ observable. We will show that
still, in such a case the additional Lorentz structures can be eliminated by the use of the equations of motion and one ends up with a rescaling of the MSSM
coupling. Based on this argument, we checked that all the loop observables we compute do not have extra divergences coming from effective terms, hence we do not
have to modify the renormalisation of the MSSM.

Having decided how we would compute observables in the BMSSM framework, it is now time to compute effectively those values and to compare them to experiments,
which is our next topic.

\section{Precision measurements}
\subsection{ElectroWeak Precision Test}
The precise measurements on the Z pole that have been carried out at LEP have shown that any new physics theory had to have very little impact on the physics at
this scale. In order to quantify possible deviations from such predictions, conventional variables have been defined : those are the $\epsilon$ variables (or
equivalently the $S,T,U$ variables). A short description of these variables can be found in Appendix C.1, or in more details in \cite{altarelli_epsilon}. Their
definition is :

\begin{eqnarray}
\epsilon_1&=&\Delta\rho\\
\epsilon_2&=&c_0^2\Delta\rho+\frac{s_0^2}{c_0^2-s_0^2}\Delta r_w-2s_0^2\Delta k\\
\epsilon_2&=&c_0^2\Delta\rho+(c_0^2-s_0^2)\Delta k
\end{eqnarray}
where the different quantities appearing have been defined in the Appendix C.1. Those $\epsilon$ variables have the property of being zero in the pure Standard
Model at tree-level, hence they strictly characterise the deviations obtained from this lowest order, either by loop corrections or new physics corrections. In
the BMSSM prediction we have an effective contribution equals to
\begin{equation}
 \delta \epsilon_1=4e^2\frac{\mw^2(\mw^2-\mz^2)}{\mz^2M^2}\left(a_{10}\tb^{-4}-a_{30}\tb^{-2}+a_{20}\right)=\1\delta\epsilon_2=\delta\epsilon_3.
\end{equation}

\subsection{The anomalous muon magnetic moment}
The anomalous magnetic moment is a famous example of a precision test. Indeed it has been measured with a very good precision for electrons and muons, and in
the muon case it yields the result
$$g_\mu=2\left(1+(1165920.80\pm0.63)10^{-9}\right).$$
This measurement reveals the structure of the vertex $\bar\mu\gamma\mu$, and in particular shows that it is away from its tree-level value which is 2. However,
this was expected since we know that loop corrections are likely to modify the strength of a coupling. What is really interesting in this observable is the
amazing precision that has been achieved on the computation of this anomalous part. In fact this result is so precise that it is quite a challenge to have a
theoretical prediction that matches the accuracy. In the Standard Model case, it implies to take the Feynman expansion up to order eight, which is probably one of the most precise calculation so far. The prediction is 
$$g_\mu^\SM=2\left(1+(1165918.90\pm0.44)10^{-9}\right).$$ 
One notices that both results differ by $2\,10^{-9}$, where the errors are $0.63\,10^{-9}$ and $0.44\,10^{-9}$ on the experimental and the theoretical side,
respectively. Hence, this may indicate the existence of new particles, however the disagreement is somehow too small to call for a discovery (we will see in the
next chapter the conventions to define a small disagreement and a significant disagreement). It is usually a test to be carried on all models beyond the
Standard model that introduce charged particles at a moderate mass scale.

\subsection{Flavour physics}
The last set of precision measurements we will use comes from flavour physics, and in particular $B$ physics, which studies the behaviour of hadrons that
contain the $b$ quark. We will in particular focus on the two rare processes $\Bsmu$ and $\Bsg$. Those processes have been measured with a significant accuracy,
and we now have the following bounds
\begin{eqnarray}
BR(\Bsmu)&<&4.7\,10^{-9}\ \text{LHC}_b\cite{lhcb_bsmu_1fb}\\
BR(\Bsg)&=&(3.55\pm0.16\pm0.09)\,10^{-4}\ \text{HFAG\footnote{Heavy Flavour Averaging Group}}\cite{hfag_bsg}
\end{eqnarray}
The computation of the predicted branching ratios in the MSSM are detailed in Appendix C.2, so I will directly present the deviations obtained in the BMSSM
framework.

\subsubsection{Prediction for $\Bsmu$}
The global picture of the calculation is exactly the same as in the MSSM : the decay is computed in the effective field theory, using Wilson coefficients
evaluated at a high scale. Since we are not introducing new particles, it would seem that we have to compute the same set of diagrams as in the MSSM case. The
situation would be different if extra vertices could also enter the process and lead to new topologies, but since all those extra vertices concern mainly Higgs
fields it does not happen (at least in the one-loop diagrams). We end up with a similar set of diagrams as in the MSSM case, up to the following differences :
\begin{itemize}
 \item The mass and mixing matrices of Higgs, squarks, charginos are changed.
 \item The weak and Yukawa couplings constant are altered.
 \item The expression for triple Higgs couplings $H^+H^-\Phi$ and Higgs to charginos $\chjm\Phi\chip$ are modified.
\end{itemize}
The last point being part of the penguin contribution. However the penguin diagram with the $H^+H^-\Phi$ vertex do not contribute at leading order in $\tb$ in
the MSSM limit, so we will discard this contribution. Most of the other changes are readily implemented since they do not alter the loop calculation but only
overall factors of gauge invariant quantities. The only difficult point lies in the $\chjm\Phi\chip$ vertex of the chargino-squark penguin loop contribution.
Indeed the MSSM expression for such a coupling simply is, at leading order in $\tb$
\begin{equation}
 \L_{C/S}\to -g_1\sqrt{2}\chjm\left(a\PL\pm b\PR\right)\chip\ \phi
\end{equation}
where $a,b$ are functions of the mass and mixing, and the $C/S$ (related to the $\pm$) show the CP-charge of the scalar Higgs. In particular the MSSM coupling
has the characteristic that it couples a left-handed chargino with a right-handed one. The BMSSM expression is more involved since we now have
\begin{eqnarray}
 \L_{C/S}\qquad\to& \chjm\left(a_{C/S}+b_{C/S}(\kslash_1-\kslash_2)+c_{C/S}\kslash_\phi\right)\PL\chip\ \phi\nonumber\\
 &\hspace*{10mm}+ (\chjm\left(a'_{C/S}+b'_{C/S}(\kslash_1-\kslash_2)+c'_{C/S}\kslash_\phi\right)\PR\chip\ \phi
\end{eqnarray}
which quite changes the picture since the coefficients are now different for CP-even and CP-odd cases, and that some derivatives have entered the coupling. Note
that the weak coupling $g_1$ is no more a common factor for the total coupling since some BMSSM operators will introduce terms that do not stem from gauge
interactions. At this point it seems difficult to use the MSSM loop calculation and just apply a rescaling since we are changing the Lorentz structure of the
coupling. However, the new derivative terms can be replaced by using the equations of motion. The fact that neither of those particles are on-shell do not
prevent us from doing so, since we can use the off-shell equations of motion. The price to pay being that, instead of replacing the derivatives by simple mass
terms, they will be traded for a non-linear function of other fields, introducing thus new vertices. Fortunately, those new vertices will not contribute at the
same order to the process we are considering. The change is then the 
following
\begin{equation}
 \dslash\chjm=-\left(\frac{m_j}{i}+f(\Phi)\right)\chjm\qquad\dslash\chip=\left(\frac{m_i}{i}+f(\Phi)\right)\chip
\end{equation}
where the $m_i$ denote the chargino masses. So we can bring ourselves back to a form similar to the MSSM one, at the difference that CP-odd and CP-even
coefficients have different expressions.
\begin{equation}
 \L_{C/S}\to \chjm\left(a_{C/S}\PL+b_{C/S}\PR\right)\chip\ \phi\\
\end{equation}
This will allow us for a simple rescaling of the MSSM amplitude, by noting that the $\PL$ and $\PR$ part can easily be told apart since they do not generate the
same loop structure. Indeed one choice will lead to the $\frac{\kslash}{k^2-m^2}$ part of the chargino propagators while the other will hit on the
$\frac{m}{k^2-m^2}$ part. After performing the loop integrals we have
\begin{eqnarray*}
 \PL &\to& m_im_j\Bloop\\
 \PR &\to& m_{\tilde{q}}^2\Bloop
\end{eqnarray*}
where the $B_2$ function is defined as in reference \cite{bobeth_bsmumu}. This leads us to the final result that the analytic formulas for the Wilson
coefficients in the MSSM can be translated in the BMSSM case by changing the coefficients of the CP-even and CP-odd Higgs contribution to their value in the
BMSSM.

\subsubsection{Prediction for $\Bsg$}
Looking at the diagrams involved in the computation of $\Bsg$, it turns out that the situation is simpler than in the previous case, since vertices with three
Higgses or one Higgs and two Higgsinos do not show up. Hence the modifications to be done are only the following :
\begin{itemize}
 \item The mass and mixing matrices of Higgs, squarks, charginos are changed.
 \item The weak and Yukawa couplings constant are altered.
\end{itemize}
The analytic formulas from MSSM can then be re-used, by plugging in the new values for masses, mixing and couplings.

\section{Dark matter constraints}
Dark Matter being one of the hot topics with important implications for supersymmetric models and therefore the work presented here, a special care will be
dedicated to such constraints. The topic being quite intricate a dedicated chapter will be be devoted to the issue, so I invite the reader to take for granted
that Dark Matter experiments are correctly taken into account, and postpone the discussion to chapter 9.

\section{Colliders Physics / Superpartners}
\begin{figure}[!h]
\begin{center}

\includegraphics[scale=1,trim=0 0 0 0,clip=true]{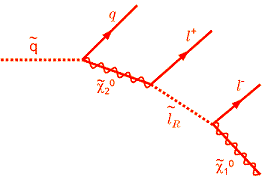}
\end{center}
\caption{\label{diag:cascade_decay}{\em Example of a typical production and decay of a squark in the R-parity conserving MSSM : the squark decays to the second
lightest neutralino plus a quark jet then the neutralino decays to a lepton and a slepton which itself decays into another lepton and the lightest neutralino
which is stable.}}
\end{figure}

The LHC being an hadronic machine, that is to say ruled by the overwhelming power of the strong interaction, squarks, obeying the very same gauge interactions
as quarks, should be copiously produced in such a collider. In fact the only unknown parameter seems to be their mass : either it lies in the kinematic reach of
the LHC and those states will be promptly detected or they will never be found. A similar statement apply also to gluinos, the superpartners of the gluons. We will see later that this statement is not exactly true, but first let us have
a look at the searches for superpartners carried out at the LHC. They are usually (though not all of them) based on the idea that if R-parity is conserved, than
a superpartner can only decay to another superpartner plus Standard Model particles, so that if a superpartner is created it will undergo a succession of decays
until it reaches the state of the lightest stable superpartner, usually a neutralino, which, due to its neutral and colourless qualities, will escape from the
detector unnoticed. This decay chain is usually 
referred to as a cascade decay, and an example is provided in figure \ref{diag:cascade_decay}.

\begin{figure}[!h]
\begin{center}
\includegraphics[scale=0.3,trim=0 0 0 0,clip=true]{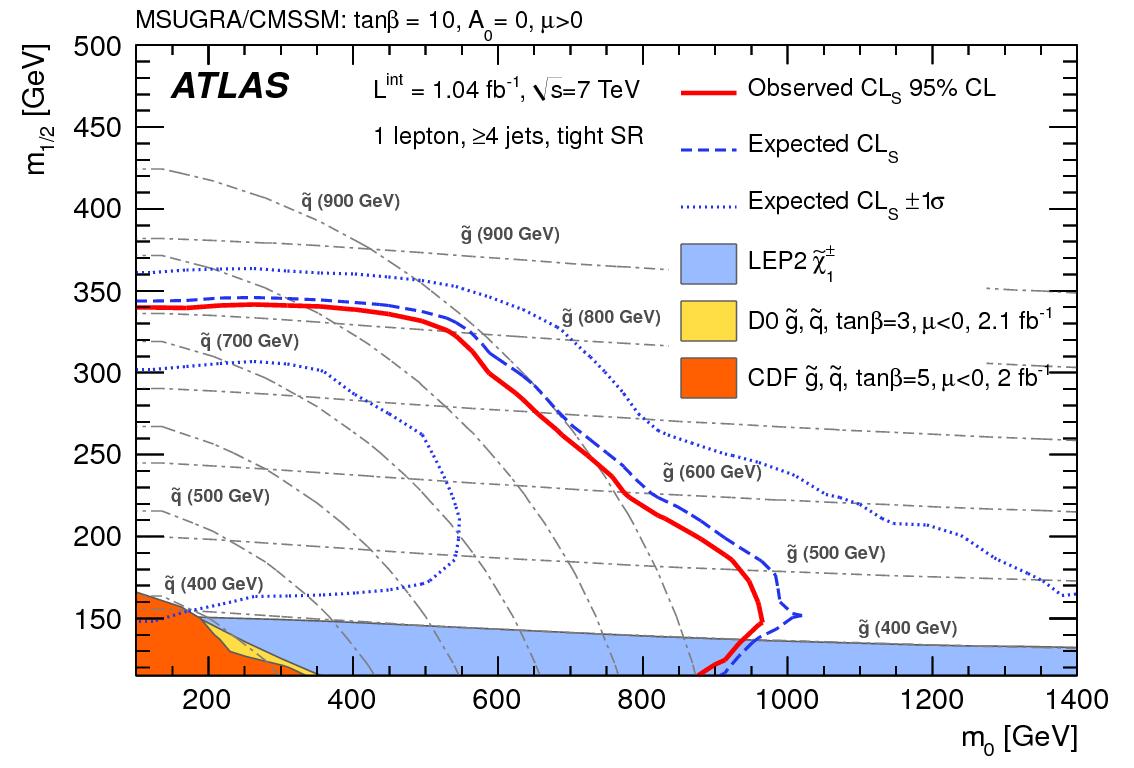}
\end{center}
\caption{\label{fig:squark_cmssm}{\em We show here the limits set by different analyses on the squark-gluino mass plane of the CMSSM, as a constraint on a
universal scalar superpartners mass $m_0$ and a universal fermionic superpartner mass $m_{\1}$. The red line is the result of the analysis lepton plus jets
performed by the ATLAS collaboration. Figure obtained from \cite{atlas_squarks}.}}
\end{figure}

The experiments will thus collect Standard Model particles and record a missing energy due to the fact that the neutralino is not observed. The momenta of the
Standard Model particles will eventually indicate the mass gap between the superpartner created and the neutralino. So far, such searches have been
unsuccessful, so experimental collaborations have been able to put lower bounds on the mass of those superpartners. Those limits in the case of a CMSSM scenario
are shown in figure \ref{fig:squark_cmssm}.

It seems at first sight to be quite contradicting our idea of a natural spectrum, that is with superpartners not too heavy in order to cure the fine-tuning
issue of the Standard Model, since those limits are bringing squarks nearly to 1 TeV. However, the discussion is not that simple, and some uncertainties will
come in the game.

\subsection{Model dependence of the limits}
So far most of the published analyses are in very specific MSSM set-up, usually the most simple one, CMSSM. This is however not enough for many theorists since
it draws extremely rigid conditions on the supersymmetry-breaking pattern. Precisely, it is now commonly agreed among the susy phenomenology community that a
natural spectrum could be obtained with light stops and a light left-handed sbottom, a gluino moderately heavy and other squarks above the TeV scale, which can
only be obtained by dropping the hypothesis of the universal squark mass. Thus, many superpartner searches are not well-suited to constrain a generic version of
the MSSM, let alone an extended version of supersymmetry. This mismatch between theory and experiment reveals an issue in current phenomenology : the
communication of results between the experimental side and the theoretical side.\\

The problem stems from the fact that one side of the community would like to have experimental results that can be interpreted in any model known or to be
invented soon while the other simply cannot analyse data without assuming a specific model. Indeed, to have an idea of the kind of final states they have to
look at, experimentalists have to take a particular model, fix the values of free parameters, compute the differential production cross-sections, generate
events accordingly and simulate the effect of those events on the detector. Only then are they able to construct a set of cuts and restrictions that will allow
the best separation between signal and background. The final part, comparing the expected signal to the actual data, is a small amount of the business involved
all along : it is just a matter of statistics. So it seems that the analysis has to be done for each model that is to be tested. This option is however not
realistic, knowing the number of models and the number of free parameters 
living in the jungle of the Beyond the Standard Model. Though there is still no general consensus, different options are explored to solve the problem : the
first one is to let theorists do the work up to (and including) the event generation and the simulation of the detector response. Hence the experiments only
have to provide the set of cuts used and the number of events recorded. A second alternative is to use simplified models at the experimental level, involving
only the particles and the interactions that are needed to account for the process studied, and treat all couplings and masses as free parameters. Then it is up
to theorists to relate those simplified parameters to their own parameters. A more detailed discussion is to be found in the Les Houches Recommendations
\cite{les_houches_reco_1203}.

\subsection{Compressed spectrum}
Another issue arises in some specific supersymmetric set-ups when the masses of superpartners get very close. Indeed, the momenta of the Standard Model
particles created all along the cascade depend on the mass splitting of the superpartners, so if this splitting gets too small, those particles will simply not
be recorded. In those specific models, also called pathological spectra, the reach of the conventional searches is very much reduced. This problem goes even
further than the previous point, since in order to record those specific events, one may have to redefine the triggering system of the experiments. The trigger
having the troublesome feature to be on-line, it means that it cannot apply to the data that is already taken.\\

Because of those different issues, we have decided to leave the superpartner searches aside by taking gluinos and squarks heavy (around 1 TeV) and only allowing
stops to be light.

\section{Colliders Physics / Higgs}
The first reason for including Higgs physics constraints in supersymmetric models is that the colliders are very sensitive to the Standard Model Higgs. By the
end of the year (2012), it will have either been found or excluded. And if the Standard Model Higgs is under pressure, this must be true for any model relying
on a Higgs mechanism for the electroweak symmetry breaking and fermion mass generation. Indeed if the Higgs particle (for a general model) generates those
masses, than its couplings to those particles have to be related somehow to the Standard Model couplings. In the MSSM for instance it imposes sum rules such as
\begin{equation}
 g_{HWW}^2+g_{hWW}^2=g_{h_{SM}WW}^2.
\end{equation}
And this general argument ensures that any model relying on the Higgs mechanism will be probed at the LHC. This is of course the case of the BMSSM. It is no
surprise that its Higgs phenomenology will be quite different from the MSSM one, since the Higgs sector was the major aim of our new operators. As we have seen
in figure \ref{fig:mh_mass}, the light Higgs mass can be raised to high values. Existing studies in the MSSM have dealt with the exploration of the MSSM
parameter space, so we will take a few scenarios as representatives of the MSSM and focus on the effective parameter space. The first case is the $\mhmax$
scenario (see \cite{carenaheinemeyer}). We chose it because it is the one that allows for maximal masses in the MSSM, so that what we compare is really the
maximal reach in term of mass. The scenario is the following. All soft masses are set to $M_{{\rm
soft}}=1$TeV, $\mu$ and $M_2$ are set to 300 GeV, $M_1$ is fixed
by the universal gaugino mass relation $M_1=\frac{5}{3}\tan^2
\theta_W M_2 \simeq M_2/2$, and $M_3=800$ GeV ($\cos^2
\theta_W=M_W^2/M_Z^2$). All trilinear couplings are set to 0,
except for $A_b=A_t=2M_{{\rm soft}}+\frac{\mu}{\tb}$ that are set
to maximise the radiative corrections to $m_h$.\\

In this scenario, the maximal mass goes from $\mh=135$ GeV in the MSSM to $\mh=250$ GeV, which of course offers brand new possibilities in the Higgs
hunting strategies. However this is not all of the story since couplings of all Higgses (not only the lightest CP-even one) will be modified from the MSSM
expectations.

\subsection{Higgs couplings}
The most relevant couplings of the Higgs at the LHC are the following
\begin{center}
 \begin{tabular}{cl}
 Production & $\gpg,\gpw,\gpz,\gpb$\\
 Decay & $\wp,\gpga,\gpw,\gpz,\gpl,\gpp$
 \end{tabular}
\end{center}
where we maintain the discussion on a generic level by denoting $\phi$ any of the three neutral bosons $h,H,A_0$. The effective operators together with the MSSM
parameters $\tb$ and $\ma$ will span a range of values for those couplings, making definite predictions not straightforward to do. However, there are
fortunately some strong correlations between the couplings : in other words, we cannot generate just any kind of coupling for each Higgs boson to any
particle.\\

\begin{figure}[!h]
\begin{center}
\begin{tabular}{cc}
\includegraphics[scale=0.35,trim=0 0 0 0,clip=true]{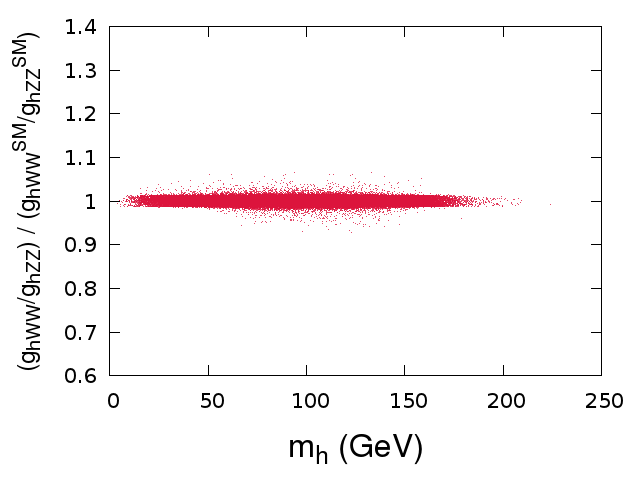}&
\includegraphics[scale=0.35,trim=0 0 0 0,clip=true]{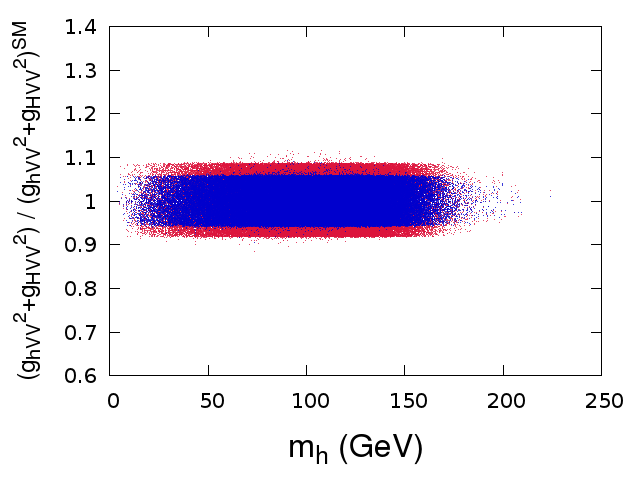}
\end{tabular}
\end{center}
\caption{\label{fig:v_couplings} {\em Couplings to the vector boson $Z$ $g_{hZZ},g_{HZZ}$, normalised to the Standard Model value. On the right plot, red and
blue points respectively correspond to $Z$ and $W$ couplings.}}
\end{figure}

\paragraph{Couplings to weak bosons :}~\\
The couplings to the $W$ and $Z$ bosons are quite constrained. First they only couple to CP-even Higgses, and then there is a correlation between the two weak
bosons. Indeed, there is an approximate custodial symmetry around in order to keep the electroweak precision tests consistent. It will impose that the ratio of
the coupling to the $Z$ boson and the $W$ boson stays the same for $h,H$ over the parameter space
\begin{equation}
\frac{\gpz}{\gpw}\approx \frac{\gpz^\SM}{\gpw^\SM}.
\end{equation}
This relation is exact in the MSSM case at tree-level, since the interaction comes from the term $\phi^\dag\,(\ggg\AA)^2\,\phi$ which impose the relation
between $W$ and $Z$. However it is slightly broken at the second order in $1/M$ by interaction terms from $\left(\phi^\dag\,(\ggg\AA)\,\phi\right)^2$ which
contributes to the $Z$ coupling but not the $W$. This ratio being however related to the mass ratio that enters the electroweak precision variable $\epsilon_1$,
it has to be small. Another correlation is the sum rule
\begin{equation}
g_{hVV}^2+g_{HVV}^2=g_{hVV\ \SM}^2\qquad (V=W,Z).
\end{equation}
Those two correlations being shown in figure \ref{fig:v_couplings}.

\begin{figure}[!h]
\begin{center}
\begin{fmfgraph*}(100,70)
\fmfleft{p1}
\fmfright{p2,p3}
\fmflabel{$h$}{p1}
\fmflabel{$b$}{p2}
\fmflabel{$\bar b$}{p3}
\fmf{dashes}{p1,v1}
\fmf{dashes,tension=0.4,label=$\tilde{b}$,label.side=left}{v2,v1,v3}
\fmf{fermion}{v2,p2}
\fmf{fermion}{v3,p3}
\fmffreeze
\fmf{photon,label=$\tilde{g}$}{v2,v3}
\fmf{fermion}{v2,v3}
\end{fmfgraph*}
\end{center}
\caption{\label{diag:dMb}{\em Example of a diagram contributing to the $\Delta m_b$ corrections : this is the vertex correction where a gluino $\tilde{g}$ is
exchanged between the two bottom quarks.}}
\end{figure}
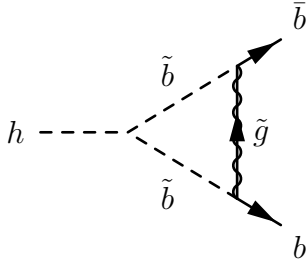

\paragraph{Couplings to fermions :}~\\
Then, since all BMSSM operators have a universal effect with respect to fermions, we will keep the fact that the Yukawa of down-type quarks and leptons are
scaled in the same way :
\begin{equation}
 \frac{\gpb}{\gpb^\SM}\approx \frac{\gpl}{\gpl^\SM}
\end{equation}
where the correlation is broken by loop effects, since there will be loops proportional to the Yukawa factor. Those loop corrections are especially relevant in
supersymmetry, where the down-type Yukawa couplings can be rather enhanced as compared to the Standard Model. In particular it is customary to include the
contribution of the diagrams in figure \ref{diag:dMb} in the variable $\Delta m_b$ and to use the following coupling to compute observables

\begin{equation}
 \gpb=\gpb^\text{tree-level}\left(1+\Delta m_b\right).
\end{equation}
The $\Delta$ correction being different for each fermion, including this contribution will break the approximate symmetry. This quantity is computed following
the prescription in \cite{micromegas2} on the loop side, plus the effective shift.\\

\begin{figure}[!h]
\begin{center}
\begin{tabular}{cc}
\includegraphics[scale=0.35,trim=0 0 0 0,clip=true]{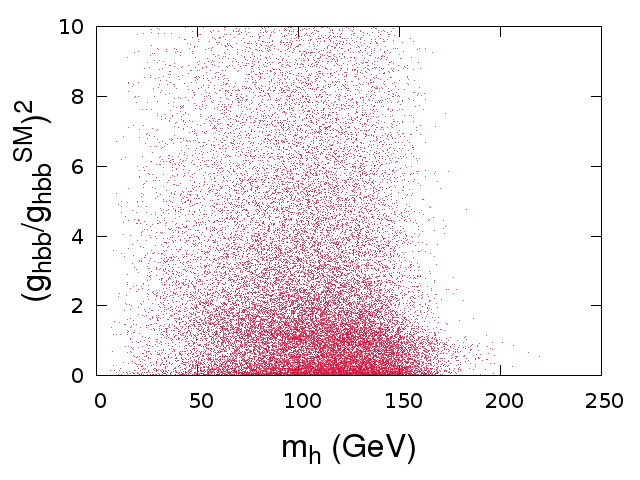}&
\includegraphics[scale=0.35,trim=0 0 0 0,clip=true]{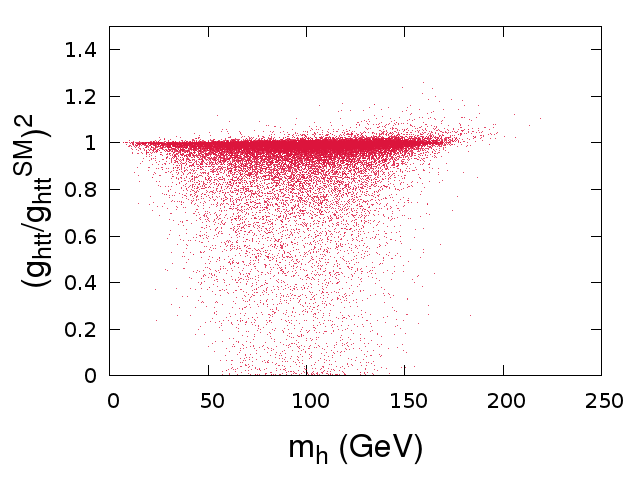}
\end{tabular}
\end{center}
\caption{\label{fig:f_couplings} {\em Couplings of the lightest Higgs to third generation fermions $b,t$ normalised to the Standard Model value.}}
\end{figure}

One can see on fig \ref{fig:f_couplings} that the effect as compared to the Standard Model can be a suppression or an enhancement : this is no new feature of
the BMSSM but a $\tb$ effect, since the Yukawa couplings are proportional to its value.

\begin{figure}[!h]
\begin{center}
\begin{tabular}{cc}
\includegraphics[scale=0.35,trim=0 0 0 0,clip=true]{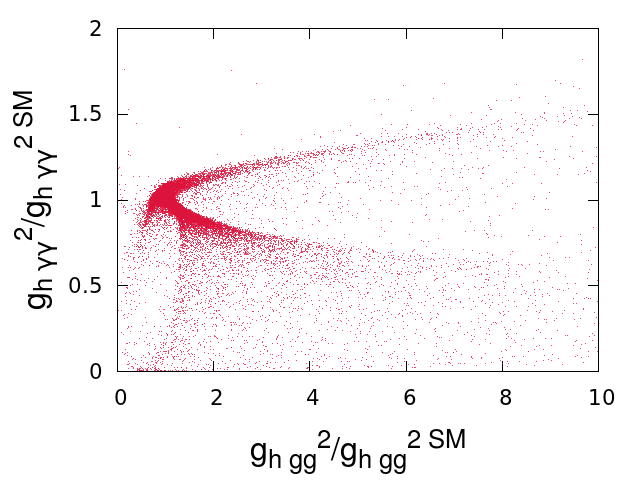}&
\includegraphics[scale=0.35,trim=0 0 0 0,clip=true]{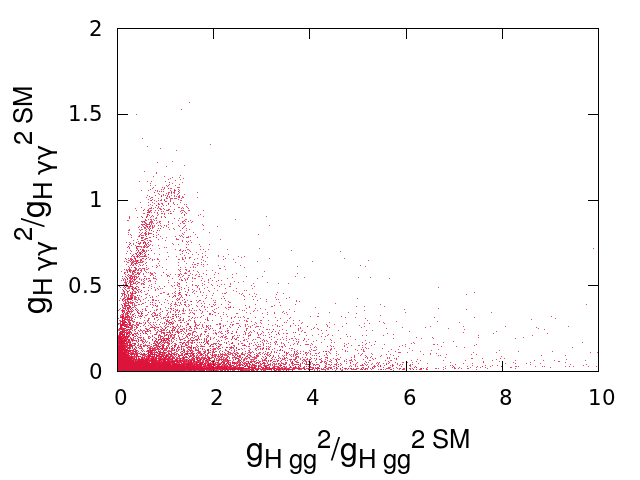}\\
\includegraphics[scale=0.35,trim=0 0 0 0,clip=true]{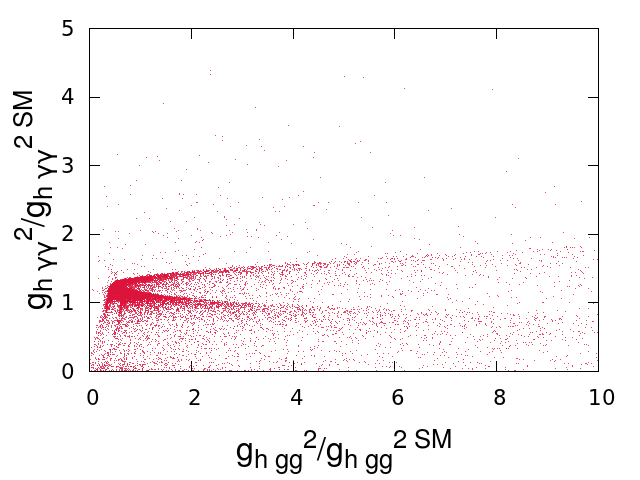}&
\includegraphics[scale=0.35,trim=0 0 0 0,clip=true]{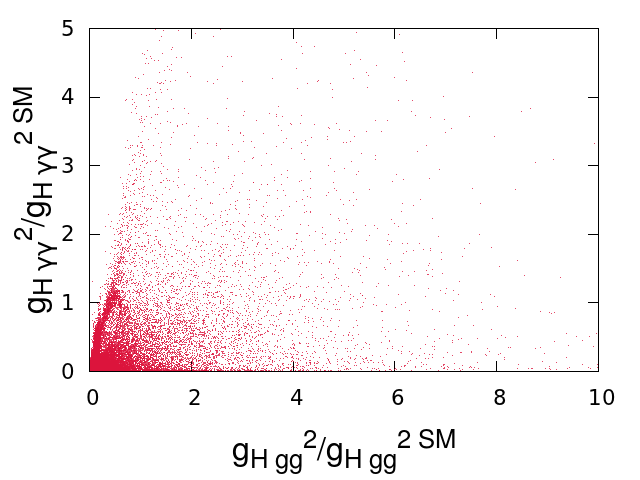}
\end{tabular}
\end{center}
\caption{\label{fig:loop_couplings} {\em Couplings of the lightest Higgs to massless gauge bosons $g,\gamma$, normalised to the Standard Model value. On the top
row is shown the case of the scenario A, and on the the bottom one scenario B. Left panels are for the lightest Higgs $h$ whereas right ones deal with $H$.}}
\end{figure}

\paragraph{Couplings to massless gauge bosons}~\\
The neutral Higgses being colourless and without charge, they do not couple at tree-level to photons and gluons. The computation of the coupling at the loop
level is however a crucial point since $gg\to \phi$ is the main production mode at the LHC and $h\to\gamma\gamma$ the most sensitive decay mode at low masses.
As compared to the standard model, there are two cases to consider : the first one is the case where the loops of superpartners are negligible, for instance if
stops are heavy. In this case, $\gpg$ is driven by top and bottom loops while $\gpga$ is mainly given by the $W$ loop. Then those two couplings are directly
correlated to $\gpb,\ \gpt$ and $\gpw$. The second case is when light superpartners of the third generation (usually stops but also staus, to a lesser extent)
come into play. In the case of stops it will allow us to decouple $\gpg$ and $\gpga$ from the three couplings $\gpb,\gpt,\gpw$ but will at the same time induce
a correlation between $\gpg$ and $\gpga$ since the same stop loop 
appears in each coupling. We show on figure \ref{fig:loop_couplings} the couplings in the $\ghg,\ghga$ plane.\\

Two implementations will be considered, in order to test these two different cases:
\begin{itemize}
\item Model A: In this scenario there is no stop mixing parameter,  $A_t=0$. 
All the soft masses of the third generation
squarks are set to $M_{{u3}_R}=M_{{d3}_R}=M_{Q_3}=400$ GeV. For these
values the masses $m_{\tilde{t}_1},m_{\tilde{t}_2}$ are around
$400$ GeV, and since the mass difference is small, the stop loop in the coupling to gauge bosons will be suppressed. This is taken as a
standard case, where stops are not too heavy and in the set up of
the BMSSM their effect is not so important.
\item Model B: A maximal
mixing scenario where one of the stop is light
$m_{\tilde{t}_1}=200$ GeV. We will take
$m_{\tilde{\tilde{t}}_2}\in[300,800]$ (GeV) and $\sin
2\theta_{\tilde t}=-1$. The heaviest stop mass is taken as a free
parameter. This will have important consequences in the production
of the Higgses and their decays. Note that, in a generic model, a
$200$ GeV stop can still escape all current collider limits.
\end{itemize}

Concerning the lightest Higgs, its coupling to gluon can be much enhanced as compared to the Standard Model, whereas the couplings to photons cannot go higher
than twice the Standard Model expectation. Furthermore, there is a correlation between the two observable : this interesting point will be discussed in the next
chapter.

\section{A short excursion on the experimental side}
\label{sec:mapping}
Despite the fact that the comparison with experiments is the very goal of phenomenology, the kind of comparison we have described so far is a very idealised
one. For one thing, it relies on the assumption that all relevant discriminations can be done at the level of the cross-section. This is a very well motivated
idea since on the theoretical part the cross-sections are unambiguously obtained from the couplings of the theory and the masses of intermediate and final
particles and on the experimental side it is encoded in the statistics of an analysis. However this assumption fails in nearly all practical cases, for the
experimental part.\\

Indeed, while we can safely affirm that cross-sections can be computed in a generic way from a given theory, since we have devoted a certain amount of pages on
that subject, we have been a bit optimistic in saying that the cross-sections could lead directly to the statistics recorded by an experiment, and vice-versa.
Let us describe (from the theorist point of view, though) how a detector works, for instance in a collider. The detector is built with a cylindrical symmetry
around the beam axis, up to a finite length from each side of the collision point. Its role is to generate interactions with the particles fleeing from the
interaction point, and through these interactions to gain knowledge on the momentum and the nature of those particles. Typically, the space momentum of a
charged particle is measured by generating a magnetic flux inside the detector and measuring the bending of its trajectory. This part is devoted to the tracker,
close to the beam axis. The energy of a particle is measured by alternating 
energy-absorbing materials, which make the energy to be transmitted from particles to the medium, and energy-measuring materials, which measure the energy of
the medium itself. This task is performed by the calorimeter. The nature of the particles is obtained from different indications stemming from the outcome of the
tracker and the calorimeter. This part is called the particle tagging and can be extremely sophisticated. The most obvious points being the following : knowing
the space momentum of a particle and the magnetic flux in the tracker, then the bending is uniquely determined by the charge. Also the transmission of energy to
the medium depends on the mass of the incoming particle : the lighter it is, the quicker it will loose energy. Of course only sufficiently stable particles can
be seen, others decaying before entering the detector. This particle tagging is exactly what hampers the relation between the cross-section and the records of
the experiment.\\

Indeed to do the mapping from the recorded events to the cross-section consistently, one must associate to each configuration of particles leaving the
interaction point the probability to be recorded exactly as it is (same particles, same momentum) by the detector, or differently, which is called a
misidentification. This probability is encoded in the efficiency and the acceptance. A first issue is that this probability is a function of the momenta of the
particles. Indeed it is clear that a particle escaping from the interaction point in a direction too close to the beam axis will not go through any piece of detector and will not be recorded at all.
On the other side, a particle going out with a too small energy will simply not be distinguishable from a particle coming from an elastic interaction in the
beam, which are numerous. This means first that experimentalists have to work with the differential cross-section, not the integrated one. But this is not the
only trouble. The efficiencies are evaluated with different tools : a reconstruction 
tool first, which aim to map the input of each cell of the detector to a set of particles that went through it. The second tool is the shower algorithm which
maps the outcome of the hard process (that is to say a handful of particles with high energies) to the set of particles that actually enter the detector,
usually several hundreds. The reason why there is so many particles arriving on the detector as compared to what the hard process produces are the QED and QCD
effects : all charged (coloured) particle will have a tendency to radiate photons (gluons) which themselves can split into charged (coloured) particles, and so
on. Fortunately, particles emitted this way have generally a small, or collinear momentum in the rest frame of the initial particle, which makes the resulting
bundle collimated in the initial direction. Those bundle are called jets. Because of the strong coupling of coloured interactions, the jets associated to quarks
are notoriously less clean than jets from leptons. Another feature of the 
shower algorithm is that it matches an input with coloured particles (quarks and gluons) to a colourless output (hadrons) : this has to be the case since QCD
confinement predicts that particles at low energy must be colourless. This is called the hadronisation process and, stemming directly from the non-perturbative
realms of QCD, it is mostly described by effective models calibrated on the data.\\

\subsection{When experiments meet theory}
The question is now the following : where can we make both ends meet? This is rather non trivial and is still widely debated. Although our short description of
the experimental set-up would indicate that the differential cross-section would be a good meeting point, it does not suit all rare processes : if one expects
no more than a couple of events over all the surface of the detector and all the measurable energy range then it makes no sense to measure the statistics at a
given energy in a given direction. Thus the most natural option is to integrate together events that differ only by their phase space variables. By doing so,
one obtains a cross-section -- called the \textit{exclusive} cross-section -- which is not the integrated theoretical cross-section, but the integration of the
theoretical cross-section convoluted with the efficiency of the detector. The trouble with this quantity is that, so far, it cannot be easily computed by one
side of the community on its own : theorists need the precise 
knowledge of the efficiencies and experimentalists the differential cross-sections for each model.\\

Unfortunately, most efforts have until now been devoted to the full calculation by each side on its own. Theorists have chosen to re-do the procedure, that is
to say, starting with the differential cross-section of the hard process to do the shower, to process the journey of particles through the detectors, and
finally to mimic the analyses of experiments in order to compare eventually with raw data. On their side, experimentalists have successfully interpreted their
data with a very narrow range of existing models and given their results as integrated cross-sections in those models. But most of those results automatically fold in the efficiency of the detector that we have discussed above, and are thus virtually impossible to recast in other models. One case stands
apart : the Standard Model, where the cooperation between both sides is extremely strong. Notably, the simulation and reconstruction tools for the Standard
Model have been tuned in a back and forth optimisation, theory improving its 
knowledge on one cross-section, than experiments comparing their data with the predictions, and so on. Particularly, the dynamics of QCD in a hadron collider
would never have reached such a precision without the interplay between both communities. And without such a precision, we would never have been able to remove
the huge background standing for all new physics searches. However there is quite a difference between the Standard Model and new physics, the first is a unique
model with so far only one unknown parameter and the other is a relentlessly growing set of theories with numerous parameters. Strictly on the manpower point of
view we cannot mobilise as many people on each new theory than what we have on the Standard Model. But on the other side, whereas the precise knowledge of the
Standard Model was a key point to estimate accurately the background, we only need to know roughly the behaviour of New Physics to get the signal right. We are
going to see in the next chapter how the issue can be treated in 
the specific case of Higgs searches applied to supersymmetry.


%% file: chapter7.tex
\chapter{Towards a non Standard Model-like Higgs signal}

\minitoc\vspace{1cm}

We are now going to interpret the results of the Higgs searches at the LHC in the BMSSM framework. This is by no means a straightforward business, indeed we
will have to refine our theoretical predictions so that they actually match the experimental quantities that are measured. The previous remarks made on the
experimental procedure are more relevant than ever, and we will see that though a quantitatively accurate match between theory and prediction is quite hopeless
with the current status, a qualitative assessment can be performed.

\section{Using LHC results}
\subsection{Experimental search : the method in a nutshell}
As was discussed at the end of the previous chapter, the experimental development of an analysis is in itself a fairly complicated business, and our aim is not
to tell whether such things are done consistently or not, but to understand how to connect the output of an analysis to a theoretical model. Basically, one
defines a final state compatible with the production and decay of a Higgs boson from two protons, then counts all events associated to this final state. After
having assessed the background related to this final state, that is to say the average number of events associated to other processes but that would still be
recorded as this final state, one subtracts this quantity to the observed number of events, and ends up with the number of events of signal. Then, knowing the
total luminosity delivered by the machine, this number can be turned into a cross-section. Two cases appear, either the cross-section is lower than the
uncertainties on the background so that there is no evidence for a signal and 
one 
can set up an upper limit for the cross-section of the Higgs signal process, or there is a signal and one can provide an experimental measure of this
cross-section, up to a certain accuracy.\\

This makes the implicit assumption that the production and decay modes of the Higgs are known, meaning that we have to set ourselves in a given model. However,
this is not a too stringent restriction since many models share the same production and decay modes. Indeed for all models where the Higgs is supposed to break
the electroweak symmetry, we can expect a coupling to the weak bosons of the order of the standard model one. And if we assume the Higgs to give mass to the
fermions, it should also couple somehow to them. This has led to the the following consensus : the search channels (that is to say the final states) are chosen
on the basis of the Standard Model expectations, but the measurements should be exploitable in any model.

\subsubsection{Likelihood function}
It seems that a sufficient description of the experimental data should be the description of the final state (particle content and phase-space specification)
and the observed number of events for signal and background $n_S$ and $n_B$ together with the uncertainty on the background estimation. This uncertainty is
fully characterised by a probability density function $f(n)$, which in most cases can be approximated by a gaussian centred on $n_B$, with mean $\sigma_B$.
However, when comparing with the expected number of events $x$ of a specific model one must take into account the probabilistic character of the measure. This
requires to construct statistical tests, for which a concise description is given in Appendix D and a detailed review in \cite{cowan, cranmer}. The idea of
those statistical tests is to quantify the compatibility of a model with the data, and answer the questions : what models are excluded by data, and which model
is the most likely, given the data? Without entering in the details, better 
explained in the appendix, let me say that the important point is that we end up with a likelihood function $L$ for each event count, from which an excluded
cross-section $\sigma^\excl$ (in the no-signal hypothesis) or a signal strength $\hat{\mu}$ (the measured cross-section) can be derived. Keeping this in mind we
will now list the most sensitive final states in the Higgs searches at the LHC.

\subsection{Describing the searches}
\begin{figure}[h!]
\begin{center}
\begin{tabular}{cc}
\hspace*{-4mm}
\includegraphics[scale=0.4,trim=0 0 0 0,clip=true]{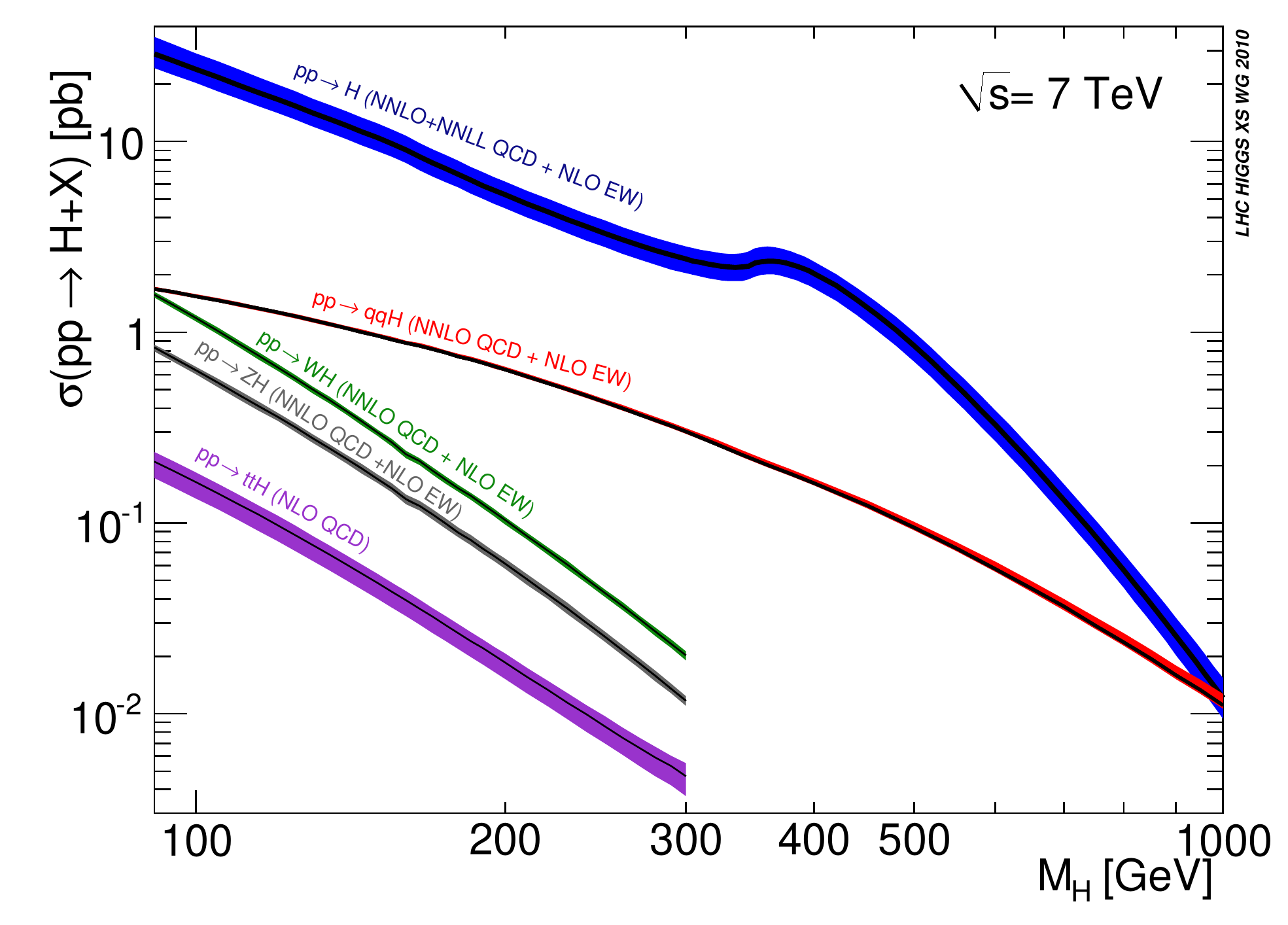}&
\includegraphics[scale=0.4,trim=0 0 0 0,clip=true]{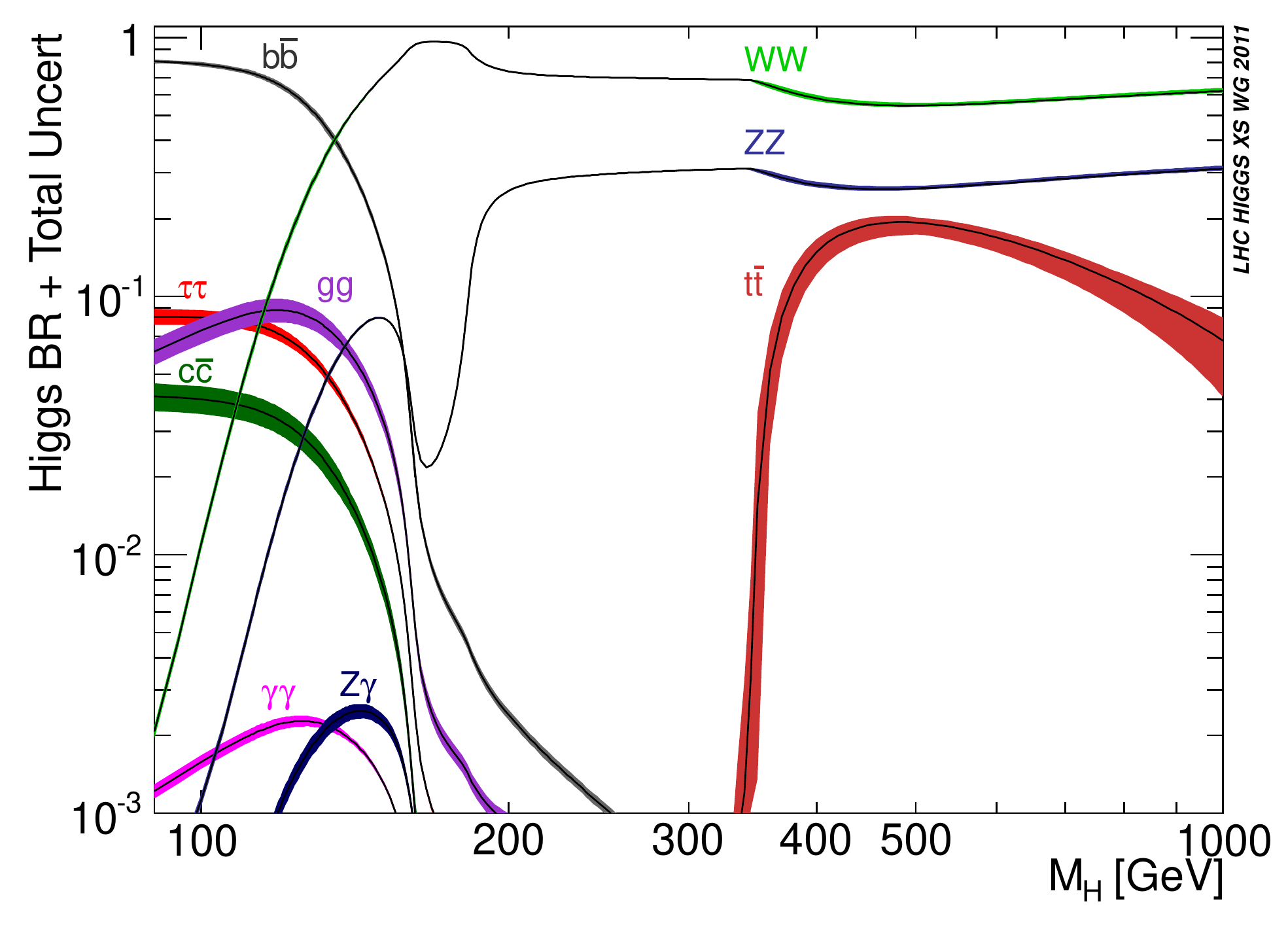}
\end{tabular}
\end{center}
\caption{\label{fig:h_prod_dec} {\em The left panel shows the cross-section of the different production mode of the Standard Model Higgs at the LHC at 7 TeV,
while the right one shows the different branching ratios of the Standard Model Higgs as a function of its mass. Plots taken from the LHC Higgs Cross Section
Working Group.}}
\end{figure}

As previously said, the definition of the search channels is mostly done on the ground of the Standard Model expectations. We show on figure
\ref{fig:h_prod_dec} the Standard Model production and decay rates, as a function of the Higgs mass.

Those expectations are the following : Higgses will be produced mainly through gluon fusion ($gg\to h$), then by vector boson fusion ($q_1q_2\to hq_3q_4$ also
known as VBF), associated vector boson production ($q_1q_2\to hV$) and heavy quark associated production ($gg\to\bar tt h$). Then, depending on its mass, it
will decay to gauge bosons and fermions. A general description of the Standard Model can be found in \cite{djouadi_review_1}, and precise predictions for LHC
have been tabulated in the LHC handbook \cite{LHC_Higgs_cs2}. The reason why many different final states are considered at the LHC is twofold, first to
improve the statistical accuracy, since it demultiplies the amount of data, and second to also be able to probe non-standard like couplings. For instance
if the Higgs couplings to gauge bosons and to fermions did not respect the Standard Model rule, we would see 
different behaviours in corresponding channels.

\paragraph{$h\to\gamma\gamma$}~\\
In the Standard Model this channel is mostly sensitive to the low mass region $\mh<150$ GeV, since the prediction drops afterwards. This channel is not a single
count analysis, it is separated into subchannels, in particular events will be divided in a low $p_T$ class and a high $p_T$ one, where $p_T$ is the transverse
momentum of the photon pair. Furthermore CMS has also defined an extra subchannel $h\to\gamma\gamma\,+\,2j$, that is requiring two extra jets on top of the
photon pair. The idea is to enhance the sensitivity towards the VBF production mode, indeed in such a mode the two outgoing quarks will yield two opposite jets
in the forward-backward regions, so by imposing this cut we will be more sensitive to the Higgses produced by VBF than those from gluon fusion. In view of
non-standard model couplings this is very interesting since the gluon fusion is driven by the $\gpg$ coupling whereas the VBF is driven by the $\gpv$
couplings which can exhibit a different behaviour in non-standard 
theories.

\paragraph{$h\to ZZ$}~\\
This channel is a category in itself, since the $Z$ bosons decay much before passing through the detectors. The associated final states are the following : $4l$
(four leptons), $2l2j$ (two leptons and two jets), $2l2\nu$ (2 leptons and two neutrinos). And also because we see only the final state, the intermediate $Z$
bosons can be off-shell, so the signal appears even at low masses $\mh<2\mz$. In particular the $4l$ mode is known as the gold-plated mode, since it has an
extremely low background. However at high masses, other modes tend to take over. The analysis is a simple counting analysis.

\paragraph{$h\to WW$}~\\
As in the previous case, different final states exist depending on the decay of the $W$, and the most sensitive one is the $2l2\nu$ final state. For the time
being the collaborations are also splitting this final state in three categories depending on the number of jets $0j/1j/2j$. As for the $\gamma\gamma$ channel, the aim is
to distinguish between different production modes.

\paragraph{$Vh\to Vbb$}~\\
It is well known in the Standard Model that a low mass Higgs $\mh<140$ GeV decays quasi exclusively (more than 80\% of the cases) to a $\bar bb$ final state.
However, such a final state is completely swamped by the QCD background of the LHC, so it cannot be used as such. The trick is to require a $Z$ boson in the
final state, which is the case if the Higgs is produced through associated vector boson production. Triggering on the $Z$ allows then to reduce tremendously the
background and to be sensitive to the signal. Because of this particular final state, this channel does not probe at all the gluon fusion mechanism of the Higgs.
At the time of writing (May 2012) this channel is still better known at Tevatron than LHC, mostly because the background is better rejected.

\paragraph{$h\to\tau\tau$}~\\
The last final state is the $\bar \tau\tau$, and since it is a unstable state, we will have to look for decay products. As for the $\bar bb$ case, it is efficient
only at low masses in the Standard Model. This channel has the interesting property of giving the opportunity for a supersymmetric analysis : indeed both ATLAS
and CMS have used it to derive limit on MSSM-like Higgses. In the MSSM case another production channel opens up (see \cite{djouadi_review_2}) : the $b$ quark fusion $\bar bb\to h$, which has
the property of being $\tb$ enhanced which make this channel particularly sensitive to the high $\tb$ MSSM parameter space.

\subsection{Combining searches}
The results on the search in each channels are regularly updated on the public website of each collaborations
(\url{https://twiki.cern.ch/twiki/bin/view/AtlasPublic/HiggsPublicResults} for ATLAS and \url{https://twiki.cern.ch/twiki/bin/view/CMSPublic/PhysicsResultsHIG}
for CMS) and one can have access to the various excluded cross-sections $\sigma_{XX}^\excl$, where $XX$ stands for the final state, as a function of the boson
mass. The results obtained in this section are based on the analysis of the 2\fb dataset
(\cite{atlas_lp_gamgam,cms_lp_gamgam,cms_lp_bb,atlas_lp_ww,atlas_lp_wwb,cms_lp_ww,atlas_lp_zz,atlas_lp_zz_2l2q,atlas_lp_zz_2l2nu,cms_lp_zz_2l2tau,
atlas_lp_tautau,cms_lp_tautau}) and the 5\fb dataset (\cite{atlas_5fb,cms_5fb}). It seems that all one has to do is compute the prediction for $\sigma_{XX}$ and
compare it to $\sigma_{XX}^\excl$. However, the naive test that would be to do this separately on each channel to decide whether the model is compatible is not
correct. Indeed, those limits are set 
with probabilistic rules, so for instance if a model has a compatibility of exactly 95\% in five independent channels, its total compatibility is much less than
95\%. Conversely, a model which is compatible at 99\% in one channel and 90\% in another could well be compatible to a level of 95\% on the whole. In particular
this means that, when dealing with two bounds coming from independent experiments, but bearing on the same quantity, one cannot just choose the most stringent
one and forget about the other. This point exemplifies the fact that what drives the compatibility of the model is the total likelihood function, so that when
dealing with multiple channels, one cannot avoid the combination.

\subsubsection{Mathematical method}
There is strictly no apparent difficulty in computing the likelihood associated to $n$ independent channels : one simply defines the combined likelihood to be
the product likelihood
$$L=\Pi_i L_i.$$
Then one follows the usual procedure, at the difference that the probability density function is now characterised by a vector $\xvec=(x_i)$, and so will be the
\pvalue\ : $p_{\xvec}$. 

\subsubsection{Exact versus approximate}
However computing such a likelihood requires an extensive access to data and is hence forbidden to theorists\footnote{Though many of them are asking the
collaborations to provide the likelihood functions for each search, see \cite{les_houches_reco_1203}.}. So we have to turn ourselves to approximate methods,
that relies only on the numbers $n_{Si},n_{Bi}$ and $x_i$. Those methods have been developed in \cite{cranmer}. The approximation, called the quadrature sum
approximation, is the following : if we define 
\begin{equation}
R=\sqrt{\sum_i \left(\frac{\sigma_i}{\sigma_i^\excl}\right)^2}
\end{equation}
where the sum runs over all channels $i$ for which $\sigma_i$ is the predicted cross-section and $\sigma_i^\excl$ the observed excluded one, then a model is
excluded if $R>1$. This approximation has the drawback that it does not account for possible correlations of nuisance parameters among channels, but has the
major advantage that it does not require more information from the experimental side than the excluded cross-section. However the question is : how precise
would it be?\\

\begin{figure}[h!]
\begin{center}
\includegraphics[scale=0.5,trim=0 0 0 0,clip=true]{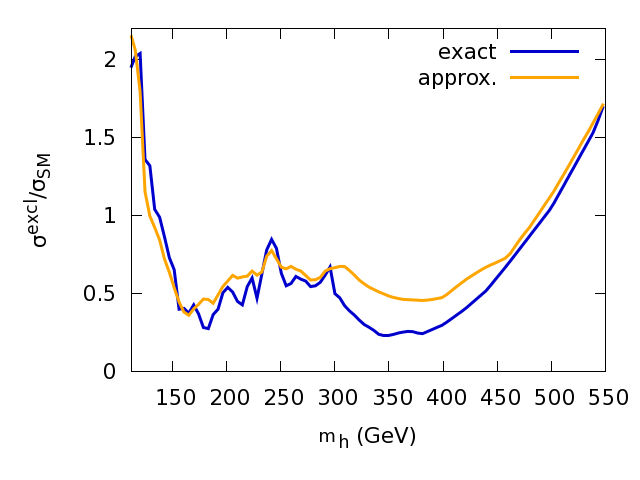}
\end{center}
\caption{\label{fig:sm_comb} {\em The figure shows the excluded
cross sections for the SM Higgs production at the 95\% confidence
level in the case of the combined analysis performed by the ATLAS
and CMS collaborations, exact, compared to the approximate
combination we perform based on the quadrature sum, approx.}}
\end{figure}

Fortunately, it is possible to put such an approximation to the test in the case of the Standard Model. Indeed, since the exact combination of ATLAS and CMS
channels in the SM
case has been published in \cite{atlas_cms_lp_comb} in fall 2011, we have
presented it as well as our own combination on the same plot, figure~ \ref{fig:sm_comb}, so one can
quantitatively weigh the discrepancy coming from the quadrature
sum approximation.
Figure ~\ref{fig:sm_comb} shows, that apart from the range
$300<m_h<450$ GeV where all $ZZ$ channels combine, the
approximation based on the quadrature sum is well justified. In
the BMSSM, this range is not reached by the lightest Higgs nor by
the CP-odd Higgs which does not couple to vector bosons. Moreover
even in the BMSSM for $m_H>300$ GeV, one is most often in the
decoupling limit where the $HWW$ and $HZZ$ are vanishingly small.
Therefore for the BMSSM the approximate combination should be
trustworthy.\\

\subsection{BSM Higgs Limits : a tricky business}
However, if it is feasible for theorists to combine approximatively different channels in their favourite model without specific access to the well protected
details of the experimental analyses, there can be two other obstacles in the determination of the compatibility of a model, which I will now describe.

\subsubsection{The inclusive/exclusive paradigm}
As was advocated at the end of the previous chapter, the experimental search is fully based on the design of cuts and other distinctions on the phase space of
the final state. As such, what is measured is not the integrated cross-section, but the convolution of the differential cross-section by the efficiency and
acceptance of the detector (this is precisely the mapping described in section \ref{sec:mapping}). This means that instead of using a production cross-section
made by simply adding all production modes as in 
\begin{equation}
\sigma_\text{incl.}=\sigma_{gg\to h}+\sigma_{VBF}+\sigma_{VH}+\sigma_{\bar bb\to h}
\label{eq:inclusive}
\end{equation}
which is an \textit{inclusive} cross-section, one has to fold in the efficiencies $\epsilon_i$ :
\begin{equation}
\sigma_\text{excl.}=\epsilon_{ggh}\sigma_{gg\to h}+\epsilon_{vbf}\sigma_{VBF}+\epsilon_{vh}\sigma_{VH}+\epsilon_{bbh}\sigma_{\bar bb\to h}
\label{eq:exclusive}
\end{equation}
in what is now an \textit{exclusive} cross-section. Two cases are then possible, the first, simple, where efficiencies are the same for all
production modes, and it can hence be factored out, so inclusive and exclusive cross-sections can be interchanged since we have
$$\frac{\sigma_\text{excl.}}{\sigma_\text{excl.}^\SM}=\frac{\sigma_\text{incl.}}{\sigma_\text{incl.}^\SM}$$
But if the efficiencies are not the same, one has to obtain them to go any further. This is somewhat problematical since collaborations do not provide them on a
public basis. The other option is to use simulation tools to generate events (such as \PY), and estimate efficiencies without input from the experiment. This
however brings a large uncertainties since most of the complicated experimental issues (how to cope with the pile-up, detector simulation, and so on) are simply
out of reach for a theorist.\\

Two direct examples of both cases are the $ZZ\to 4l$ and $WW\to 2l2\nu$ channels. In the first one there are practically no cuts, one simply requires the decay
products to have an outgoing direction that actually leads to the detector and to have enough energy to trigger the detector response. In such a case, the
efficiency is nearly the same on all production modes, so the \textit{inclusive} ratio can be used. On the latter, the final states are discriminated on the
number of extra QCD jets : $0j/1j/2j$. Then the efficiencies are quite different, for instance the VBF production mode will go mostly in the $2j$ bin, while the
gluon fusion will mainly go in the $0j$ bin. In such a case, the \textit{exclusive} ratio has to be used.

\subsubsection{There is no model independent combination}
The second issue is related to combinations done internally by the experiments. Indeed, since they currently focus on the Standard Model Higgs (which is not a
totally disreputable strategy), they only need to compute the \pvalue s along the Standard Model direction, that is with $\xvec=\mu\xvec_\SM$, where $\mu$ is a
real parameter. That is why most of the results for Higgs searches are given either as $\mu^\excl$ in the no-signal case or $\hat{\mu}$ in the signal case. But
if the prediction of a model lies in a direction $\xvec'$ that would not be proportional to $\xvec_\SM$, then the combined result can simply not be used. The
reason why a proportionality factor would be accepted is that, in the quadrature sum approximation, such a factor would factorize the whole expression.

This emphasizes the fact that, every time one attempts a combination of different channels, one has to assume a model, or in other words, that there is no model
independent combination. This conclusion leads us to ask for experimental papers where, on top of the Standard Model combination (which would be the primary
aim), one could find the excluded cross-sections per channels or sub-channel, before any combination. This is for instance the case of the diphoton channel : it
is indeed discriminated on the transverse momentum of the photon pair, with a low $p_T$ bin and a high $p_T$ bin. Since the different production modes do not
populate the two bins in the same way, their relative sensitivity will hence vary from one model to another. However, no separate excluded cross-sections are
quoted, only the combination of the two in the case of the Standard Model is given. A workaround for this issue would be that collaborations would give for each
of their analyses the excluded cross-section obtained by merging 
all subchannels together. The result would hence be model-independent. However this may significantly weaken some limits, so the ideal case would of course to
have excluded cross-sections for each subchannel.

\section{The no-signal case}
The first analysis is to interpret the exclusion bounds from both collaborations ATLAS and CMS in the BMSSM framework. It led to a publication,
\cite{gdlr_higgs_1112}, which allowed us to draw some conclusions that will be presented here. We will now fix an MSSM scenario in order to reduce the number of
free parameters, and define the region of the effective parameter space to be explored. Then I will show two different interpretations of the LHC limits,
depending on the amount of information one is willing (or able) to extract from the experimental notes : a special point will be made in differentiating the
BMSSM behaviour from the MSSM one. Finally we will slightly change the MSSM parameter space to allow for a configuration where the light Higgs boson can become
extremely elusive at the LHC, and we will conclude on the constraining power of the Higgs searches.

\subsection{Parameter space}

In this first try, we go back to our first scenario : $m_{h\ max}$, with heavy stops. We vary $\tb$ and $\ma$ in the range
$$\tb\in[2,40]\qquad\ma\in[50,450].$$

Since we are dealing with a large parameter space (22
dimensional), performing a satisfactory scan is a crucial issue.
We attempted first a search with random Markov chains, but it
ended up to be limited by the frequentist character of the
technique : indeed a Markov chain will stay in regions depending
on the number of allowed points in it. Because our model exhibits
regions that are extremely more populated than others, the Markov
chains showed a tendency to stagnate. Besides, we are not
interested in the density of points, but in disentangling what
lies in the reach of such a generic susy model, and what is
incompatible with it. In particular one of our first motivations,
in accordance with the main reason for considering such scenarios,
is to explore regions where $m_h$ is much heavier than what is in
the usual MSSM. So, after an exploratory random scan we aim at
populating regions giving largest values for $m_h$. We first carry
a blind random scan on all parameters, then we pick  up a point
exhibiting a large mass and scan again by perturbing around this
point. In fact we perturb only around the $\grandO{1/M^2}$
$a_{ij}$ values rescaling them by a common factor while rescanning
on the $\grandO{1/M}$ which give the leading order effect in the
increase of $m_h$. In the scans that we will show later, for
$\tb=2$, we perturb around
$$\begin{array}{|cc|cc|cc|cc|}
& & a_{10} & 0.168605 & a_{11} & -0.55814 & &\\
a_{12}& 0.511628 & a_{20} & 0.0465116 & a_{21} & 0.639535 & a_{22} & 0.802326\\
a_{30} & 0.151163 & a_{31} & 0.744186 & a_{32} & 0.284884 &  a_{40} & 0.238372\\
a_{41} & 0.383721 & a_{42} & 1. & a_{50} & 0.848837 & a_{51} & -0.133721\\
a_{52} & -0.732558 & a_{60} & 0.598837 & a_{61} & 0.575581 & a_{62} & 0.331395
\end{array}$$
We will refer to this combination as $c_1$. We have then carried
out reduced scans on the parameter space by randomly choosing a
triplet $(x,\zeta_{10},\zeta_{11})$ in the cube $[-1,1]^3$,
and associate it to the point
\begin{equation}
 p=(\zeta_{10},\zeta_{11},x\times c_1)
\end{equation}
This choice strongly relies on the fact that most of the
phenomenological change are brought by the order 5 operators, even
though the order 6 ones are essential in raising $m_h$ further.

\paragraph{Constraint from $t \to H^+  b$}~\\
Because there is no charged Higgs boson in the Standard Model, the searches for $H^+$ are somehow less numerous than the ones for the neutral bosons.
Constraints from not too heavy charged Higgses are imposed
exploiting the search of charged Higgs boson in top decays, done
by CMS (\cite{cms_eps_Hp}). The latter explores the channel
$t\rightarrow H^+ b$, $H^+\rightarrow\nu\tau^+$. Special care was
taken in the computation of the branching ratio of $t\rightarrow
H^+ b$, since it can be affected both by QCD corrections and by
supersymmetric-QCD corrections. The first have been included using
the \texttt{HDecay} code, and the second by including the $\Delta
m_b$ correction following\cite{deltamb_spira,micromegas2}. To end
up with the correct branching ratio, QCD corrections were also
taken into accounts for $t\rightarrow W^+ b$ using
\texttt{HDecay}.\\

Concerning neutral Higgses, the direct searches at colliders are taken into account
by comparing the ratio $\sigma/\sigma_{SM}$ at the 95\% CL
exclusion value for each analysis by LEP, TEVATRON and the LHC.
This is automated via HiggsBounds \cite{HB} for LEP and Tevatron. We
must also account for the case where two of them get degenerate : in
this case the two cross-sections must be added. We define two
Higgs bosons to be degenerated when their mass difference is less
than 10 GeV for hadron colliders (LHC and Tevatron) and 2 GeV for
LEP. Concerning the LHC analyses, and as stated in the previous section, it is not straightforward to include them consistently, let us see first a naive
implementation.

\subsection{First LHC implementation : the inclusive}

In this first analysis all Higgs search data from from both ATLAS
and CMS as presented at {\em Lepton Photon 2011} are used ranging from
$1$fb$^{-1}$ to 2.3 fb$^{-1}$. For short we will sometimes refer to this analysis as $2$fb$^{-1}$ data,
\begin{itemize}
 \item $H\rightarrow\gamma\gamma$, done by ATLAS (\cite{atlas_lp_gamgam}) and CMS (\cite{cms_lp_gamgam}).
 \item $VH\rightarrow V\oo{b}b$, done by CMS (\cite{cms_lp_bb}).
 \item $H\rightarrow WW$, done by ATLAS on different final states ($l\nu l\nu$ \cite{atlas_lp_ww}, $l\nu qq$ \cite{atlas_lp_wwb}) and CMS ($l\nu l\nu$
\cite{cms_lp_ww}).
 \item $H\rightarrow ZZ$, done by ATLAS on different final states ($4l$ \cite{atlas_lp_zz}, $2l2q$ \cite{atlas_lp_zz_2l2q} and $2l2\nu$
\cite{atlas_lp_zz_2l2nu}) and CMS ($4l$ \cite{cms_lp_zz}, $2l2q$ \cite{cms_lp_zz_2l2q}, $2l2\nu$ \cite{cms_lp_zz_2l2nu} and $2l2\tau$ \cite{cms_lp_zz_2l2tau}).
 \item $H\rightarrow\tau\tau$, done by ATLAS (\cite{atlas_lp_tautau}) and CMS (\cite{cms_lp_tautau}).
\end{itemize}

A priori, all these analyses are dedicated to the SM, however we
can still try to compare \textit{exclusive} cross-sections : we will hence compute the total cross-section by adding all production modes, as stated in eq
\ref{eq:inclusive}. All Standard Model cross-sections have been taken from the
LHC Higgs cross-section working group
(\cite{LHC_Higgs_cs1,LHC_Higgs_cs2}) except for the $b$ quark
fusion, computed with \texttt{bbh@NNLO} (\cite{bbh_at_nnlo}).\\
Another subtlety related to the quadrature sum combination that we use is that, for analyses that exist only in one of the
collaborations ($V b \bar b$ in CMS, $WW\to l\nu qq$ ATLAS, $ZZ
\to ll \tau \tau$ in CMS), we make up for the lack of the
corresponding analysis by including it in our analysis through a
scaling factor $\sqrt{2}$ to the corresponding ratio. This
approach is followed in Ref.\cite{carena_bmssmhiggs_1111} also.
The test applies separately to   the three neutral Higgses (though
in the CP-odd case, some analyses like $H \rightarrow WW$ do not
apply) and rejects all points were at least one Higgs fails to
pass the test.\\

\subsection{Differences between MSSM and BMSSM}

\begin{figure}[!h]
\begin{center}
\begin{tabular}{cc}
\includegraphics[scale=0.3,trim=0 0 0 0,clip=true]{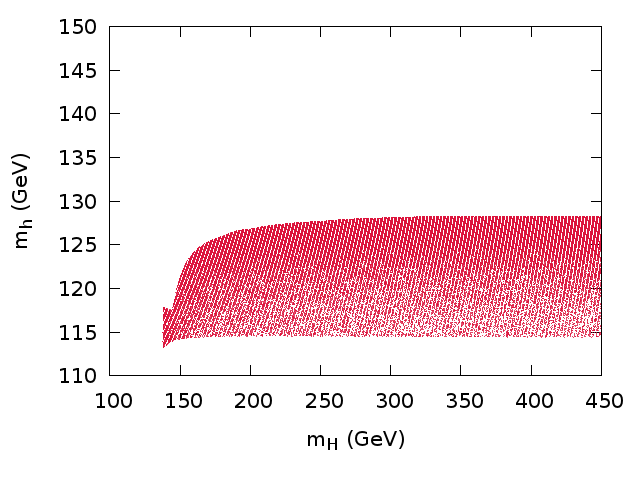}&
\includegraphics[scale=0.3,trim=0 0 0 0,clip=true]{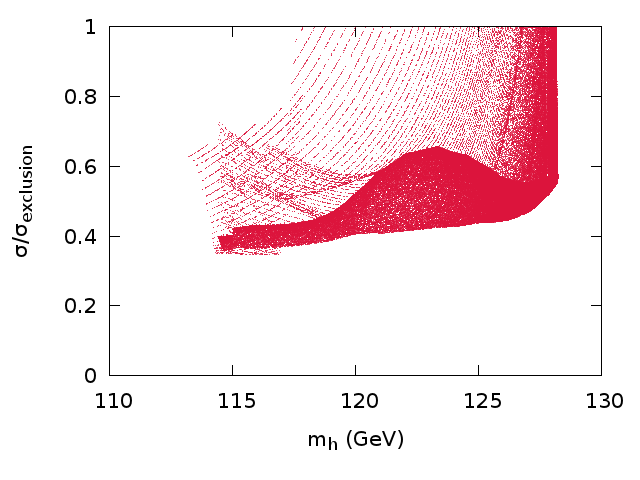}
\end{tabular}
\end{center}
\caption{\label{fig:mh_MSSM} {\em The allowed range in the
$m_h-m_H$ plane for our reference MSSM model is shown in the left
panel. The right panel shows the ratio $\sigma/\sigma^\excl$ as a function of $m_h$. The bulk corresponding to the high point density region corresponds to 
cases where the highest signal comes from $h$, whereas thin stripes correspond to points where the highest signal comes from $H$ or $A^0$.}}
\end{figure}

\begin{figure}[h!]
\begin{center}
\hspace*{-0.8cm}
\begin{tabular}{cc}
\includegraphics[scale=0.36,trim=0 0 0 0,clip=true]{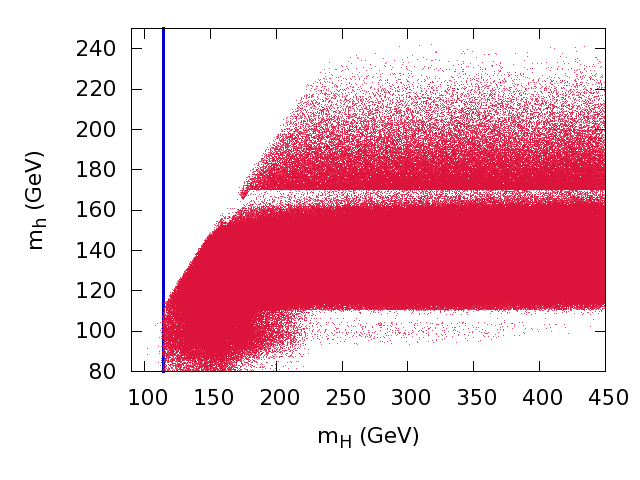}&
\includegraphics[scale=0.36,trim=0 0 0 0,clip=true]{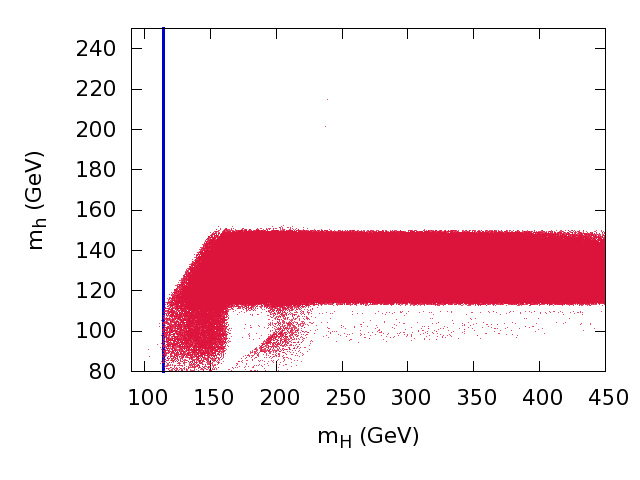}
\end{tabular}
\end{center}
\caption{\label{fig:mh_BMSSM} {\em BMSSM predictions before (left)
after (right) applying the LHC constraints in the $(m_H,m_h)$
plane. The vertical blue line shows the LEP SM bound $m_\Phi=114$
GeV.}}
\end{figure}

We show in fig.~\ref{fig:mh_MSSM} and \ref{fig:mh_BMSSM} the
allowed points obtained with the analysis we have just described,
either in the MSSM and BMSSM cases. In the MSSM case in the particular scenario we have chosen,
fig.~\ref{fig:mh_MSSM}, the light Higgs mass is distributed in
between the LEP bound (114 GeV) and the maximum of the radiative
corrections (about 130 GeV). We have also plotted here the ratio
$R_\sigma=\sigma/\sigma^\excl$ of each point (it is not
necessarily a $h$ signal, but can be any of the three Higgs
bosons) against the mass of the lightest Higgs. We notice that
this MSSM scenario is largely unaffected by the current experimental results
since the ratio between the predicted production rate to the
excluded production rate can be as small as 0.4. There are of course points, especially some  with the highest $m_h$
predicted in this model  which require much less
luminosity increase to be excluded or discovered.
\\

In the case of the BMSSM, the fact that $m_h$ can be raised to values as
high as $250$ GeV changes the picture quite drastically as compared to
what was allowed before the LHC data (LEP and Tevatron data are
included in both sets). Fig.~\ref{fig:mh_BMSSM} shows that with
just about $2$ fb$^{-1}$ of collected data the $m_h-m_H$
plane has shrunk considerably due to the fact that a rate 2 times
smaller than the SM for $m_h>160$ GeV is excluded. This shows in
particular that $m_h > 150$ GeV is now excluded. Therefore the main
{\em raison d'\^{e}tre} of such models that aimed at raising the
lightest Higgs mass considerably is now gone. Only an extra
$15$ GeV increase for the lightest Higgs compared to the maximal
value attained in the usual MSSM framework is still allowed. Therefore the majority
of models that survive have $114<m_h<150$ GeV, but we do find some regions with smaller values of $m_h$.\\

Indeed, while we find that the heaviest CP-even Higgs is above the LEP
limit, $m_H>114$ GeV, in the range $114 < m_H < 220$ GeV we find
models where the lightest Higgs is lighter than the LEP limit of $114$ GeV, we even find that models with $m_h<M_Z$
are still possible. In these configurations the lightest Higgs is far
from being SM-like. We have seen that the $hWW$ coupling can be drastically reduced. In this
case it is $H$ that picks up almost the totality of the $HWW/HZZ$
coupling, which explains why $m_H>114$ GeV (LEP constraint). The
configuration with $m_h<100$ GeV consists of two separate scenarios as
fig.~\ref{fig:mh_BMSSM} shows. One notices a region that
corresponds to $m_H> 2 m_h$ starting at $m_H=160$ GeV. Here the
branching ratio $H\mapsto hh$ can be as high as 0.6, with $h$
decaying almost exclusively to $b$ quarks, making such scenarios
difficult to probe at the LHC. For $114< m_H < 160$ (GeV), some
scenarios are still viable because they correspond to $gg \to H$
that can go down to $50\%$ the value of the SM. Since this
reduction is limited to no more than $50\%$, such scenarios will
eventually be excluded by a luminosity increase. Other scenarios
in this mass range have a $\tb$ enhanced $bbH$ coupling, which
is constrained through $VH\rightarrow V\oo{b}b$ and
$H\rightarrow\tau\tau$ which includes $b\bar b \to H$ (since the BR to $\tau$ is also
significantly enhanced). As the luminosity will increase so will
the sensitivity of these last two channels.

During the process of writing the article reference ~\cite{carena_bmssmhiggs_1111} has carried at the same time a
very similar analysis. Our results are in very good agreement with
theirs apart from the region with $m_h<100$ GeV which is more
populated in our case. Note that although we carry a similar LHC
analysis, we differ in the choice of the MSSM reference point and
more importantly in the scan over parameters. One example is the
scan in $\tb$ that is covered uniformly in the range $2$ to $40$
in our case, whereas in Ref.~\cite{carena_bmssmhiggs_1111} the
emphasis was on $\tb=2$ and $\tb=20$  with a sparse scan in
between. In fact we have verified that our models that survive the current LHC Higgs constraints have
$\tb$ in the range $5$ to $15$.

\subsection{Second LHC implementation : the exclusive}
As was stated on the initial warning at the beginning of the chapter, the use of the \textit{inclusive cross-section} is not guaranteed to give correct results.
In fact this depends on whether the analysis can
differentiate between the different production modes of the Higgs,
which would allow to fold in the weight of the different channels
in the analysis. As an example if one tries to interpret the $WW\to 2l2\nu$ channel, which is separated in the $0j/1j/2j$ bins with a model with
gluon fusion dominating all other production modes, like a heavy
4th generation (and to a certain extent the SM), the exclusion
will be purely driven by the 0-jet subchannel. If one takes a
fermiophobic model, the gluon fusion vanishes  and the exclusion
is given by the 2-jets subchannel. It is clear that the \textit{exclusive}
obtained in the 4th generation model or the fermiophobic model
with the same \textit{inclusive} cross-section is not the same and will not
lead to the same exclusion limits. In such a case one need to know the efficiencies of each production mode, or at least their relative ratios.\\

\begin{figure}[h!]
\begin{center}
\begin{tabular}{cc}
\includegraphics[scale=0.3,trim=0 0 0 0,clip=true]{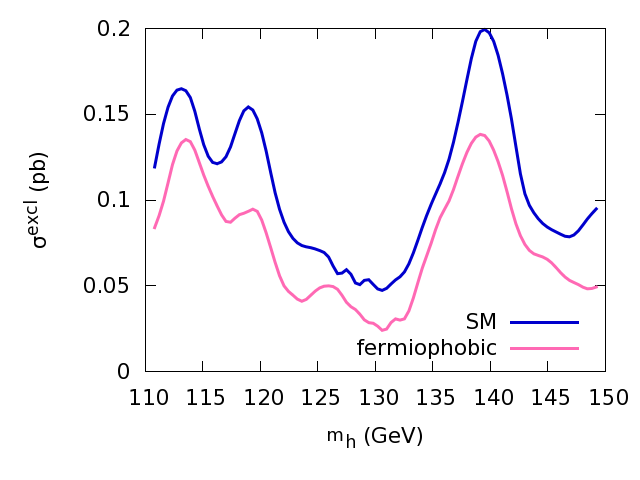}&
\includegraphics[scale=0.3,trim=0 0 0 0,clip=true]{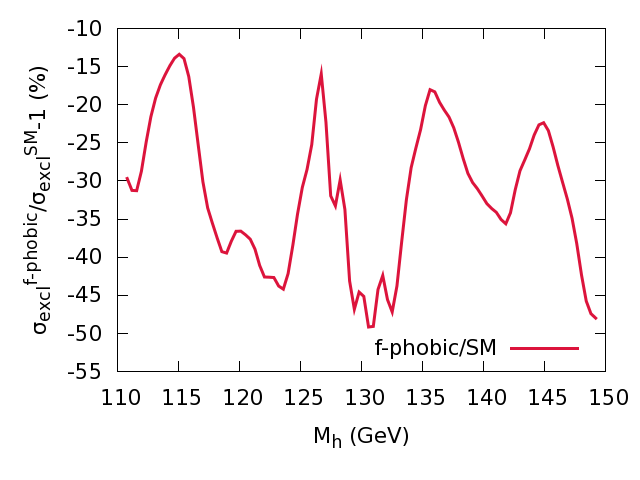}
\end{tabular}
\end{center}
\caption{\label{fig:gamgam} {\em The panel on the left shows the
SM and fermiophobic excluded cross-sections in the $H\rightarrow
\gamma\gamma$ CMS analysis, these plots are extracted from
Ref.~\ref{fig:gamgam}. Cross sections are given  in picobarns. On
the right is shown the relative difference between fermiophobic
and SM analyses, in percent units.}}
\end{figure}

For $H \to \gamma \gamma$ the separation in the bins high/low $p_T$ could be very useful and efficient in, again,
the case of a fermiophobic Higgs whose $gg$ induced cross section
is vanishing, in sharp contrast to the SM Higgs, since the $p_T$ spectrum of the VBF and associated vector production is harder than the gluon fusion one. This
particular
model can be used as an example, though perhaps extreme since
one important SM channel is absent, to quantify the difference one
gets from an inclusive (in this case merging all $p_T^{\gamma
\gamma}$ regions) compared to an exclusive search or
exclusion limit  for each $p_T^{\gamma
\gamma}$ region. Such an approach has been performed by CMS
\cite{cms_lp_gamgam}\footnote{Note that very recently a similar
analysis was also released by the ATLAS collaboration
\cite{atlas_hcp_gamgam}, which we however do not consider here}.
In that analysis the classification is done according to
$p_T^{\gamma \gamma}>40$ GeV for enhancing the fermiophobic signal
over background. If we consider the inclusive cross-section, in
this case no $p_T^{\gamma \gamma}$ separator, to set the limits on
$\sigma\times BR$, the limit is model-independent since there is no combination. Using the
$p_T^{\gamma \gamma}>40$ GeV as a separator gives a much more
powerful limit, though model dependent since it is a combination, as shown in fig.~\ref{fig:gamgam}.
Unfortunately CMS does not provide separate exclusion limit for each region in $p_T^{\gamma \gamma}$, which prevents us to derive ourselves the combined
exclusion. In fact CMS gives the values obtained for $\sigma^\text{excl}$ using both models, in the inclusive and the exclusive analyses for  $m_h=120$ GeV.
$$\begin{array}{ccc}
 & \sigma_\text{merged}^\excl (pb) & \sigma_\text{split}^\excl (pb)\\
 \text{SM} & 0.1308 & 0.1104\\
  \text{fermiophobic} & 0.1303 & 0.0696\\
\end{array}
$$
One notices first that the values of $\sigma_\text{merged}^\excl$ are nearly the same in the two models, which was expected.  We note that the gain in the SM is
approximately 20\%, and in the fermiophobic model nearly 50\%.

This means that if one had used the SM limit with the inclusive
approach in the context of a fermiophobic Higgs, one would have
lost a factor 2 in sensitivity compared to a more refined
exclusive analysis.\\
Unfortunately at present the details of the analyses performed by
ATLAS and CMS do not provide all the needed information and
efficiencies that we require for an exclusive approach. At present
in a phenomenological analysis like ours the best that can be done
is to simulate the experimental analysis through a Monte-Carlo with the caveat
that some detector issues contributing to the efficiencies may be lost.

\subsubsection{Refining the analysis}
Improving the analysis means that we will attempt to exploit those channels where separators leading to exclusive observables
have been conducted. Of course the situation is different from the case of the fermiophobic model in the $\gamma \gamma$ signature
where only one channel is selected. Moreover the fermiophobic model is well defined, the $gg \to h$ cross section is vanishing. In the scans we perform in the BMSSM
case one is in fact considering many models where a given Higgs mass, $m_h$,
corresponds to models with very different properties. Let us first go through all the channels we have used in the previous analysis and comment on how
one could, for some of them, take into account the exclusive nature of a particular final state.

\begin{figure}[h!]
\begin{center}
\includegraphics[scale=0.3,trim=0 0 0 0,clip=true]{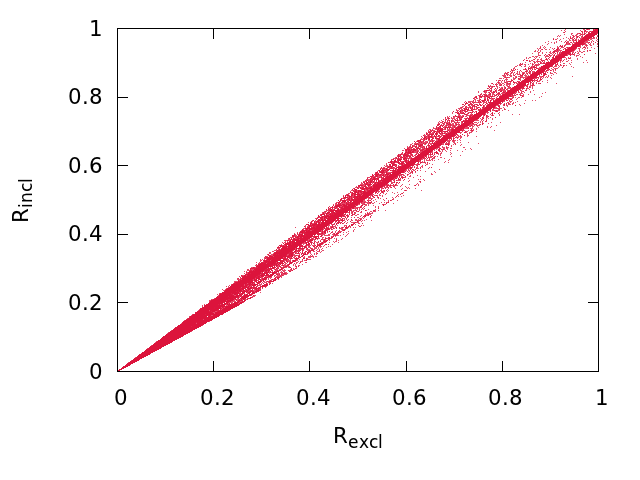}
\end{center}
\caption{\label{fig:ww_incl} {\em We show the exclusion power based on the inclusive analysis compared to the refined analysis, see text, applied to searches
for the heaviest CP-even Higgs}}
\end{figure}

\begin{itemize}
 \item $VH\rightarrow V\oo{b}b$. In this case there is only one
 production mode, the vector boson associated production. Although
 it is strictly speaking two modes, the $Z$ and $W$, the scaling
 factor from the SM is nearly exactly the same, which simplifies
 the analysis. Here we can safely use inclusive cross-sections.

\item $H\rightarrow \tau^+\tau^-$.
This channel is of interest in the MSSM and BMSSM for high $\tb$. $H$ is produced
either through $gg$ fusion of $bb$ fusion.
The ATLAS analysis (\cite{atlas_lp_tautau}) presents excluded cross-sections
for each of these two production modes.  This is most useful when analysing a new model as we can weigh each sub-channel separately.  This piece of information
is extremely helpful since
it gives the efficiency in a very handy way : one has just to
compute the ratio of each production cross section to its excluded
value, sum them and compare to 1. Indeed as we deal with a
counting experiment, this is adding events from each production
mode and compare it with the excluded number of events, which, in the
approximation of no theoretical systematics, is justified.

 \item $H\rightarrow ZZ\rightarrow 4l$.
Unlike the $WW$ signature where an analysis including 0-jet, 1-jet and 2-jet is performed, for the $ZZ$ channel one only has at the moment
a fully inclusive analysis.

\item $H\rightarrow \gamma\gamma$.
We have just seen in the fermiophobic Higgs search that CMS, and similarly ATLAS, divide the phase space
according to $p_T^{\gamma \gamma}$, thus allowing to give different exclusion limits if one assumes a fermiophobic
model rather than the SM. As we have just argued,  the efficiencies in the two regions are not given. Our procedure here is to correct
the exclusive analysis of CMS \cite{cms_lp_bb} by $20\%$ to  recover the fully inclusive limit. Although this scaling was derived for $m_h=120$ GeV,
considering the narrow range of the $\gamma \gamma$ channel we assume this scale factor to be roughly constant. This is a conservative approach, but a precise
analysis requires the exclusion cross section for each subchannel (here the $p_T^{\gamma \gamma}$ regions) and the efficiencies of each mode.

 \item $H\rightarrow WW\rightarrow l\nu l\nu$. Both ATLAS and CMS split the channel according to the number of
 recorded jets, which allow to gain sensitivity to specific
 production modes ($gg \to H$ or VBF). Fortunately enough, ATLAS provides exclusion
 limits for the 0-jet and 1-jet subchannel. Providing the 2-jet that would select the VBF would
 be extremely useful. Once again though the weight of the 0-jet and 1-jet in the ATLAS analysis are folded in, these weights are not provided.
Simulating the ATLAS analysis one could in principle calculate these weights or efficiencies. We have run
\texttt{PYTHIA} for a SM Higgs boson through
gluon fusion, VBF or $b$ quark fusion and extracted  the
efficiency of each production mode. Although this may seem far too
naive since full detector simulation is not applied we are only interested in the relative efficiencies, say
the ratio between the VBF  and gluon fusion. One  expects that a full detector simulation does not affect these ratios much.
The ratios we calculated were validated by the ATLAS collaboration\footnote{The VBF ratio to gluon fusion was in
very good agreement. Private communication.}. $b$ fusion could not be checked since it is not included in a SM Higgs analysis.  We were then able to fold in
these ratios
within a refined exclusive analysis.  We show in fig
\ref{fig:ww_incl} the relative difference between the
inclusive and exclusive, defined in eq~\ref{eq:inclusive},\ref{eq:exclusive}. This
relative correction is mainly positive, up to $30$\% which can be
traced back to the fact that the $b$ fusion efficiency is higher
than the gluon fusion one.
\end{itemize}

\begin{figure}[h!]
\begin{center}
\begin{tabular}{cc}
\includegraphics[scale=0.3,trim=0 0 0 0,clip=true]{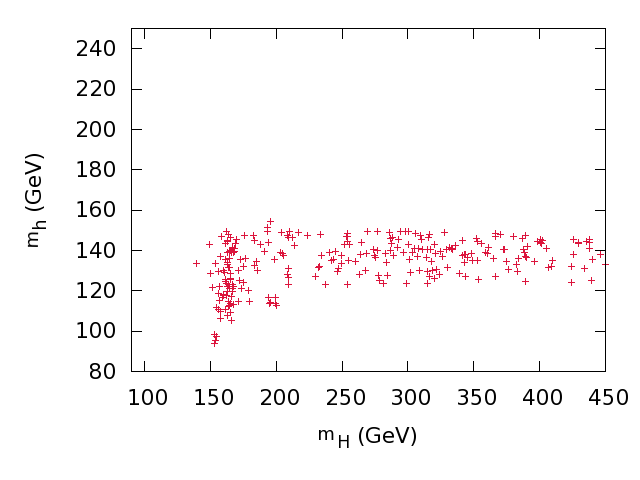}&
\includegraphics[scale=0.3,trim=0 0 0 0,clip=true]{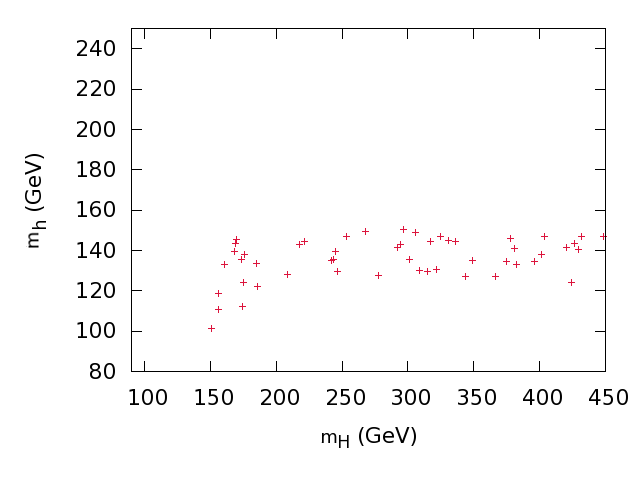}
\end{tabular}
\end{center}
\caption{\label{fig:mh_subset} {\em Taking a small subset of models, we apply an inclusive analysis, left panel, and compare it to the result of a refined
analysis, right panel. }}
\end{figure}

To summarise, we see that for the moment, the refinement concerns only two channels and may seem a modest improvement, but it is important to send a
request  to the collaborations so that details of the analyses with the weight and efficiencies of all channels and sub-channels be released. It is important to
stress that what we call the refined analysis is our approach to arrive at what we think is a better treatment of such models, nonetheless with the inclusive
analysis this allows to compare and quantify the assumptions. It should also be clear that the refined analysis does not necessarily mean that it is more
constraining than the inclusive one.
Before turning
to the final results taking into account these refinements and in order to understand their impact when scanning over a large set of parameters, we compare the
exclusion power in terms of the inclusive approach compared to the refined analysis, eq.~\ref{eq:exclusive}, applied to the heaviest CP-even Higgs for
illustration. The comparison is shown in fig.~\ref{fig:ww_incl} corresponding to the luminosity $2$fb$^{-1}$ (Lepton-Photon 2011). Note that we only display
values with $R_{{\rm incl}}=\sigma_\text{incl.}/\sigma^\excl<1, R_{{\rm excl}}=\sigma_\text{excl.}/\sigma^\excl<1$ corresponding to models that are still
viable. When the luminosity increases, the condition $R_{{\rm incl}}<1, R_{{\rm excl}}<1$ can be read from the plot, but would correspond to smaller $R$ values.
~
The figure shows that there is, unfortunately, little spread around $R_{{\rm incl}}=R_{{\rm excl}}$, the largest differences attaining about $20\%$ for $R<0.3$.
 A scan over the entire parameter set, taking all constraints on all Higgses, showed practically not much difference between the refined and inclusive approach
when projected on the $m_h-m_H$ plane. So we will not show such plots. However to illustrate that the two analyses do exclude different sets of models, we have
generated a well chosen subset of models\footnote{The subset has $R_{{\rm excl}}>0.99$ applied  to all three Higgses. In the refined analysis $R_{{\rm excl}}<1$
is imposed while in the inclusive analysis $R_{{\rm incl}}<1$.}
 and passed them through the two  analyses, inclusive and refined. In this (biased) chosen subset of models, we see in fig.~\ref{fig:mh_subset} that the refined
analysis excludes many more models.  Had we performed a full scan, the differences in the projection on the plane $m_h-m_H$ would hardly be visible.

\subsection{Elusive Higgs : the case of invisible decay}

We present now the preliminary results of the consequence of the
Higgs decaying to invisible particles\footnote{The earliest
mention  of an invisible Higgs and its connection to dark matter
that we are aware of is made in a simple extension of the standard model
\cite{zee_darkons_1985}.}. Despite numerous advantages, the $m_{h\
max}$ scenario does not cover the full diversity of the MSSM nor
the BMSSM, in particular it does not cover cases where the Higgs
can decay to neutralinos, in particular the lightest ones. The
latter are good dark matter candidates and therefore these decays
of the lightest Higgs are into invisibles. In order to have a
sufficient branching ratio to the neutralino one must have a
neutralino which is light enough, $M_{\tilde{\chi}^0_1}<m_h/2$. We
do not wish here to conduct a thorough analysis of the BMSSM
Higgses into invisibles and review all the constraints from dark
matter, we leave this to a more focused study. Dark matter issues
within the BMSSM taking into account the dim-5 operators were
conducted in
\cite{cheung_bmssmdm_0903,gondolo_bmssmdm_0906,bernal_bmssmcosmo_0906,goudelis_bmssmdm_0912}.
Though succinct our implementation includes dim-6 operators
automatically. In the recent approach of
\cite{romagnoni_bmssmgoldstino_1111} which can be related to a
BMSSM implementation, decays are into invisible light scalars.

In this exploratory study we consider $M_{\tilde{\chi}^0_1}<80$
GeV. In order to achieve this while taking into account LEP limits
on the chargino mass, such light neutralinos are dominantly
bino-like. However in order to couple to the Higgs efficiently
there must be a higgsino component that is not too negligible, see
for example \cite{boudjema_higgs_inv}. One should therefore have
$M_1$, the bino mass, and $\mu$ not too far apart. We will set
$M_1=50$ GeV to have a light neutralino and $\mu=200$ GeV to have
enough mixing. The alert reader will have noticed that this value
of $\mu$ is smaller than what we have been using so far. In order
that our previous results are not much affected so that we can compare
with what an invisible decay brings, one should remember  that the
phenomenology without invisibles is not much changed  if one keeps the
ratios $\mu/M,\ m_s/M$, that governs the effective expansion in
$1/M$, identical to what
was stated in eq.~\ref{eq:ratio}. Very small differences are due to a change in the small contributions of order 
$v^2/M^2$.\\

\begin{figure}[h!]
\begin{center}
\includegraphics[scale=0.4,trim=0 0 0 0,clip=true]{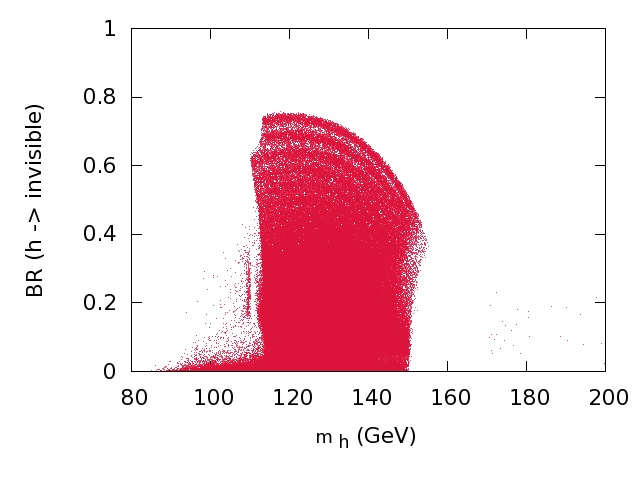}
\begin{tabular}{cc}
\includegraphics[scale=0.3,trim=0 0 0 0,clip=true]{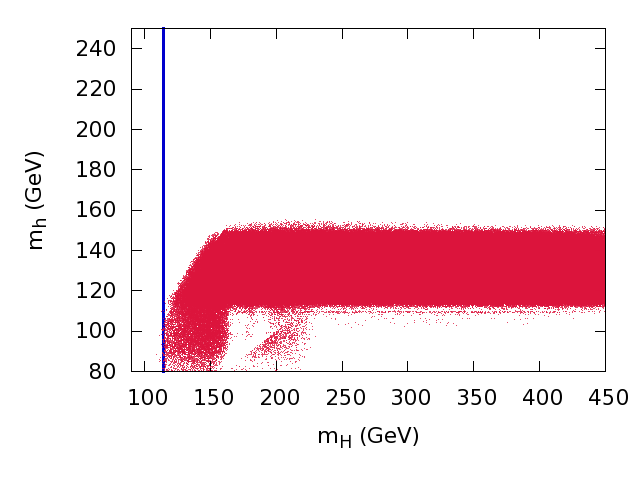}
&
\includegraphics[scale=0.3,trim=0 0 0 0,clip=true]{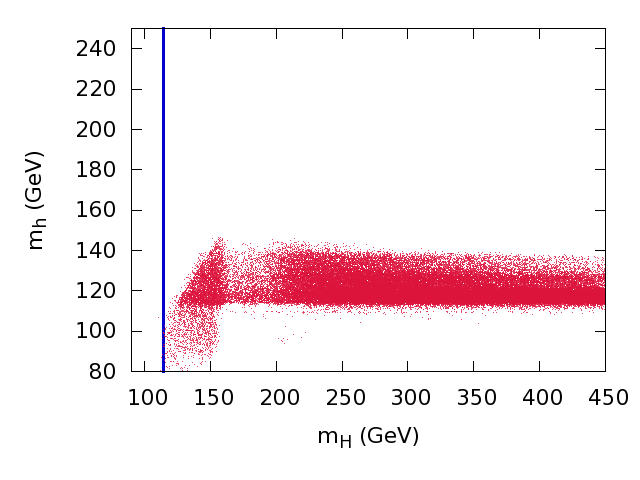}\\
\end{tabular}
\end{center}
\caption{\label{fig:BMSSM_inv} {\em Higgs decays to invisible neutralinos. The first graph shows the branching ratio of the
light Higgs to the lightest neutralinos, see text for details on the parameters of the neutralinos. In the second row, the first panel shows the allowed
$m_h-m_H$ space taking into account
the present LHC constraint ($2$fb$^{-1}$). The graph on the right is for a luminosity of $15$fb$^{-1}$, assuming no signal appears.}}
\end{figure}

Fig~\ref{fig:BMSSM_inv} shows the branching ratio of the light
Higgs to the lightest neutralinos. Between $m_h=120$ GeV and
$m_h=150$ GeV, the branching ratio is substantial ranging from
$~80\%$ to $40\%$ for $m_h=150$ GeV, at which point it drops
precipitously to almost $0\%$ because of the opening of the $WW$
channel. When the branching  into the lightest neutralinos is
large it reduces all the usual branchings and leads to a much
reduced sensitivity of the Higgs signal. Fig.~\ref{fig:BMSSM_inv}
shows how the picture changes when decays to invisibles are
allowed. With the current data  ($2$ fb$^{-1}$) it is difficult to
see that changes have occurred. This is not surprising since our
invisible scenario can only cut in the $m_h$ range $120-150$ GeV .
With the present luminosity this range is still very much viable
even without Higgs decays as we have seen.  With the luminosity at
$15$ fb$^{-1}$, we clearly see the {\em damaging} effect of the
invisible decays. More models with Higgs masses up to $m_h=140$ GeV
survive compared to the case where no invisible Higgs decays are
allowed.

\subsection{Conclusion}

It is quite clear by now that Higgses in supersymmetric models
that go beyond the MSSM are very much constrained by the LHC
searches, even though the primary goal of those searches is the
Standard Model Higgs. We have first shown that the higher-order
terms appearing in the effective Lagrangian  alter  the Higgs
phenomenology quite significantly, in particular by raising the
lightest Higgs mass to values up to 250 GeV. This feature alone
was the main motivation of the BMSSM. We have shown that with the
advent of the LHC and the data collected so far, experiments no
longer allow a lightest supersymmetric Higgs to have a mass beyond
$150$ GeV even in these BMSSM set ups and even if we allowed for
decays into invisibles as provided by the lightest neutralinos.
With the increase of the luminosity most of the remaining models
at $15$ fb$^{-1}$ are within a thin layer in lightest Higgs mass,
with $114<m_h<140$ GeV with a concentration around $m_h\sim
120$ GeV, apart from an island with $m_h<100$ GeV for $m_H$ low
enough, $m_H<150$ GeV. Invisible decays  allow more models with
$m_h \sim 140$ GeV. Within this picture, set in terms of
exclusions, and with $15$ fb$^{-1}$ of data in the no-signal case, a similar conclusion
in terms of masses applies to the MSSM, the BMSSM lightest Higgs
is allowed to be less than about $10$ GeV heavier that what it can
be in the MSSM, whereas before the advent of the LHC masses for
the lightest BMSSM Higgs up to $250$ GeV were possible.  Still the
phenomenology of the two models are quite different. Although our
philosophy in this paper has been towards constraining the BMSSM
models in the pessimistic prospect of no Higgs signal, it would be
very interesting to revisit the models in case of a signal. If the
density of allowed models that we have found is any indication for
where a possible signal may be hiding and if the possible slight
excess in the latest data from the LHC is confirmed,  it would be
extremely interesting to check whether the signals are better
described by a BMSSM Higgs with $m_h=125$ GeV and what the
properties of the latter are. Could one always tell it apart from
a MSSM one or even a standard model one? We have not addressed
this issue here. What we have addressed however, though perhaps
partially, is how to exploit LHC results made for the SM Higgs in
the context of other models that can have quite different
properties. We have made a request that the collaborations should
provide more details about the weight of the different
sub-channels that are used in their analyses.

\section{The signal case}
The second analysis, following the exciting piece of data released at the end of 2011, explores the eventuality of a signal around $m_\phi=125$ GeV. The
interpretation of this would be signal was done in two Higgs doublet model (\cite{Ferreira:2012my}), in the MSSM
(\cite{Heinemeyer:2011aa,gunion_higgs_125,carena_higgs_125,arbey_higgs_125,muhlleitner_higgs_125,maiani_higgs_125,Desai:2012qy,draper_higgs_125,hall_higgs_125})
, the NMSSM (\cite{ellwanger_higgs_125,cao_higgs_125,kang_higgs_125}) and with an effective lagrangian
(\cite{grojean_higgs_125,contino_higgs_125,falkowski_higgs_125}). In the case of the BMSSM, we have just seen that the allowed parameter space could afford
either a light Higgs or a heavy one at such a mass. Though for the sake of the example the analysis will use the best fit values for the signal strength
recorded by both collaboration (which are, so far, subject to large experimental uncertainties), the idea is to estimate what kind of non Standard Model like
signals we can accommodate in 
the BMSSM framework. Indeed, as we have anticipated in the introduction, there are correlations between couplings, so that it is not possible to reproduce any
kind of signature. After having worked out the possibility of a light Higgs generating the signal or a heavy Higgs doing so, we will also study the prospect for
signals in other channels that so far present no excesses.

\subsection{Parameter space}
Considering the impact of the third family on Higgs physics, we
decided to allow some flexibility in the stop sector, as compared to our first scenario. First, since we are not looking for a particularly heavy Higgs, we will
drop the $m_{h\ max}$ scenario to a more natural spectrum, with lighter stops, and we will hence consider scenario A and B presented in the previous chapter. In
scenario B, the largest value of the heaviest stop
$m_{\tilde{t}_2}=800$ GeV that we allow in the scan should be
regarded an extreme example, not only from the point of view of
naturalness but also because it is not far from the new
scale $M=1.5$ TeV. Furthermore the
heavy scale $M$ can be enhanced with little change to our results
provided one keeps fixed the  ratios $m_s/M,\mu/M$.\\

\subsection{Signal features}
The data that is most indicative of a possible signal is the
following (uncertainties correspond to the $1\sigma$ band)
\renewcommand{\labelitemi}{$\star$}
\renewcommand{\labelitemii}{$\bullet$}
\begin{itemize}
\item ATLAS\cite{atlas_5fb}:\\
The ATLAS collaboration records a combined (all channels) signal
strength of $1.5$~${}_{-0.5}^{+0.6}$ at $m_h=126$ GeV. It may be
considered as most revealing in channels with best resolutions on
the Higgs mass:
\begin{itemize}
\item The inclusive $\gamma\gamma$ channel  where the signal
strength is $2^{+0.9}_{-0.8}$ (see \cite{atlas_5fb_gamgam}\footnote{see the additional plots on
\url{https://atlas.web.cern.ch/Atlas/GROUPS/PHYSICS/PAPERS/HIGG-2012-02/}})
\item $ZZ \to 4l$ channel  where the
signal strengths is $1.2^{+1.2}_{-0.8}$ compatible with $WW \to ll
\nu \nu$, though the $WW$ channel  has a worse mass resolution.
\end{itemize}
\item CMS collaboration\cite{cms_5fb} reports a combined signal strength of $1.2^{+0.3}_{-0.4}$ at $m_h=124$ GeV
\begin{itemize}
\item In the $\gamma \gamma$
channels, the first CMS release with 4.9 fb$^{-1}$ was based on an
analysis with four subchannels that gave a signal strength of $1.7\pm0.8$ at $m_h=123.5$
GeV (see ref (\cite{cms_5fb_gamgam_b})). The updated release added
a dijet-tagged subchannel $\gamma\gamma+\text{ 2 jets}$ yielding by itself a signal
strength of $3.8^{+2.4}_{-1.8}$. The combination of the five subchannels yield a signal strength of $2.1^{+0.8}_{-0.7}$ (\cite{cms_5fb_gamgam}).

\item For the $ZZ
\to 4l$, the signal strength is $0.5^{+1.0}_{-0.7}$. Note that the mean is low, moreover
the mean value for $m_h$ is at 126 GeV.
\item the $b\bar b$ and
$\tau^+ \tau^-$ channels analysed by CMS\cite{cms_5fb_gamgam_b} in the mass range $122-128$ GeV have so much uncertainty
that they are of little use in the present analysis.
\end{itemize}
\end{itemize}

Let us emphasise again that there is still much
uncertainty in these results, some of which may not help in
drawing a  coherent picture, execpt perhaps in the $\gamma \gamma$
channel. The signal strengths are compatible with a Standard Model
Higgs, however it is tempting and in any case educative to
entertain the idea that some non standard Higgs scenario is
emerging. What is very interesting is that the different channels
and subchannels will allow, when measured with better precision,
to discriminate between different models and implementations of
the BMSSM. Most probably a first step in this discrimination in
this mass range will be performed with $\gamma \gamma, VV, \gamma
\gamma + 2\text{jets}$ perhaps also with the incorporation of the
$\oo{\tau}\tau$ channel. In the case of a multi-Higgs system this will
be done in parallel with searches for other Higgses. In the rest
of the paper we will investigate what kind of correlations between
these observables are possible within the BMSSM, for example
whether enhancements in all channels are possible.

\subsection{Light Higgs case : $h$}

\paragraph{Model A:}~\\

\begin{figure}[h!]
\begin{center}
\includegraphics[scale=0.3,trim=0 0 0 0,clip=true]{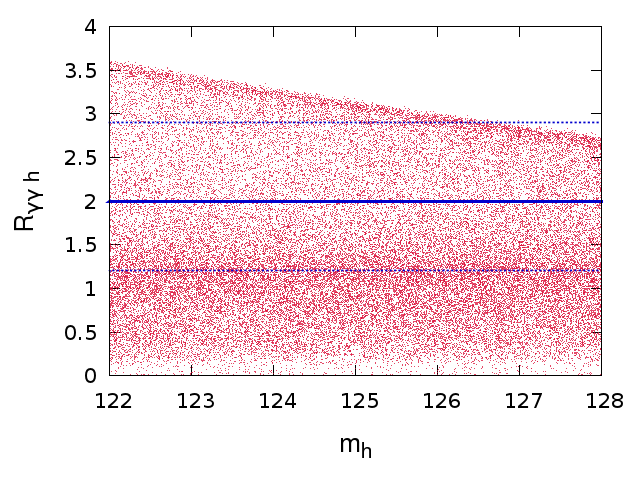}
\end{center}
\caption{\label{fig:h_Rgam} {\em Allowed region in the plane
$m_h,\Rgam$. The blue line represents the ATLAS best fit for the
signal strength, and the dotted lines are the one sigma deviations
from this value in model A.}}
\end{figure}
\begin{figure}[h!]
\begin{center}
\begin{tabular}{cc}
\includegraphics[scale=0.3,trim=0 0 0 0,clip=true]{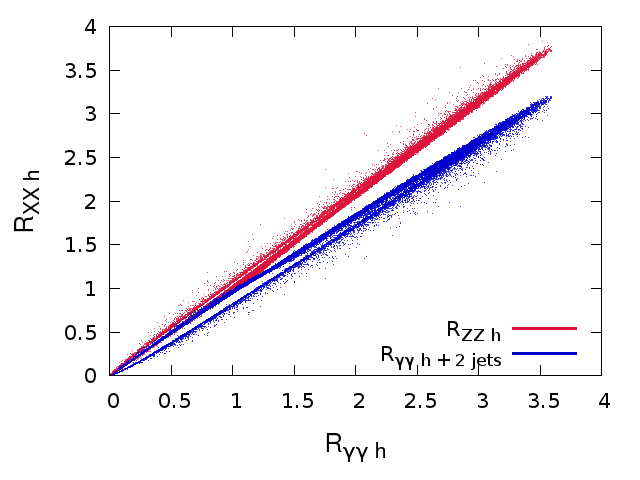}&
\includegraphics[scale=0.3,trim=0 0 0 0,clip=true]{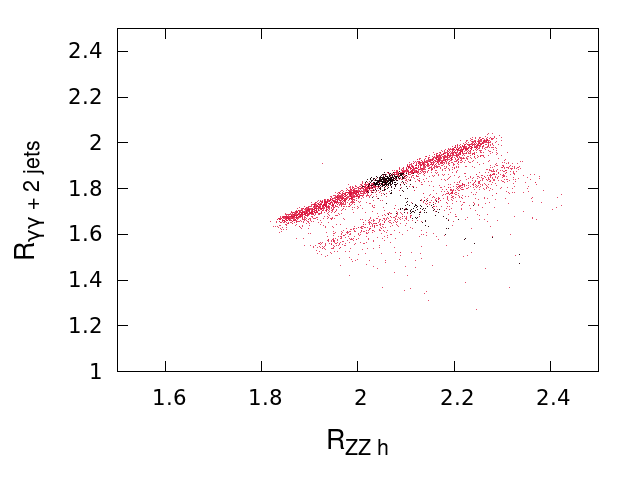}
\end{tabular}
\end{center}
\caption{\label{fig:nostop} {\em Left panel: correlation between
$\Rgam$, $\Rz$ and $R_{\gamma\gamma+\text{2 jets}}$ for $122< m_h
< 128$ GeV. Right panel: Imposing $\Rgamh=2.0\pm10\%$ (points in
red) and  $\Rgamh=2.0\pm1\%$ (points in black) we show the
correlation in the plane $\Rz$ and $R_{\gamma\gamma+\text{2
jets}}$. Both figures are for model A.}}
\end{figure}

Fig.~\ref{fig:h_Rgam} shows that with the current data, the BMSSM
yields a production rate in the inclusive $pp\rightarrow
h\rightarrow\gamma\gamma$ that can be quite small (as small as
$0.1$), and hence unobservable with the current luminosity or in
the very near future. More interestingly there is however no
difficulty in finding a signal in this channel that is up to $3.5$
times that of the SM. There is a very strong correlation with the
signatures in the other promising channels, namely $VV \equiv ZZ \to 4l$
and the $2\gamma \; + \;2 \; jets$, see fig \ref{fig:nostop}. With
small differences we have $R_{\gamma\gamma} \simeq R_{ZZ} \sim
R_{\gamma\gamma+\text{2 jets}}$. Rates above those of the SM are
mostly driven by reduction in the width of to $b \bar b$ which
increases all channels. This is trivially seen for the $2\gamma$
{\it versus} $ZZ$ channel. In the case of the
$\gamma\gamma$/$\gamma\gamma+\text{2 jets}$ correlation, when the
rates are above those of the SM, the inclusive channel is higher
by 20\% or so : this is related to the contribution of the $b$ quarks. Therefore  a
configuration with $R_{ZZ \to 4l}=1, R_{\gamma \gamma}=2,
R_{\gamma \gamma \; + \; 2 \; jets}=3$ is very much disfavoured in
Model A.

It is important to stress that the characteristics we find in
these scenarios occur for all values of $\tb$, even if statistically,
with a simple scan, the population with smaller $\tb$ is larger.

\begin{figure}[h!]
\begin{center}
\includegraphics[scale=0.3,trim=0 0 0 0,clip=true]{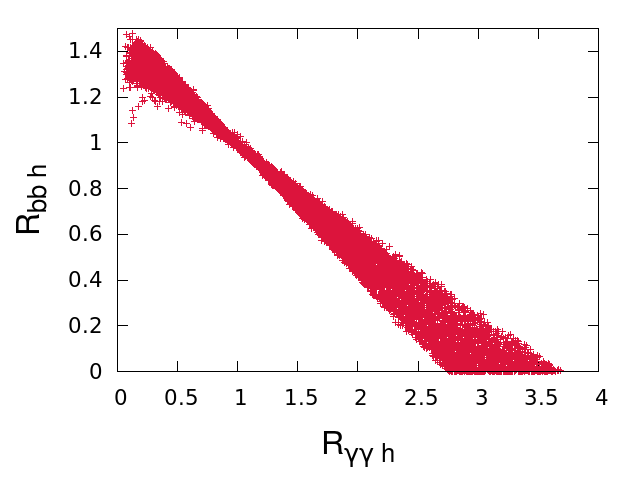}
\end{center}
\caption{\label{fig:tev_bb} {\em Correlation between diphoton
channel ($R_{\gamma\gamma}$) and the $VH\to V\oo{b}b$
($R_{\bar{b}b}$) in Model A.}}
\end{figure}

\noindent\subparagraph{Tevatron and the $\bar{b} b$ channel}~\\
While this work was being finalised, the Tevatron Collaborations
released new analyses \cite{tevatron_10fb} pointing out to a
possible  signal in $VH\to V \bar b b$ channel with a rate that
could be  compatible with the Standard Model expectation and with
a mass that could correspond to where the excesses are seen at the
LHC. This would seem at first sight to disfavour a scenario where
$g_{h\bar{b}b}$ is very much reduced. However,  one must keep in
mind that since the decay $H \to \bar{b}b$  dominates for
$m_h=125$ GeV, a suppression of the coupling by a factor two does
not imply a suppression of the branching ratio by a factor two.
The suppression is much more modest and there can still be a
significant enhancement of the diphoton channel without
suppressing too much the $VH\to V\bar{b}b$ channel. It must be
stressed that a more precise  measurement of the latter process
would really be helpful. Indeed, there exists also a correlation
between the diphoton (inclusive) channel and this channel, as
shown in fig \ref{fig:tev_bb}.

\paragraph{Model B:}~\\

It has been known for some
time\cite{Djouadi:higgsstopreduce,boudjema_light_stop} that, within
the MSSM, light stops endowed with a large mixing can drastically
reduce the $gg$ induced production. Even if this is accompanied by
an increase in the decay width to photons, the combined effect can
be a large drop in $gg \to h \to \gamma \gamma$. This effect is
encapsulated in the coupling of the stops to the Higgs. The
coupling of the lightest stop, $\tilde{t}_1$,
$g_{h\tilde{t}_1\tilde{t}_1}$  reads in the large $\ma$ limit
\begin{equation}
\label{eq:ght1t1}
 g_{h\tilde{t}_1\tilde{t}_1}\simeq\frac{g}{M_W}\left(\sin^2(2\theta_{\tilde{t}})\frac{m_{\tilde{t}_1}^2-m_{\tilde{t}_2}^2}{4}+m_t^2+O(M_Z^2)\right)
\end{equation}
$\theta_{\tilde{t}}$ is the mixing angle of the stops. The
$\tilde{t}_2$ coupling is obtained through $\tilde{t}_1
\leftrightarrow \tilde{t}_2$. The non mixing term $m_t^2$ {\em
adds} up with the top contribution, whereas the mixing term
interferes {\em destructively} with the top. For large mixing with
large enough gap between the two stops masses  this means that a
reduction in $gg \to h$ occurs but accompanied with a more modest
increase in the $h \to \gamma \gamma$ due to the fact that the dominant contribution, the $W$ loop, remains constant. Of
course the $Br(h \to \gamma \gamma)$ can be much more efficiently
increased if a drop in $h \to b \bar b$ occurs as within the
BMSSM.
Therefore we see that by letting light stops jump into the game
and keeping a ratio in the $\gamma\gamma$ channel higher than the
standard model, the correlations between the different channels
will change.

\begin{figure}[h!]
\begin{center}
\includegraphics[scale=0.3,trim=0 0 0 0,clip=true]{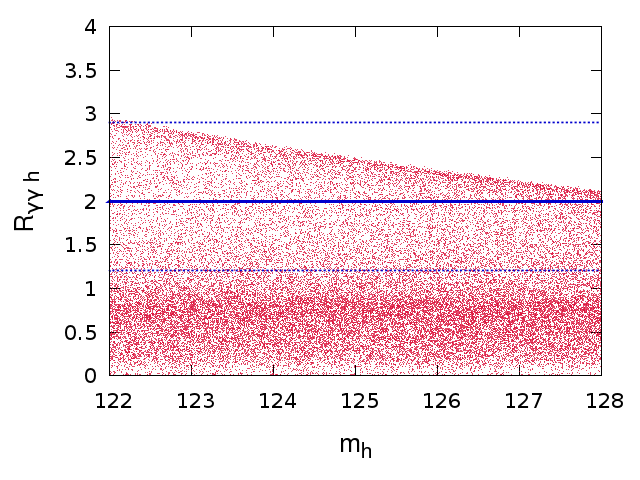}
\end{center}
\caption{\label{fig:h_Rgam2} {\em Allowed region in the plane
$m_h,\Rgam$. The blue line represents the ATLAS best fit for the
signal strength, and the dotted lines are the one sigma deviations
from this value in model B with maximal mixing and  with
$m_{\tilde{t}_2}=600$ GeV.}}
\end{figure}

\begin{figure}[h!]
\begin{center}
\begin{tabular}{cc}
\includegraphics[scale=0.3,trim=0 0 0 0,clip=true]{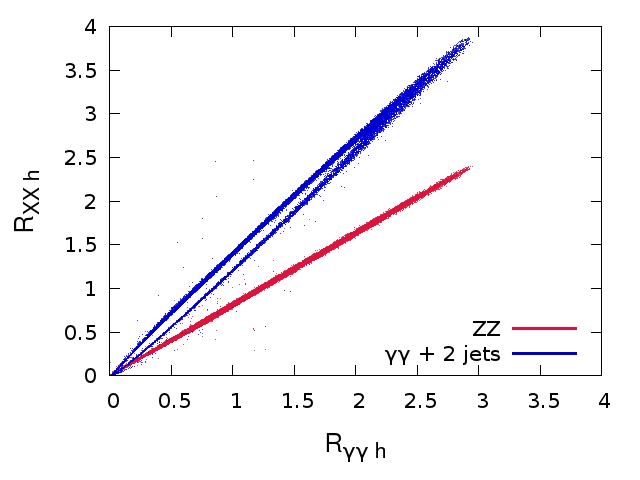}&
\includegraphics[scale=0.3,trim=0 0 0 0,clip=true]{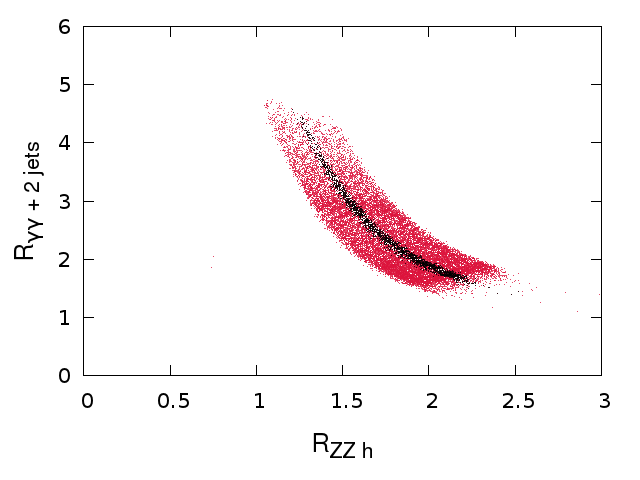}
\end{tabular}
\end{center}
\caption{\label{fig:corrstop} {\em Left panel: correlations
between $\Rgam$, $\Rz$ and $R_{\gamma\gamma+\text{2 jets}}$ for
$122< m_h < 128$ GeV in the maximal mixing scenario of model B with
$m_{\tilde{t}_2}=600$ GeV. Right panel is a subset after imposing
$\Rgamh=2.0\pm10\%$ (points in red) and $\Rgamh=2.0\pm1\%$ (points
in black) in the plane $\Rz$ and $R_{\gamma\gamma+\text{2 jets}}$
in model B scanning in the range $m_{\tilde{t}_2}\in[300,1000]$
(GeV)}}
\end{figure}

We first note, see fig.~\ref{fig:h_Rgam2}, that in the maximal
mixing case $\sin^2(2\theta_{\tilde{t}})=1$ and with
$m_{\tilde{t}_2}=600$ GeV, $\Rgam$ is reduced somehow compared to
model A, however one still obtains
enhancements of a factor $2$ (and more) compared to the SM.
However, now the $\gamma\gamma+\text{2 jets}$ can be much higher
than the $\gamma\gamma$ channel, whereas previously we had
$R_{\gamma\gamma+\text{2 jets}}=1.5$ for $\Rgam=2$, now for the
same value of $\Rgam$ $R_{\gamma\gamma+\text{2 jets}}=2.5$, see
fig.~\ref{fig:corrstop}. Moreover the weight between
$R_{\gamma\gamma+\text{2 jets}}$ and $R_{ZZ}$ has been inverted,
we now have $R_{\gamma\gamma+\text{2 jets}} > R_{ZZ}$.  Scanning
over $m_{\tilde{t}_2}$ from 300 GeV to 1 TeV will open up more
possibilities for the correlations between these channels. The
results of this scan are shown in the right panel of
fig.~\ref{fig:corrstop}. For example imposing that
$\Rgam=2.0\pm10\%$ one can obtain $R_{\gamma\gamma\text{ + 2
jets}}=3.8$ together with $\Rz=1.3$. We can therefore recover
values that correspond to the best fits for these observables
obtained by the two collaborations. We stress again that this is
illustrative and shows how much flexibility in the model can be
introduced. While in the case of no trilinear mixing term in the stop sector (Model A) all
channels seemed to have nearly trivial correlations, raising the $A_t$
mixing term will in most cases raise the $\gamma\gamma\text{ + 2 jets}$
channel compared to the $\gamma\gamma$ channel, and also decrease
the $ZZ\rightarrow 4l$ channel with respect to the $\gamma\gamma$
one.

\subsection{Flavour constraint}
The implication of constraints from flavour physics was not present in our first paper \cite{gdlr_higgs_1203} as well as in the work presented so far. We have
then decided to compute the prediction for some of the flavour observables in order to assess how much this would change the picture in the Higgs sector. The
first one is given by $\Bsmu$, its effect is mostly to disfavour low values of $\ma$. This is not surprising since, as in the MSSM, we expect the supersymmetric
contribution to be important when $\tb$ is high and $\ma$ low. Note that the limit is not very effective since high values are already ruled out by the
$A_0\to\bar\tau\tau$ search. This constraint does not change the feature of the signal produced by the lightest
Higgs.\\

\begin{figure}[h!]
\begin{center}
\begin{tabular}{cc}
\includegraphics[scale=0.3,trim=0 0 0 0,clip=true]{figure/h_Rgamh_RXh_stop.png}&
\includegraphics[scale=0.3,trim=0 0 0 0,clip=true]{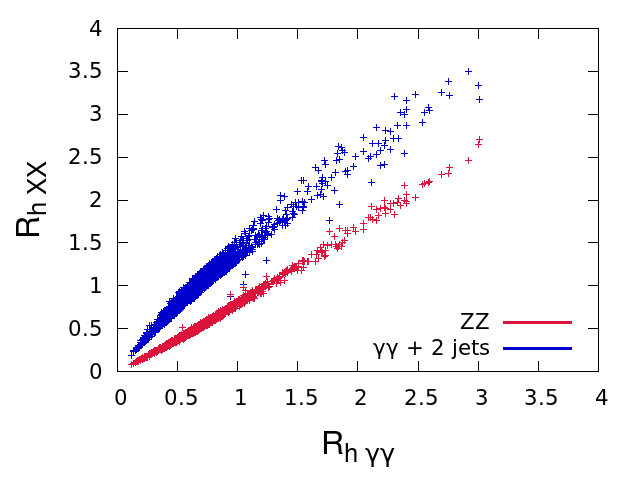}
\end{tabular}
\end{center}
\caption{\label{fig:flavour_on} {\em Correlations
between $\Rgam$, $\Rz$ and $R_{\gamma\gamma+\text{2 jets}}$ for
$122< m_h < 128$ GeV in the maximal mixing scenario of model B with
$m_{\tilde{t}_2}=600$ GeV. On the left are point \textit{without} flavour constraints, and on the right points \textit{with} flavour constraints, and by
relaxing furthermore $s_{2\theta_t}=1$ to $0.8<s_{2\theta_t}<1$.}}
\end{figure}

As concerns $\Bsg$, the situation is a bit different : in the scenario A it will disfavour region with small $\ma$ (say $\ma<200$ GeV) and the scenario B is even more affected. This feature comes from the
fact that a major contribution to this observable is given by a stop-chargino loop which is proportional to
$s_{2\theta_t}(m_{\tilde{t}_2}^2-m_{\tilde{t}_1}^2)\tb$. In
model B, we are looking for non-zero $s_{2\theta_t}(m_{\tilde{t}_2}^2-m_{\tilde{t}_1}^2)$ since it also mediates the stop loop going in gluon fusion and
$\gamma\gamma$ decays. Thus the experimental bounds on $\Bsg$ will impose a low value of $\tb<5$. However we have seen that the combination of a high $\ma$ and
a small $\tb$ was driving the BMSSM to a decoupling limit, so the suppression of the $\ghb$ coupling will be less efficient. This feature is shown in figure
\ref{fig:flavour_on}.

In the next section, which deals with the case of a heavy Higgs boson generating the signal, we will however leave the flavour issue aside.

\subsection{Heavy Higgs case : $H$}
\label{case_hh} As fig.~\ref{fig:mh_BMSSM} makes clear, the
BMSSM is compatible with a scenario where it is the heavier of the
two CP even Higgses, $H$, which is in the range $122-128$ GeV and may thus be
responsible for a signal,  while the lightest Higgs $h$ has so far gone
undetected. Such possibility, even though restrained, has also
been evoked in the case of the MSSM \cite{Heinemeyer:2011aa}. We
review such a possibility in the case of the BMSSM both in a
scenario with degenerate moderate stop masses and a scenario with large stop mixing
and a light stop.

\paragraph{Model A:}~\\
\label{case_hhmodela}
\begin{figure}[h!]
\begin{center}
\begin{tabular}{cc}
\includegraphics[scale=0.3,trim=0 0 0 0,clip=true]{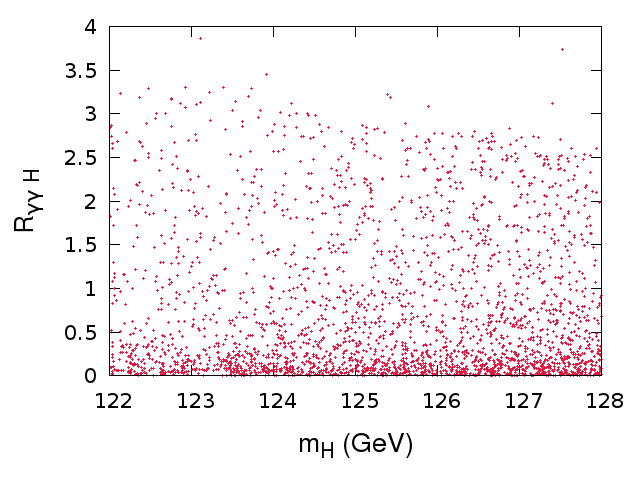}&
\includegraphics[scale=0.3,trim=0 0 0 0,clip=true]{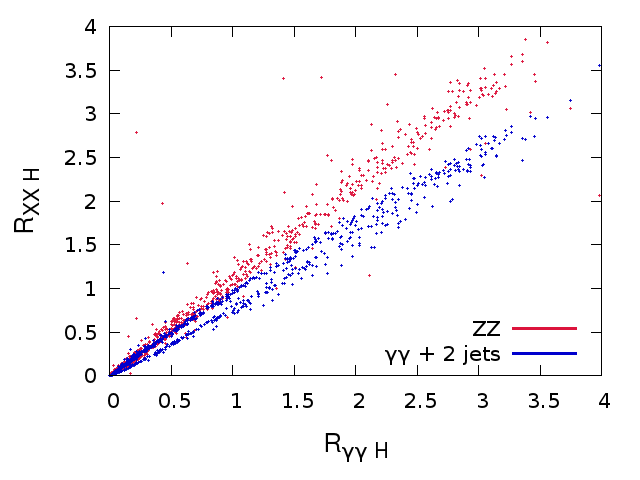}
\end{tabular}
\end{center}
\caption{\label{fig:H_Rgambis} {\em We show here the allowed
region in the plane $m_H,\Rgam$ (left panel) and the associated
correlations between $\Rgam$, $\Rz$ and $R_{\gamma\gamma+\text{2
jets}}$ for $122< m_H < 128$ GeV (right panel) in Model A.}}
\end{figure}
The statement we have just made can be made more quantitative.
Solutions with $122<m_H<128$ GeV correspond to a situation where
all three Higgses are light in the sense of being all three below
the $WW$ threshold, $m_h<120$ GeV $\ma < 160$ GeV.  We find that some features, for the signal observables,
are to a large extent similar to what we have found in the case of
$h$. In a way $h$ and $H$ have swapped their role as to which
is SM-like. A SM-like state is defined with respect to the  strength of the $VVH/h$ coupling. Indeed,
this is illustrated in fig.~\ref{fig:H_Rgambis}. $\Rgam$ can still
reach values as large as 3.5, there are  correlations between
$\Rgam$, $\Rz$ and $R_{\gamma\gamma+\text{2 jets}}$ with  $\Rz >
R_{\gamma\gamma+\text{2 jets}}$ in most cases, but not all as was
the case for $122<m_h<128$ GeV. In this case, there is some spread in the
correlations between  $\Rz$ and $R_{\gamma\gamma+\text{2
jets}}$, see fig.~\ref{fig:H_Rgambis}.\\

\paragraph{Model B:}~\\

\begin{figure}[h!]
\begin{center}
\begin{tabular}{cc}
\includegraphics[scale=0.3,trim=0 0 0 0,clip=true]{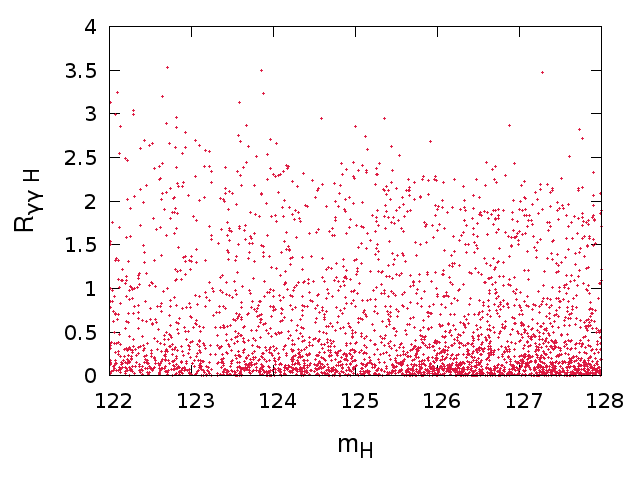}&
\includegraphics[scale=0.3,trim=0 0 0 0,clip=true]{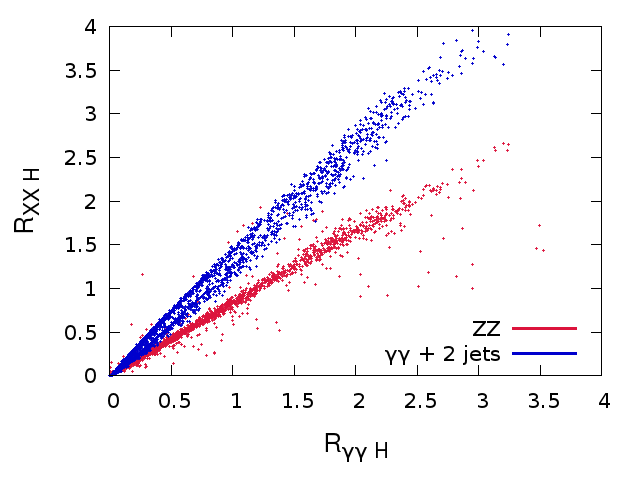}
\end{tabular}
\end{center}
\caption{\label{fig:H_Rgam2bis} {\em We show here the allowed
region in the plane $m_H,\Rgam$ (left panel) and the associated
correlations between $\Rgam$, $\Rz$ and $R_{\gamma\gamma+\text{2
jets}}$ for $122< m_H < 128$ GeV (right panel) with maximal mixing
and $m_{\tilde{t}_2}=600$ GeV.}}
\end{figure}

The most noticeable change is
the correlation between $\Rz$ and $R_{\gamma\gamma+\text{2
jets}}$, see fig.~\ref{fig:H_Rgam2bis}. We now easily find
$R_{\gamma\gamma+\text{2 jets}} > \Rz$. The spread in this
correlation has increased. One can find scenarios with $R_{ZZ}<1$
even for $\Rgam>2$. For $\Rgam \sim 2$, $R_{\gamma\gamma+\text{2
jets}}> 2$ is attained.

\subsection{Prospect for other Higgses}
\paragraph{Case of the light Higgs $h$ as signal}~\\
Although an unambiguous disproof of the SM would be, in the
case where the signal at $m_h=125$ GeV were confirmed, a precise
determination of the signal strength to be different from the SM
expectation, such a precision may require some time. At the same
time as the luminosity increases other channels and signatures may
become sensitive in corroborating the signals with $m_h \sim 125$
GeV. These channels could either be other channels where  the same
Higgs with mass 125 GeV takes part or channels affecting  the
other Higgses of the model. In the first case, the other allowed
decay modes are $\oo{\tau}\tau$ and $\oo{b}b$ final state, however
if the trend towards an increase in the $2\gamma$, $ZZ$ and
$2\gamma + 2 jets$ is reinforced requiring a reduced $hb\bar b$
(and consequently $h\oo{\tau}\tau$) in the BMSSM, the $\oo{\tau}\tau$
and $\oo{b}b$ whose current sensitivity in the SM is quite low will
require substantial increase in the luminosity.

To pursue this investigation about the prospects of signals in
other channels, we keep for the sake of illustration those models
compatible with
\begin{equation}
 1.2<\Rgam<2.9\qquad\&\qquad 0.5<\Rz<2.4,
 \label{eq:signal}
\end{equation}
which is the one sigma band obtained by the ATLAS collaboration
and show the different $R_{XX}^{{\rm exclusion}}$. Again,
eq.~\ref{eq:signal} is an arbitrary choice, taken for the sake of
concreteness. One should keep in mind that as more data is
collected, this requirement will
become either stronger or perhaps even totally irrelevant.\\

\subparagraph{Model A}~\\
\begin{figure}[h!]
\begin{center}
\includegraphics[scale=0.3,trim=0 0 0 0,clip=true]{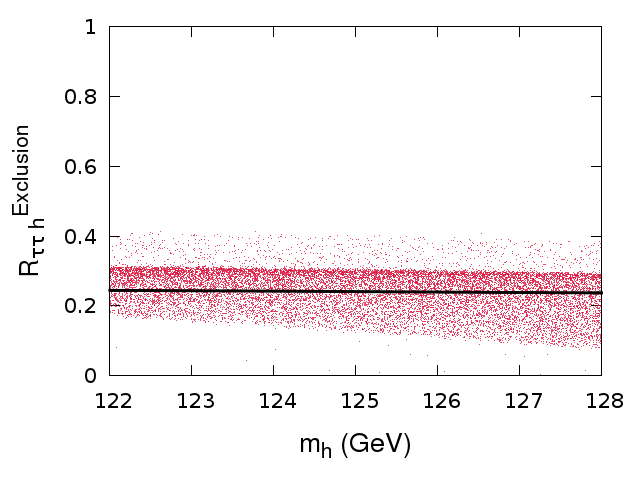}
\end{center}
\caption{\label{fig:htau} {\em Discovery perspective for the channel $h\rightarrow\oo{\tau}\tau$. The line is  black corresponds to the SM.}}
\end{figure}

We start with the $\oo{\tau}\tau$ signal of $h$. We see in fig. \ref{fig:htau} that $R_{h\tau
\tau}^{{\rm exclusion}}$ is always below $0.4$, with a
concentration below $0.3$ that corresponds also to the expectation from a SM Higgs, therefore a luminosity in excess of $30\ 
{\rm fb}^{-1}$ is needed in the most favourable cases. Most cases
will require much more luminosity, up to 500 fb$^{-1}$ in the worst case. Incidentally we note that this channel, despite the reduced
$h \oo{\tau}\tau$ coupling, can be above that of the SM, which shows that a reduced $h\oo{\tau}\tau$ does not mean a large drop in the $\oo{\tau}\tau$
branching ratio, moreover the production cross section can be larger than in the SM.\\

\begin{figure}[h!]
\begin{center}
\begin{tabular}{cc}
\includegraphics[scale=0.3,trim=0 0 0 0,clip=true]{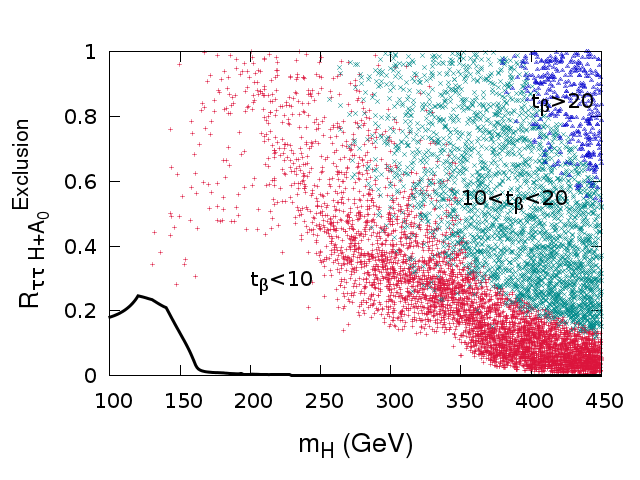}&
\includegraphics[scale=0.3,trim=0 0 0 0,clip=true]{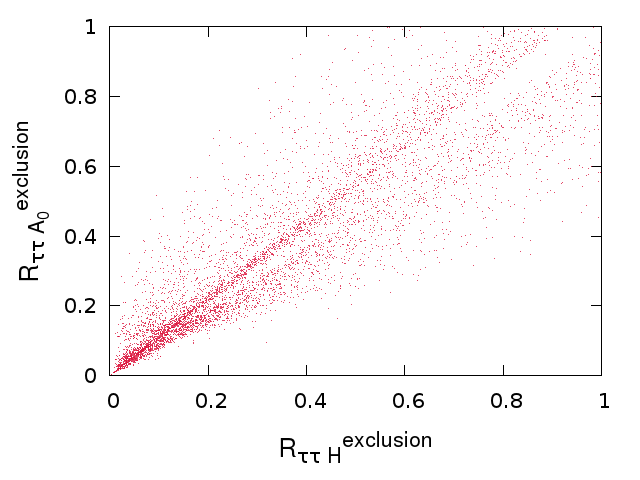}
\end{tabular}
\end{center}
\caption{\label{fig:other2} {\em Discovery perspective in the
$\oo{\tau}\tau$ channel through the heavier Higgses: in the case
where $A^0$ and $H$ are degenerate within 10 GeV (left panel) and
when  $A^0$ and $H$ are not degenerate (right panel). In the
latter the correlations between the two signal is shown. In the
panel on the left, the SM case is shown in black. The different
shades for the BMSSM correspond (from left to right) to cases with
$\tb<10$(red) , $10<\tb < 20$ (green) and $\tb>20$(blue).  }}
\end{figure}
Would the other Higgses be more sensitive? The answer  can be
drawn from fig.~\ref{fig:other2}. Some scenarios can be probed
with little increase in the present luminosity. Generically, high
$\tb$ ($\tb>20$) will be probed within the next 30 fb$^{-1}$,
while low $\tb$ ($\tb<10$) could be quite hopeless if the heavier
Higgses are heavier than 400 GeV. We find that
$R^\text{exclusion}>0.9$ are reached in cases where $A^0$ and $H$
are close enough in mass to be degenerate ($|m_{A^0}-m_H|<10 $
GeV), yielding thus a single signal.   $R^\text{exclusion}>0.9$ is
reached also when the degeneracy is lifted, in which case one
expects both signals to be revealed with roughly the same
luminosity, see the correlation in  fig.~\ref{fig:other2}. Models
with $m_H \sim \ma <250$ GeV (degenerate case) show a ratio
$\RtauE>0.4$, which means that the region where the decoupling is
not complete between light and heavy Higgses could be probed with
about 30 fb$^{-1}$ . In the non-degenerate case, there is of
course a loss of a factor two, but there is still a lower limit to
the exclusion ratio in this mass range. But in many models we will have
$R^\text{exclusion} <0.4$. Consequently $A^0$ and $H$ will go undetected even
with a luminosity in excess of $30 {\rm fb}^{-1}$. This discussion
shows that studying the $\tau$ channel in Higgs physics is
crucial. Not only can it deliver new signals but it can give
important information on the parameters of the model.
\\

\begin{figure}[h!]
\begin{center}
\begin{tabular}{cc}
\includegraphics[scale=0.3,trim=0 0 0 0,clip=true]{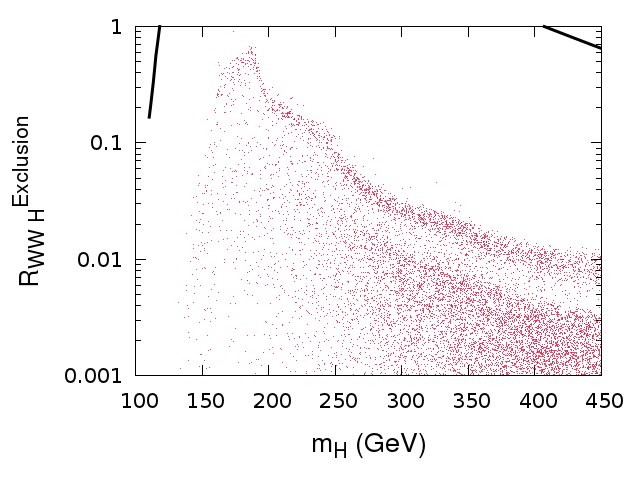}&
\includegraphics[scale=0.3,trim=0 0 0 0,clip=true]{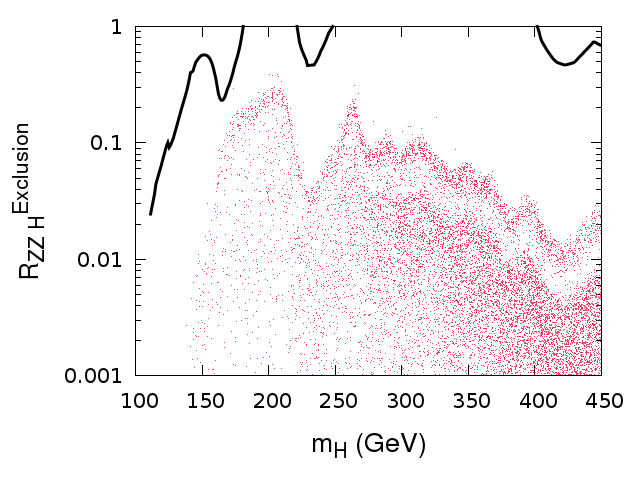}
\end{tabular}
\end{center}
\caption{\label{fig:other1} {\em Discovery perspective for other
signals in Model A: $H\rightarrow ZZ$ on the left and $H\rightarrow WW$ on
the right. The curve in black  is the SM Higgs hypothesis. For $WW$, the curve is out of the bounding box, this confirms that for Higgs masses above the $WW$
threshold  this channel is very constraining.  }}
\end{figure}
Other channels offer little prospects, apart if $M_H \sim 180$ GeV
where the search sensitivity in the clean $WW$ and somehow also
the $ZZ$ channel is high, despite the fact that the $HWW$ is quite
small, see fig.~\ref{fig:other1}.

\noindent\subparagraph{Model B: Maximal mixing and a light stop}~\\
\begin{figure}[h!]
\begin{center}
\includegraphics[scale=0.3,trim=0 0 0 0,clip=true]{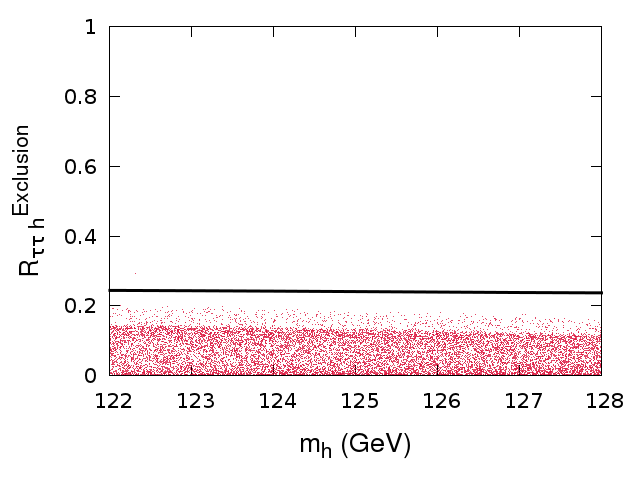}
\end{center}
\caption{\label{fig:htau2} {\em Discovery perspective for the
channel $h\rightarrow\oo{\tau}\tau$ in the maximal mixing scenario. The
line in black  represents the SM.}}
\end{figure}
In the maximal mixing case, with $m_{\tilde{t}_2}=600$ GeV, there
are few differences. The drop in $gg \to h$ is the reason behind the drop in sensitivity. Subsequently
the $\oo{\tau}\tau$ channel of $h$ will be even less sensitive, as can
be seen in fig.~\ref{fig:htau2}. $R_{h\oo{\tau}\tau}^{{\rm
exclusion}}$ is now below 0.2.
\\

As concerns the heavier Higgses, the changes are marginal compared
to model A with both stop masses almost degenerate. The best prospects are in the $\tau$
channels and in the $WW$ channel if $m_H \sim 180$ GeV. The
corresponding figures are similar to those shown for Model A and we therefore do not display them here.

\paragraph{Case of the heavy Higgs $H$ as signal}~\\
\noindent\subparagraph{Model A :}~\\
\begin{figure}[h!]
\begin{center}
\includegraphics[scale=0.3,trim=0 0 0 0,clip=true]{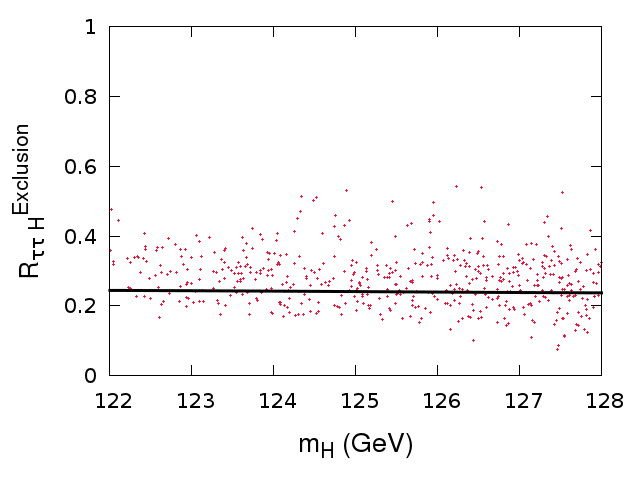}
\end{center}
\caption{\label{fig:mass_H} {\em $R_{\oo{\tau}\tau H}^{Exclusion}$ for
$122<m_H<128$ GeV in model A. The
line in black is the SM.}}
\end{figure}
Another mode where $H$ could be observed is the $H\to \oo{\tau}\tau$
channel. Fig.~\ref{fig:mass_H} suggests that prospects here might
be better than for $h$ giving a signal in the range $122<m_h<128$
GeV. Indeed, there are solutions with $R_{\oo{\tau}\tau
H}^{Exclusion}=0.5$
that would need about $20 {\rm fb}^{-1}$ to be uncovered.\\
\begin{figure}[h!]
\begin{center}
\begin{tabular}{cc}
\includegraphics[scale=0.3,trim=0 0 0 0,clip=true]{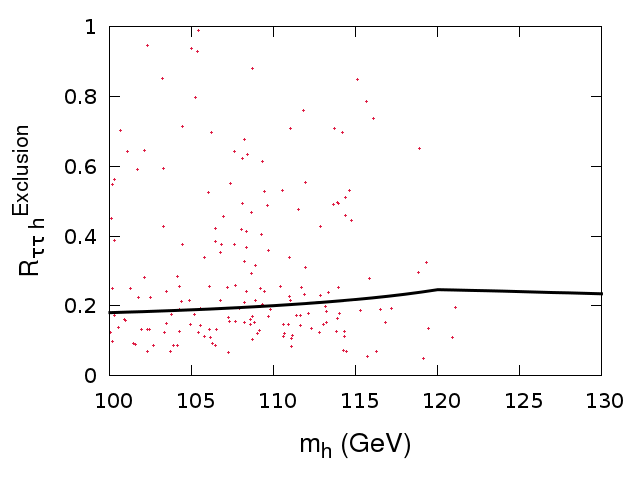}&
\includegraphics[scale=0.3,trim=0 0 0 0,clip=true]{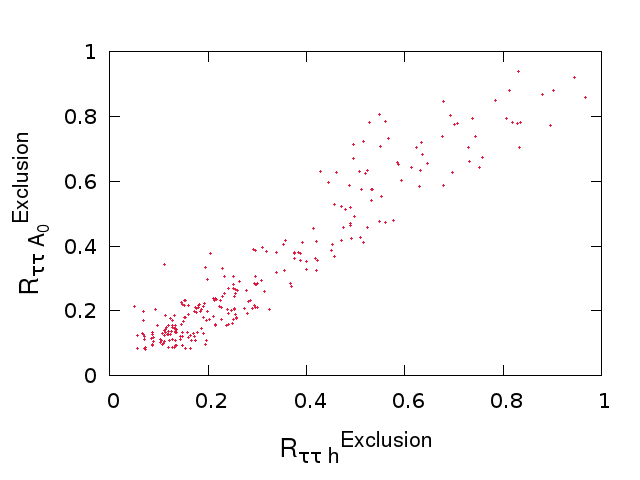}
\end{tabular}
\end{center}
\caption{\label{fig:H_other} {\em $R^\text{exclusion}_{\oo{\tau}\tau
h}$ as a function of $m_h$ for $122< m_H < 128$ GeV in the case
where $A^0$ and $h$are not degenerate within $10$ GeV (left panel).
The right panel shows the correlations between the $h$ and $A^0$
in the $\tau$ channels in model A.
The line in black in the left panel represents the SM.}}
\end{figure}
Observability of the other Higgses shows, in many cases, very good
prospects, gain in the $\oo{\tau}\tau$ channels, $R_{\oo{\tau}\tau
h}^{Exclusion}>0.6$ are obtained, see fig.~\ref{fig:H_other} .
Therefore it is worth pursuing searches of $h$, for $m_h<120$ GeV
in the $\oo{\tau}\tau$ channel. $A^0$ could also be uncovered with
the same luminosity, in fact fig.~\ref{fig:H_other} shows the
correlation between $h$ and $A^0$ in the $\oo{\tau}\tau$ channel.
There, of course, remains also many situations with $R_{\tau
\tau}^{Exclusion}<0.2$ that would be difficult to decipher.\\

\noindent\subparagraph{Model B :}~\\
\begin{figure}[h!]
\begin{center}
\includegraphics[scale=0.3,trim=0 0 0 0,clip=true]{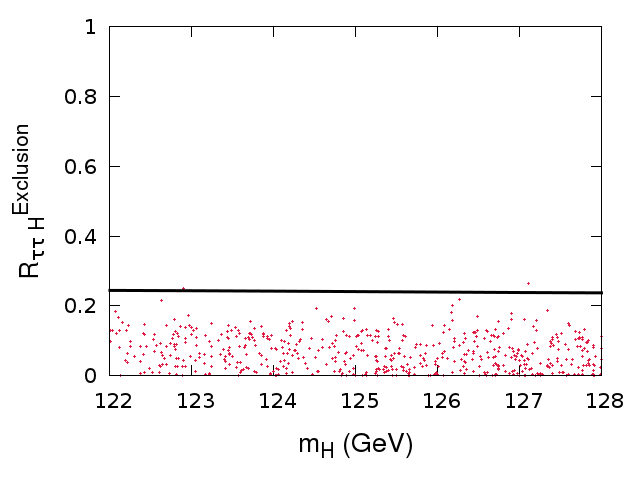}
\end{center}
\caption{\label{fig:mass_H2} {\em $R_{\oo{\tau}\tau H}$ in the same
mass $122<m_H<128$ GeV in the case of maximal stop mixing and
$m_{\tilde{t}_2}=600$ GeV. The line in black is the SM. }}
\end{figure}
We now turn to the maximal mixing case
and restrict  ourselves to $m_{\tilde{t}_2}=600$ GeV
($m_{\tilde{t}_1}=200$ GeV). Compared to the previous case, Model A,  one notes that there is a reduction in $R_{\oo{\tau}\tau H}$.
This is mainly driven by  the drop in
$gg \to H$, see fig.~\ref{fig:mass_H2}, very low values are also due to quite small $h \tau\bar \tau$ couplings.\\

The visibility of $A^0$ and $h$ is little affected by the trilinear stop mixing parameter. Our conclusions are little changed. Again it is very
important to pursue the search in the $\oo{\tau}\tau$ channel.

\subsection{Conclusion}
\label{sec_conclusions} Despite the fact that no sign of
supersymmetry has been found so far, the BMSSM framework is a very
efficient set up that extends the realm of the MSSM in a most
natural way as concerns the realisation of the Higgs. In the MSSM
framework there is some tension with naturalness for a Higgs mass
of $125$ GeV that requires heavy stops, in the BMSSM this is not an
issue. Although one must exercise extreme care with the so called
tantalising hints for a Higgs signal around this mass, 125 GeV, it
is extremely important to scrutinise the properties of the Higgs
with such a mass in many models, in particular the BMSSM which
represents an effective implementation of a variety of
supersymmetric models having the same field content as the MSSM.
Despite the uncertainty in the measurements of both ATLAS and CMS, 
these tantalising hints have come also with the temptation of attributing to the signals in the
inclusive $2\gamma$, the $2\gamma + \; {\rm jets}$ and
perhaps in the $ZZ \to 4l$ channels, values that are higher  than what is expected from
the SM. Such scenarios are very difficult to attain in
the MSSM without a certain amount of tuning. It is therefore very
important to find out whether some configurations, especially
those leading to enhancements in these most important channels can
be realised in the BMSSM. As important is to find out how these
enhancements or signals are correlated and how different kinds of
correlations can be realised. We have shown that a vanilla BMSSM
where stops are at very moderate masses and almost degenerate
easily allows enhancements in all these channels for $m_h \sim
125$ GeV with the constraint that the rate $ZZ \to 4l$  would
generally be higher than the rate $\gamma \gamma \;+ 2 {\rm
jets}$. A light stop with large mixings in the stop sector offers
more possibilities especially as concerns correlations between
these three important channels. Our study also reveals that
although it is easier to have such realisations work for the
lightest Higgs of the BMSSM, solutions where it is the heaviest
Higgs that has a mass around 125 GeV also exist. Once a signal at
$125$ GeV has been confirmed a better measurement of the rates, in
particular the $2\gamma$ (inclusive and exclusive) as well as the
$4l$ would narrow considerably the parameter space of the BMSSM.
At the same time as more precision is  achieved and more
luminosity is gathered one can constrain the models through the
other Higgses (those outside the 125 GeV window) but also through
other channels of the Higgs at 125 GeV. Our study reveals that in
both cases it is crucially important and telling to investigate
the $\oo{\tau}\tau$ channel. We have not folded in the possible
constraints from dark matter as we have argued
that this introduces some model dependencies (including those from
cosmology) but it is clear now that we have entered a fascinating
era. The study we have conducted is an example which shows that
even before any new direct signal of New Physics is discovered,
the study of the Higgs, once confirmed, will give important clues
on the New Physics. We eagerly await more data and analyses from
the experiments and we urge, once more, our colleagues to provide
as much information as possible on the data so that one can gain
access to the different individual subchannels that make up an
inclusive channel.


%% file: chapter8.tex
\chapter{Dark Matter I : a supersymmetric candidate}

\minitoc\vspace{1cm}

\section{Dark Matter}
The evidence for a new kind of matter in our universe are numerous : there is the apparent modification of gravitational motion, the measured value of the relic density and possibly some excesses in cosmic rays. Counter evidence are the direct detection experiments, which have shown hints but no actual signal and the LHC searches. Concerning the two first effects, that are the rotational velocity of stars in a galaxy and the relic density, though there is no experimental evidence that these two effects have the same cause, it is quite interesting to consider that it might be the case, simplifying thus our description of the universe. The last piece of evidence is currently a hot topic at the experimental level : it is the fact that dark matter particles may annihilate to produce Standard Model particles that we could detect on earth, leading thus the close scrutiny of cosmic ray fluxes to be a major hope in assessing the nature of dark matter. In contrast, the LHC is 
tackling the issue the other way round since one starts with Standard Model particles and ends up with dark matter particles if the latter is sufficiently coupled to the Standard Model sector. Last, we are also trying to detect interactions of the dark matter particles going through earth in the so-called direct detection experiments. So far, no significant and consistent signal has been reported (though there have been several false alarms, none of them has achieved a sufficient compatibility with other experiments to be a plausible dark matter signal), thus imposing bounds on the properties of such particles.\\

The appealing aspect of dark matter is that it implies at least one new particle in addition to the Standard Model. Less appealing aspects are the facts that it can be pretty much any kind of particle as long as it is neutral and long-lived, and that all astrophysical measurements include uncertainties on astrophysical and cosmological models as well as the cosmic ray propagation throughout the universe. This explains why most of the searches are done with effective models that make the fewest possible hypotheses on the particle physics side. It may hence seem a bit curious to try to constrain a very high level model such as supersymmetry, since it has a very specific kind of dark matter. However this would be true if the only purpose of supersymmetry was to account for dark matter, but as we have already seen, they are many other motivations for introducing supersymmetry. So the idea is to try, from the purely particle physics point of view, to derive constraints for observables in the dark matter sector 
for supersymmetry.\\

We are now going to study what predictions of dark matter we have in supersymmetry. Let us see what are those predictions and how they are currently computed
\begin{itemize}
 \item Rotational velocity of stars. It depends on the history of the universe and structure formation, and is pretty much independent of supersymmetric parameters, or even the presence of supersymmetry itself.
 \item Relic density. It relies on the history of the universe, cosmology and particle physics for interactions with the Standard Model particles. Concerning cosmology the standard cosmological model is mainly used, though some deviations of it have also been studied (see \cite{arbey_0906}, for instance). Concerning the particle physics in the supersymmetric landscape the MSSM\footnote{In the following there will be plenty of different MSSM frameworks, depending on the number of free parameters they assume, and whether they can be taken as real or complex.}, NMSSM, U(1)'MSSM and other supersymmetric extensions have been studied. The relic density turns out to be a very powerful constraint, given the impressive experimental accuracy, and can rule out entire regions of the parameter space. However its power can be quite reduced : first if one assumes that there is more than one dark matter particle, the required range transforms into an upper bound only. Then the prediction is also likely to be significantly 
modified when playing with the cosmological scenario. Last, the theoretical uncertainty on the computation can be much higher than the experimental one, reducing thus the apparent sensitivity of the observable.
 \item Indirect detection. The same tools as for the relic density are used, even though the observables are a bit different. We now have to deal with the trouble of propagation of cosmic rays in some cases (as for positrons), which tends to diminish the constraining power on the particle physics at hand. This being said, not looking at those observables would be a mistake since it may be the most unambiguous kind of signal we could expect : the academic example being the case of a gamma ray line, that is to say a cosmic ray of photons at a given energy that would have been produced by annihilation of dark matter. This possibility has in particular been recently highlighted with the interpretation of the Fermi data on a possible gamma ray line around 130 GeV, see \cite{fermi_gamma_130}.
 \item Direct detection. The direct detection experiments are basically sensitive to two quantities, the dark matter density in the solar system, and the interaction cross-section with nuclei. They have recently been on the spot since they just started to probe interesting regions, as for instance with the results on spin-independent cross-section from XENON100 (\cite{xenon_100_2011}), but also with the first searches at the LHC\footnote{Although the LHC itself cannot discover Dark Matter since it cannot probe whether those particles are stable or not, it can be used to derive constraints on specific models where the LSP is recorded as missing energy.}(\cite{lhc_dm_2011}).
\end{itemize}

This research area has experienced a full boom recently, and for each of the observables listed above, several directions for improvement are undertaken, which cover a wide range in particle physics, astrophysics, cosmology and astronomy. Being more concerned by the particle physics issues, I have devoted my study to the improvement of the computation of the relic density in supersymmetric models. As we will see, this is a quantity already known to a very good accuracy by experimentalists, but for which theoretical predictions are much less precise, so any gain of precision on the theoretical side would mean a better understanding of the models.

\section{Computation of the relic density}
\subsection{Definition of the \RD}
In cosmology the quantity used to describe the number of particles of a given species $\phi$ is its abundance, given by
\begin{equation}
 \Omega_\phi=\frac{8\pi G}{3H_0^2}\rho_\phi
\end{equation}
where $G$ is the gravitational constant $H_0$ the Hubble rate and $\rho$ the density of the species $\phi$. The density of a species is a parameter (together with the pressure $P$) of its stress-energy-momentum tensor $T_{\mu\nu}$
\begin{equation}
 T^{\mu\nu}=(P+\rho)u^{\mu}u^{\nu}-Pg^{\mu\nu}.
\end{equation}

It turns out that we can measure very precisely the abundance of some classes of species by analysing the CMB energy spectrum. This spectrum is a function of all abundances of particles at the time of the decoupling of photons in the universe. Specifically, particles will have different contribution to the photon spectrum whether they are charged and whether they are massive. This implies that the spectrum can be parametrised by the five following abundances
$$\Omega_\gamma,\Omega_b,\Omega_\nu,\Omega_h,\Omega_\Lambda$$
where $\Omega_b$ stands for all charged matter, $\Omega_\nu$ for all neutral massless matter, $\Omega_h$ for neutral massive matter and $\Omega_\Lambda$ is unrelated to matter since it is the contribution from the cosmological constant. Since each class of species has a different shape in the photon spectrum, one can fit the combination of all five categories to the CMB spectrum to derive numbers for each abundance. Therefore, one ends up with a measure for $\Oh$ which is precisely the amount of dark matter we expect in the universe. In practice one combines the results of  the 7-year {\tt WMAP} data
\cite{Jarosik:2010iu}, the
baryon acoustic oscillations from {\tt SDSS}\cite{Percival:2009xn}
and the most recent determination of the Hubble
constant (\cite{Riess:2009pu}) one  arrives at
\begin{equation}
 \Oh=0.1126\pm0.0036\qquad\cite{wmap_7}.
\end{equation}
An even more accurate measure is expected soon from the PLANCK satellite, launched in 2009.\\

However, one quickly realises that the density of a species is not a quantity intrinsic to a particle physics model, but depends also on the history of the universe, so if we want to use $\Oh$ as a measure of the particle physics parameters, we have to work out this history.

\subsection{Evolution of the density of a species}
Let us start by parametrising the evolution of the universe itself. Up to a good approximation the universe seems to have a black body radiation : one consequence being that the temperature $T$ of the photons in the universe is a monotonous function. This is why we usually consider the history of universe in term of $T$ rather than time. The behaviour of a species follows roughly two steps from the dawn of time to the present :
\begin{itemize}
 \item A period of thermal equilibrium, where we have $\rho_\phi\propto e^{-m_\phi/T}$
 \item The occurrence of the freeze out, which turns the variation to $\rho_\phi\propto a^{-3}(T)$
\end{itemize}
where $a(T)$ is the universe scale factor. The freeze out defines the photon temperature at which the characteristic time of interactions between particles of the species is roughly equal to the expansion rate of the universe : in other words from this point on, the species is too sparse to be considered as self-interacting. The exponential behaviour of the density during thermal equilibrium makes this transition sharp, so it is a very good approximation to take two different behaviour on each side of the freeze out. What we are interested in is the value $\rho(T_0)$, that is the density at the time of the photon decoupling, corresponding to the appearance of the CMB. Since we know the evolution of the scale factor, we only need to compute the freeze-out time $T_f$ and the initial density. The initial density is obtained from the primordial entropy in the high-energy limit :
\begin{equation}
 \rho_\phi=\frac{7}{8}\frac{\pi^2}{15}\frac{g_\phi}{2}T^2
\end{equation}
where $g$ counts the number of degrees of freedom of $\phi$. And the evolution can be obtained from the Bolztman equation :
\begin{equation}
 \frac{dn}{dt}=-3Hn-<\sigma v>(n^2-n_{eq}^2)
\end{equation}
where $n$ is the number of particles of the species and $<\sigma v>$ the cross-section convoluted with the \Moller\ velocity, which is a relative velocity defined so that $vn^2$ is Lorentz invariant. The cross-section we are referring to is the one of the process $\tilde{\chi}\tilde{\chi}\to X$, where $\tilde{\chi}$ is the dark matter candidate and $X$ a state with any number of Standard Model particles. The way to compute the relic density is hence the following : first determine $<\sigma v>$, and then integrate numerically the Boltzmann equation until the freeze-out condition is met. The distribution of the \Moller\ velocity $v$ being determined on astrophysical grounds, the role of particle physics is then to establish the prediction for the cross-section, which is what we will now see.

\subsection{Coannihilation}
So far we have been considering that the contribution to the relic density was coming from a single dark matter particle, which is a good approximation in the sense that any other particle which is more massive will undergo a higher reduction during the thermal equilibrium. Hence it would seem that the relic density is mainly given by the lightest dark matter particle, which we often call the LSP, for Lightest Stable Particle. This stems from the fact that a heavier particle would mostly decay to this LSP. However this assumption turns out to be wrong if we have another particle nearly mass-degenerate with the dark matter candidate. Such particle would then be called the NLSP, for Next-to-Lightest Stable Particle. Supersymmetry is accustomed to such a spectrum, indeed we will see that some very generic choices of parameters predict a lightest neutralino and a lightest chargino very close in mass. In this case the relevant quantity is no more $<\sigma v>$, but 
\begin{equation}
 <\sigma v>\to\sum_{i,j}<\sigma_{ij}v_{ij}>
\end{equation}
where $\sigma_{ij}$ now stands for the process $\tilde{\chi}_i\tilde{\chi}_j\to X$ ($i$ running on the dark matter particles close to the LSP mass). This implies that we need to keep track of the behaviour of each species.

\paragraph{Neutralino as a Dark Matter candidate :}~\\
One advantage of supersymmetry is that is has a natural candidate for Dark Matter with the lightest neutralino. Indeed the neutralino corresponds exactly to the WIMP (Weakly Interactive Massive Particle) definition : it is neutral, colourless and massive (its mass depending on $M_1,M_2$ and $\mu$). It can furthermore become stable if R-parity is conserved. This symmetry, first introduce to forbid a fast proton decay, is the discrete symmetry that can be realised by assigning to each Standard Model particle a unit charge and an opposite charge to all their superpartners and requiring the product of the charge to be conserved. This has the consequence that all cross-sections must have an even number of superpartners or, in other words, that we cannot have a superpartner decaying to Standard Model particles only. Thus, since the lightest superpartner can not decay to any other superpartners it must be stable. But as we will see, though the neutralino is in principle a good candidate for dark matter, the 
prediction for the relic density will very 
much depend on the values of the MSSM parameters.

\section{Neutralino annihilation : a tree-level study}
\begin{figure}[!h]
\begin{center}
\begin{tabular}{lcr}
\begin{fmfgraph*}(100,70)
\fmfleft{p2,p1}
\fmfright{p4,p3}
\fmflabel{$\chii$}{p1}
\fmflabel{$\chii$}{p2}
\fmflabel{$f$}{p3}
\fmflabel{$\bar f$}{p4}
\fmf{fermion}{p1,v1,p3}
\fmf{fermion}{p2,v2,p4}
\fmf{scalar,label=$\tilde{f}$}{v1,v2}
\end{fmfgraph*}
&
\hspace{5mm}\begin{fmfgraph*}(100,70)
\fmfleft{p2,p1}
\fmfright{p4,p3}
\fmflabel{$\chii$}{p1}
\fmflabel{$\chii$}{p2}
\fmflabel{$f$}{p3}
\fmflabel{$\bar f$}{p4}
\fmf{fermion}{p1,v1,p2}
\fmf{fermion}{p3,v2,p4}
\fmf{dashes,label=$A_0$}{v1,v2}
\end{fmfgraph*}\hspace{5mm}
&
\begin{fmfgraph*}(100,70)
\fmfleft{p2,p1}
\fmfright{p4,p3}
\fmflabel{$\chii$}{p1}
\fmflabel{$\chii$}{p2}
\fmflabel{$W^+$}{p3}
\fmflabel{$W^-$}{p4}
\fmf{fermion}{p1,v1,p2}
\fmf{fermion}{p3,v2,p4}
\fmf{photon,label=$Z$}{v1,v2}
\end{fmfgraph*}
\end{tabular}
\end{center}
\caption{\label{fig:rd_tree_main}{\em Example of diagrams contributing to the relic density in the MSSM. All show a different process of the annihilation of two neutralinos : the left one is a sfermion t-channel to a fermionic final state, the middle one the s-channel of the CP-odd Higgs to the same final state and the right one the $Z$ s-channel to a final state with two $W$.}} 
\end{figure}
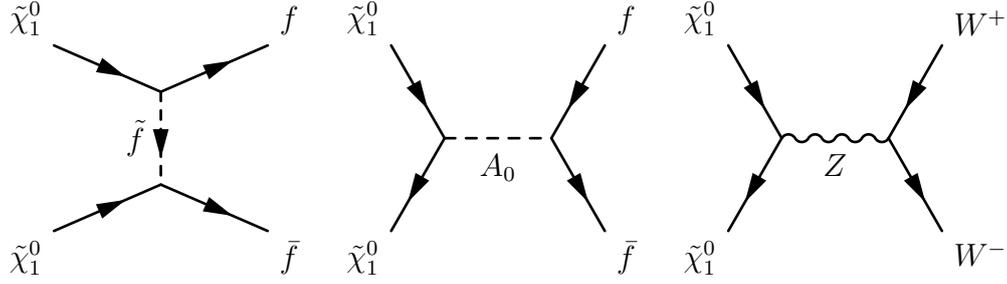

One intricacy of the calculation of $<\sigma_{ij}>$, as compared to colliders or direct detection observables is that we have to add many different processes : namely all with the LSP or the NLSP (or even more particles) in the initial state, and any number of Standard Model particles in the final state. That means that computations are usually automated (as in \momegas \cite{micromegas1,micromegas2}, \darksusy\ \cite{darksusy} or \texttt{SuperIso-Relic} \cite{superiso_relic} for instance). While this does not introduce dramatic complications\footnote{apart from a computing time issue} since all processes are $2\rightarrow 2$ or $2\rightarrow 3$, it makes the interpretation in terms of supersymmetric parameters a bit harder. The good news is that it is verified that on most of the parameter space no more than a handful of processes actually contribute to 90\% of the \RD, and simply a bit more at 99\%, which is enough compared to the experimental accuracy. This does not mean that a given set of processes will 
dominate everywhere on the parameter space, indeed different regions will usually lead to different channels, as we will soon see. I show on figure \ref{fig:rd_tree_main} some of the main processes contributing to the relic density in the MSSM : the task is now to determine the dominant ones.\vspace{5mm}\\

There are two major factors driving the dominance of a channel : the kinematical reach and the strength of the process. The first depends on the LSP mass, if it is lower than half the sum of the masses of the products then the channel will be closed. This important point comes from the fact that we expect the dark matter particles to behave non-relativistically, so that the reaction occurs approximatively at rest. The strength of a given process is mainly driven by the strength of the couplings, and then by the mass of mediator particles. The Standard Model couplings being unchanged by supersymmetry, what we have to look at are the couplings from neutralinos and charginos to the Standard Model spectrum. Those couplings are the following :
\begin{equation}
 \t{h}\t{h}gA,\ \t{A}\t{A}gA,\ hg\t{A}\t{h},\ y\t{h}f\t{f}\text{ and }\t{f}g\t{A}f.
\end{equation}
where $h,A,g,y$ are generic notations for Higgs fields, gauge fields, gauge couplings and Yukawa couplings respectively. One notices that gauge couplings involve one or two neutralinos, hence they will appear both in s and t-channels. Yukawa couplings will appear only in t-channel. However since the physical states -- neutralinos and charginos -- are combination of higgsinos and gauginos, the couplings will be rescaled by the coefficients of the mixing matrix. The nature of the LSP (or the NLSP) is used to describe the mixing elements of higgsino/gaugino parts. One usually labels $Z_n$ the unitary mixing matrix 
\begin{equation}
 \t{\psi}^{0}=Z_n\t{\chi}^0
\end{equation}
where $\t{\psi}^0$, defined in chapter 4, labels neutral higgsino and gaugino. The nature of the LSP (in the case where it is indeed the neutralino) will be defined as bino if $|Z_{n11}|^2>0.99$, and so on for each nature. Note that for higgsino, there are two states $\tilde{h}_1$ and $\tilde{h}_2$ so the higgsino part is given by $|Z_{n13}|^2+|Z_{n14}|^2$ The rest will be denoted the mixed cases.\\

\noindent Hence the relevant information to determine the dominant processes are
\begin{itemize}
 \item the LSP mass $\mchi$
 \item the coefficients $|Z_{ni1}|^2$
\end{itemize}
and, to a lesser extent
\begin{itemize}
 \item $\ma,\tb$ for Higgs exchange in the s-channel
 \item $M_{\t{f}}$ for sfermions exchange in t-channel
\end{itemize}
We will now see what are the dominant processes for each pure case : bino, wino and higgsino.

\subsection{Parameter space}
Since we will be studying different compositions of the
neutralinos we will take different values for the set
$M_1,M_2,\mu$. On the other hand we will fix some default parameters in the rest of the MSSM parameter space, starting with the Higgs sector :
\beqn
M_{A^0}=1{\rm TeV} \quad \quad \tb=4.
\eeqn
The sfermion sector is specified by a rather heavy spectrum (in
particular within the limits set by the LHC for
squarks\cite{cms_susy}). All sleptons left and
right of all generations have a common mass which we take  to be
different from the common mass in the squark sector. All
tri-linear parameters $A_f$ (including those for stops and
sbottom) are set to 0. The default values for the sfermion masses
are
\beqn
M_{{\tilde l}_{R}}&=&M_{{\tilde l}_{L}}=500\ {\rm GeV},\nonumber \\
M_{{\tilde u}_{R}}&=&M_{{\tilde d}_{R}}=M_{{\tilde Q}_{L}}=800\ {\rm
GeV},\nonumber \\
A_f&=&0 \,.
\eeqn
The choice for squarks and gluinos to be at $800$ GeV might be considered as in tension with the direct search of superpartners at LHC. We will however vary the squark masses up to 3 TeV in our study, so this somewhat small value of 800 GeV must be considered as a simple default value. We  will focus on relatively light neutralinos (around
100 GeV) scattering with a relative velocity $v=0.2c$. Note that this parameter space differ slightly from the one we used in the Higgs constraints, since $A^0$ is now heavy, this choice is made in the purpose of studying the behaviour of the relic density without the possibility of resonances. The fact that the lightest Higgs is still light is not an issue in that respect, indeed a system with two Majorana fermions at non-relativistic velocities behave in a CP-odd way, so that a coupling to CP-even Higgs bosons is suppressed. The relic density is computed with \momegas-2.4, with a model processed by \lanHEP-3.1.

\subsection{The pure cases}
\subsubsection{Bino case}
\begin{figure}[!h]
\begin{center}
\begin{tabular}{cc}
\includegraphics[scale=0.3,trim=0 0 0 0,clip=true]{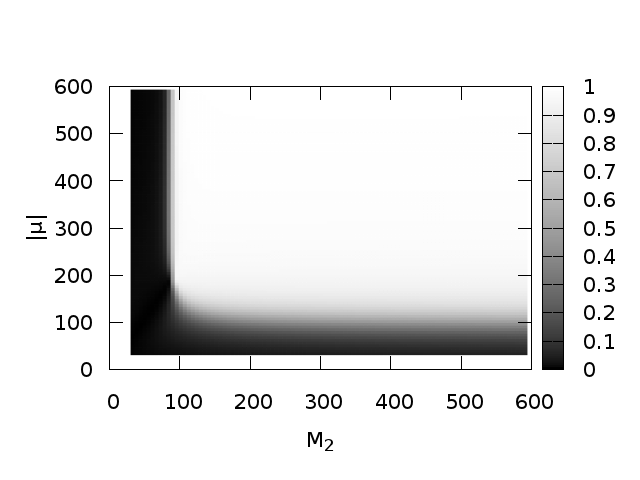}
&\raisebox{2.5cm}{$|Z_{n11}|^2$}
\end{tabular}
\end{center}
\caption{\label{fig:bino_case} {\em We show here the bino part of the lightest neutralino when varying $M_2,\mu$ with a fixed $M_1=90$ GeV. One notices a sharp transition between wino and bino cases (vertical line on $M_2\sim90$ GeV) and another still quick between bino and higgsino (horizontal tline).}}
\end{figure}

This case is generically obtained for $M_1<M_2,\ M_1<|\mu|$. It turns out that we do not need a large hierarchy between the parameters. Indeed on figure \ref{fig:bino_case} one observes the value of $|Z_{n11}|^2$ with $M_1=90$ GeV, varying $M_2$ and $\mu$. It is clear that the transition between pure cases is extremely sharp. 

In this case we have approximatively $\mchi\sim M_1$. The only non-vanishing coupling is $\t{f}g\t{A}f$ hence the dominant final state will be $f\bar f$, and there is no coannihilation since other charginos or neutralinos are heavier. On table \ref{tab:bino_case} are plotted the relative contributions to the \RD\ on a parameter point 
\begin{equation}
 M_1=90,\quad M_2=200,\quad \mu=-600\qquad\text{(GeV)}
\end{equation}
which will be our benchmark for the bino case.\\

\begin{figure}[!h]
\begin{center}
\begin{tabular}{c|c}
$\Oh$&6.68517\\\hline
$\chii\chii\to\bar\tau\tau$ & 28\%\\
$\chii\chii\to\bar\mu\mu$ & 28\%\\
$\chii\chii\to\bar ee$ & 28\%\\
$\chii\chii\to\bar cc$ & 3\%\\
$\chii\chii\to\bar uu$ & 3\%\\
others & 10\%
\end{tabular}
\end{center}
\caption{\label{tab:bino_case} {\em Contribution of the different processes to the relic density in the pure bino case. The dominant process is thus $\chii\chii\to\bar ff$, mediated by gauge couplings.}}
\end{figure}

This is quite independent from $\ma,\tb$ (since no Higgs exchange contribute) and raising $\Mf$ lowers the cross-section, which stems from the fact that it appears in the t propagator. Because of the nature of the couplings the contributions are directly related to the hypercharge, except for the top that is kinematically excluded. Note that since on this point the annihilation cross-section is quite low, the relic density will end up too high (more than one order of magnitude higher than the measured value). The options to enhance the cross-section in order to obtain a correct relic density value is either to lower the slepton masses in order to enhance the t-channel exchange or to allow for a small but non-zero higgsino component in order to open the s-channel of gauge bosons, as we will see in mixed cases.

\subsubsection{Higgsino case}
\begin{figure}[!h]
\begin{center}
\begin{tabular}{cc}
\includegraphics[scale=0.3,trim=0 0 0 0,clip=true]{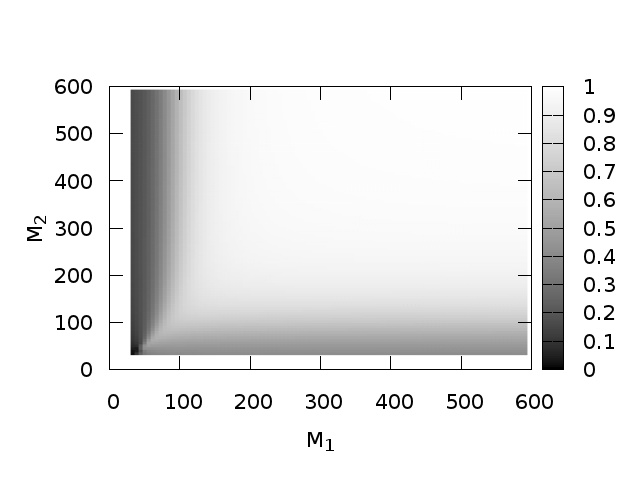}
&\raisebox{2.5cm}{$|Z_{n31}|^2+|Z_{n41}|^2$}
\end{tabular}
\end{center}
\caption{\label{fig:hino_case} {\em We show here the bino part of the lightest neutralino when varying $M_1,M_2$ with a fixed $\mu=-100$ GeV. In this case the transition between different cases are less direct, $\mu$ needs to be significantly lower than $M_1$ and $M_2$ to be in a pure higgsino case.}}
\end{figure}

This case is obtained with $|\mu|$ smaller than $M_1,M_2$. The transition to this case along those parameters is less sharp than in the previous case, as can be seen in figure \ref{fig:hino_case}.

Note that what we call the higgsino component is actually the quantity $|Z_{n31}|^2+|Z_{n41}|^2$ since there are two of them. The LSP mass is then $\mchi\sim|\mu|$ and the non vanishing couplings are $gV\t{h}\t{h}$ and $y\t{h}\t{f}f$. However, except for the top quark, the Yukawa coupling will be small compared to the Higgs gauge coupling, hence the dominant final states will be $WW$ and $ZZ$ through a chargino/neutralino t-channel. We now have a coannihilation channel opening with the lightest chargino as NLSP, and this channel will also mostly proceed through gauge interaction, that is to say through the $s$-channel of a $W$ boson and a fermionic final state. We show on table \ref{tab:hino_case} the relative contribution of each channel on our following higgsino benchmark :
\begin{equation}
 M_1=500\qquad M_2=600\qquad \mu=-100\qquad\qquad\text{(GeV)}
\end{equation}

\begin{figure}[!h]
\begin{center}
\begin{tabular}{c|c}
$\Oh$&0.00460804\\\hline
$\chii\chii\to W^+W^-$ & 30\%\\
$\cha\chii\to\bar sc$ & 16\%\\
$\cha\chii\to\bar du$ & 16\%\\
$\chii\chii\to ZZ$ & 8\%\\
$\cha\chii\to\bar \tau\nu_\tau$ & 5\%\\
$\cha\chii\to\bar \mu\nu_\mu$ & 5\%\\
$\cha\chii\to\bar e\nu_e$ & 5\%\\
others & 15\%
\end{tabular}
\end{center}
\caption{\label{tab:hino_case} {\em Contribution of the different processes to the relic density in the pure higgsino case. The dominant process is here the annihilation of neutralinos to weak bosons, byt we also have an important part coming from coannihilation with the lightest chargino.}}
\end{figure}

Involving no Higgs nor sfermions exchanges, the result for the relic density is independent of $\ma,\tb$ and $\Mf$. We have here the case where the cross-section is quite sizeable and produce a very low relic density. Such a point would thus be excluded if we require dark matter to reproduce exactly the measured value for $\Oh$ recorded by WMAP, but would be still viable if we take it to be only an upper bound.

\subsubsection{Wino case}
\begin{figure}[!h]
\begin{center}
\begin{tabular}{cc}
\includegraphics[scale=0.3,trim=0 0 0 0,clip=true]{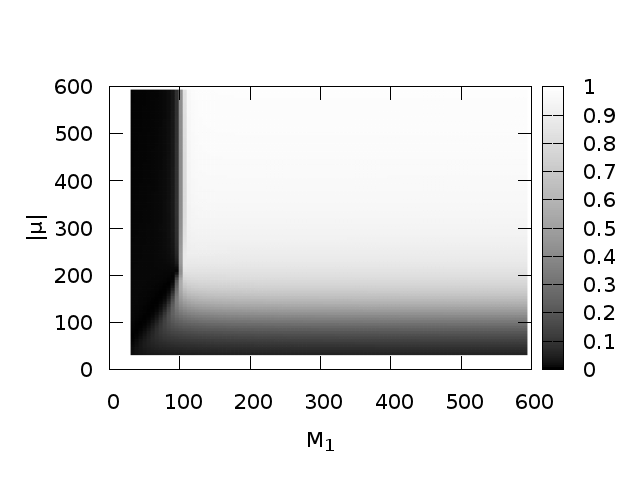}
&\raisebox{2.5cm}{$|Z_{n12}|^2$}
\end{tabular}
\end{center}
\caption{\label{fig:wino_case} {\em We show here the wino part of the lightest neutralino when varying $M_1,\mu$ with a fixed $M_2=100$ GeV. As can be guessed from the previous cases, the transition is sharp towards bino cases and smooth to the higgsino ones.}}
\end{figure}

The last case is generically obtained with $M_2$ smaller than $M_1,|\mu|$, with once more a sharp transition as seen on figure \ref{fig:wino_case}.\\

We will have $\mchi\sim M_2$, but since we also have $m_{\t{\chi}_1^+}\sim\mchi$ the co-annihilation will have a large contribution. The relevant couplings are $\t{V}\t{V}gV$, $d\t{f}g\t{V}f$, and it turns out that the main process is $\chii\t{\chi}_1^+\rightarrow ff'$. In table \ref{tab:wino_case} are shown the relative contributions to the \RD\ on a wino benchmark, taken as 
\begin{equation}
 M_1=500\qquad M_2=100\qquad \mu=-600\qquad\qquad\text{(GeV)}
\end{equation}

\begin{figure}[!h]
\begin{center}
\begin{tabular}{c|c}
$\Oh$&0.000335223\\\hline
$\cha\chii\to ff'$ & 65\%\\
$\cha\chii\to W^+Z$ & 6\%\\
$\chii\chii\to W^+W^-$ & 5\%\\
$\cha\cha\to\bar W^+W^+$ & 5\%\\
others & 19\%
\end{tabular}
\end{center}
\caption{\label{tab:wino_case} {\em Contribution of the different processes to the relic density in the pure wino case. Now the co-annihilation with a chargino is the dominant process, one also notes that it is very efficient since the relic density has a very small value.}}
\end{figure}

Because of the large superpartner masses in our set-up, the main process $\cha\chii\to ff'$ will predominantly operate through a $W$ exchange.

\subsection{The Mixed cases}
\begin{figure}[!h]
\begin{center}
\begin{tabular}{cc}
\includegraphics[scale=0.3,trim=0 0 0 0,clip=true]{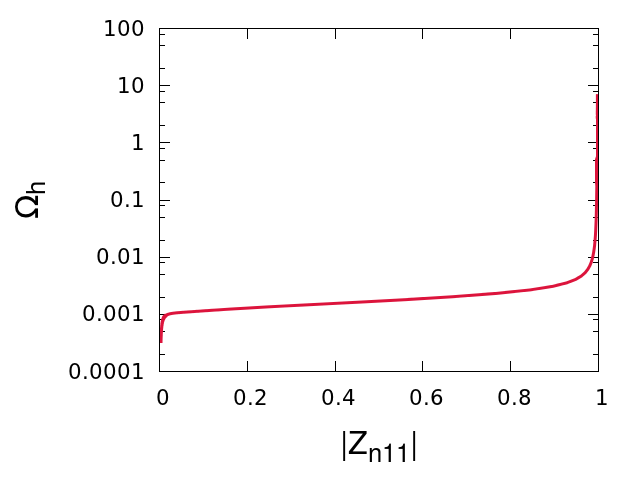}&
\includegraphics[scale=0.3,trim=0 0 0 0,clip=true]{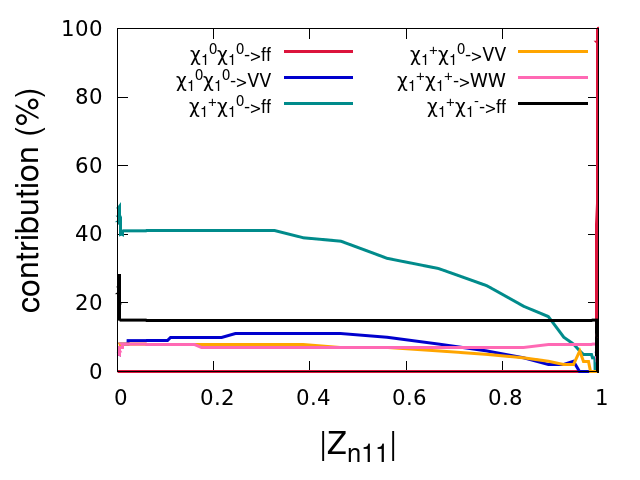}\\
\includegraphics[scale=0.3,trim=0 0 0 0,clip=true]{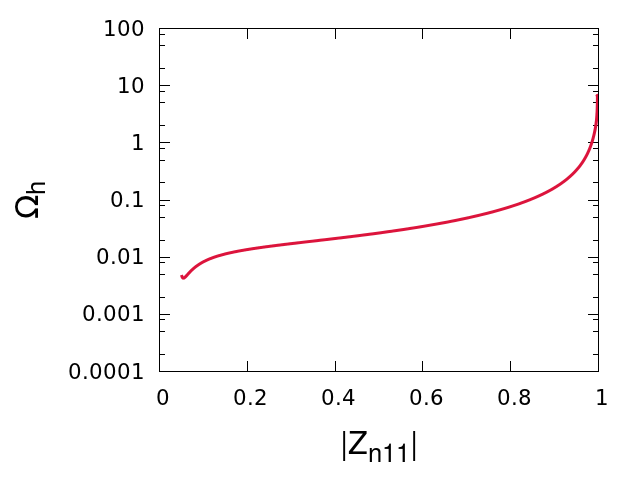}&
\includegraphics[scale=0.3,trim=0 0 0 0,clip=true]{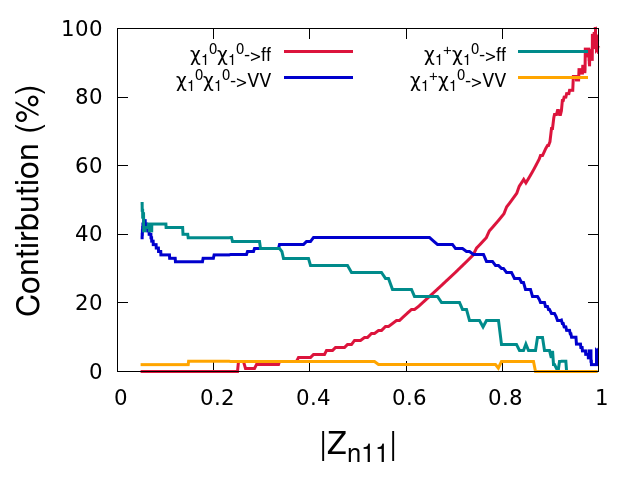}\\
\includegraphics[scale=0.3,trim=0 0 0 0,clip=true]{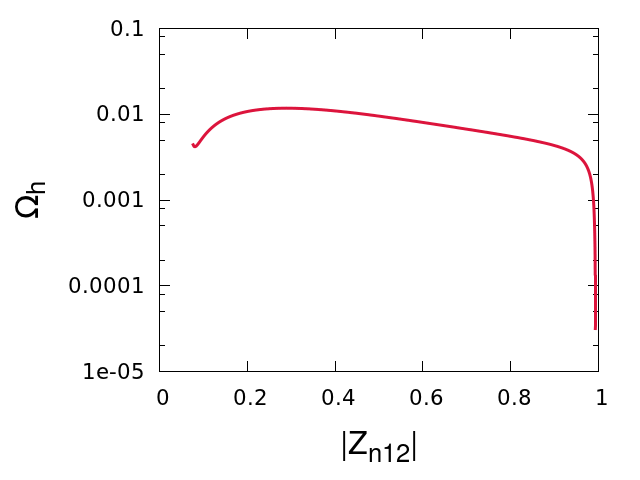}&
\includegraphics[scale=0.3,trim=0 0 0 0,clip=true]{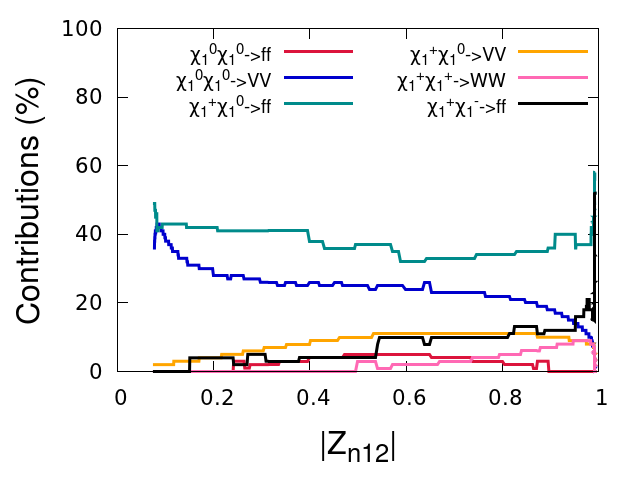}
\end{tabular}
\end{center}
\caption{\label{fig:mixed_case} {\em We show here three interpolations. On the top from the wino point to the bino point, where we see that the neutralino annihilation to fermions dominates only the pure bino case : for intermediate cases the co-annihilation with the lightest chargino is the most important process. Moreover the annihilation of two charginos can also contributes and since it does not depend on $M_1$ it does not change much along the mixed cases. The middle plot is the interpolation from higgsino to bino. Now the annihilation channel to fermions is still an important channel in mixed cases : this occurs thanks to the s-channel of a gauge boson that opens with a bino and a higgsino in the initial state. For mainly higgsino states the annihilation to weak bosons and the co-annihilation to fermions are the most important processes. The bottom plot goes from wino to higgsino and is mostly dominated by co-annihilation to fermions and weak bosons final states.}}
\end{figure}

When interpolating between two pure cases one meets cases where the LSP is a mixture of different natures, which is the mixed region. When going from one case to the other some process strengths will increase while others will decrease. The change in the relative contribution can be extreme since the total value of $\Oh$ is likely to change quickly, which will affect the relative contribution of a channel even if its strength stays constant. I have carried out simple  interpolations in between the three pure cases. In order to understand fully what happens, care has been taken so that $\tb,\ma$ and $\Mf$ stay the same. Hence all changes in the relative contribution should be seen with the $|Z_{ni1}|^2$ quantities. I show in figure \ref{fig:mixed_case} the variation of $\Oh$, and of each relative contribution. 

In particular one notices that when going away from the bino point, the cross-section rises nearly instantaneously, hence the relic density drops at once : this is because the channels of the bino case are much less efficient than the channels of the other cases. Note that in the mixed higgsino-bino case, because we are allowing for a coupling bino-higgsino-higgses, the process $\chii\chii\to\bar ff$ can still be an important contribution.\\

As a concluding remark, let me emphasize that I have only sketched some possible configurations in the supersymmetric parameter space for the relic density. In particular, our choice for Higgs and superpartners masses to be very high (1 TeV) has excluded the possibility of Higgs resonances in the various processes. Such a feature would of course dramatically change the value of the relic density and the relative contribution, but the aim was to catch the gross picture of the \RD\ along the MSSM parameter space, and in particular the role of the nature of the LSP. We are now going to see that, when turning to finer study, the effects of the other parameters are to be taken into account.


%% file: chapter9.tex
\chapter{Dark Matter II : the call for precision}

\minitoc\vspace{1cm}

The impressive accuracy of the WMAP measurement for the relic density (the last
release (\cite{wmap_7}) showing a 3\% uncertainty) should underline that we now
have to answer the following question : can we content ourselves with the
zeroth-order prediction or do we need to go beyond? On the BMSSM side, the point
is then to know whether extra particles can have a sizeable effect on the LSP
phenomenology. At the time being, this hypothesis is not considered to be so
crucial, and dedicated studies are scarce. On the loop expansion side it is
first not obvious that considering radiative corrections to the relic density is
relevant : indeed it is a purely electroweak process at the tree-level hence
such corrections are expected to be small. But as small as they are, they may
still be important compared to the precision we are aiming at and furthermore,
we could expect sizeable contribution from superpartners loops even in the case
of a heavy spectrum, constraining thus the susy spectrum also at high masses.
This 
would be particularly interesting in the view of an interplay with the LHC,
which will only probe moderately heavy superpartners. This situation is akin to
the precision electroweak observables and their sensitivity to the top and Higgs
mass. For example, non decoupling effects termed in analogy with electroweak SM
observables, super-oblique corrections have been revealed in one-loop
calculations of supersymmetry
observables\cite{feng_nondecoupling,randall_nondecoupling,nojiri_nondecoupling}.
Computing the \RD\ at the first order of radiative corrections -- i.e. the
one-loop order -- has previously been proven to be feasible (see
\cite{baro07,Freitas-relic-qcd,Klasen-relic-qcd,boudjema_chalons1,
Bjorn_review_rc}, among others). Results from those have strongly advertised the
necessity to go beyond the tree-level, however, the full one-loop computations
is far from being a standard. The reason for this is that the full
computation may be too thorny and time-consuming. 
Those two issues have called for another approach to the radiative corrections,
which will be described in this chapter : the effective approach.

\section{Corrections to the cross-section}

\subsection{New Physics corrections}
Before going to the one-loop computations, that will be relevant in all
supersymmetric models, let us first dwell on the case of the possible
corrections brought by extra physics beyond the MSSM. Precisely, since we have
developed in the previous chapters a formalism to account for effects of the New
Physics through effective operators, the question is to assess the consequences
of those operators on the computation of the relic density. Hence we will for a
moment go back to the parameter space that was used for the Higgs study. It
turns out that higher order corrections do not affect the Dark Matter relic
density as strongly as they affect the scalar Higgs sector. This is partly due
to the fact that the gauginos are not directly modified by the effective
operators, since those are written in term of Higgs superfields. Concerning the
neutralino mass matrix, it is changed in the higgsino mass terms and the
higgsino-gaugino mixing. At the level of couplings we will have shifts to the
following couplings
$$\t{h}\t{h}gA,hg\t{A}\t{h},\ y\t{h}f\t{f}$$
Although we know from chapter 3 how to derive the mixing matrix and the masses
at any order of the perturbation theory, the analytic expressions tend to be
quite lengthy in the case of the 4 by 4 neutralino mass matrix. We will
therefore keep the discussion at the level of numerical considerations.

\begin{figure}[!h]
\begin{center}
\includegraphics[scale=0.35,trim=0 0 0 0,clip=true]{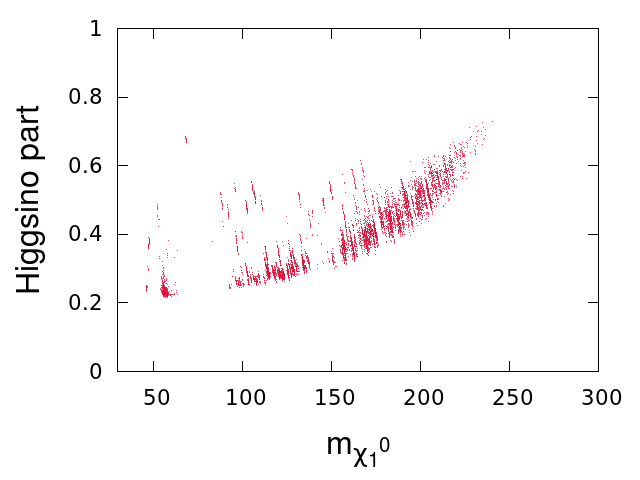}
\end{center}
\caption{\label{fig:res_1} {\em Allowed regions when adding the dark matter
constraints in terms of the higgsino fraction of the lightest neutralino, that
is $\sqrt{|Z_{n\,31}|^2+|Z_{n\,41}|^2}$ versus its mass.}}
\end{figure}

\paragraph{BMSSM features}~\\
It turns out that the relative corrections are small in most of the parameter
space. What is more interesting is then to see whether there is an interplay
between the constraint on the Higgs searches at the LHC and the dark matter
observables. To study such a correlation we will first impose the requirement
that the relic density $\Oh$ is totally accounted for by the model. The other
constraint comes from the direct detection experiments. We have used XENON 100
limits (from \cite{xenon_100_2011}) on the spin independent cross-section of the
dark matter candidate on the nucleus. The latter being also computed using
\momegas (\cite{micromegas2}). As in the usual MSSM case, the observables
associated to dark matter will strongly depend on the nature of the lightest
neutralino, that is to say whether its dominant contribution is bino, wino or
higgsino, or a mixture of different species. In order to encompass all the
different possibilities we have extended our MSSM scenario to let $M_1,M_2$ vary
freely in the 
range
\begin{equation}
M_1,M_2 \in[50,600]\ \text{GeV}.
\end{equation}
Since $\mu$ is fixed at $\mu=300$ GeV, this will generate all the possible
hierarchies between the three parameters. Having extended our search in such a
manner, we have performed a scan taking into account constraints from Higgs
physics and the ones from flavour physics. Among those points, the lightest
neutralino will be mostly a mixture of bino and higgsino, ranging from the case
of an equal mixing (50\% bino and 50\% higgsino) to a mostly bino case (95\%
bino). One can see in fig.\ref{fig:res_1} what are the masses allowed for the
dark matter candidate, and to which part of higgsino species they correspond.\\

\begin{figure}[!h]
\begin{center}
\includegraphics[scale=0.35,trim=0 0 0 0,clip=true]{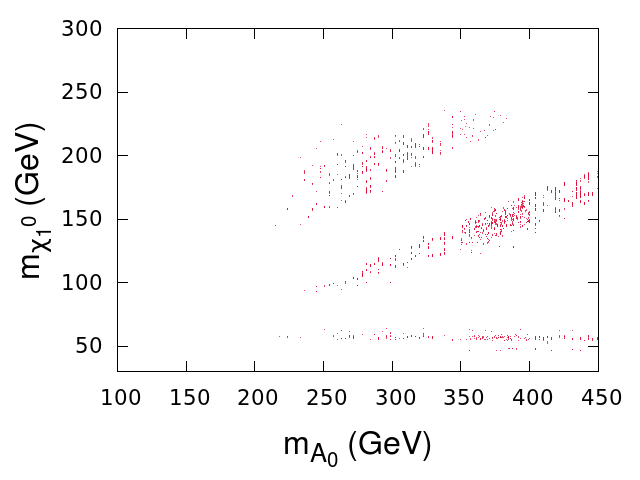}
\end{center}
\caption{\label{fig:res_2} {\em Correlation between $\ma$ and $m_{\chii}$ for
points where the relic density is driven by the process $\chii\chii\to f\bar f$.
One notices a lower band, which is due to the $Z$ resonance, and two strips that
are on each side of the line $\ma=2m_{\chii}$ : points in-between them have a
too strong $A_0$ resonance and the relic density ends up to be too small.}}
\end{figure}

\noindent We can categorize the main channels for the relic density as follows :
\begin{itemize}
\item $\chii\chii\to f\bar f$ : this is the most frequent case. Though the cross
section of this process is usually too small to respect the relic density
constraint, it can be enhanced by an $A_0$ resonance characterised by
$2\mchi\sim\ma$. Lower masses can benefit from a $Z$ resonance. This point is
highlighted in fig.\ref{fig:res_2}, where we show the correlation between the
mass of the lightest neutralino and the pseudo-scalar mass, for points where the
dominant contribution comes from such a channel.
\item $\chii\chii\to WW/ZZ$ : at a given mass for the neutralino, this channel
corresponds to the points with the highest higgsino part.
\item $\chii\chii\to WH/ZH$ : this channel only opens for high masses, that is
$m_{\chii}>200$ GeV. This is simply a kinematical restriction since this channel
requires $2\mchi>m_{H^+}+\mw$ to open.
\end{itemize} 

\begin{figure}[!h]
\begin{center}
\includegraphics[scale=0.4,trim=0 0 0 0,clip=true]{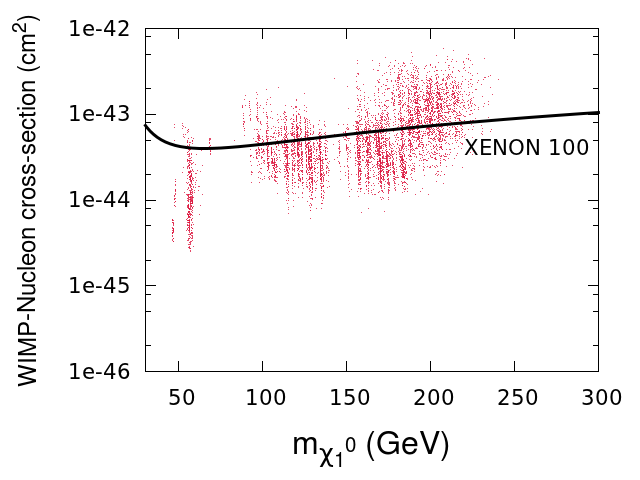}
\end{center}
\caption{\label{fig:res_3} {\em WIMP-Nucleus cross section, for the Xenon
nucleus, as a function of the WIMP mass $m_{\chii}$. In black is shown the upper
bound derived by XENON 100 (\cite{xenon_100_2011}) experiment.}}
\end{figure}

\paragraph{Direct detection}~\\
As is shown on figure \ref{fig:res_3}, the points that respect the relic density
constraint are quite constrained by the bounds obtained from the XENON 100
(\cite{xenon_100_2011}) experiment. Not only an important part of the points are
excluded by the  search, but the remaining points stay close to the bound and
could thus be probed with the following upgrade. One notices that the highest
masses are the most affected, since the predicted cross-section is somewhat
higher than for low mass.\\

\subsection{Quantum corrections}
\begin{figure}[!h]
\begin{center}
\begin{tabular}{cc}
\begin{fmfgraph*}(80,50)
\fmfleft{p1}
\fmfright{p2}
\fmf{fermion}{p1,v1}
\fmf{fermion}{v2,p2}
\fmf{dashes,left,tension=0.4}{v1,v2,v1}
\end{fmfgraph*}
&
\begin{fmfgraph*}(80,50)
\fmfleft{p1}
\fmfright{p2}
\fmfv{decor.shape=cross}{v1}
\fmf{fermion}{p1,v1,p2}
\end{fmfgraph*}\\
Self energy & Field renormalisation \\
\begin{fmfgraph*}(80,50)
\fmfleft{p1,p2}
\fmfright{p3}
\fmf{fermion}{p1,v1}
\fmf{fermion}{v2,p2}
\fmf{fermion,tension=0.4}{v1,v2}
\fmf{dashes,tension=0.4}{v1,v3,v2}
\fmf{dashes}{v3,p3}
\end{fmfgraph*}
&
\begin{fmfgraph*}(80,50)
\fmfleft{p1,p2}
\fmfright{p3,p4}
\fmf{dashes}{p1,v1}
\fmf{dashes}{p2,v2}
\fmf{dashes}{p3,v3}
\fmf{dashes}{p4,v4}
\fmf{fermion,tension=0.4}{v1,v2,v4,v3,v1}
\end{fmfgraph*}\\
Vertices & Boxes\\
\end{tabular}
\caption{\label{diag:category} {\em Categories of loop corrections}}
\end{center}
\end{figure}
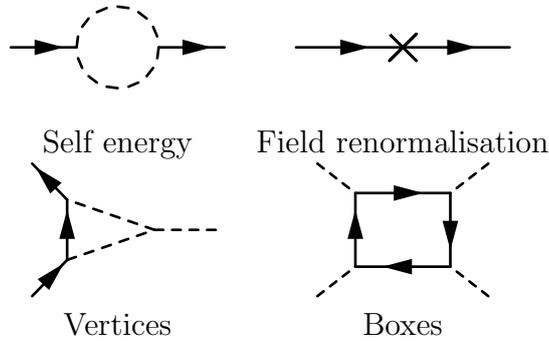

We will now switch to the other source of corrections to the theoretical
prediction : the loop corrections. Although we already know from chapter 3 how
to compute an observable at the one-loop level, from chapter 4 how to apply this
to the MSSM and from chapter 5 how to automatise fully the computation, it is
not totally irrelevant to have a look at some analytical features of the
computations first. Indeed, as warned in chapter 5, one cannot compute numerical
results blindly and then try to see some physical effects in them, instead one
has to go the other way round : first guess some physical dependencies and then
verify them numerically. As an example it is usually quite hard to determine
which contribution dominates a whole process, and to do so numerically one has
to separate the total amplitudes in different contributions. The one-loop order
radiative corrections to the tree-level process $\chii\chii\to XY$ are formally
categorised upon the number of propagators running in the loop. To wit, those 
categories are shown in figure \ref{diag:category}.

Note that the two point functions (corresponding to the first row) are purposely
separated in two parts, the last one corresponding to wave-function
renormalisation. Since we use the $OS$ scheme, the one-loop part of the
propagator vanishes or, in other words, the loop diagrams in the Field
renormalisation category automatically cancel the one-loop part of the mixing of
the fields. At this point there is no way of knowing whether one category is
more important than another. Besides the question does not really make sense :
indeed the categories have been decided on a purely pictorial point and not by
relating them to some observables and, as it usually happens in such cases, the
different amplitudes turn out not to be gauge invariant and to contain
divergences. One could argue that the boxes diagrams are not divergent (since we
cannot have 4 fermions in the loop, having already two in the initial state).
While this is true, they are not always gauge-invariant when considering gauge
particle in the final 
state.\\

To those corrections one must add the bremstrahlung corrections : however it
affects only the final states since neutralinos are neutral. As an example, I
show in Table \ref{tab:loop_ff} the size of the full one-loop correction to the
process $\chii\chii\to\bar \mu\mu$ for the three pure cases discussed in the
last chapter.

\begin{table}[!h]
\begin{center}
\begin{tabular}{lc}
Nature & Relative correction\\
Bino & 19.6\%\\
Wino & 20\%\\
Higgsino & -7.5\%
\end{tabular}
\end{center}
\caption{\label{tab:loop_ff}{\em Relative correction of the full one-loop
cross-section with respect to the tree-level cross-section of the process
$\chii\chii\to\bar\mu\mu$ in the three pure cases.}}
\end{table}

It is hence clear that the relative correction brought by the one-loop diagrams
have quite different origins, depending on the nature of the LSP. This
interesting feature being developed later, we are left with a crucial issue of
the full one-loop calculations : they are CPU time-consuming.

\subsubsection{Issue : the computing time}
The number of Feynman diagrams needed to obtain the result of figure
\ref{tab:loop_ff} are of the order of several hundreds at least. This feature is
unfortunately quite generic to supersymmetry since the theory includes a large
number of particles. This is also the reason why the computation of the relic
density at one loop in the MSSM heavily relies on automated and numerical tools.
Precisely, those numbers were obtained by using the \SL\ program presented in
chapter 5. However whereas the use of this tool is quite efficient on a given
set of benchmarks, it is much less suited to the exploration of the vast
parameter space of the soft masses of the MSSM. Indeed the computing time of a
single point is not negligible since it requires to evaluate a large number of
loop integrals. To give an idea, the comparison with the tree-level computation
exhibits at least 2 orders of magnitude in the computing time. This is an issue
for the different studies that aim at scanning the MSSM parameter space. One
solution to 
this 
issue is the introduction of an effective Lagrangian which would mimic the
loop-effects. This idea has led to a publication (see \cite{gdlr_dm_1108}),
and I will present our results in the following section.

\section{Effective approach for quantum corrections}
The idea being to introduce effective couplings in order to retain the one-loop
contribution, we will simplify the task by considering only one process, instead
of the full relic density. The process will be the following
\begin{equation}
\chii\chii\to\bar\mu\mu
\end{equation}
This process will be estimated on the same parameter set as in our tree-level
study, and the idea is to compare the full one-loop computation with the
effective calculation.

\subsection{Running of the electromagnetic coupling}
A first example of an effective coupling is that if one includes the universal
correction to the electromagnetic coupling $e$, one improves considerably the
tree-level computation. In our bino case the running of $\alpha_{em}$, which we
will call $\alpha_\text{eff}(Q)$ where we have used $Q=2\mchi$, yields a 14\%
contribution to the process we consider, hence the major part of the full
one-loop correction ($\sim 20$\%). The definition of such a running is simply a
choice of renormalisation scheme : instead of requiring that the one loop
correction to the electromagnetic vertex $\gamma f\bar f$ vanishes at
$p_\gamma\sim0$, we set this for $|p_\gamma|= Q$ resulting in an enhancement in
the value of $e$. One can relate directly this to the shift of the photon self
energy from $p^2=0$ to $p^2=Q^2$. Since $e$ does not only enter in $\gamma$
vertex but in any electroweak gauge coupling, this will affect nearly all
vertices involved in the \RD\ computation. We define hence the $\alpha_\eff$
cross-section
\begin{equation}
\sigma_{\alpha(Q)}=\sigma^0(\alpha=\alpha(2\mchi))
\label{eq:alpha_eff}
\end{equation}
 
\subsection{Effective vertices}
However this running does not contain all of the radiative corrections, so we
may be interested in other effective couplings to get closer to the full
one-loop corrections. Among the full set of those one-loop corrections one can
in particular construct a
finite subset that is not specific to the final state, that is to say the muons.
This subset will
be involved in all processes involving neutralinos. For example,
the vertex correction to $\neuto \neuto Z$ is obviously
independent of the muon being in the final state, a similar
statement can be said for $\neuto \neuto h/H/A$. Also, all
occurrences of the wave function renormalisation of the neutralino
(including transitions between neutralinos) and the $Z$ are
process independent. The same can be said  also of the
counterterms to the gauge couplings and the vacuum expectation
values or in other words $v,\tb$. On the other hand the wave
function renormalisation of the muon is specific to the muon final
state. The box contributions, as well as the QED correction are also specific to
the
process. The construction of the universal correction for the
effective coupling $\neuto f \tilde{f}$ from $\neuto \mu
\tilde{\mu}$ is different from that of $\neuto \neuto Z$, since in
the latter all three particles can be considered as universal. For
example the full correction to the vertex $\neuto \mu \tilde{\mu}$ consists of
a 1-PI 3-point function vertex correction (triangle) which is
muon specific and that does not need to be calculated to build up
the effective coupling. It also contains wave function
renormalisation of the neutralinos as well as  counterterms for
the gauge couplings and for other universal quantities such as
$\tb$ which must be combined to arrive at the universal
correction for the $\neuto f \tilde{f}$ vertex. The aim is therefore to extract
these process independent
contributions and define effective vertices for the LSP
interactions. This is akin to the effective coupling of the $Z$ to
fermions where universal corrections are defined. Describing the
bulk of the radiative corrections in terms of effective couplings
has been quite successful to describe for example the observables
at the $Z$ peak. Although not describing most perfectly the effect
of the full corrections for all observables (for example $Z b \bar
b$ receives  an important triangle contribution due to the large
top Yukawa coupling) one must admit that the approach has done quite a
good job. Most of the effective corrections were universal,
described in terms of a small set of two-point functions of the
gauge bosons. \\
The other benefit was that such approximations were sensitive to
non decoupling effects that probe higher scales (top mass and Higgs mass). The
set of two-point functions, and for $\neuto \neuto Z$ three-point
functions, should of course lead to a finite and gauge invariant
quantity. Loops involving gauge bosons have always been problematic (even in
the case of the $Z f \bar{f}$) in such an approach
since it is difficult to extract a gauge independent value. The aim is to
consider
the couplings of the neutralinos as would be needed for
approximating their annihilation cross section independently of
the final state. Therefore one would expect that apart from the
rescaling of the gauge couplings which can be considered as an
overall constant, the mixing effect between the different
neutralinos should be affected. One can in fact re-organise a few
of the two point functions (that can be written also as
counterterms) to define an effective coupling for the neutralino.
One should of course also correct in this manner the $Z \mu^+
\mu^-$ coupling. Let us stress again that in this first
investigation we will primarily take into account the effects of
fermions and sfermions in the universal loops. For the $\neuto
\neuto Z$ effective we also attempt to include the virtual
contribution of the gauge bosons especially that for the
higgsino-like the coupling to $W$ and $Z$ are not suppressed.

\paragraph{$\chisff$}~\\
To find the process independent corrections to this coupling, we
recall that in the basis $(\tilde{B}^0,\tilde{W}^{0},
\tilde{H}_1^0,\tilde{H}_2^0)$ before mixing and for both $f_{L,R}$
the couplings for the two chiral Lorentz structures reads as
\beqn
\label{eq:wbh_fsf} \frac{1}{\sqrt{2}} \big( g^\prime Y_f, g
\tau^3_f, y_{1,f}, y_{2,f} \big)=\frac{1}{\sqrt{2}} \big( g^\prime
Y_f, g \tau^3_f, \frac{g m_u}{M_W c_\beta},  \frac{g m_d}{M_W
s_\beta}\big) \rightarrow \big( g^\prime , g , \frac{g }{M_W
c_\beta}, \frac{g }{M_W s_\beta}\big),
\eeqn

$Y_f,\tau_3^f$ are the isospin and $SU(2)$ charges of the
corresponding fermion/sfermions. The last term, on the right, is not the exact
coupling, but its universal part, where universal is meant with respect to the
fermions in the final state. The two higgsinos couple
differently to the up and down fermions with a coupling that is
proportional to the Yukawa coupling. Though  this is not universal
we can still isolate a universal part where there is no  reference
to the final fermion/sfermion. This is what is meant by the last
expression in Eq.~\ref{eq:wbh_fsf} where the explicit mass of the
corresponding fermion masses has been dropped. The
variations/counterterms on these parameters have to be implemented
before turning to the physical basis. In the case of effective coupling of
neutralinos,  this is
achieved by defining an effective mixing matrix such that $N \ra
N+\Delta N^{\chi f \tilde{f}}$ in all couplings of the neutralino.
The $\Delta N^{\chi f \tilde{f}}$ read as

\beqn
\label{eq:app_hollik}
\Delta N_{i1}^{\chi f \tilde{f}}&=&\frac{\delta g^\prime}{g^\prime} N_{i1} +
\frac{1}{2}\sum_{j}N_{j1}\delta Z_{j i}, \nonumber \\
\Delta N_{i2}^{\chi f \tilde{f}}&=&\frac{\delta g}{g}N_{i2} + \frac{1}{2}
\sum_{j}N_{j2}\delta Z_{j i}, \nonumber \\
\Delta N_{i3}^{\chi f \tilde{f}}&=&\left(\frac{\delta g}{g}-\1\frac{\delta
M_W^2}{M_W^2}-\frac{\delta
c_{\beta}}{c_{\beta}}\right)N_{i3}+\frac{1}{2}\sum_{j}N_{j3}\delta Z_{j i},
\nonumber \\
\Delta N_{i4}^{\chi f \tilde{f}}&=&\left(\frac{\delta
g}{g}-\1\frac{\delta M_W^2}{M_W^2}-\frac{\delta
s_{\beta}}{s_{\beta}} \right)N_{i4} + \frac{1}{2}
\sum_{j}N_{j4}\delta Z_{j i}.
\eeqn
where $j$ runs from 1 to 4 and for the LSP , $i=1$.

All the counterterms above are calculated from self-energy
two-point functions and are fully defined in
~\cite{Sloops-higgspaper,baro09}. $\delta g/g=\delta e/e-\delta
s_W/s_W,\delta g^{\prime}/g^{\prime}=\delta e/e-\delta c_W/c_W$.
$\delta s_\beta/s_\beta=c_\beta^2 \delta \tb/\tb$.
Eq.~\ref{eq:app_hollik} agrees with what was suggested in
\cite{Hollik_susyeff}. Let us stress again that in these
self-energies no gauge bosons and therefore no neutralinos and
charginos are taken into account but just sfermions and fermions, otherwise this
would not be finite.
For a bino-like, self-energies containing gauge and Higgs bosons
(with their supersymmetric conterparts) are not expected to
contribute much. This is not necessarily the case for winos and
higgsinos.

\subsubsection{$\chii\chii Z$}
Since all particles making this vertex can now be considered as
being process independent (as far as neutralino annihilations are
concerned), all counterterms including wave function
renormalisation of both the $Z$ and $\neuto$ must be considered.
The price to pay now is that the genuine triangle vertex
corrections  $\neuto \neuto Z$ must also be included. It is only
the sum of the vertex and the self-energies that renders a finite
result. When correcting this vertex one must also correct the $Z
\mu^+ \mu^-$ vertex keeping within the spirit of calculating the
universal corrections. This can be implemented solely through
self-energy corrections (excluding the muon self-energies) and
there is no need to calculate here the genuine vertex corrections.
An exception would be the production of the $b$ and to some extent
the top where genuine vertex corrections are important. Talking of
heavy flavours, when computing the correction to
the $\neuto \neuto Z$ with the Z off shell with an invariant mass $Q^2$, one
should also include the $\neuto \neuto
G^0$ vertex, where $G^0$ is the neutral Goldstone boson. In our
case we restrict ourselves to almost massless fermions. The case
of the top and bottom final states will be addressed elsewhere
together with the potential relevant
contribution of the Higgses in the $s$-channel. \\
Since one is including the genuine 1-PI vertex correction, it is
important to inquire whether this correction generates a new
Lorentz structure beyond the one found at tree-level. The
contribution to the tree-level Lorentz structure is finite after
adding the self-energies and the vertex. Any new Lorentz structure
will on the other hand be finite on its own. General arguments
based on the Majorana nature of the neutralinos backed by our
numerical studies show that no new Lorentz structure is generated
for neutralinos. First of all, at tree-level one has only one
structure
\beqn
{\cal {L}}_{\neuto \neuto Z}=\frac{g_Z}{4}
\Big(N_{13}N_{13}-N_{14} N_{14} \Big) \neuto \gamma_\mu
\gamma_5\neuto Z^\mu, \quad \quad g_Z=\frac{e}{c_W s_W}.
\eeqn
The overall strength is a consequence of the fact that the
coupling emerges solely from the higgsino with a gauge coupling.
Indeed in the $(\tilde{B}^0,\tilde{W}^{0},
\tilde{H}_1^0,\tilde{H}_2^0)$ basis the coupling is $\propto g_Z
(0,0,1,-1)$. Only the Lorentz structure $\gamma_\mu \gamma_5$
survives as a consequence of the Majorana nature. With $p_1,p_2$
denoting the incoming momenta of the two $\neuto$, at one-loop a
contribution $(p_1^\mu-p_2^\mu)$ does not survive symmetrisation,
whereas $(p_1^\mu+p_2^\mu)$ will not contribute for massless
muons. We calculate this correction  for a $Z$ with an invariant
mass $Q^2$, in the application this $Q^2$ will be set to the
invariant mass of the muon pair.  This vertex contribution is
denoted $\Delta g_{\neuto \neuto Z}^{\bigtriangleup}(Q^2) \equiv
\Delta g_{\neuto \neuto Z}^{\bigtriangleup}$. The contribution of
the coupling counterterms defining $g_Z$ {\bf and} the $Z$
wave function renormalisation define the universal correction to
the $Z$ coupling strength $g_Z^{\rm eff}=g_Z (1+\Delta g_Z)$,
with $\Delta g_Z/g_Z= \delta g_Z/g_Z+ \delta Z_{ZZ}/2$. $\delta
Z_{ZZ}$ is the wave function renormalisation of the $Z$. We of
course have to add the wave function renormalisation of the
$\neuto$ like what was done for the $\neuto f \tilde{f}$ vertex.
We improve on this implementation by taking into account the fact
that the $Z$ is off-shell and therefore the wave function
renormalisation through $\delta Z_{ZZ}=\Pi_{ZZ}^{\prime}(M_Z^2)$ is
only part of the correction that would emerge from the correction
to the complete $Z$ propagator in the $s$-channel contribution with invariant
mass $Q^2$.
Note that here there is no need for including a $Z\gamma$
transition since photons do not couple to neutralinos. Collecting
all these contributions, the
effective vertex is obtained by making \\
$g_Z \to g_{\neuto \neuto Z}^{{\rm eff}}$ and $N_{i 1} \to N_{i 1}
+ \Delta N_{i 1}^{\neuto \neuto Z}$ with
\beqn
g_{\neuto \neuto Z}^{{\rm eff}}&=&g_Z (1+ \Delta g_Z(Q^2) + \Delta
g_{\neuto \neuto Z}^{\bigtriangleup}(Q^2)); \\ \quad \Delta
N_{ij}^{\neuto \neuto Z}&=&\frac{1}{2} \sum_{k} N_{kj} \delta
Z_{ki}, \;\; (i,j,k)=1\dots 4.
\eeqn
Explicitly
\def\sw2{s_W^2}
\def\cw2{c_W^2}
\def\MZ2{M_Z^2}
\def\MW2{M_W^2}
\beqn
\label{delgzuni} \Delta
g_Z&=&\1\left(\Pi'_{\gamma\gamma}(0)-2\frac{s_W}{c_W}\frac{\Pi_{\gamma
Z}(0)}{\MZ2}\right)+\1\left(1-\frac{\cw2}{\sw2}\right)
\left(\frac{\Pi_{ZZ}(\MZ2)}{\MZ2}-\frac{\Pi_{WW}(\MW2)}{\MW2}\right)
\nonumber \\
&-&\frac{1}{2} \left(
\frac{\Pi_{ZZ}(Q^2)-\Pi_{ZZ}(M_Z^2)}{Q^2-\MZ2} \right)\ .
\eeqn
At the same time for the fermion with charge $q_f$ we correct the
$Z f \bar f$ $\propto g_Z (\gamma_5 +(1-4 |q_f| s_W^2))\gamma_\mu$
by effectively making $g_Z \to g_Z (1+\Delta g_Z)$ with $\Delta
g_Z$ defined in Eq.~\ref{delgzuni} and $s_W^2$ to
\beqn
\label{delsweff} \Delta
\sw2=\frac{\cw2}{\sw2}\left(\frac{\Pi_{ZZ}(M_Z^2)}{\MZ2}-\frac{\Pi_{WW}(M_W^2)}{
\MW2}\right)+
\frac{c_W}{s_W}\frac{\Pi_{\gamma Z}(k^2)}{k^2}\ .
\eeqn
By default we include only the fermions and sfermions in the
virtual corrections described by
Eqs.~\ref{delgzuni}-\ref{delsweff}. For the $\neuto \neuto Z$ one
expects the contribution of the gauge bosons and the
neutralinos/charginos to be non negligible especially for the
higgsino case. In fact, including such contributions still gives
an ultraviolet finite result for $g_{\neuto \neuto Z}^{{\rm eff}}$
in Eq.~\ref{delgzuni} which is a non trivial result. Moreover this
contribution is gauge parameter independent in the class of
(linear) and non-linear gauge fixing conditions\cite{grace-1loop}.
To weigh up the gauge/gaugino/higgsino contribution we will
therefore also compare with this generalised effective $g_{\neuto
\neuto Z}^{{\rm eff}}$ including all virtual particles. Observe that in
Eq.~\ref{delgzuni} we have
the contribution $\Pi_{\gamma Z}(0)$ which vanishes for fermions
and sfermions but which is essential for the contribution of the
virtual $W$. In any case including gauge bosons in the
renormalisation of electromagnetic coupling requires the inclusion
of the $\Pi_{\gamma Z}(0)$ in Eq.~\ref{delgzuni} for gauge
invariance to be maintained\cite{grace-1loop}. We stress that we
will present the effect of the generalised effective coupling
$g_{\neuto \neuto Z}^{{\rm eff}}$ as an indication of the gauge
boson contribution while keeping in mind that this result may lead
to unitarity violation. Indeed through cutting rules, the $W$ loop
can be seen as made up of  the scattering $W^+W^- \to Z \to \mu^+
\mu^-$ that needs a compensation from the cut in the box. For the effective $Z
\mu^+ \mu^-$ coupling we only include the fermion/sfermion
contribution in Eqs.~\ref{delgzuni}-\ref{delsweff}, adding the
gauge bosons would require part of the 1-PI triangle contribution
to $Z \to \mu^+ \mu^-$.

\subsection{Robustness of the effective operators}

To analyse consistently the efficiency of effective corrections we
will refer to the following quantities :

\beqn
\label{Delta_U} \Delta_{\rm eff}=\frac{\seff-\s0}{\s0} \,.
\eeqn
Here $\seff$ is the cross section calculated with the effective
couplings that include, by default, universal process independent particles
excluding gauge bosons and gauginos/higgsinos. We will explicitely specify when
including all virtual particles in those corrections, referring to it as
$\DU^{W}$. This correction
will  be compared to the correction solely due to the running of
the electromagnetic coupling, see Eq.~\ref{eq:alpha_eff}. To
see how well the correction through the effective couplings
$\neuto f \tilde{f}$ and $\neuto \neuto Z$ reproduces the full
one-loop correction we introduce
\beqn
\Delta_{NU}=\frac{\sol-\seff}{\s0} \,, \nonumber \\
\Delta_{\rm full}=\frac{\sol-\s0}{\s0} \;.
\eeqn
with $\sol$  the full one-loop cross section, $\Delta_{NU}$
measures what we will refer to as the non-universal corrections
although strictly speaking this measures the remainder of all the
corrections that are not taken into account by the effective
vertices approach. $\Delta_{\rm full}=\Delta_{\rm
eff}+\Delta_{NU}$ is the full one-loop correction.

\paragraph{Bino case}~\\
\subparagraph{Effective vs full corrections}~\\
We start with our bino point : $(M_1,M_2,\mu)=(90,200,-600)\ GeV$ which yields a
lightest  bino-like neutralino (the bino composition is 99\%) with
mass $\mneuto=91\ GeV$. At tree-level the cross section for
relative velocity $v=0.2$ is $\sigma_{\mu^+
\mu^-}^{\tilde{b}}=6.75 \times 10^{-3}{\rm pb}$. Note for further
reference that this is an order of magnitude larger than
annihilation into a pair of $W$'s: $\sigma_{W^+
W^-}^{\tilde{b}}=4.51 \times 10^{-4}{\rm pb}$. The annihilation
proceeds predominantly through the $t$-channel, binos coupling to
$Z$ vanishing. This leads to the following set of
corrections
\beqn
\DU=17.52\% (\DA=14.56\%) \quad \quad \DNU=2.06\%(\Delta_{\rm
full}=19.58\%).
\eeqn

\begin{figure}[htbp]
\begin{center}
\hspace*{-1.4cm}
\begin{tabular}{cc}
\includegraphics[width=0.56\textwidth]{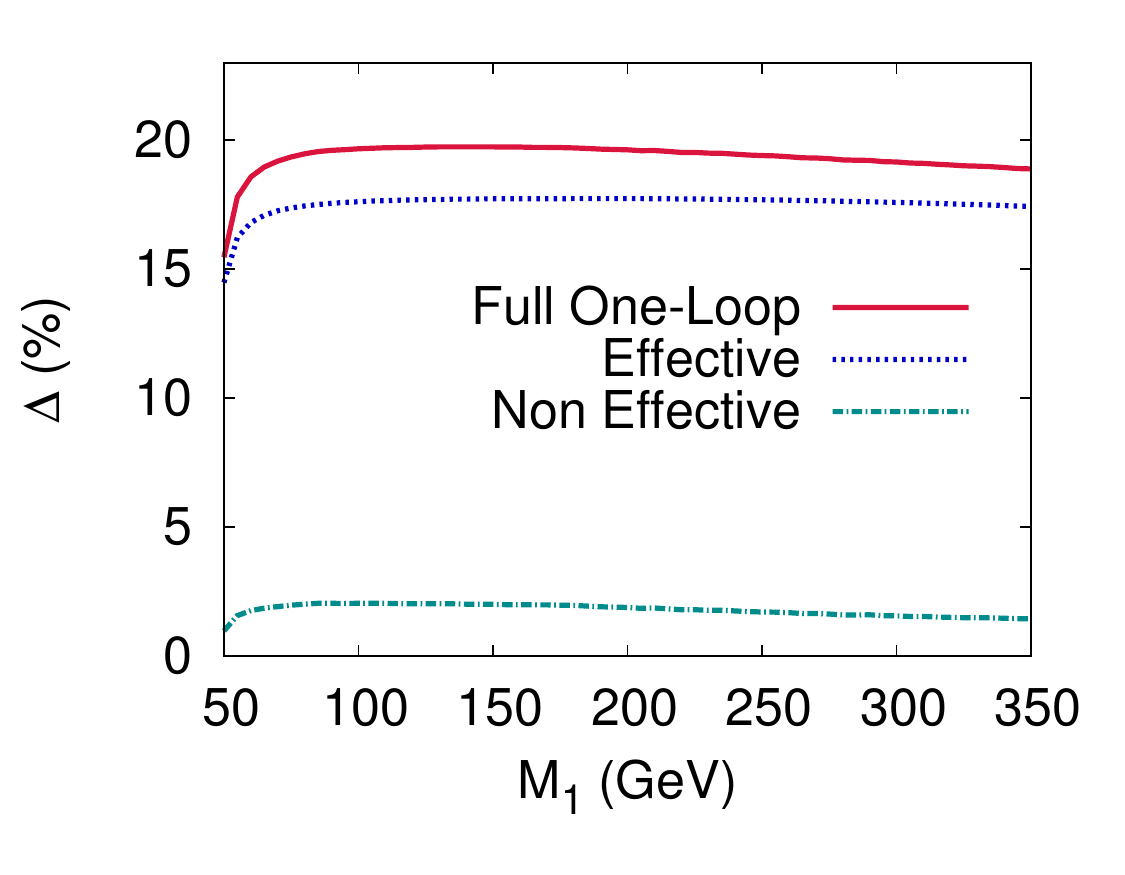}
&
\includegraphics[width=0.56\textwidth]{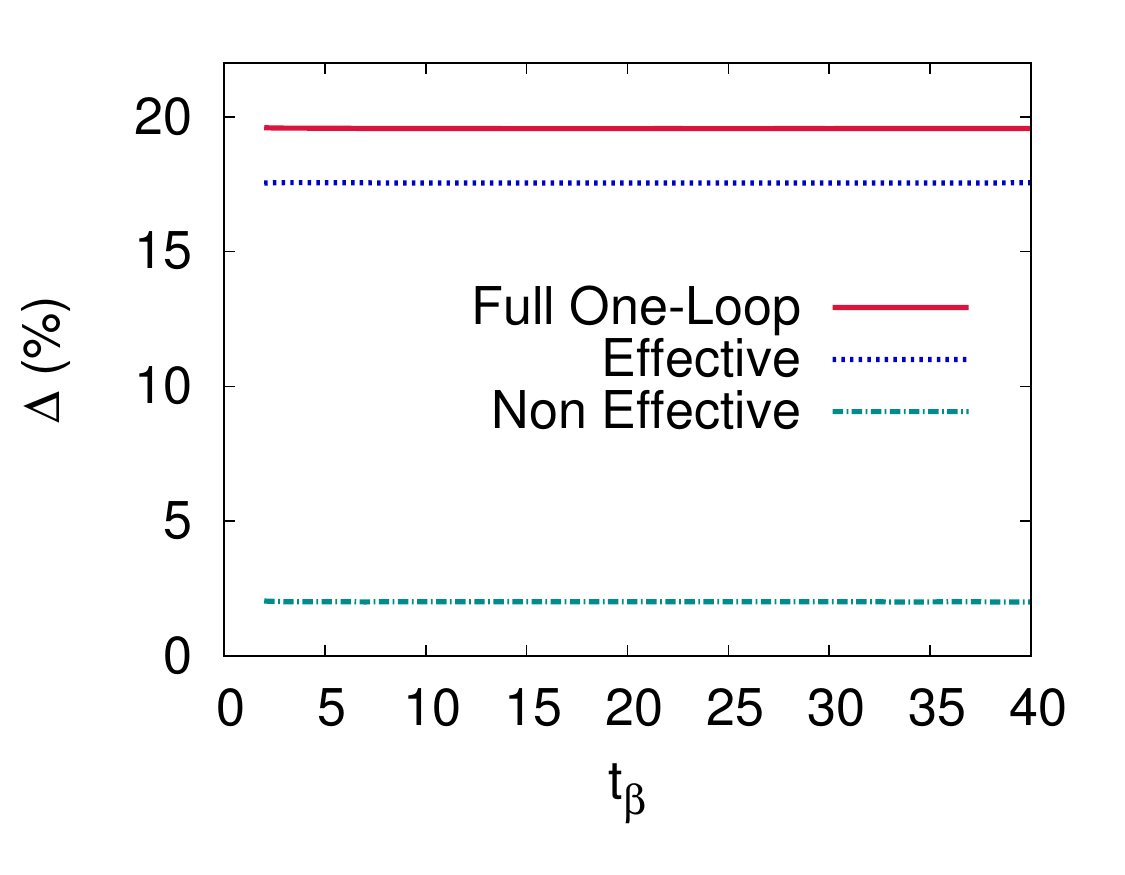}
\end{tabular}
\caption{{\em Corrections to the tree-level cross-section for the
process $\chii\chii\rightarrow \mu^+ \mu^-$ in the bino case as a
function of $M_1$ (left panel) and $\tb$ (right panel). We show the
full one-loop, the effective correction and the difference which
we term non effective. $M_2=500, \mu=-600\ GeV$.}} \label{fig:bino2}
\end{center}
\end{figure}

For our first try  the effective universal coupling does
remarkably well falling short of the full calculation by only $2\%$. Note that
although the most naive
implementation through a running of the electromagnetic coupling
fares also quite well it is nonetheless  $5\%$ off the total
correction, therefore the effective correction through the
effective couplings performs better. It must be admitted though
that
the bulk of the correction is through the running of $\alpha$. \\
To see how general this conclusion is we scanned over the set
$(M_1,M_2,\mu)$ while maintaining $\neuto$ with a 99\% bino like
component. This is simply obtained by taking $M_2=500,
\mu=-600\ GeV$ and scanning up to $M_1=350\ GeV$. We also checked how
sensitive our conclusion is depending on $\tb$ by varying $\tb$
from 2 to 40. The suspersymmtery breaking sfermion masses were
first kept at their default values. As Fig.~\ref{fig:bino2} shows,
our conclusions remain quantitatively unchanged. There is no
appreciable dependence in $\tb$, we arrive at the same numbers as
our default $\tb$ value. As for the dependence in $M_1$ it is very
slight, for $M_1\sim 50\ GeV$ there is perfect matching with our
effective coupling implementation, then as $M_1$ increases to
$350\ GeV$, the non universal corrections remain negligible, below
$2\%$.

\begin{figure}[htbp]
\begin{center}
\includegraphics[width=0.8\textwidth]{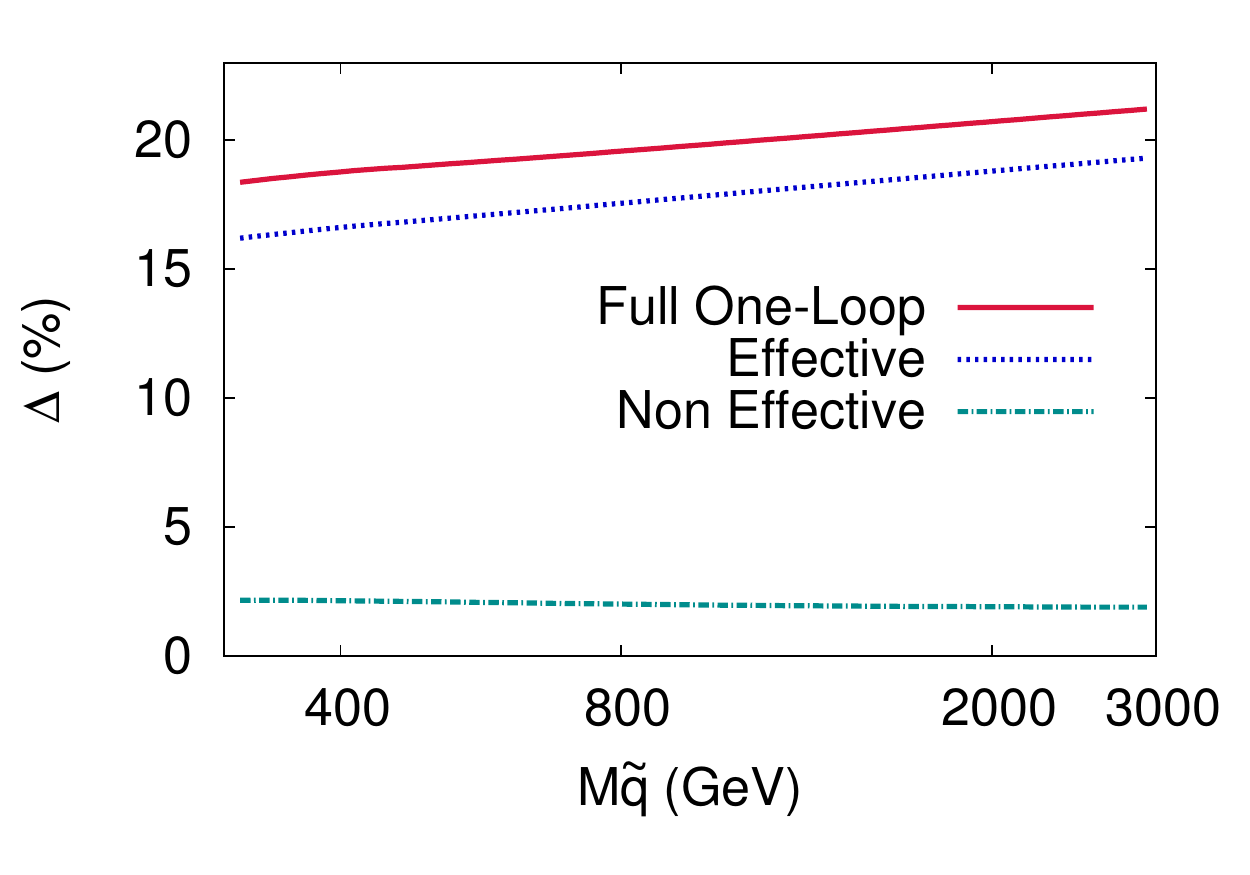}
\caption{{\em Corrections to the tree-level cross-section for the
process $\chii\chii\rightarrow\mu^+ \mu^-$ in the bino case
($M_1=90,M_2=200,\mu=-600\ GeV$) as function of the common soft
supersymmetry breaking squark mass.}} \label{fig:bino_nondecoup}
\end{center}
\end{figure}

The annihilation of neutralinos and hence the relic density  is a
very good example of the non decoupling effects of very heavy
sparticles, a remnant of supersymmetry breaking. The variation in
the fermion/sfermion masses is all contained in the effective
couplings that we have introduced. Leaving  the dependence on the
smuon mass at tree-level, and the very small (see below)
contribution of the smuon to the 1-PI vertex $\neuto \mu
\tilde{\mu}$, the bulk of the smuon mass dependence is within the
effective coupling. Fig.~\ref{fig:bino_nondecoup} shows how the
correction increases as the mass of the squarks increases from
$400\ GeV$ to $3\ TeV$, we take here a common mass for the
supersymmetry breaking squark masses (both right and left in all
three generations). The non universal correction of about $2\%$ is
insensitive to this change in squark masses whereas both $\DU$ and
$\DNU$ show the same logarithm growth that brings a $3\%$ change
as the squark mass is varied in the range $400\ GeV$ to $3\ TeV$.
This result also confirms that genuine vertex corrections and box
corrections are very small.\\
We have also extracted the individual
contribution of each species of fermions to the total
non-decoupling effect of sfermions. To achieve this we numerically
extracted the logarithm dependence of the non decoupling effect
for each species of sfermions. We have parameterised the effective
correction as
\beqn
\label{fit_delta_nd_eff} \Delta^f=a_{\tilde{f}} \ln
m_{\tilde{f}}/Q - a_f \ln m_f/Q + b_f \quad {\rm with} \quad
Q=2\mneuto \quad \tilde{f}=\tilde{d}_R+\tilde{u}_R+\tilde{Q}_L
\eeqn
The coefficients of the fit are given in
Table~\ref{table_af_bino}. As expected the fit to $a_f$ is
extremely well reproduced by the running of $\alpha$, {\it i.e},
$a_f=N_c q_f^2 \frac{4 \alpha}{3\pi}$. We also find
$a_{\tilde{e}}=a_{\tilde{\tau}}=a_{\tilde{\mu}},
b_e=b_\mu=b_\tau$. The fit to $a_f$ is made to validate the fit
procedure.
\begin{table}[htbp]
\begin{center}
$$
\begin{array}{|c|c|c|c|c|c|}
\cline{2-6}
\multicolumn{1}{c|}{} & a_{\tilde{Q}_L} & a_{\tilde{u}_R} & a_{\tilde{d}_R} &
a_f & b_f\\
 \hline
e & 0.0010 & - & 0.00231 & 0.00310 & 0.15\% \\
(u,d)  & 0.000575 & 0.00236 & 0.000698 & (0.00413,0.00103) & 0.15\% \\
(t,b)  & -0.00406 & 0.00838 & 0.000661 & (0.00413,0.00103) & 0.16\%\\
\hline
\end{array}
$$
\end{center}
\caption{{\em Coefficients of the $\ln (m_f)$ (running couplings)
$\ln (m_{\tilde{f}})$(non decoupling effects) in $\DU$. $(c,s)$
give very similar results to $(u,d)$.}\label{table_af_bino}}
\end{table}
The most important observation is that the stops behave
differently, this is due to the Yukawa coupling of the top and
mixing. If there were not a compensation between left and right
contribution of the stops (compare to $\tilde{u}$) the
contribution of the stops would be even more important and would
dominate. Considering the different contributions and the scales
that enter our calculations it is difficult to attempt at giving an
analytical result, but leaving the stop aside the different
contributions to $a_{\tilde{f}}$ can be roughly approximated by
$y_f^2 N_c N_d/8/c_W^2$, $N_d=2$ for doublets and $1$ for singlet
of $SU(2)$. $y_f$ is the hypercharge, corresponding to the
couplings of the sfermions to the bino component.

\paragraph{Scheme dependence in the bino case}~\\
We have compared
the full correction to an approximate effective implementation and
observed that the approximation is quite good. However, even the
full correction, being computed at one-loop, is potentially dependent on the
renormalisation scheme chosen. As
discussed earlier we analyse the $\tb$ scheme dependence and the
$M_1$ scheme dependence. For $\tb$ we obtain the following
corrections:
$$ 19.58\% (DCPR), \quad 19.79\% (\oo{DR}), \quad 19.51\% (MH).$$ This confirms
that
the $\tb$ scheme dependence is very negligible. For the bino case
it is natural to reconstruct $M_1$ from the LSP, nonetheless
analysing the $M_1$ scheme dependence one chooses another
neutralino, say $\neutt$ which in our example is a wino-like. This
introduces more uncertainty or error since with this scheme the
corrections attain $24.08\%$, more than $4\%$ compared to the
usual scheme.
\newpage
\paragraph{Higgsino case}~\\
\subparagraph{Effective versus full corrections}~\\
We now switch to the higgsino point (600,500,-100) which gives a LSP with
$M_{\chii}=95\ GeV$ with a  99\% higgsino content. The sfermion
parameters are the default values. In the higgsino case the cross
section is dominated by the exchange of the $Z$ in the
$s$-channel, so the bulk  of the corrections through the effective
couplings will be through the effective $\neuto \neuto Z$. For
further reference note that the tree-level cross section for
annihilation into muons is $\sigma_{\mu^+ \mu^-}^{\tilde{h}}=2.58
\times 10^{-3}{\rm pb}$, tiny and totally insignificant especially
compared to annihilation into $W$, $\sigma_{W^+
W^-}^{\tilde{h}}=18.83\ {\rm pb}$. This is an observation we will
keep in mind. The one-loop corrections we find for $\sigma_{\mu^+
\mu^-}^{\tilde{h}}$ are
\beqn
( {\rm for} \; \mu=-100{\rm GeV}) \quad 
\DU=13.55\%(\DA=14.62\%) \quad \DNU=-21.09\%(\Delta_{full}=-7.54\%)
\eeqn

This result is in a quite striking contrast to the bino case. The
effective coupling does not reproduce at all the full correction
and is off by as much as $21\%$. It looks like, at least for this
particular choice of parameters, that  going through the trouble
of implementing the effective $\neuto \neuto Z$ was in vain since
this correction is, within a per-cent, reproduced by the naive
running of $\alpha$. As we will see both these conclusions depend
much on the parameters of the higgsino and even the squark masses.
For example consider $\mu=-50\ GeV$, leaving all other parameters the
same. Of course this is a purely academic exercise, since in this
case, the charginos with mass $m_{\chi^\pm_1}=55\ GeV$ are ruled out
by LEP data. Nonetheless, in this case
\beqn
( {\rm for} \; \mu=-50{\rm GeV}) \quad  \DU=10.7\%(\DA=12\%)
\quad \DNU=-6.9\%(\Delta_{full}=3.8\%).
\eeqn
Had we included all particles in the effective vertex, we would get a correction
$\DU^W=4.4\%$ improving thus the agreement with the one-loop correction for this
particular
value of $\mu$ up to $0.6\%$. At the same
time a correction in terms of
a running of $\alpha$ will be off by more than 8\%. \\

\begin{figure}[htbp]
\begin{center}
\begin{tabular}{cc}
&
\includegraphics[width=0.8\textwidth]{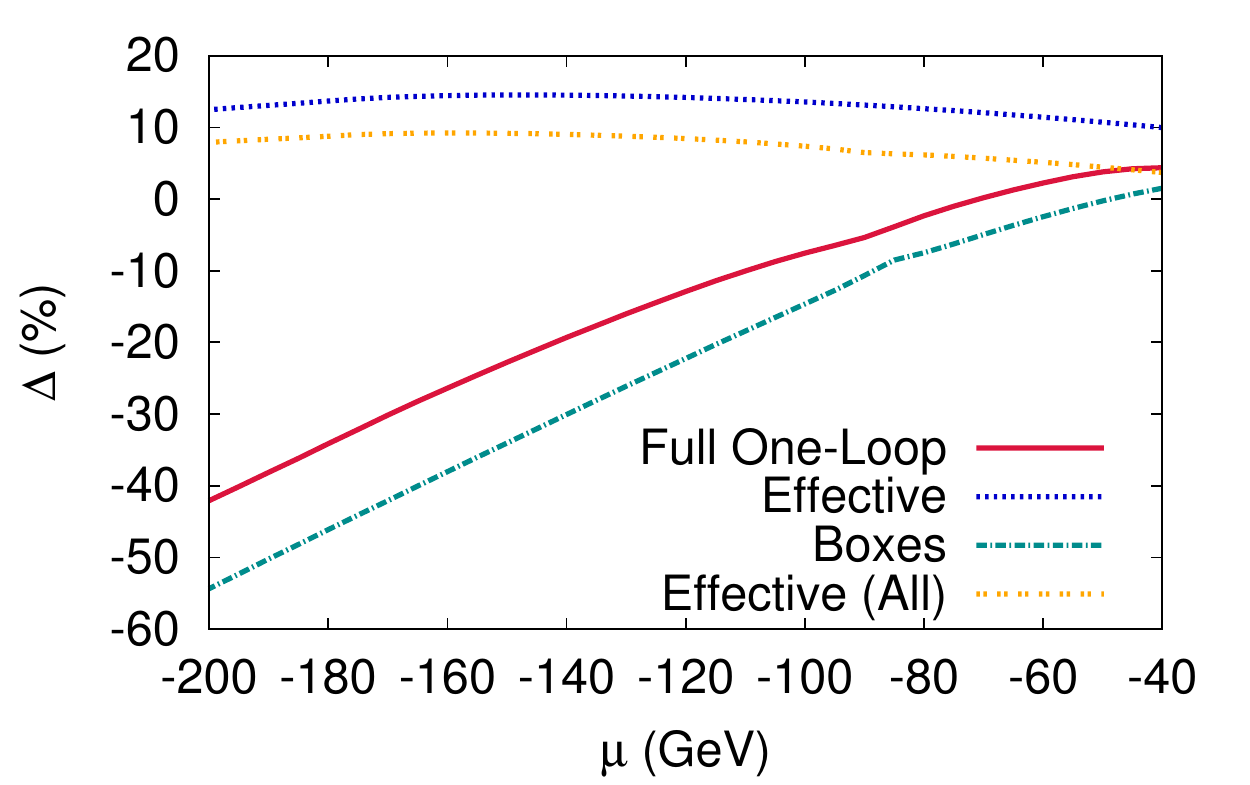}
\end{tabular}
\caption{{\em Corrections to the tree-level cross-section for the
process $\chii\chii\rightarrow \mu^+ \mu^-$ in the higgsino case
as a function of $\mu$. Shown are the effective vertex correction
(Effective, with only fermions/sfermions in the loops), the
effective $\neuto \neuto Z$ coupling  including all particles
denoted  Effective (All), the non QED boxes (Boxes) and the full
correction. $M_2=500, \mu=-600\ GeV$.}} \label{fig:hino2}
\end{center}
\end{figure}
\begin{figure}[htbp]
\begin{center}
\begin{tabular}{cc}
&
\includegraphics[width=0.7\textwidth]{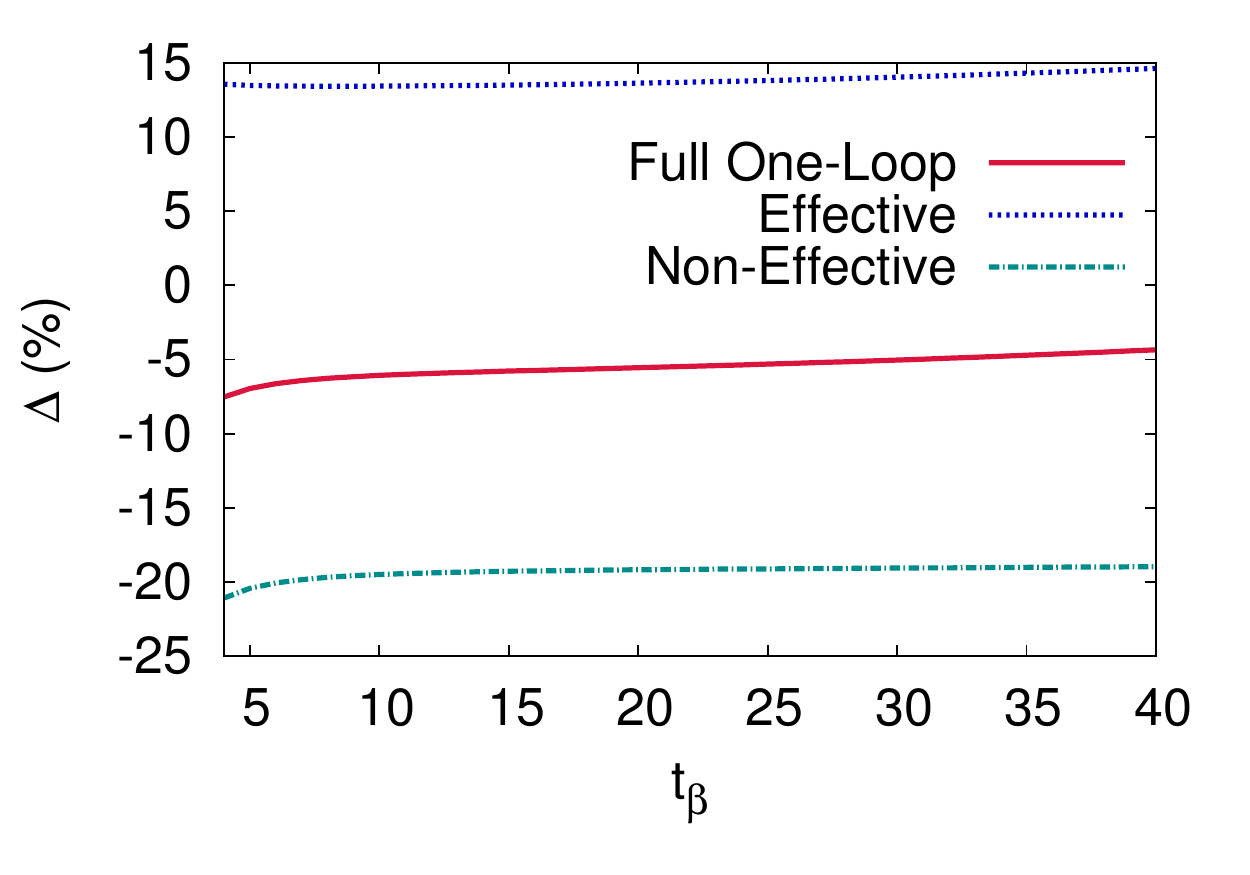}
\end{tabular}
\caption{{\em Corrections to the tree-level cross-section for the
process $\chii\chii\rightarrow \mu^+ \mu^-$ in the higgsino case
as a function of $\tb$. We show the full one-loop, the effective
correction and the remainder (Non-effective). $\mu=-100, M_1=500,
M_2=600\ GeV$. }} \label{fig:hinotb}
\end{center}
\end{figure}

These two examples show that one can not, in the higgsino case,
draw a general conclusion on the efficiency of the effective
coupling as what was done in the bino case. Let us therefore look
at how the corrections change with $\mu$, and therefore with the
mass of the LSP,  while maintaining its higgsino nature. We have
varied $\mu$ from $-200\ GeV$ to $-40\ GeV$. Fig.~\ref{fig:hino2}
shows that the full correction is extremely sensitive to the value
of $\mu$. For $\mu=-200\ GeV$ the full one-loop correction is as much
as $-42\%$, casting doubt on the loop expansion. The effective
coupling corrections with only fermions/sfermions on the other hand is much
smoother and
positive bringing about $10\%$ correction. Including all particles
in the effective $\neuto \neuto Z$ vertex brings in an almost
constant  reduction of about $6\%$. Therefore as the value of
$|\mu|$ increases the effective one-loop corrections in the case
of the higgsino case can not  be trusted. The same figure shows
that the behaviour and the increase in the corrections is due
essentially to the contribution of the boxes. Here the boxes mean
the non QED box (involving an exchange of a photon which are
infrared divergent before including the real photon emission\footnote{The
contribution of the QED box + real photon emission is only 0.1\%}). The
large contribution of the boxes can be understood by looking at
the box diagram. Indeed, as argued
previously, cutting through the box reveals that it represents
$\neuto \neuto \to W^+ W^-$ production that rescatter into $\mu^+ \mu^-$. Both
these
processes have very large cross sections compared to the
tree-level $\neuto \neuto \to \mu^+ \mu^-$. Our conclusion is
therefore that the effective vertex approximation is inadequate as
soon as the channel $\neuto \neuto \to W^+ W^-$ opens up. When
this occurs, in practical calculations of the relic density, the
channel $\neuto \neuto \to \mu^+ \mu^-$ is irrelevant and must
rather analyse the loop corrections to $\neuto \neuto \to W^+
W^-$. This process was studied in\cite{boudjema_chalons1,baro07} and will
be investigated further through an effective approximation in a
forthcoming study. \\
On the other hand, the dependence of the relative correction on
$\tb$ is quite modest even though there is certainly more
dependence than in the bino case, especially at lower values of
$\tgb$. This is shown in Fig.~\ref{fig:hinotb}.

\begin{figure}[htbp] \begin{center}
\hspace*{-1.2cm}
\mbox{\includegraphics[width=0.55\textwidth]{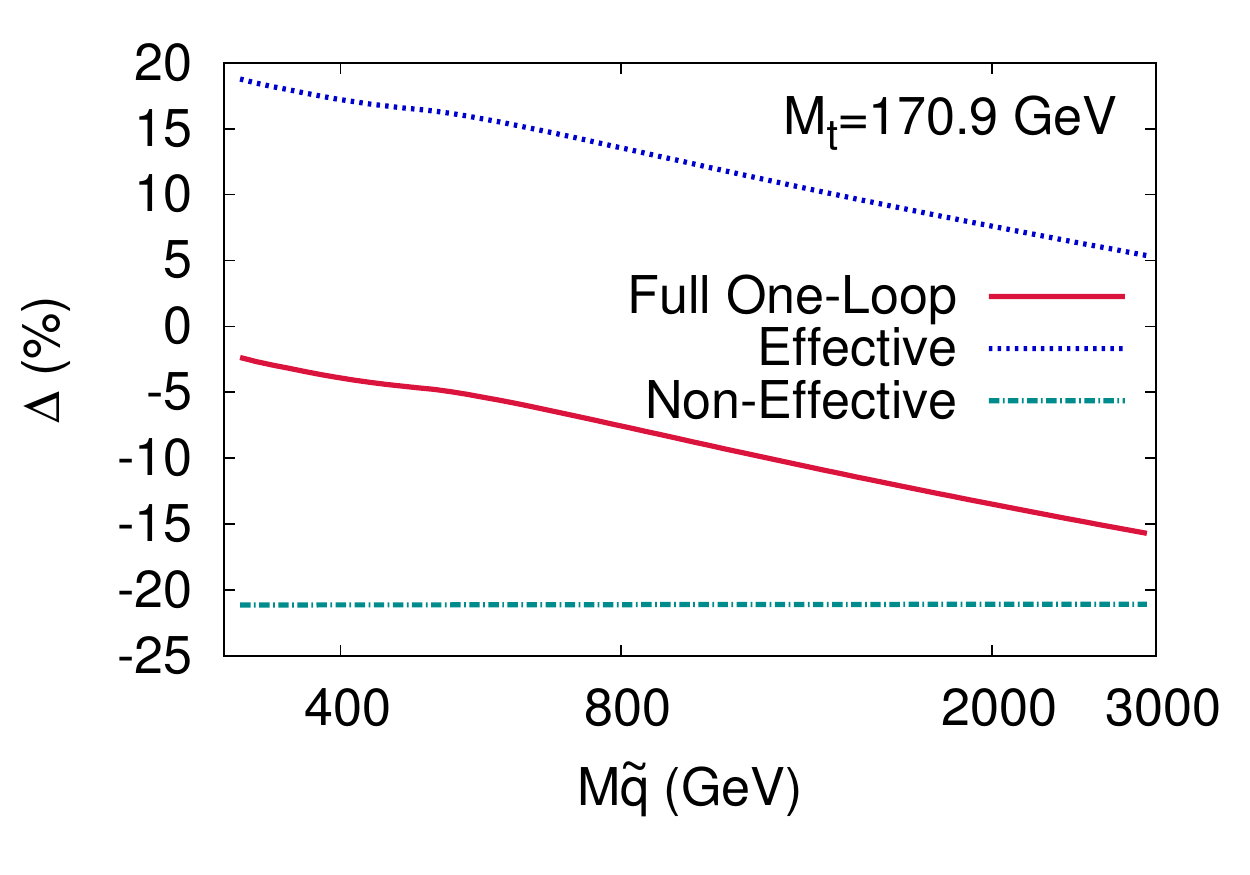}
\includegraphics[width=0.55\textwidth]{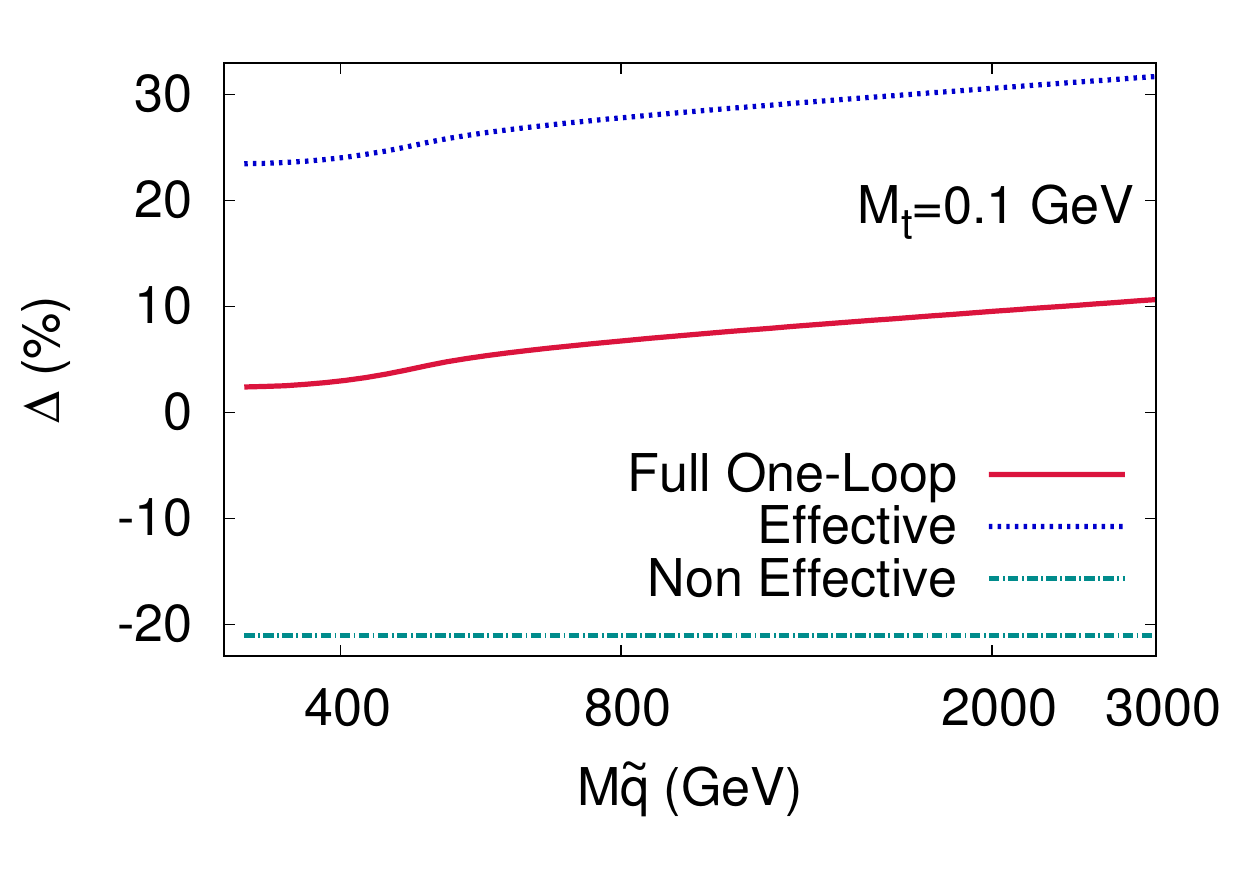}}
\caption{Corrections to the tree-level cross-section for the
process $\chii\chii \to \mu^+ \mu^-$ in the higgsino case
($M_1=600,M_2=500,\mu=-100$) as function of the common squark
mass. The right panel illustrates the case $m_t=0.1\ GeV$.}
\label{fig:higmsq}
\end{center}
\end{figure}

We now investigate the non-decoupling of very heavy squarks (and
heavy sfermions in general). Since we are in a Higgsino scenario
we expect the Yukawa of the fermions to play a more prominent role
than what was observed in the bino case. This is well supported by
our study. Fig.~\ref{fig:higmsq} shows how the effective (with
only fermions and sfermions) and the full correction gets modified
when the common mass of all squarks (all generations, left and
right) increases from $400\ GeV$ to $3\ TeV$. To
better illustrate the important effect of the Yukawa of the
top/stop sector we plot the corrections also for $m_t=0.1\ GeV$. For
$m_t=170.9\ GeV$, the correction drops by about $13\%$ when the mass
of the squarks increase from $400\ GeV$ to $3\ TeV$. This is much
more dramatic than in the bino case where we observed a 3\%
increase in the same range. Observe that for our default squark
mass of $800\ GeV$, the  effective correction including
sfermions/fermions is such that it almost accidently coincides
with the running of $\alpha$. If one switches off the top quark
mass, instead of a 13\% decrease we observe an 8\% increase for
$m_t=0.1\ GeV$! Observe that the difference one sees for
$m_{\tilde{Q}}=400\ GeV$ between $m_t=170.9\ GeV$ and $m_t=0.1\ GeV$ is
due essentially to the running  of $\alpha$ with very light top
that accounts for  $3\%$.

The special role played by the top can be seen even more clearly
from each individual contribution of the
fermion/sfermions and the fit of the contribution according to
Eq.~\ref{fit_delta_nd_eff} as was done for the bino case.
\begin{table}[htbp]
\begin{center}
$$
\begin{array}{|c|c|c|c|c|c|}
\cline{2-6}
\multicolumn{1}{c|}{}& a_{\tilde{Q}_L} & a_{\tilde{u}_R} & a_{\tilde{d}_R} & a_f
& b_f\\
 \hline
e  & 0.00304 & - & 0.000366 & 0.00309 & -0.12\% \\
(u,d)  & 0.00861 & 0.000489 & 0.000122 & (0.00414,0.00101) & -0.13\% \\
(t,b) & -0.0701 & 0.000826 & 0.000108 & (0.00414,0.00101) & 0.13\%\\
\hline
\end{array}
$$
\end{center}
\caption{{\em Coefficients of the $\ln (m_f)$ (running couplings)
$\ln (m_{\tilde{f}})$(non decoupling effects) in $\DU$. $(c,s)$
give very similar results to $(u,d)$. Higgsino
case.}\label{table_af_higgsino}}
\end{table}
The contribution of the stop is clearly (especially through
$\tilde{Q}_L$) an order of magnitude larger than for all other
sfermions, see Table~\ref{table_af_higgsino}. It is the only one
that brings a negative contribution. Since this effect is in the
universal $\neuto \neuto Z$ it will show up in many processes
where the higgsino contributes.

\paragraph{Scheme dependence in the higgsino case}~\\
We analyse here the $\tb$ scheme dependence and the $M_1$ scheme
dependence. For $\tb$ we obtain the following corrections:
$$ -7.5\% (DCPR), \quad -12.4\% (\oo{DR}), \quad -4.76\% (MH).$$
As expected and in line with the behaviour of the corrections with
respect to $\tgb$, Fig.~\ref{fig:hinotb}, we see that the
corrections though larger than in the bino case are nonetheless
within 5\%. On the other hand, expectedly the choice of $M_1$ has
less impact than in the bino case where the reconstruction of
$M_1$ is essential to define the LSP. In the case of the higgsino,
changing the $M_1$ scheme turns the full correction from -7.5\% (in DCPR
scheme for $\tb$) to -10.7\%, a 3\% uncertainty.

\subsection{Conclusion}
Very few analyses have been done taking into account the full
one-loop corrections to the annihilation cross sections entering
the computation of the relic density despite the fact that this
observable is now measured within 3\% precision. In supersymmetry
radiative corrections have been known to be important, yet
practically all analyses that constrain the parameter space of
supersymmetry are performed with tree-level annihilation cross
sections. Taking into account the full one-loop corrections to a
plethora of processes is most probably unrealistic. On the other
hand one must incorporate, if possible simply and quickly, a
parameterisation of the theory error or implement the corrections
through effective couplings of the neutralino, in the case of
supersymmetry. This is what we have attempted in this study for
two of the most important couplings of the neutralinos $\neuto f
\tilde{f}$ and $\neuto \neuto Z$. In order to look more precisely
at the impact of each of these effective couplings we take as a
testing ground a most simple process, $\neuto \neuto \to \mu^+
\mu^-$ and select a neutralino that is either almost pure bino or
pure higgsino. We do not strive at finding a scenario with the
correct relic density since our primary task is to study this
vertices and the approximations in detail. In this exploratory
study taking a final state involving gauge bosons would only
confuse the issues. Nonetheless, the impact of the gauge bosons is
studied. Indeed, we have shown how the construction of the
effective $\neuto \neuto Z$ is quite different from that of the
$\neuto f \tilde{f}$. For the latter the effective coupling
involves self-energy corrections, whereas for the former the
one-particle irreducible vertex correction must be added. These
examples and the construction of the effective coupling already
pave the way to a generalisation to the effective couplings
$\neuto \neuto h,H,A$ and $\neuto \chi^+ W$ which we will address
in forthcoming publications with applications to different
process, including gauge boson final states. Even with the
effective couplings we have derived, we could generalise the study
of $\neuto \neuto \to \mu^+ \mu^-$ to cover not only pure winos,
but also mixed scenarios and also heavy fermions. \\
Our preliminary study on the simple process $\neuto \neuto \to
\mu^+ \mu^-$ is already very instructive. To summarise the bino
case, we can state that the effective couplings approach is a very
good approximation that embodies extremely well the non decoupling
effects from heavy sfermions, irrespective of many of the
parameters that are involved in the calculation, as long as one is
in an almost pure bino case. The effective coupling implementation is
within $2\%$ of the full one-loop calculation. Here, this reflects
essentially the correction to the $\neuto f \tilde{f}$ coupling.
The scheme dependence from $\tb$ is very small, this result
stands for large $M_1$ masses as long as the neutralino is more
than $90\%$ bino like. In particular for higgsino-like
LSP in excess of $90$ GeV as imposed by present limits on the
chargino, the effective coupling implementation in the
annihilation $\neuto \neuto \to \mu^+ \mu^-$ fails. It worsens as
the mass increases due to the importance of a large box
contribution corresponding to the opening up of $\neuto \neuto \to
W^+ W^-$ which would in any case be  the dominant process to take
into account when calculating the relic density. The large Yukawa
of the top has a big impact on the radiative corrections and in
particular on the non-decoupling contribution of a very heavy
stop. Although this is an example which shows, in principle, the
failure of the effective approach apart from correctly reproducing
 the non-decoupling effect of very heavy squarks, we need
 further investigation on the dominant processes, in this case
 annihilations into $W,Z$, to see if these dominant processes
 could on the other hand be reproduced by an effective coupling
 approach. If the effective approach turns out to be efficient
 for the dominant processes, where and if the box corrections are tamed, the
effective coupling could still be a good alternative for the
 calculation of the relic density with high precision. We leave
 many of these interesting issues to further analyses.


%% file: chapter10.tex
\chapter*{Conclusion}
We have developed in this thesis a method to shed some light on the features of supersymmetry in view of Higgs physics and observables pertaining to dark matter : the effective approach. We have thus investigated the BMSSM framework, an extension of the MSSM that encompasses many different extensions of the MSSM. It turns out that allowing for extra-physics that affect the Higgs sector of the MSSM produces a much richer Higgs phenomenology compared to the MSSM. This flexibility means however that at the level of the Lagrangian many new terms are introduced and we have then shown how to deal with those effective corrections with the usual tools for phenomenology, namely by the modification of \lanHEP\ and \HD. As a first consequence of the effective operators we have seen that the lightest Higgs mass could be significantly raised at tree-level, alleviating thus the fine-tuning issue of the MSSM. Comparing our model to the LHC results obtained in mid 2011, it appears that the cases where the mass is too much 
enhanced (up to 250 GeV) are now ruled out : we are left with a light Higgs that has to be less than 150 GeV. But the effective operators can also alter significantly the signal and the associated rates of this light Higgs boson and we have exemplified the point with the case of a 125 GeV Higgs. Some channels can be enhanced as compared to the Standard Model expectation, in particular this is the case of the diphoton channel. This feature is all the more interesting since the hint of a signal that was recorded by ATLAS and CMS actually shows some discrepancies, though not significant with the 2011 dataset only, but that may eventually lead to a non-standard like Higgs boson. We have shown that not only the increase in the diphoton channel, which is seen both by ATLAS and CMS, could be explained by a reduction of the coupling to $b$ quarks but also that the hierarchy between the $ZZ$, $\gamma\gamma$ and $\gamma\gamma+2j$ channels could be reproduced by the mean of the stop loop. We have then worked out the 
consequences for other experimental data, namely the electroweak precision tests, the flavour physics and dark matter observables. The flavour physics, in particular the computation of the $\Bsmu$ observable has highlighted some new structures appearing in the BMSSM Lagrangian that can be reduced by the use of the equations of motion. It turns out that in most of the cases there is a very tight interplay between Higgs physics and flavour observables since the experimental constrain on $\Bsmu$ has disfavoured our low $\ma$ region while the $\Bsg$ one has restricted the effect of the light stops loop contribution to the Higgs processes to be in the low $\tb$ region.\\

An important development that we have carried out in the aim of exploiting the LHC performance in the BMSSM framework is the recasting of the Standard Model analyses to BSM theories. Precisely, we have evaluated the accuracy of some approximations as for instance the quadrature sum of different signals to combine the statistical significances and the use of the inclusive predicted cross-sections instead of the exclusive ones. We have also seen that limits on the cross-sections that are obtained by a combination of different subchannels are generically model-dependent : this is the case of the diphoton channel for instance, where the combination in the case of the Standard Model Higgs hypothesis and the fermiophobic Higgs hypothesis are quite different. We have seen what could be done to improve the current status of this reinterpretation, namely the communication of efficiencies per production mode for each search channel and have been in close contact with experimentalists along our work to advocate for the 
availability of experimental results in a way better suited to new physics interpretation.\\

We have then turned to another set of constraints on supersymmetric theories that consists in the dark matter observables. Our work has focused on the precise computation of the relic density in the MSSM. We have decided to introduce once again an effective approach, but whereas the one implemented in the BMSSM aimed at accounting for extra physics beyond the MSSM, this specific one was built to account for radiative corrections brought by MSSM particles. We have performed the implementation of different effective vertices ($\chii\tilde{f}f$, $\chii\chii Z$ and $Z\bar ff$) and assessed the robustness of the approach in the case of annihilation of neutralinos to fermions. We have found that the full one-loop result was very well approximated in the case of a bino-like neutralino, where the discrepancy between both calculations was found to be less than 2\%. We have also discussed the case of the Higgsino-like neutralino with an improved effective vertex $\chii\chii Z$, and have concluded that the bulk of the 
corrections was taken by box corrections, and as such escaped from an universal effective coupling parametrisation.\\

Those studies naturally lead to a well-defined continuation : first in the case of the Higgs searches, it is of a crucial importance to maintain our discussion and collaborations with the experimental community to achieve the goal of a better communication of the results. This year (2012) is particularly important in that respect since it will eventually lead to the discovery or exclusion of the Standard Model Higgs boson. Either in the no-signal case or if the couplings of such a Higgs boson appear to be non standard, we will have to recast those results in different BSM parametrisations. Since on the direct searches, very few improvements are expected before the LHC upgrade, the Higgs sector may well be the place to test supersymmetry, making it worthwhile to pursue our investigation of the BMSSM framework. It is then interesting to relax our assumption that higher-order operators may include only Higgs superfields : what other operators could stem from other sectors, for instance the flavour sector in 
general or the stop sector in particular? What would this change for direct stop searches? Enlarging thus our set of effective operators would bring us a more general description of the UV completion which, when compared to experimental data, would turn in a more precise constraint of supersymmetry.\\

Finally, our effective approach to the precision computation of the relic density has led to many questions and improvements that have still to be performed : we need to extend our study to other processes (in particular Higgs resonances), other initial and final states (as the co-annihilation mechanism with in particular $W,Z$ and Higgs production). The aim being to include the implementation as a package to \momegas, so that any theory built on the MSSM could account for the MSSM radiative corrections in an efficient way.\\

Although we have a supersymmetric framework all along our work, many of the different techniques that we have exploited, applied or developed either at the theoretical level or at the level of experimental analyses can be extended outside the realm of supersymmetry. It is therefore all the more exciting to scrutinize the data to come for any deviations from the Standard Model in the hope that, perhaps, we will be able to catch a glimpse of what New Physics really is.

\chapter*{(Français) Conclusion}
Nous avons développé dans cette thèse une méthode pour tester certaines caractéristiques de la Supersymmétrie dans la perspective des recherche du Higgs ainsi que des observables reliées à la matière sombre : l'approche effective. Nous avons ainsi étudié le modèle BMSSM, une extension générique du MSSM qui recouvre de nombreuses extensions spécifiques. Il s'avère que l'ajout de nouvelle physique au secteur du Higgs du MSSM amène à une phénoménologie du Higgs bien plus riche. Cette flexibilité vient cependant avec de nouvelles complications au niveau du Lagrangien puisque de nombreux nouveaux termes sont introduit, nous avons ainsi montré comment intégrer ces corrections effectives aux outils standards de la phénoménologie, en l'occurrence par la modification de codes comme \lanHEP\ ou \HD. Un première conséquence de ces opérateurs est d'augmenter la masse du Higgs léger à l'arbre, permettant ainsi de réduire le problème de fine-tuning du MSSM. En  comparant notre modèle aux résultats obtenus au LHC à l'été 
2011, il apparait que la possibilité d'augmenter considérablement cette masse (jusqu'à 250 GeV) est désormais exclue : nous sommes donc restreint à un Higgs léger de masse inférieure à 150 GeV. Mais les opérateurs effectifs peuvent aussi altérer significativement un éventuel signal et les taux de production associés et nous avons pris pour exemple le cas d'un signal de Higgs à 125 GeV. Certains canaux peuvent ainsi être augmentés vis à vis de la prédiction du Modèle Standard, c'est en particulier le cas du canal en diphoton. Cette caractéristique est d'autant plus intéressante que les possibles signaux enregistrés par les collaborations ATLAS et CMS ont effectivement de telles déviations, qui bien que très imprécises pour le moment pourraient dans un futur proche (les données 2012) révéler un boson de Higgs non-standard. Nous avons montré que non seulement l'augmentation du canal en diphoton, observé également par ATLAS et CMS, pouvait être expliquée par une réduction du couplage au quark 
b, mais aussi que la hiérarchie entre les signaux $ZZ$, $\gamma\gamma$ et $\gamma\gamma+2j$ pouvait être reproduite par l'effet de la boucle de stop léger. Nous avons ensuite déterminé les conséquences d'un tel Higgs dans d'autres expériences, à savoir les tests de précision électrofaible, la physique de la saveur et la matière sombre. La physique de la saveur et en particulier le calcul de la désintégration $\Bsmu$ a mis en avant de nouvelles structures de Lorentz apparaissant dans le Lagrangien du BMSSM, qui peuvent être simplifiées par l'utilisation d'équations du mouvement. Il s'avère que dans la plupart des cas la contrainte expérimentale sur $\Bsmu$ met en danger les régions de paramètres avec un faible $\ma$ alors que la contrainte venant de la désintégration $\Bsg$ a permis de restreindre l'effet du stop léger dans les processus du Higgs à une région de faible $\tb$.

Un important développement que nous avons mené à terme dans le but d'exploiter les données du LHC dans le cadre du BMSSM est la ré-interprétation des analyses du Modèle Standard dans des théories BSM. Concrètement nous avons évalué la précision de certaines approximations comme par exemple la somme en quadrature des différents signaux pour combiner les significations statistiques, ou encore l'utilisation de sections efficaces inclusives à la place des sections efficaces exclusives. Nous avons aussi observé que les limites obtenues par une combinaison de différents canaux sont toujours dépendantes du modèle : c'est ainsi le cas de l'analyse en diphoton où la combinaison dans le cadre du modèle Standard donne un résultat très différent de la combinaison dans le cadre d'un modèle fermiophobique. Nous avons vu ce qui pouvait être fait pour améliorer l'état actuel de ces ré-interprétations, en l'occurrence pas la communication des efficacités par mode de production du Higgs pour chaque 
canal et sommes restés en contact proche avec des expérimentateurs tout au long de nos recherches pour défendre l'idée de la mise en commun des résultats expérimentaux dans un format plus adapté aux interprétations en modèles de nouvelle physique.

Nous nous sommes ensuite tournés vers un autre ensemble de contraintes de nouvelle physique que sont les observables de matière sombre. Notre travail s'est focalisé sur le calcul de précision de la densité relique dans le cadre du MSSM. Nous avons décidé d'introduire  à nouveau une approche effective, mais alors que dans le cas du Higgs les opérateurs effectifs avaient pour but de reproduire les effets d'une nouvelle physique au delà du MSSM, cette approche spécifique a pour but de reproduire les corrections radiatives issues des particules du MSSM. Nous avons effectué l'implémentation de divers vertex effectifs ($\chii\tilde{f}f,\ \chii\chii Z,\ Z\bar ff$) et estimé la performance de l'approche effective dans le cas particulier de l'annihilation de neutralinos en fermions. Nous avons trouvé que la contribution totale des diagrammes à une boucle était très bien approximée par l'approche effective dans le cas d'un neutralino de type bino, puisque la déviation entre les deux calculs est inférieure à 2\%. Nous 
avons ensuite discuté le cas d'un neutralino de type Higgsino et amélioré le vertex effectif $\chii\chii Z$, pour enfin conclure que la partie la plus importante des corrections venait des diagrammes de boîtes et échappait par là même à une paramétrisation effective universelle.

Ces études définissent naturellement une suite logique : premièrement dans le cas des recherches du Higgs, il est d'un importance cruciale de maintenir notre discussion et notre collaboration avec la communauté expérimentale pour parvenir à une meilleure communication des résultats. Cette année (2012) est particulièrement importante à cet égard puisqu'elle nous permettra de découvrir ou d'exclure le Higgs du Modèle Standard. Que ce soit dans le cas où nous ne verrions pas de signal ou si les couplages du signal venaient à être non standards, nous aurions à ré-interpréter ces résultats dans de différents modèles BSM. Puisque du côté des recherches directes de superpartenaires, les améliorations ne devraient être que marginales en attendant l'upgrade du LHC, le secteur du Higgs pourrait être le meilleur endroit pour tester la supersymmétrie, justifiant ainsi la poursuite de nos investigations du côté du BMSSM. Il est alors intéressant d'aller au delà de la restriction des nouveaux opérateurs au secteur du Higgs 
: 
quels autres opérateurs pourrait-on considérer, par exemple dans le secteur de la saveur en général ou dans le secteur du stop en particulier? Quels en seraient les conséquences pour la recherche directe de stops? En augmentant ainsi notre ensemble d'opérateurs effectifs nous aurions une description plus générale de la théorie à haute énergie, qui donnerait au niveau des expériences des contraintes plus précises sur la supersymmétrie.

Enfin notre approche effective au calcul de la densité relique a débouché sur de nombreuses questions et maintes améliorations peuvent être faites : nous avons besoin d'étendre notre étude à de nouveaux processus (en particulier les résonances de Higgs) et d'autre états initiaux et finaux (comme le mécanisme de co-annihilation avec des productions de Higgs, de W et de Z). Le but est de pouvoir inclure cette implémentation comme un paquet du code \momegas, de façon à permettre à chaque théorie bâtie sur le MSSM de pouvoir calculer les corrections radiatives de MSSM de manière efficace.

%% file: appendix.tex
\chapter{Appendix}

\section{Perturbative Linear Algebra}
\label{eff_diag}
The relations between initial fields and parameters and physical ones, which are the key for phenomenology, are obtained through relations of linear algebra, and since phenomenology is built on perturbative expansions, it is natural to look for results combining both approaches. The three operations that we will need and that are not straightforward are the following
\begin{itemize}
 \item diagonalisation of a hermitian matrix
 \item Singular valued decomposition of a matrix
 \item Takagi diagonalisation of a symmetric matrix
\end{itemize}
All matrices being complex and the perturbative parameter is not specified, it can be either the $1/M$ of an effective expansion or a $\epsilon$ of a loop expansion.

\subsection{Diagonalisation}
We will assume that the matrix has the form
$$M^2+\dA$$
where $M^2$ is a real diagonal matrix at zeroth order and $\dA$ an hermitian matrix at order equal or higher than 1. We can always turn any hermitian matrix to this form by doing a zeroth order diagonalisation. The aim is now to find an anti hermitian matrix $\dP$ and a real diagonal matrix $\dM^2$ so that
\begin{equation}
(1+\dP)^\dag(M^2+\dA)(1+\dP)=M^2+\dM^2
\label{eq:pert_diag_1}
\end{equation}
The requirement of the antihermicity of $\dP$ is equivalent to the requirement that $1+\dP$ is hermitian since
$$(1+\dP)^\dag(1+\dP)=1+\dP^\dag+\dP$$
at first order. Note that when going to higher order, the expression is a bit more subtle
\begin{eqnarray}
\dP^{(1)}+\dP^{(1)\,\dag}&=&0\\
\dP^{(2)}+\dP^{(2)\,dag}+\dP^{(1)\,\dag}\dP^{(1)}&=&0
\label{eq:dP_herm}
\end{eqnarray}
Writing eq.\ref{eq:pert_diag_1} at first order and using the previous relation we obtain
$$-\dP M^2+M^2\dP+\dA=\dM^2$$
which, by evaluating off diagonal term and diagonal terms separately, yields
\begin{eqnarray}
\dP_{ij}&=&\frac{\dA_{ij}}{m_j^2-m_i^2}\\
\dM^2_{i}&=&\dA_{ii}
\end{eqnarray}
where the first equation applies to $i\neq j$ only. Note that $\dP_{ii}$ has only the requirement to be imaginary : we will by simplicity take it equal to zero.

\subsection{Singular valued decomposition}
The matrix to be decomposed is
$$M+\dA$$
where both are complex matrices, and $M$ has real non-vanishing elements only on the diagonal. We will then look for $\dU,\dV$ antihermitian matrices and $\dM$ with diagonal real values. There obey the decomposition equation
\begin{equation}
(1+\dU)^T(M+\dA)(1+\dV)=M+\dM
\end{equation}
The first order equation is then
\begin{equation}
\dU^TM+M\dV+\dA=\dM
\label{eq:svd_eff}
\end{equation}
This is solved by the system
\begin{eqnarray}
\dU_{ij}&=&\frac{m_i\dA_{ji}+m_j\dA_{ij}^*}{m_j^2-m_i^2}\\
\dV_{ij}&=&\frac{m_i\dA_{ij}+m_j\dA_{ji}^*}{m_j^2-m_i^2}\\
\dM_i&=&Re(\dA_{ii})\\
\dU_{ii}+\dV_{ii}&=&-\frac{Im(A_{ii})}{m_i}
\end{eqnarray}
where the first two equations apply on $i\neq j$ and the two last equations are derived from the diagonal part of eq.\ref{eq:svd_eff} using the fact that $\dM$ is real and $\dU,\dV$ imaginary on the diagonal.

\subsection{Takagi diagonalisation}
This diagonalisation deals with complex symmetric matrices
$$M+\dA$$
where $M$ is real positive diagonal and $\dA$ complex symmetric. We look for $\dP$ antihermitian and $\dM$ real positive diagonal so that
\begin{equation}
(1+\dP)^T(M+\dA)(1+\dP)=M+\dM
\end{equation}
which turns at first order to
\begin{equation*}
\dP^TM+M\dP+\dA=\dM
\end{equation*}
solved by
\begin{eqnarray}
Re(\dP_{ij})&=&\frac{Re(\dA_{ij})}{m_j-m_i}\\
Im(\dP_{ij})&=&-\frac{Im(\dA_{ij})}{m_j+m_i}\\
\dM_i&=&Re(\dA_{ii})
\end{eqnarray}
where the second equation also applies to $i=j$.

Each case can be enhanced to the second order by using eq.\ref{eq:dP_herm}. We note the appearance of singularities for the mixing when the zeroth order mass $m_i$ and $m_j$ get degenerated. In this case $\dP$ stop being small since the mixing will purely be driven by $\dA$.

\section{Application of the \SL\ program}

\subsection{Generation of the model}
This generation is mostly done by \lanHEP, but assisted in different ways by functions (that, for practical reasons, are run in \MA) which amongs others will compute masses and mixing.\\
With a bit of work on the \lanHEP\ language, one can slim down the required input to the gist of supersymmetric models : that is the name of all vector superfields and chiral superfields with their charges (note that the matrix form of the generators of the gauge group have to be specified for each representation by the user), the expression of K\"ahler potential, superpotential and supersymmetric breaking terms. Finally the non zero vacuum expectation values have to be specified. Such a model file looks like\\

\begin{DDbox}{\linewidth}
\begin{verbatim}
% Initial theory
% Gauge sector
parameter g1,g2,g3.
vector_superfield W:(0,3,1),B:(0,1,1),G:(0,1,8).

% Matter sector
chiral_superfield L:(1,0,-1)...

%Potentials
let K = anti(X)*expV(ycharge(X),wcharge(X),scharge(X))*X
             where X in [L,...]
let W = mu H1 * H2.
let L_sb = m_1 * tilde(H1) * tilde(H1)+...

Option InfiOrder=0.
read get_lagrangian
\end{verbatim}
\end{DDbox}

The output will be highly unphysical, since all fields are kept in the (unbroken!) gauge basis, so they have no observable meaning. However it is enough to extract the quadratic part of the lagrangian, given the procedure we have outlined, and we can now have tree-level expression for masses and mixing. At this point it is handy that the user specifies the names he or she wants to use for the physical fields and mixing matrices, which implies to define a convention on the mixing. The whole process can be synthetized neatly via personal routines. With the one I have created the things look like\\

\begin{DDbox}{\linewidth}
\begin{verbatim}
Physical$Particles={"A","Z","W+","h",...};
Physical$Mixing={{ {"B","W3"}->{"A","Z"} , R[theta_w] , {0,MZ} },
                { {"h2","h1"}->{"h","H"} , R[alpha] , {Mh,MH} },
                { {"~B","~W3","~h1","~h2"}->{"~o1","~o2","~o3","~o4"} ,
                             Zn , {MNE1,MNE2,MNE3,MNE4} },...

Calcul$Type=Analytical;
l=Physical$Particles;
{M$scalar,M$fermion,M$boson}=GetQuadraticLagrangian[l];
M$fermion=GauginoRotation[M$fermion];
M$scalar=MomentumRotation[0,M$scalar];
M$fermion=MomentumRotation[1/2,M$fermion];
M$boson=MomentumRotation[1,M$boson];
{Z,M}=GetMassesMixing[M$scalar,M$fermion,M$boson]
WriteMixing["mixing.mdl",Z];
WriteMasses["masses.mdl",M];
\end{verbatim}
\end{DDbox}

The output of those routine is a \lanHEP\ model file that specifies relations between initial fields and physical fields and which basically looks like

\begin{DDbox}{\linewidth}
\begin{verbatim}
% Masses
% If Calcul$Type=Analytical
Mh=(g1^2+g2^2)*().....

% If Calcul$Type=Numerical
Mh=MassMatr(Matr_h_cp_even,1).
\end{verbatim}
\end{DDbox}

\begin{DDbox}{\linewidth}
\begin{verbatim}
% Mixing
let h1=ca*h+sa*H.
let h2=-sa*h+ca*H.
\end{verbatim}
\end{DDbox}

\noindent The only missing part being the relations between physical and initial parameters. They are obtained via another set of routines.\\

\begin{DDbox}{\linewidth}
\begin{verbatim}
Initial$Parameters={"g1","g2","v1",...};
Physical$Parameters={"MZ","MW","EE",...};
Physical$Definition:=Block[{},
        MZ=GetMass["Z"];MW=GetMass["W"];
        EE=GetCouplage["e","e","B"]*GetMixing["B","A"]
          +GetCouplage["e","e","W3"]*GetMixing["W3","A"];
        ... ];

F = Physical$Definition;
G = Inverse[F];
WriteParameter["param.mdl",Initial$Parameters,Physical$Parameters,G];
\end{verbatim}
\end{DDbox}

So we have now created the missing part of the \lanHEP\ model files and we caqn now express the lagrangian in terms of physical quantities. To this aim we just have to run again our first \lanHEP\ input file, but we modify the particles/parameter description, to the following result\\

\begin{DDbox}{\linewidth}
\begin{verbatim}
% Physical theory
% Gauge sector
parameter MZ=91.1954,
	  MW=80,823,
	  EE=0.343 ...
read param.mdl,masses.mdl.

vector A:(gauge),Z:(mass MZ),'W+':(mass MW),...
spinor e:(mass Me),'~o1'/'~o1':(mass MNE1),...
scalar h:(mass Mh),'~e1'/'~E1':(mass MSe1),...
read mixing.mdl.
\end{verbatim}
\end{DDbox}

This will force \lanHEP\ to consider the physical fields as the particles (that is, the one appearing in the final Feynman rules), whereas initial fields ($B$,$~eL$,...) are now internal variables. At this point everything is fine to generate the Feynman rules without the loop contribution, and this is the last thing to work. As we have seen the expression for $\dZ$ terms are quite generic and ae automatically written by \lanHEP. They mostly relies on the fact that \FA/\FC will be able to compute the loop part of $\Geff$, indeed we have\\

\begin{DDbox}{\linewidth}
\begin{verbatim}
%Correspondence
Gamma_loop_boson(X,Y,k^2) = SelfEnergy[{prt["X"]}->{prt["Y"]},k^2]
Gamma_loop_fermion_S(X,Y,k^2) = SelfEnergy[{prt["X"]}->{prt["Y"]},k^2]
Gamma_loop_fermion_L(X,Y,k^2) = SelfEnergy[{prt["X"]}->{prt["Y"]},k^2]
Gamma_loop_fermion_R(X,Y,k^2) = SelfEnergy[{prt["X"]}->{prt["Y"]},k^2]
\end{verbatim}
\end{DDbox}

However the definition of $\delta P_I$ must have an input fed by the user : it depends on the renormalisation scheme. For the $P_I$ that are extracted from masses, one can again use the generic formulaes that we have already found, but for other observables, they have to be worked out by the user. So the inclusion of loop contribution on the Feynman rules is done by an additional \lanHEP\ file which looks like\\

\begin{DDbox}{\linewidth}
\begin{verbatim}
%One-loop corrections
infinitesimal dZAA=SelfEnergy[{prt["A"]}->{prt["A"]},0],
	      dZAZ=SelfEnergy[{prt["A"]}->{prt["Z"]},0].
	      
infinitesimal dEE.

transform A->(1+dZAA)*A+dZAZ*Z.
transform EE->EE+dEE
\end{verbatim}
\end{DDbox}

Generating the full set of Feynman rules is now only amtter of running the \lanHEP again, by fixing InfiOrder=1. this time. One can then choose to which output this can be connected, either to \FA/\FC or to \CH/\momegas. Let us now go the secund part.

\subsection{Computing the process cross-section}
Having generated the Feynman rules of our model in the \FA/\FC\ format, we can now go to the computation itself. The program is however far from complete due to the fact that \FA/\FC\ match exactly my definition of codes : they provide a library of function that can be used in calculating a given process in a given model. The aim of our program -- \texttt{SloopS} -- is then to combine the functions of these libraries in a convenient, still general but much more automated way. The first part is to generate the expression of the amplitude of a process and write in a \FO\ routine. This is done via a set of routines from \FA/\FC, which have been merged together by the commands

\begin{DDbox}{\linewidth}
\begin{verbatim}
%Creating process amp
Start["o1o1WW",100];
DoProcess[{"~o1","~o1"}->{"W+","W-"},1,{0,0,1,-1}];
\end{verbatim}
\end{DDbox}

The next step is to set up a link to the fortran code, so that the outputted cross-section can be used in scans and with plotting facilities. \MA is very suited as a front-end since it can plot really easily the output of scans. The scans are prepared and launched with the following commands

\begin{DDbox}{\linewidth}
\begin{verbatim}
%Performing scans
PrepareScan[100,"o1o1WW",{M1,M2,mu},{XS}]
Pin={{90,200,-600}.{100,200,-600},...};
Pout=DoScan[100,"o1o1WW",Pin];
\end{verbatim}
\end{DDbox}

\section{Statistics}
Statistics pertain most areas of high energy physics, and in particular in the search for the Higgs boson : (see \cite{cowan,cranmer} for a detailed review), where we will define the likelihood of a model versus the data as the quantity
\begin{equation}
 L(x)=\frac{(n_B+x)^{n_B+n_S}e^{-(n_B+x)}}{(n_B+n_S)!}\,L(\theta)
\end{equation}
which is simply a Poisson law between the expected number of events $n_B+x$ and the observed one $n_B+n_S$. $L(\theta)$ is called the nuisance function : it represent the auxiliary measurements that are done to determine $n_B$. It is used to construct the test statistic $t_x$ :
\begin{equation}
 t_x=-2\ln\frac{L(x)}{\hat{x}}
\end{equation}
where $\hat{x}$ is the quantity that maximise $L$. Hence $t_x$ is a positive quantity, which indicates the compatibility of the model to the data by yielding high values to less compatible models. One constructs then a quantity called the \pvalue\ :
\begin{equation}
 p_x=\int_{t_{x\ \text{observed}}}^{+\infty}f(t_x|x)dt_x
 \label{eq:pvalue}
\end{equation}
where $f(t_x|x)$ is the probability density function of the variable $t_x$ computed on an probabilistic data constructed in the hypothesis of an expected signal $x$. In other words, it is the area under the tail of the probability density function $f$, starting at $t_{x\ \text{observed}}$. One can says that $p_x$ is the probability to observe, if we were to do the experiment again and that the data would be distributed according to the model, something less compatible than what we have observed. Although in realistic cases this function is not a gaussian, it is conventional to re-express $p_x$ as $Z_x$, the tail of a mean one gaussian:
$$p_x=\int_{Z_x}^{+\infty}e^{(y-x)^2}dy$$
At this point, one use different variables if one is trying to obtain an upper bound on the signal cross-section or to quantify the deviation from the no-signal hypothesis.

\section{Precision Test}
\subsection{Electroweak Precision Variables}
Recommending to the interested reader the review \cite{altarelli_epsilon}, I will now provide a brief description of those variables. Being related to the physics of the $Z$ boson those variables will in particular impose constraints on the electroweak symmetry breaking mechanism. They are obtained from the following experimental quantities
\begin{equation}
\frac{\mw}{\mz},\Gamma_l,\Gamma_b,A_l^{FB}
\end{equation}
where $\Gamma_l$ is the leptonic width of the $Z$, that is to say the partial width averaged over $e,\mu$ and $\tau$ and $\Gamma_b$ is the partial width to $\bar bb$. $A_l^{FB}$ is the leptonic forward backward asymmetry, that is the asymmetry between events where the fermion $f$ goes in the forward direction, and those where $\bar f$ goes in the forward direction (remember that since we are the Z pole, $f$ and $\bar f$ are produced back to back). In the case of the Standard Model taken at tree-level it seems that all four measurements are entirely correlated, indeed we have seen that the mass ratio and the weak couplings only depended on the initial parameters $g_1$ and $g_2$, and since we also know the value of the electromagnetic constant $e$, there is only one free parameter in the game. The situation is a bit relaxed when we include the radiative corrections, since this will bring a running of the electromagnetic coupling (which is defined at vanishing energy whereas we are now sitting on the Z pole) 
and QCD corrections in the case of the partial width to $b$ quarks. In order to concentrate on the deviation from the tree-level expectation of the Standard Model, we trade those quantities to the $\epsilon$ variables :
\begin{eqnarray}
\epsilon_1&=&\Delta\rho\\
\epsilon_2&=&c_0^2\Delta\rho+\frac{s_0^2}{c_0^2-s_0^2}\Delta r_w-2s_0^2\Delta k\\
\epsilon_2&=&c_0^2\Delta\rho+(c_0^2-s_0^2)\Delta k
\end{eqnarray}
where the $\Delta x$ quantity are themselves extracted from
\begin{eqnarray*}
\Delta\rho&=&-4\left(g_A+\1\right)\\
\Delta_k&=&\frac{1-\frac{g_V}{g_A}}{4s_0^2}-1\\
\Delta r_W&=&1-\frac{\pi\alpha(\mz)}{\sqrt{2}G_F\mw^2}\left(1-\frac{\mw^2}{\mz^2}\right)^{-1}
\end{eqnarray*}
$g_A$ and $g_V$ are the axial and vector weak couplings, extracted from
\begin{eqnarray*}
\Gamma_l&=&\frac{G_F\mz^3}{6\pi\sqrt{2}}(g_V^2+g_A^2)\left(1+\frac{3\alpha(\mz)}{4\pi}\right)\\
A_l^{FB}&=&\frac{3g_V^2g_A^2}{g_V^2+g_A^2}
\end{eqnarray*}
and $s_0$ and $c_0$ the sinus and cosinus of an angle defined by
$$s_0^2c_0^2=\frac{\pi\alpha(\mz)}{\sqrt{2}G_F\mz^2}.$$

\begin{figure}[!h]
\begin{center}
\includegraphics[scale=0.3,trim=0 0 0 0,clip=true]{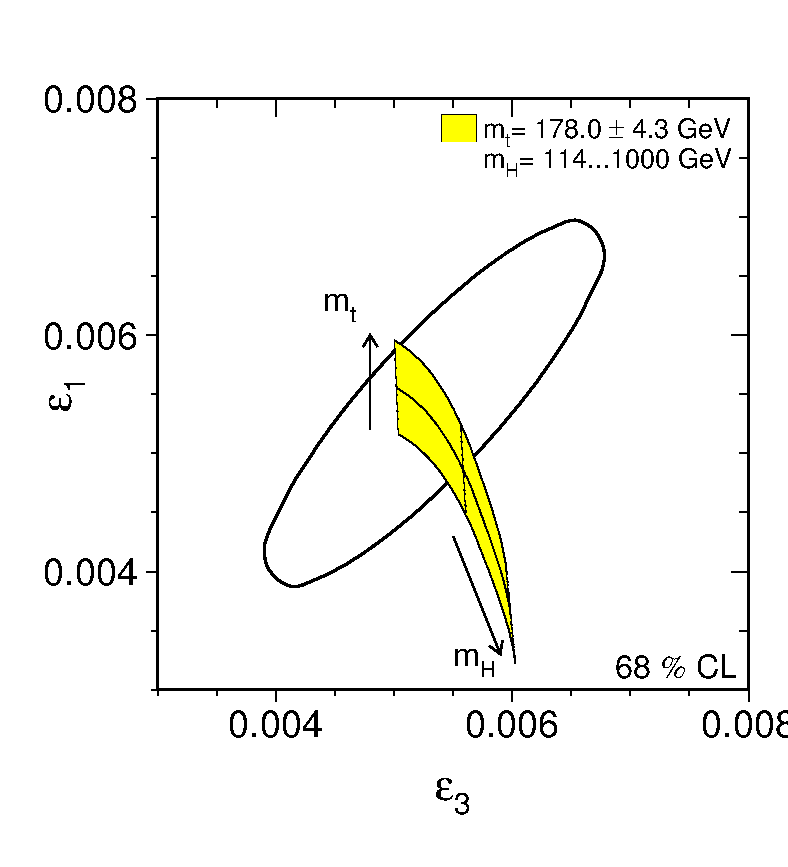}
\end{center}
\caption{\label{fig:ewpt} {\em The allowed range in the
$\epsilon_1,\epsilon_3$ plane by the precision tests at LEP (plot taken from \cite{lep_epsilon}). The black ellipse is the $2\sigma$ contour, and the yellow region is the Standard Model prediction.}}
\end{figure}

We have then to turn the experimental constraints in the $\epsilon$ space. For instance, one can see on figure \ref{fig:ewpt} the allowed space when projected onto the $\epsilon_1,\epsilon_3$ plane, a result taken from the LEP Electroweak Working Group (\cite{lep_epsilon}).

\subsection{Flavour Physics}
As said in the introduction, baryons composed with quarks from the second and third generation are unstable, and hence allow one to gather information by studying the characteristics of their decays. Those characteristics being the width, the branching ratios and the asymmetries between baryons and anti-baryons, among others. The choice of the heavy flavour is mainly motivated by theorists. Indeed, the particular behaviour of the strong interaction, that gets infinitely attractive at low energies, has the consequence to make the Feynman expansion non-perturbative under a certain energy. The technique used in such a region is called lattice QCD, for the reason that it is based on a grid discretisation of the spacetime, and has the drawback that it cannot be used in the same systematic way as the Feynman expansion. It turns out that the critical energy where one has to switch from a theory to another is approximately around the mass of the quarks of the second generation, which implies that the only baryons 
that can be treated in the perturbative regime are the one from the heavy flavours, hence the relevance of the $B$ physics. An interesting side effect in the context of this thesis is that heavy quarks are also the most sensitive ones to the Higgs physics, because of their large Yukawa couplings, and moreover in the special case of supersymmetry the Yukawa coupling of the $b$ quark is $\tb$-enhanced, so if we are to look for evidence for supersymmetry, $B$ physics is a clever guess.\\

The baryon $B_s$ being composed of a $b$ quark and an anti $s$ quark, this decay can be seen as the $b\bar s\to\bar\mu\mu$ process. It seems at first sight a rather clumsy process since we have no interactions that connect two fermions from different families. This fact is however not really in agreement with real experiments where one notices that the quark basis for weak interactions is not the same as the mass basis. This implies in particular that, as for neutrinos, a weak interaction can connect two different mass generations together. On the theoretical side, such a feature is easily accounted for by introducing Yukawa terms that connect generations together : this will cause the mass matrix to be non-diagonal in the gauge basis, so that when turning to the mass basis, the gauge interactions will be non-diagonal. Those trans-generation interactions are fully parametrised by an unitary matrix $V$ called the Cabbibo-Kobayashi-Maskawa (CKM) matrix. This matrix has the property of being close to the unity 
matrix, with off-diagonal elements quite small. In particular, the highest contribution to the $\Bsmu$ observable will be proportional to
$$\Bsmu\propto|V_{tb}V_{ts}|^2$$
since it connect $b$ and $s$ quarks.\\

\begin{figure}[!h]
\begin{center}
\begin{tabular}{ccc}
\begin{fmfgraph*}(100,80)
\fmfleft{p3,p1}
\fmfright{p4,p2}
\fmflabel{$b$}{p1}
\fmflabel{$s$}{p2}
\fmflabel{$\mu$}{p3}
\fmflabel{$\mu$}{p4}
\fmf{fermion}{p1,v1}
\fmf{fermion}{v2,p2}
\fmf{fermion}{v3,p3}
\fmf{fermion}{p4,v4}
\fmf{fermion,tension=0.3,label=$u,,c,,t$}{v1,v2}
\fmf{photon,tension=0.3,label=$W^-$}{v1,v3}
\fmf{photon,tension=0.3,label=$W^-$}{v4,v2}
\fmf{fermion,tension=0.3,label=$\nu$}{v4,v3}
\end{fmfgraph*}
&
\begin{fmfgraph*}(100,80)
\fmfleft{p3,p1}
\fmfright{p4,p2}
\fmflabel{$b$}{p1}
\fmflabel{$s$}{p2}
\fmflabel{$\mu$}{p3}
\fmflabel{$\mu$}{p4}
\fmf{fermion}{p4,v4,p3}
\fmf{fermion}{p1,v1}
\fmf{fermion}{v2,p2}
\fmf{fermion,tension=0.3,label=$u,,c,,t$,label.side=left}{v1,v2}
\fmf{photon,tension=0.3,label=$W^-$,label.side=right}{v1,v3,v2}
\fmf{dashes,tension=0.5,label=$H$}{v3,v4}
\end{fmfgraph*}
&
\begin{fmfgraph*}(100,80)
\fmfleft{p3,p1}
\fmfright{p4,p2}
\fmflabel{$b$}{p1}
\fmflabel{$s$}{p2}
\fmflabel{$\mu$}{p3}
\fmflabel{$\mu$}{p4}
\fmfblob{4mm}{v1}
\fmf{fermion}{p4,v2,p3}
\fmf{fermion}{p1,v1,p2}
\fmf{scalar,tension=0.3,label=$h$}{v1,v2}
\end{fmfgraph*}
\end{tabular}
\end{center}
\caption{\label{fig:bsmu_diag}{\em Categories of one-loop diagrams contributing to the $\Bsmu$ decay : boxes, penguins and counterterms.}}
\end{figure}
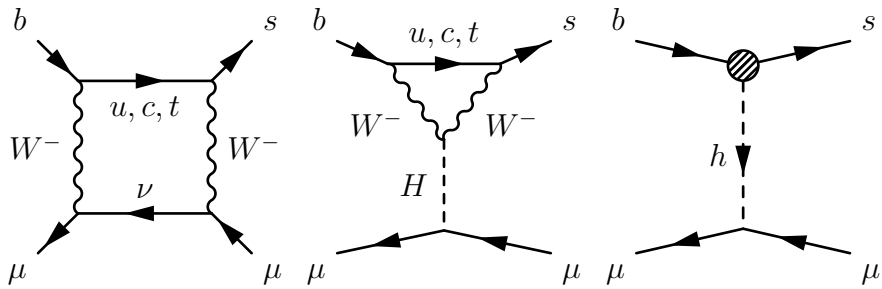

Without going in too much of the details of the calculation (which can be found in \cite{bobeth_bsmumu,huang_bsmumu}, among others), the different steps are the following : first one writes the loop effective action ($\Gamma_\looop$) containing all operators likely to contribute to the process $b\bar s\to\bar\mu\mu$ up to a given order (usually dimension 6 operators), then one obtains the coefficients of those operators by computing the associated loop diagrams at a high scale ($\mw$), the coefficients are finally evolved down to the $B_s$ scale and the process simply evaluated with the effective action. The determination of the effective action is made relevant by the fact that there exists a plethora of processes in flavour physics, so instead of working out the cross-sections for each process and having to cope each time with the lengthy loop computation, one does all loop integrations once to derive the coefficients of the effective action, and then any process can be computed straightforwardly. 
Furthermore, 
in the calculation of the coefficients, different sectors of a theory will simply add their contributions, which eases the task : for instance once the Standard Model part of the coefficients has been computed, any extension of the Standard Model can use this result and add only diagrams where new particles appear. Some of the diagrams contributing to the operators associated to $\Bsmu$ are shown in figure \ref{fig:bsmu_diag}.\vspace{1cm}\\

